\documentclass [PhD] {uclathes}
\newcommand{ \be }{\begin{equation}}
\newcommand{ \ee }{\end{equation}}
\newcommand{ \bea }{\begin{eqnarray}}
\newcommand{ \eea }{\end{eqnarray}}
\newcommand{ \bf }{\begin{figure}[htpb]}
\newcommand{ \ef }{\end{figure}}
\newcommand{ \bmn }{\begin{minipage}}
\newcommand{ \emn }{\end{minipage}}
\newcommand{ \bt }{\begin{table}[htpb]}
\newcommand{ \et }{\\end{table}}
\newcommand{ \la }{\langle}
\newcommand{ \ra }{\rangle}

\newcommand{ \rar }{\rightarrow}

\newcommand{ \as }{$\alpha_{s}$ }
\newcommand{ \pT }{$p_{T}$ }

\newcommand{ \dAu }{$d$ + Au }
\newcommand{ \pp }{$p+p$ }
\newcommand{ \ppbar }{$p+\bar{p}$ }
\newcommand{ \pA }{$p$ + A }
\newcommand{ \AuAu }{Au + Au }
\newcommand{ \sNN }{$\sqrt{s_{NN}}$ }
\newcommand{ \s }{$\sqrt{s}$ }
\newcommand{ \Jpsi }{$J/\psi$ }

\usepackage{amssymb}
\usepackage{amsthm}
\usepackage{revsymb}
\usepackage{rotate}
\usepackage{epsfig}
\usepackage{eepic}
\usepackage{epic}
\usepackage{graphics}
\usepackage[center]{caption2}
\usepackage{wrapfig}
\usepackage{doublespace}
\usepackage{multicols}
\usepackage{multirow}
\usepackage{amsmath}
\usepackage{indentfirst}

\begin{document}
\ifx\href\undefined\else\hypersetup{linktocpage=true}\fi 



\title{
Single Electron Transverse Momentum and Azimuthal Anisotropy
Distributions: \linebreak Charm Hadron Production at RHIC }

\author         {Xin Dong}
\department {Physics}
\supervisor{Ziping Zhang, Nu Xu}
\degreeyear     {2005}
\copyrightyear{2005}
\nosignature


\dedication{ \textsl{Dedicated to my dear mother}
}


\acknowledgments{ This thesis would have never come out without
the support and contribution from lots of people. I would like to
express my gratitude to those listed in the following and many
others I might not mention.

Firstly, I would thank Prof. Ziping Zhang and Prof. Hongfang Chen
for introducing me into this field and offering me lots of
freedom. I am grateful for their continuous supervision and
support in the last six years. I thank Dr. Hans-Georg Ritter and
Dr. Nu Xu for offering me the opportunity to work with the great
RNC group in LBNL. I am greatly thankful for Dr. Nu Xu's guidance
and tons of fruitful discussions in the last two years. I would
especially thank Dr. Zhangbu Xu from BNL for guiding me all the
analysis details throughout the thesis and plenty of help on
living when I was in BNL.

I would like to thank my classmate Dr. Lijuan Ruan. She helped me
a lot in BNL and we had an enjoyable cooperation on work and
abundant helpful discussions. Thanks should go to Dr. Jian Wu, Dr.
Ming Shao for their help on the calibrations of TOF detectors and
my living in BNL.

I appreciate many assistances from RNC group members in LBNL,
especially Dr. Kai Schweda and Dr. Paul Sorensen. They also
offered lots of valuable discussions on physics. I would give my
special thanks to Dr. Paul Sorensen for his elegant careful
wording corrections on my thesis. I also thank Prof. Huan.Z. Huang
from UCLA for many helpful suggestions on physics. I thank Dr. An
Tai from UCLA and Dr. Haibin Zhang from BNL for valuable
discussions on the charm physics.

Thank Dr. Jerome Lauret and STAR software group for their
continuous support on software. Thank all other STAR collaborators
for obtaining beautiful detector performance and data. I would
particularly thank STAR TOF group for their super efforts on
making this new detector function well.

I would thank my friends Dr. Tao Huang from BNL and Yi Zheng from
MIT for their kindly help when I was in Brookhaven and Berkeley.

Finally, I express my deep gratitude to my family. I'll never
forget tens of years of unceasing self-giving sacrifices from my
mother. I'll never forget continuous support and understanding
from my brother.

 }




\abstract {\par {\em Quantum Chromodynamics} (QCD) is a basic
gauge field theory to describe strong interactions. Lattice QCD
calculations predict a phase transition from hadronic matter to a
deconfined, locally thermalized {\em Quark-Gluon Plasma} (QGP)
state at high temperature and small baryon density. Plenty of
exciting results from RHIC experiments in the first three years
have demonstrated that a hot dense matter with strong collective
motion which cannot be described with hadronic degrees of freedom
was created at RHIC. Charm quarks are believed to be mostly
created from initial gluon fusion in heavy ion collisions. Since
they are massive, charm hadrons are proposed to be ideal probes to
study the early stage dynamics in heavy ion collisions.

We provide here an indirect measurement of charm semi-leptonic
decay. Single electron transverse momentum ($p_T$) distributions
from 200 GeV \dAu, \pp collisions and 62.4 GeV \AuAu collisions,
and single electron azimuthal anisotropy ($v_2$) from 62.4 GeV
\AuAu collisions are presented.

Electron identification is performed with the combination of the
prototype Time-Of-Flight detector (TOFr) and the ionization energy
loss ($dE/dx$) in the {\em Time Projection Chamber} (TPC).
Photonic background electrons are subtracted statistically by
reconstructing the invariant mass of the tagged $e^{\pm}$ and
every other partner candidate $e^{\mp}$. Partner track finding
efficiency is estimated from Monte Carlo simulations. More than
$\sim 95\%$ of photonic background (photon conversion and $\pi^0$
Dalitz decay) can be subtracted through this method. The
non-photonic electron \pT spectrum is extracted. In \AuAu
collisions at \sNN = 62.4 GeV, due to low yield of charm quarks,
only inclusive electron and photonic electron \pT spectra are
presented to illustrate the feasibility of this method in \AuAu
collisions. Electron azimuthal anisotropy in \AuAu 62.4 GeV is
calculated from the event plane technique. Photonic background in
each $\Delta\phi$ bin is reconstructed using the invariant mass
method. $v_2$ of inclusive electrons and photonic electrons is
presented.

The non-photonic electron \pT spectrum is consistent with the
direct open charm reconstruction spectrum in \dAu collisions. The
total charm production cross section per nucleon-nucleon collision
is calculated from the combined fit of the $D^0$ spectrum and
non-photonic electron spectrum in \dAu collisions. The result is
$d\sigma^{NN}_{c\bar{c}}/dy=0.30\pm0.04$(stat.)$\pm0.09$(syst.)
mb, which is significantly higher than the Next-to-Leading-Order
perturbative QCD (pQCD) calculations extrapolated from low energy
data points. The total charm cross section measurement is crucial
to investigate $J/\psi$ production mechanism in \AuAu collisions,
which may be a robust signature of QGP formation. Not only the
total yield is larger, but also the \pT spectrum is harder than
those from pQCD predictions, indicating a possible unusual charm
fragmentation function at RHIC energy. In \AuAu 62.4 GeV, the
photonic electron \pT spectrum agrees with the inclusive electron
spectrum well. The $v_2$ of photonic electrons also agrees with
that of inclusive electrons. These are consistent with the
expected low charm yield at this energy. The success of this
technique in \AuAu 62.4 GeV data makes us confident about
extracting both the non-photonic \pT spectrum and $v_2$ in the
coming large 200 GeV \AuAu data set.

Many open charm measurements can be carried out with a full barrel
TOF and heavy flavor tracker upgrade. These upgrades will allow us
to study the details of open charm collective motion, elliptic
flow, correlations and so on with much better statistics. These
two sub-detector upgrades will also allow us to reconstruct vector
mesons ($\omega$, $\phi$ {\em etc.}) from the low mass di-electron
invariant mass distributions, which may offer us indications of
possible chiral restoration issues during a possible phase
transition at RHIC.

 }


\graphicspath{plots/}

\makeintropages         
\chapter{Introduction: Quantum Chromodynamics and Heavy Ion Collisions}

\section{Quantum Chromodynamics}

From the lepton-nucleon {\em Deep Inelastic Scattering} (DIS)
experiments in the late 1960's, it was formed that hadrons have
structure and the quark parton model was exposed. The introduced
sub-nucleon structures include quarks which are constituents of
hadrons, and gluons which propagate interactions between partons.
This kind of interaction, namely the strong interaction, is one of
the four fundamental interactions in the nature. {\em Quantum
ChromoDynamics} (QCD) has been established since the 1970's to
describe strong interaction, and together with the unification of
electroweak theory, composes the {\em Standard Model} (SM), which
is a reasonably successful model to describe all interactions
except gravity.

QCD~\cite{QCDbook} is based on the gauge group $SU(3)_{C}$, with
gauge bosons - color octet gluons for factors and a unique group
coupling constant $g_s$. The subscript $C$ denotes the quantum
number - color, which is an exact symmetry. Quarks belong to a
color triplet representation in this symmetry, but hadronic states
are assumed to be color singlets in QCD. Owing to the non-abelian
character of the color group, the invariant QCD
Lagrangian\footnote{See Appendix A for QCD Lagrangian} requires
gauge (gluon) self-interactions, which do not appear in {\em
Quantum ElectroDynamics} (QED) - the gauge theory describing
electromagnetic interaction.

Generally speaking, QCD is a non-perturbative gauge theory in most
cases. It can be calculated using a computer-assisted method -
{\em Lattice QCD}~\cite{LQCDintro}. In this calculation, the
spacetime is discretized and replaced by a lattice with lattice
spacing equal to $a$. The quark fields are only defined at the
elements of the lattice and the gauge fields are defined on the
links of the lattice. The action is rewritten in such a way that
the limit $a\rar 0$ formally gives the original continuous action.
Lattice QCD has been widely used for reliable QCD calculations.
The precision of Lattice QCD calculations are limited by the
lattice spacing or the computing power.

\subsection {QCD running coupling constant \as}
The renormalized QCD coupling shows a scale dependent coupling
$\alpha_{s}(\mu)$ (running coupling), similar to that in QED.
However, the QED running coupling increases with energy scale,
while the gluon self-interactions lead to a complete different
behavior in QCD. $\alpha_{s}(\mu)$ can be written as:
\be \label{equas}
\alpha_s(\mu)\equiv\frac{g_s^2(\mu)}{4\pi}
\approx \frac{4\pi}{\beta_{0}\ln(\mu^2/\Lambda_{QCD}^2)}
\ee When $\beta_0>$ 0, this solution illustrates the {\em
asymptotic freedom} property: \as$\rar$ 0 as $\mu\rar\infty$,
which means QCD can be calculated perturbatively in high momentum
transfer or short distance approach. On the other hand, this
solution also shows strong coupling at $\mu\sim\Lambda_{QCD}$, so
QCD is non-perturbative in this case. \as needs to be determined
from experiment. The world averaged \as at the fixed-reference
$\mu_0=M_Z$ is $\alpha_{s}(M_{Z})=0.1187\pm0.002$~\cite{PDG}, and
the QCD scale $\Lambda_{QCD}\sim200$ MeV. Fig.~\ref{alphas} shows
the measured $\alpha_s$ at different momentum transfer scale $\mu$
compared with Lattice QCD calculations.

\bf \centering\mbox{
\includegraphics[width=0.60\textwidth]{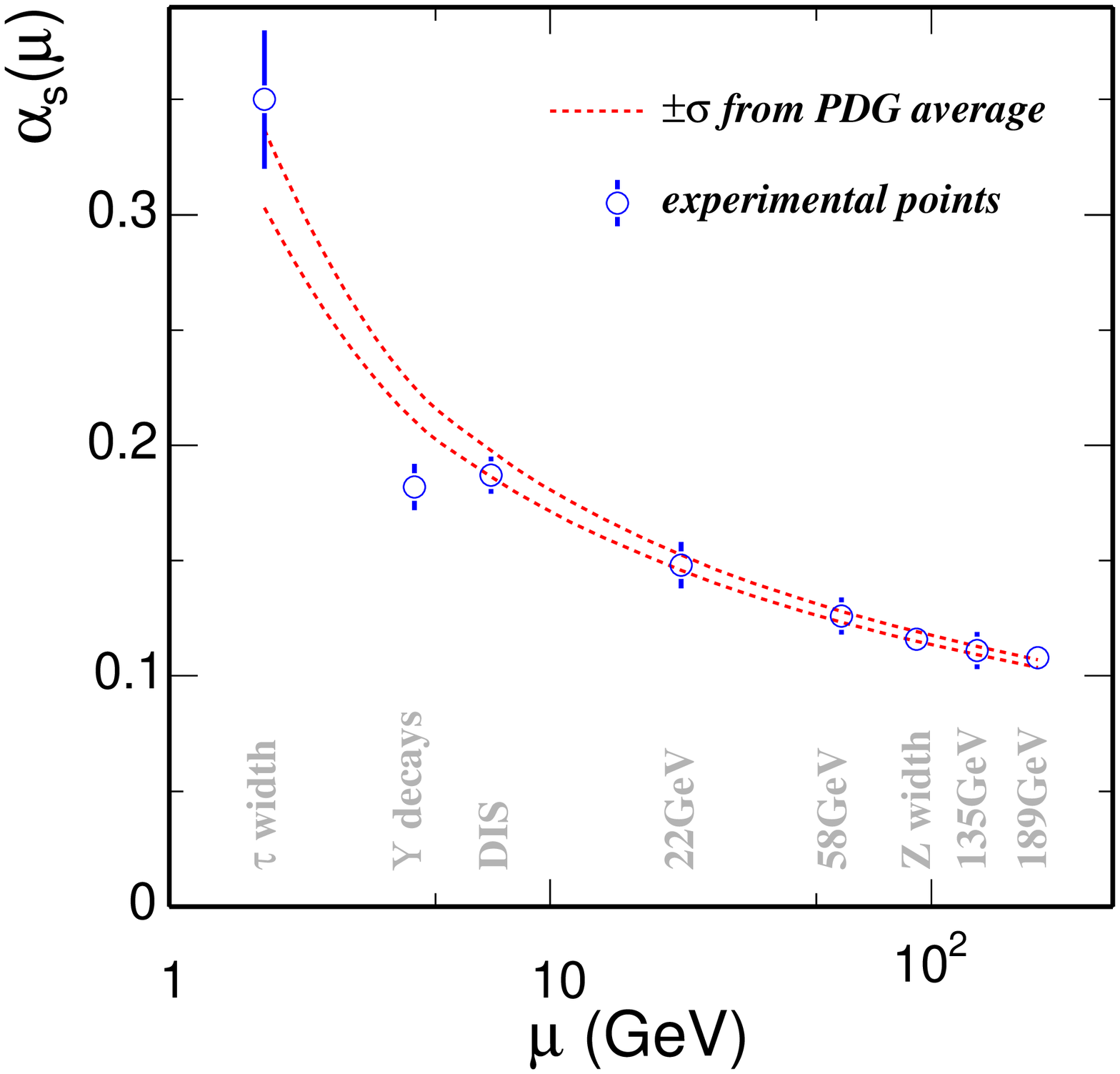}}
\caption[QCD running coupling constant $\alpha_s$]{Measured QCD
running coupling constant $\alpha_s$ from different experiments
compared with Lattice QCD calculations.} \label{alphas} \ef

\subsection {Perturbative QCD (pQCD)}
At sufficiently high $\mu$, where \as is sufficiently small,
physics quantities, such as cross sections, can be calculated to a
truncated series, known as {\em Leading Order} (LO), {\em
Next-to-Leading Order} (NLO) {\em etc.}. There are plenty of
experiments on high energy processes which offer quantitative
tests of pQCD. Due to the complexity of hadronic processes, the
test for pQCD has more difficulties than that for QED.

Assuming factorization, the cross section of a process $A + B \rar
C + ...$ can be written as: \be \label{Xsecfun}\sigma_{AB\rar C}=
f_{a/A}(x_a,\mu^2_{F})f_{b/B}(x_b,\mu^2_F)\otimes
\hat{\sigma}_{ab\rar c}(\hat{s},\mu^2_F,\mu^2_R,\alpha_s) \otimes
D_{c\rar C}(z,\mu^2_F) \ee Only the middle term
$\hat{\sigma}_{ab\rar c}$ can be calculated in pQCD from Feynman
diagrams. The first term $f_{a/A}(x_a,\mu^2_{F})$ or
$f_{b/B}(x_b,\mu^2_{F})$ is the hadron {\em Parton Distribution
Function} (PDF) and the last term $D_{c\rar C}(z, \mu^2_F)$ is the
{\em Fragmentation Function} (FF) that describes the transition
from a parton to a hadron. For leptons, these two terms do not
contribute in this formula. Hence, we can measure PDFs through
lepton-nucleon DIS interactions and FFs through high energy
$e^+e^-$ collisions. $\mu_R$ is the renormalization scale,
originating from the need to regularize divergent momentum
integrals in calculating high order diagram loops. $\mu_F$ is the
factorization scale, at which the parton densities are evaluated.
$\hat{s}$ is the partonic center of mass energy squared. From this
formula, we can see it is complicated to determine the expected
hadron production cross section in hadron-hadron collisions.

Heavy quark ($c,b$) production, due to large masses, is believed
to match to pQCD prediction better than light quark production.
And because they cannot be produced through the initial light
hadron fragmentation, the FF part is irrelevant to the total
production cross section of heavy quarks. Hence the measurement of
the total heavy quark production cross section offers a powerful
test of pQCD.

Let us take the calculation of heavy quark pair $Q\bar{Q}$ cross
section in \pp collisions as an example~\cite{vogtXsec}. At LO,
heavy quarks are created by $gg$ fusion and $q\bar{q}$
annihilation, while at NLO, $qg$ and $\bar{q}g$ scattering is
included. To any order, the partonic cross section can be
calculated as an expansion in the power of $\alpha_s$, assuming
$\mu_R=\mu_F=\mu$: \be \hat{\sigma}_{ij}(\hat{s}, m_{Q}^2, \mu^2)
= \frac{\alpha_s^2(\mu)}{m^2}
\sum_{k=0}^{\infty}(4\pi\alpha_s(\mu))^2
\sum_{l=0}^{k}f_{ij}^{(k,l)}(\eta) \ln^l(\frac{\mu^2}{m_Q^2}) \ee
where $\eta=\hat{s}/4m_Q^2-1$ and $f_{ij}^{(k,l)}(\eta)$ is called
the scaling function, obtained from the calculation of Feynman
diagrams for each order. The total production cross section can be
obtained from the above partonic cross section: \be
\sigma_{pp}(s,m_Q^2)=\sum_{i,j=q,\bar{q},g}\int_{\frac{4m_Q^2}{s}}^{1}
\frac{d\tau}{\tau}\delta(x_ix_j-\tau)f_{i/p}(x_i,\mu^2)f_{j/p}(x_j,\mu^2)
\hat{\sigma}_{ij}(\tau,m_Q^2,\mu^2) \ee

Numerical results of theoretical calculations show strong
dependence on the truncation of the series, the selection of
parameters $\mu_F$, $\mu_R$ and $m_Q$ {\em
etc.}~\cite{manganoXsec,vogtXsec,raufeisenXsec}. Hence,
theoretical predictions still have large uncertainties.

Experimentally, measurement of charmed hadrons is difficult due to
their short lifetime ($c\tau(D^{0})=124$ $\mu$m), low production
rates, and large combinatoric background. Both direct
reconstruction through the hadronic channel and indirect
measurement of the semi-leptonic decay of charmed hadrons were
preformed in the previous measurements. In the low center-of-mass
energies ( $^{<}_{\sim}$ 40 GeV ), measurements were done on fixed
targets~\cite{charmreview,NA32,E743,E653,E769}. At \s$\sim52-63$
GeV (ISR), the measurements were done from $D\Lambda_{c}$
production and $ee$, $e\mu$ pair correlation, and the results
showed inconsistency between different
publications~\cite{charmreview}. At higher energies, the UA2
experiment measured single electron distributions at 630 GeV
\ppbar collisions~\cite{UA2}. However, due to large uncertainties
of the decay from charmed hadron to electrons, these measurements
included large errors. The CDF II collaboration made a direct
measurement of open charm hadrons at \s = 1.96 TeV \pp
collisions~\cite{CDFII}, but the spectrum only covers the high \pT
region. Recent pQCD calculations seem to reproduce the spectrum at
CDF. No charm measurement in elementary \pp collisions at RHIC
energy was done and theoretical predictions differ significantly
at this energy due to the large uncertainties of many parameters.
So charm production at the baseline elementary \pp collisions at
RHIC energy is helpful.

\subsection {Confinement and chiral symmetry breaking}
Since quarks have color quanta, while hadrons are color-neutral to
us, quarks must be confined within hadrons. This can also be
explained from the QCD coupling $\alpha_{s}$. When two quarks are
separated to a large distance, which corresponds to a small energy
scale, the coupling becomes strong, {\em i.e.} intuitively, more
and more self-coupled gluons hold the quarks not to be isolated.
It is quite different in QED because there is no self coupling
between photons, so that we can observe isolated electric charges.

In the absence of quark masses, the QCD Lagrangian can be split
into two independent sectors: the left- and right-handed
components~\cite{QCD}. This Lagrangian is invariant under chiral
symmetry transformations then. This symmetry, which is the
extension of classical $SU(3)$, is a global $SU_L(n_f)\times
SU_R(n_f)$ symmetry for $n_f$ massless quark flavors. However, it
is spontaneously broken in the vacuum in the Nambu-Goldstone's way
to realize this symmetry and this breaking gives rise to
$(n_f^2-1)$ massless Goldstone particles. Thus we can identify the
$\pi$, $K$, $\eta$ with the Goldstone modes of QCD: their small
masses being generated by the quark-mass matrix which explicitly
breaks the global chiral symmetry of the QCD
Lagrangian\footnote{See Appendix A for chiral symmetry and
effective lagrangian approach.}.


\subsection {QCD Phase transition}
QCD matter is mostly observed as nuclei or hadron gas in our
current condition. With sufficient temperature and energy density,
QCD predicts a phase transition to a new matter, named {\em Quark
Gluon Plasma} (QGP), with new (color) {\em degrees of freedom}
(d.o.f.). In this new phase, quarks and gluons are liberated from
hadrons, and can move around in a larger distance rather than
confined in hadrons, which is called {\em deconfinement}.
Meanwhile, the broken chiral symmetry in normal QCD matter will be
restored and consequently, masses of scalar mesons and vector
mesons will decrease~\cite{LQCDchiral}. Lattice QCD calculations
provide quantitative predictions on this phase transition: the
critical temperature of this phase transition is $T_c\sim150-180$
MeV, and the energy density at the critical point is
$\varepsilon_c(T_c)\sim1-3$ GeV/fm$^3$ ($\sim0.17$ GeV/fm$^3$ for
nuclear matter)~\cite{LQCDreview}. The appearance of these color
d.o.f. can be illustrated by a sharp increase in pressure with
temperature, shown in Fig.~\ref{LQCDpressure}~\cite{LQCDpressure}.

\bf \centering\mbox{
\includegraphics[width=0.60\textwidth]{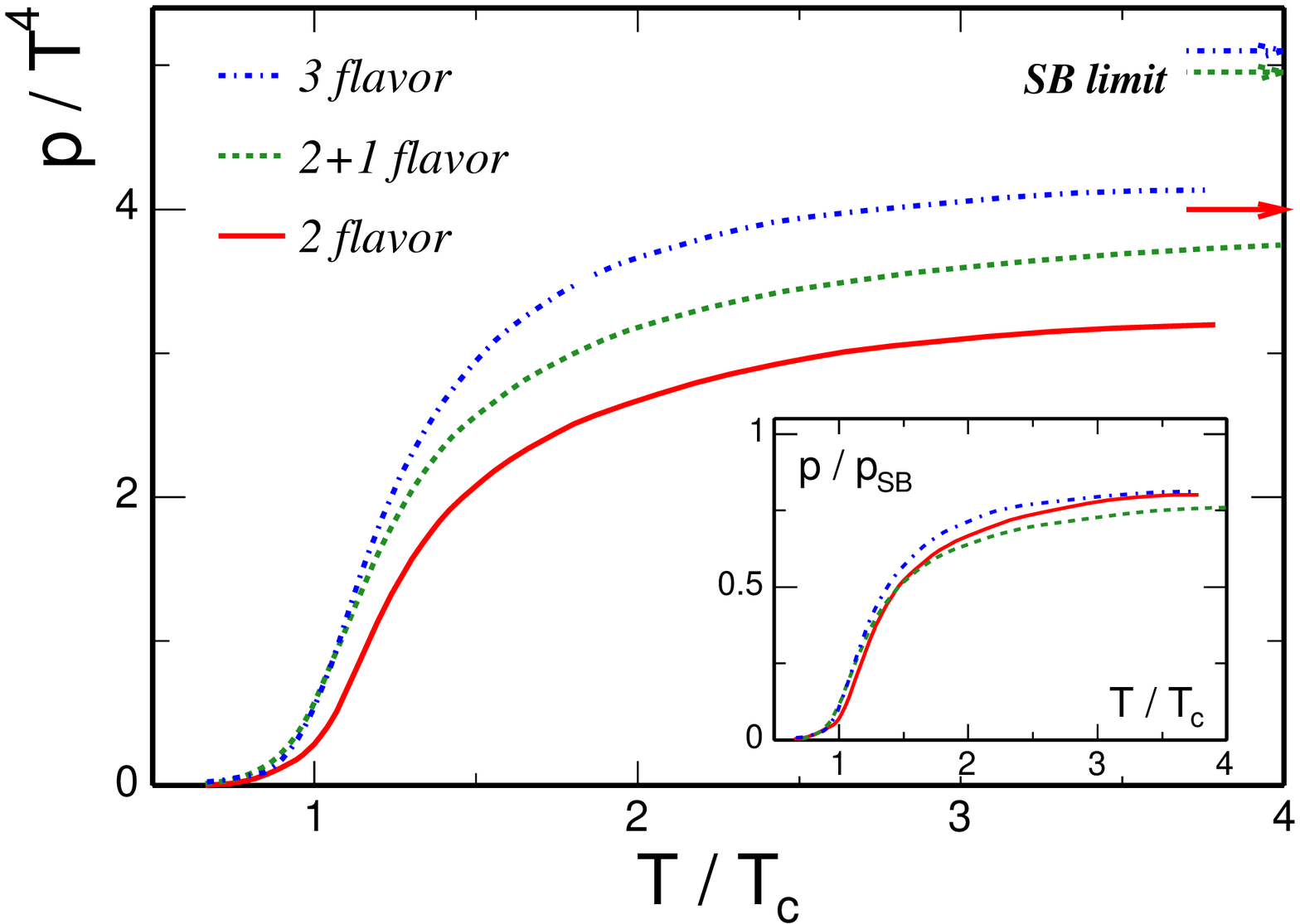}}
\caption[LQCD calculation for pressure]{The evolution of $p/T^4$
with the increase of temperature $T$ for 3 different flavor
configurations. The arrows indicate the SB limit for each case.
The insert plot shows the ratio of $p/p_{SB}$ with function of
$T$. } \label{LQCDpressure} \ef

The arrows indicate the Stefan-Boltzmann limits, which are for the
systems with massless, non-interacting quarks and gluons. The
similarity of the three curves in the insert plot of
Fig.~\ref{LQCDpressure} illustrates that besides the effect of
quark masses, there should be interactions in the newly formed
system, which is different from the original QGP scenario (weakly
interacting), but is already demonstrated by experimental results
(see next section).

Lattice QCD calculations of the potential between two heavy-mass
quarks also offer evidence of deconfinement.

\bf \centering\mbox{
\includegraphics[width=0.60\textwidth]{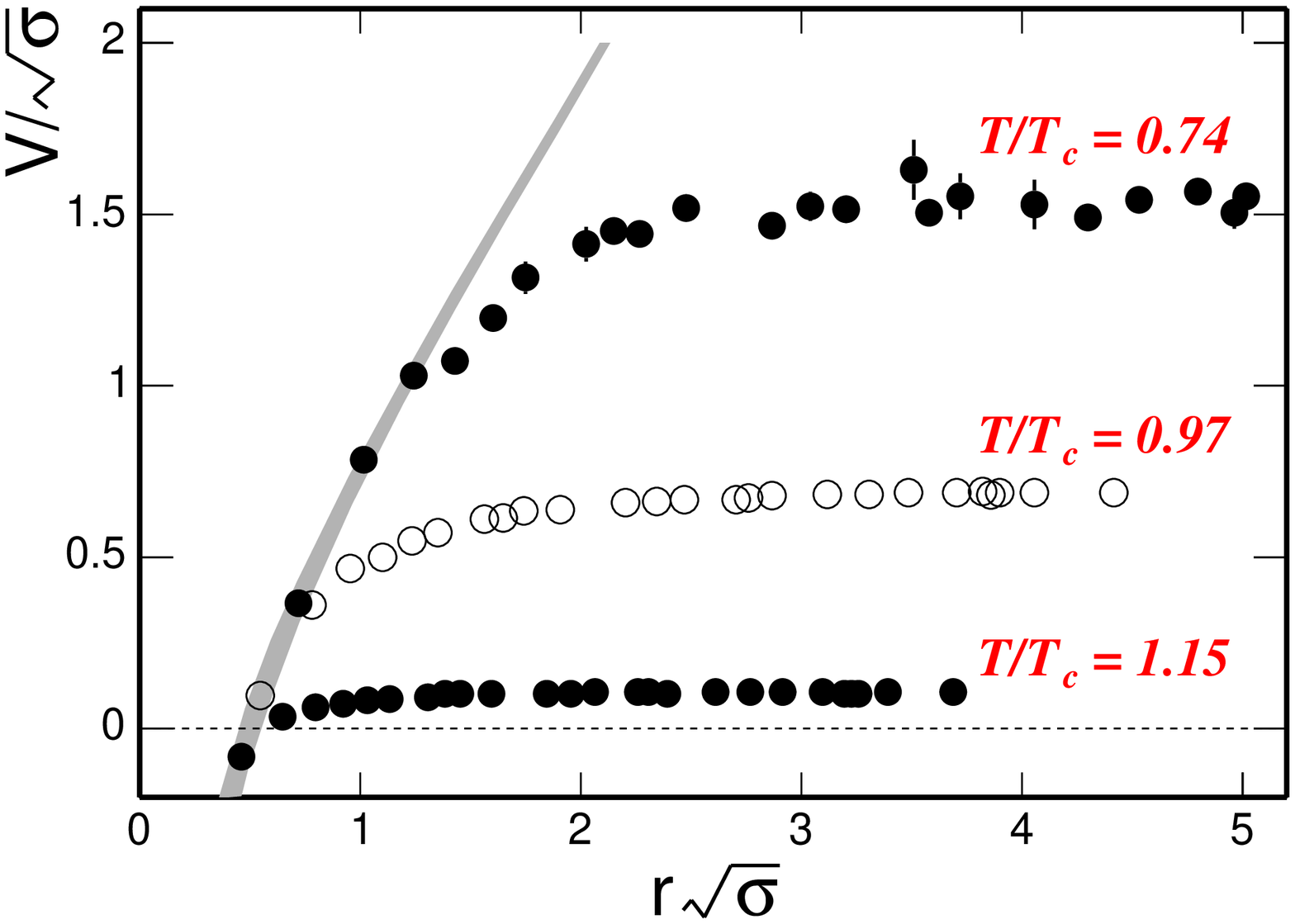}}
\caption[LQCD calculation for heavy quark potential]{Lattice
calculations for the heavy-mass quark potential in different
temperature cases. The band depicts the Cornell potential of
$V(r)=-\alpha/r+\sigma r$ with $\alpha=0.25\pm0.05$.}
\label{potential} \ef

Fig.~\ref{potential} shows a recent calculation of the heavy-mass
quark-antiquark pair Cornell potential in different temperature
conditions~\cite{LQCDpressure}: with the increase of temperature,
the rampart of the potential between two quarks, which causes
confinement, will bend down and thus liberate quarks from the
trap. In addition, the continuous bending without sudden change
indicates a crossover transition at high temperature and vanishing
net quark density.

\section{Heavy Ion Collisions}
Experimentally, to search for this new kind of matter, a large
amount of energy needs to be packed into a limited space volume.
Heavy ion collisions have been proposed as a more effective way
because the initial energy density increases as a power law
function with the atomic number while only logarithmically with
collision energy~\cite{LinThesis}. Since the 1970's, physicists
from BEVALAC at LBL, SIS at GSI, AGS at BNL, SPS at CERN and RHIC
at BNL {\em etc}~\cite{QM04review} have been trying to reach and
cross the phase transition boundary in the laboratory through
relativistic heavy ion collisions.

\bf \centering\mbox{
\includegraphics[width=0.60\textwidth]{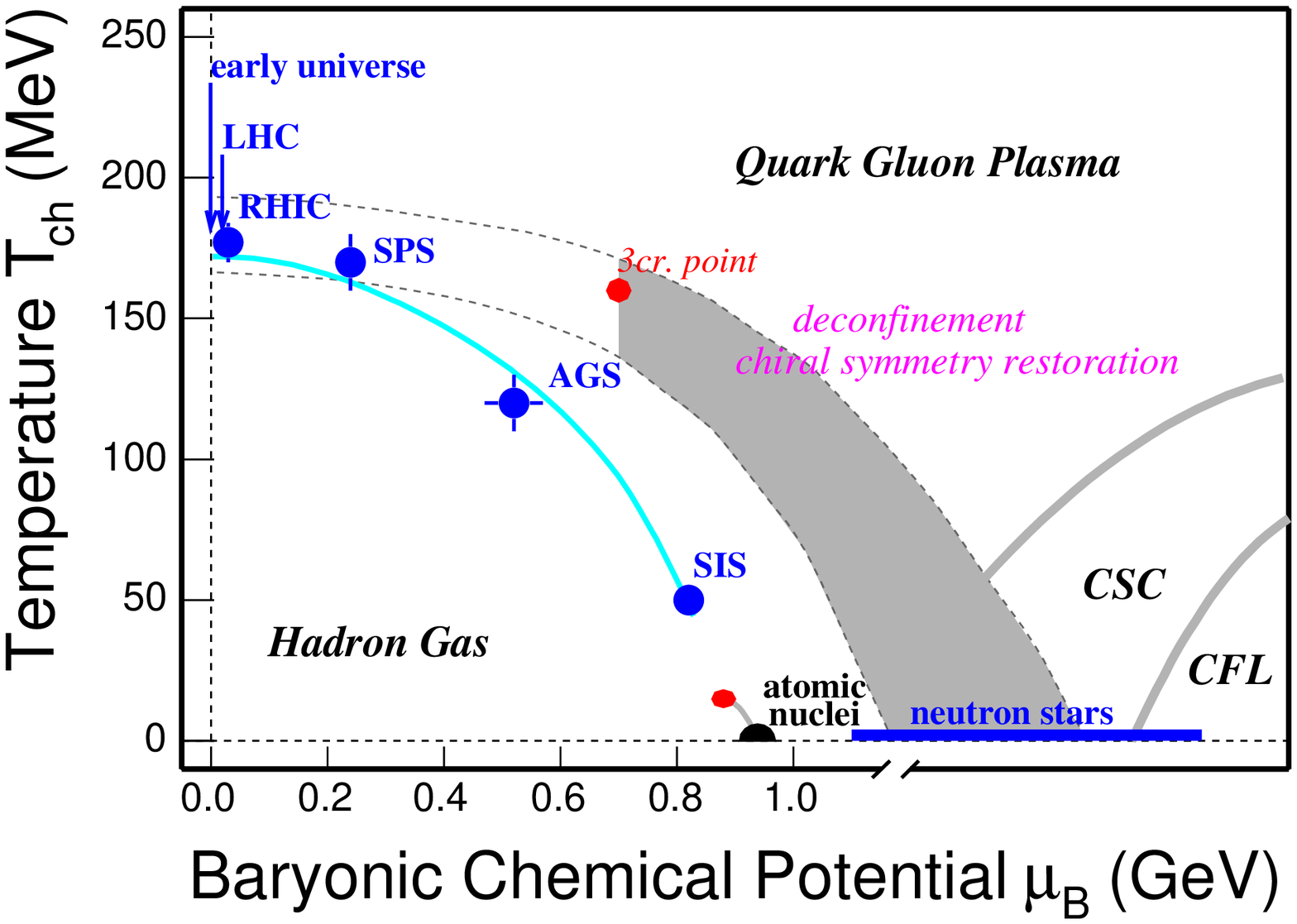}}
\caption[QCD phase diagram]{QCD phase diagram. The grey shadows
depict first-order phase transition boundaries. The red dots
depict the critical points and "3cr point" is calculated from
Lattice QCD~\cite{LQCDcriP}. The blue dots depict the positions of
several colliders from statistical fit~\cite{PBMreview}.}
\label{phasediagram} \ef

Fig.~\ref{phasediagram} shows the QCD phase diagram and
approaching transition boundary of the colliders.

For the past four years the {\em Relativistic Heavy Ion Collider}
(RHIC) at {\em Brookhaven National Lab} (BNL) has conducted very
successful runs. Plenty of exciting physics results reveal that
the matter created at RHIC is quite different from what we
observed before: It cannot be described by hadronic degrees of
freedom and demonstrates many of the signatures from a QGP
scenario. These measurements provide strong hints for the
discovery of QGP~\cite{phenixwhitepaper}. Some of the key
measurements will be discussed in the following
sections\footnote{Useful kinematic variables are defined in
Appendix B.}.

\subsection{Nucleon stopping power and initial energy density}
One of the most fundamental quantities we need to investigate is
whether the initial deposited energy is enough to cross the energy
density threshold for QGP formation. Because baryon number is
conserved and the rapidity distributions are only slightly
affected by rescattering in the late stage of collisions, the
measured net baryon ($B-\bar{B}$) distribution can reveal the
energy loss of initial participants and allow us to estimate the
degree of nucleon stopping power. Fig.~\ref{yloss} shows rapidity
loss $\la\delta y\ra = \la y\ra - y_p$ for different energies from
AGS to RHIC~\cite{brahmsStop}. Using the data point at RHIC, one
can estimate that $73\pm6^{+12}_{-26}$ GeV of the initial 100 GeV
per participant is deposited and available for excitations.

\bf \centering\mbox{
\includegraphics[width=0.6\textwidth]{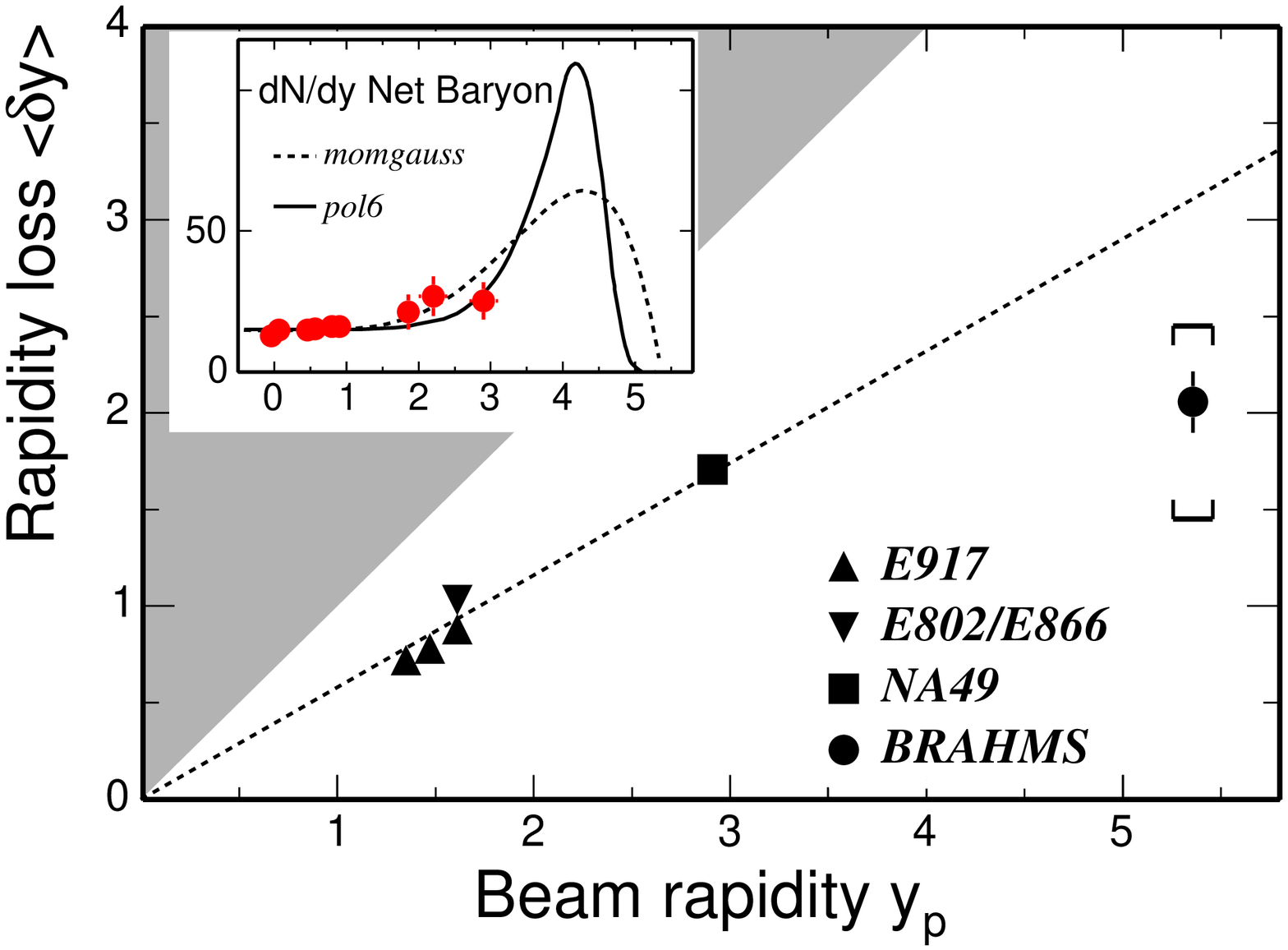}}
\caption[Rapidity loss in heavy ion collisions]{The rapidity loss
as a function of beam rapidity. The shadow indicates an unphysical
region and the dashed line depicts the phenomenological scaling
$\la\delta y\ra=0.58y_p$. The insert plot shows the measured data
points of net baryon distribution from BRAHMS and different
extrapolations to full rapidity.} \label{yloss} \ef

The initial Bjorken energy density~\cite{bjorken} can be
calculated using: \be \epsilon_{Bj} =
\frac{1}{A_{\bot}\tau}\frac{dE_{T}}{dy} \ee where $\tau$ is the
formation time and $A_{\bot}$ is the nuclei transverse overlap
region area. The PHENIX $dE_{T}/dy$ measurement~\cite{phenixEt}
indicates an initial energy density of $\sim 5$ GeV/fm$^3$
($\tau\approx$ 1 fm/c, $A_{\bot}=\pi R^2$, $R\approx 1.2A^{1/3}$
fm) for central \AuAu collisions at RHIC, well above the expected
critical energy density $\epsilon_{c}\sim 1$ GeV/fm$^3$.

Thus, the initial condition in \AuAu collisions at RHIC is
believed to be capable of forming the new partonic matter - QGP.

\subsection{Jet quenching}
In heavy ion collisions, high \pT ($p_T>\sim5$ GeV/c) particles
are believed to be produced mainly from the initial QCD
hard-scattering processes~\cite{highpt130}. These energetic
particles can be used as unique probes by studying their
interactions with the medium. Experimentally, the nuclear
modification factor, the difference between the spectrum in A+B
collisions {\em with respect to} (w.r.t.) a \pp collision
reference, is extensively used. It is defined as: \be \label{Rab}
R_{AB}(p_{T}) =
\frac{d^{2}N_{AB}/dp_{T}dy}{T_{AB}d^{2}\sigma_{pp}/dp_{T}dy} \ee
where $T_{AB}=\langle N_{bin} \rangle/\sigma_{pp}^{inel}$ is the
nucleus overlap function, calculated from a Glauber
model~\cite{HICintro}. $N_{bin}$ represents the number of binary
collisions in a nucleus-nucleus collision. The experimental
result~\cite{dAuhighpt} in Fig.~\ref{RabSTAR} shows that there is
a strong suppression relative to the binary scaling expectation at
high \pT in the central \AuAu collisions $-$ jet quenching. But
this suppression is not seen in \dAu collisions, the control
experiment, which suggests the suppression in central \AuAu is due
to the final state interactions rather than initial state effect
and thus a very dense matter must be created in central \AuAu
collisions at RHIC.

\bf \bmn[b]{0.47\textwidth} \centering\mbox{
\includegraphics[width=1.0\textwidth]{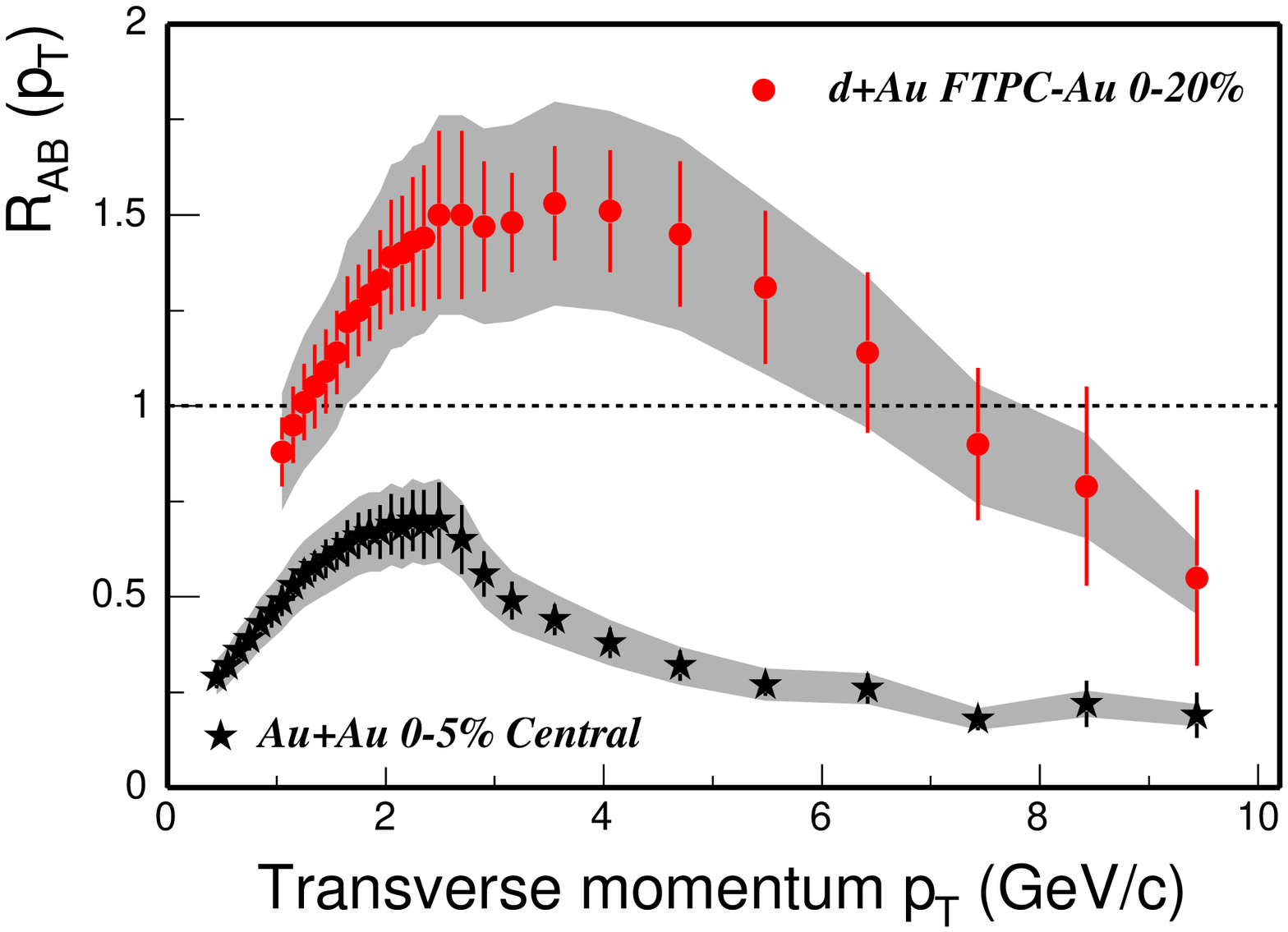}}
\caption[$R_{AB}$ of \AuAu and \dAu]{Nuclear modification factor
in central \AuAu and \dAu collisions.} \label{RabSTAR} \emn
\bmn[b]{0.47\textwidth} \centering\mbox{
\includegraphics[width=1.0\textwidth]{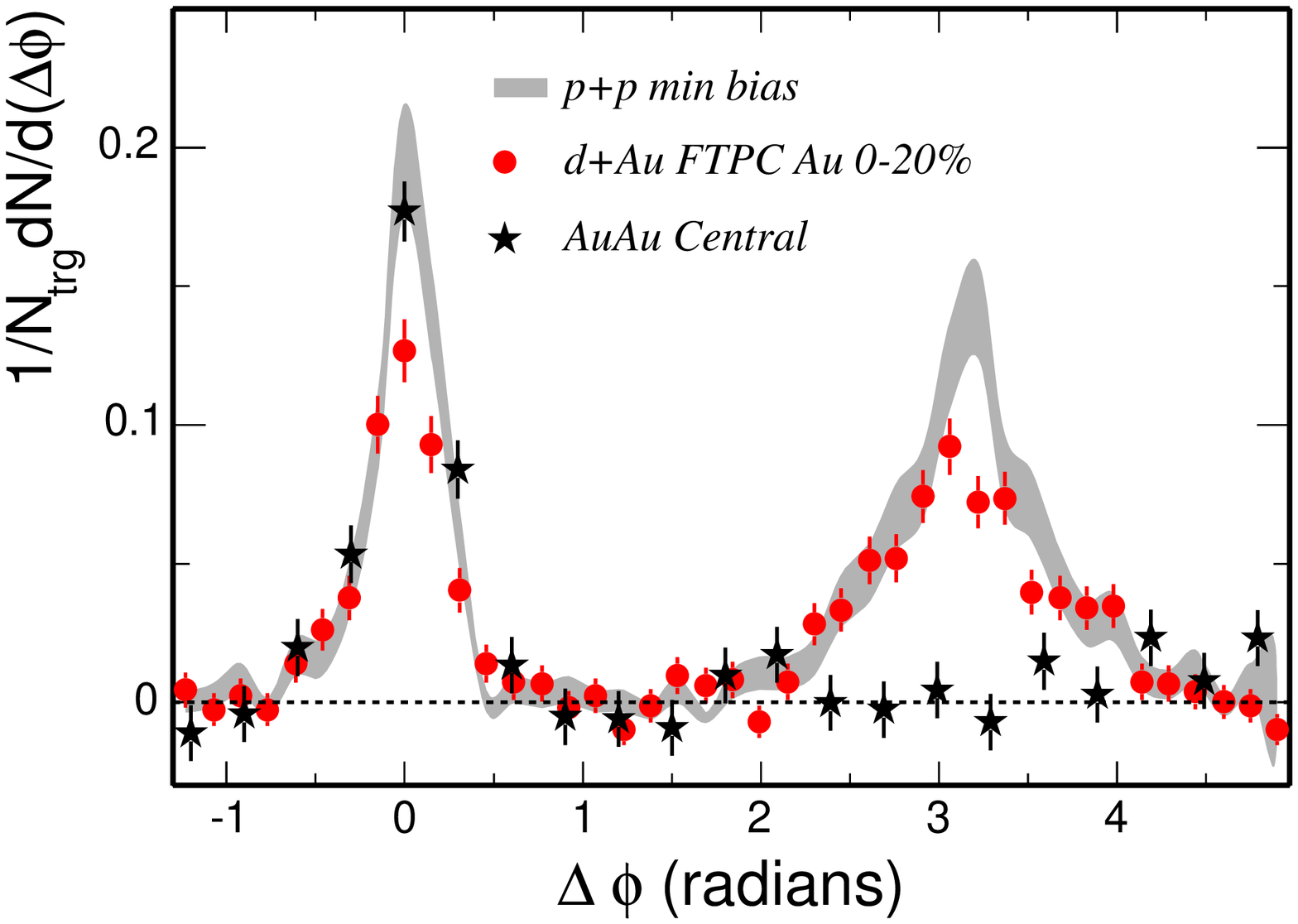}}
\caption[Dihadron correlations]{Dihadron azimuthal angle
correlation at high \pT in \pp, \dAu and \AuAu collisions.}
\label{dihadron} \emn \ef

This result has also been illustrated by a dihadron azimuthal
angle correlation study. Fig.~\ref{dihadron} shows the associated
hadrons ($p_T>2$ GeV/c) azimuthal distribution relative to a
triggered hadron ($p_T>4$ GeV/c). The enhanced correlation at
$\Delta\phi\sim0$, which means the pair is from a single jet, was
observed in \pp, \dAu and \AuAu collisions. The pair from the
back-to-back jet correlation at $\Delta\phi\sim\pi$ only appears
in \pp and \dAu while it almost completely disappears in central
\AuAu collisions.

\bf \centering\mbox{
\includegraphics[width=0.4\textwidth]{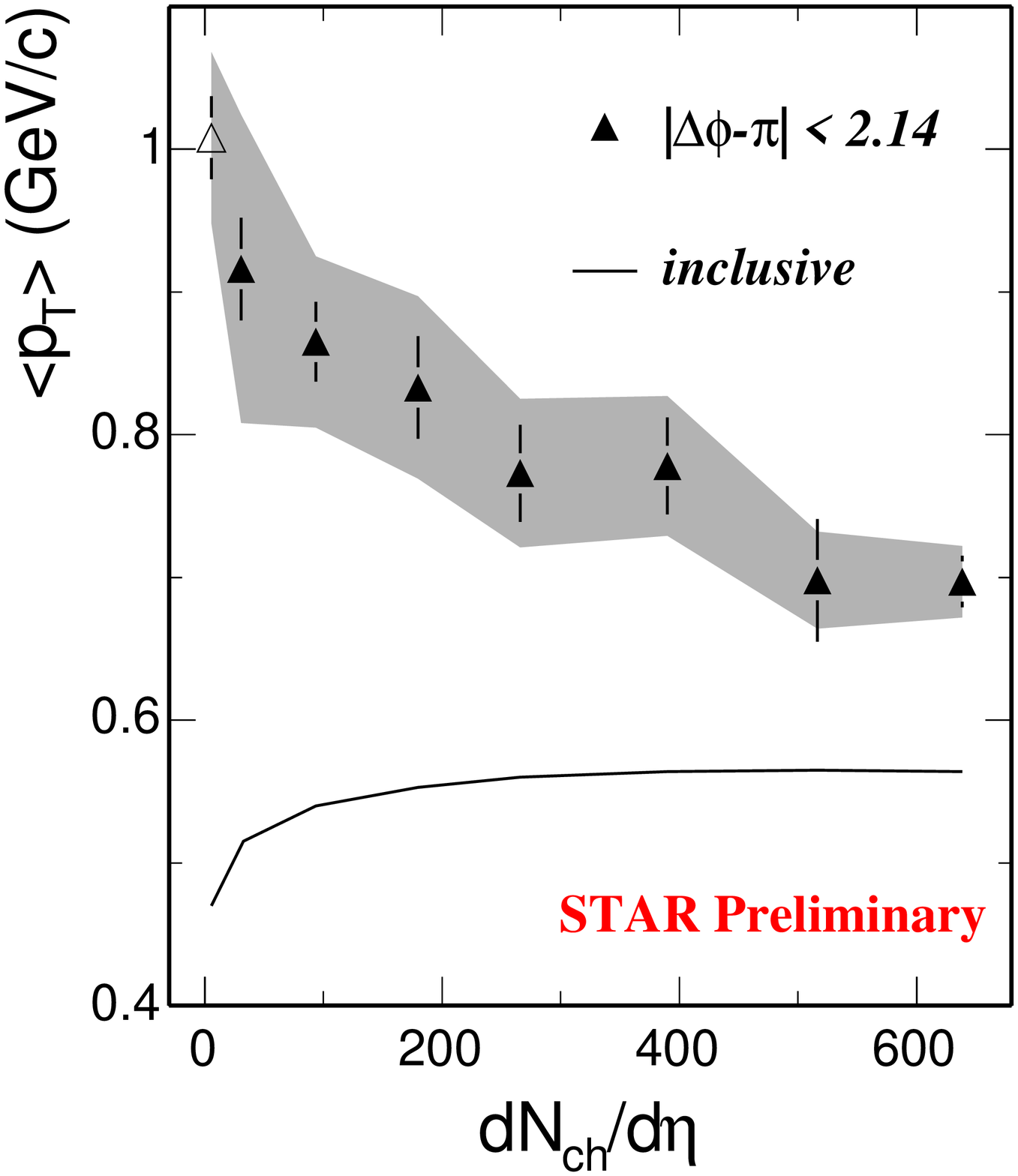}}
\caption[$\la p_T\ra$ of away side jet]{The $\la p_T\ra$ of away
side jet evolves with the change of multiplicities in \pp (open
triangle) and \AuAu (solid triangles) collisions.}
\label{meanPtaway} \ef

Due to momentum conservation, the disappearance of away-side fast
partons must result in an excess of softer emerging hadrons.
Fig.~\ref{meanPtaway} shows the centrality dependence of the
$\langle p_{T}\rangle$ of the associated away-side charged hadrons
(threshold lowered to 0.15 GeV/c), compared with that of inclusive
hadrons, illustrating the above point~\cite{jetspectra}. This also
offers a hint of the attainment of thermalization via the frequent
soft parton-parton interactions in the early collision stages. But
how strong those partons interact is still a crucial open
question, that needs to be answered quantitatively to address the
evidence of early thermalization of the system.

\subsection{Collective motion}
Hadron spectra are useful tools to study the properties of the
bulk system and specified particles can be used to probe different
stages after the heavy ion collisions. Fig.~\ref{particlespecta}
shows the identified particle \pT spectra measured in central
Au+Au collisions~\cite{pikpspectra,k0slambda,phispectra,xiomega}.

\bf \centering\mbox{
\includegraphics[width=0.48\textwidth]{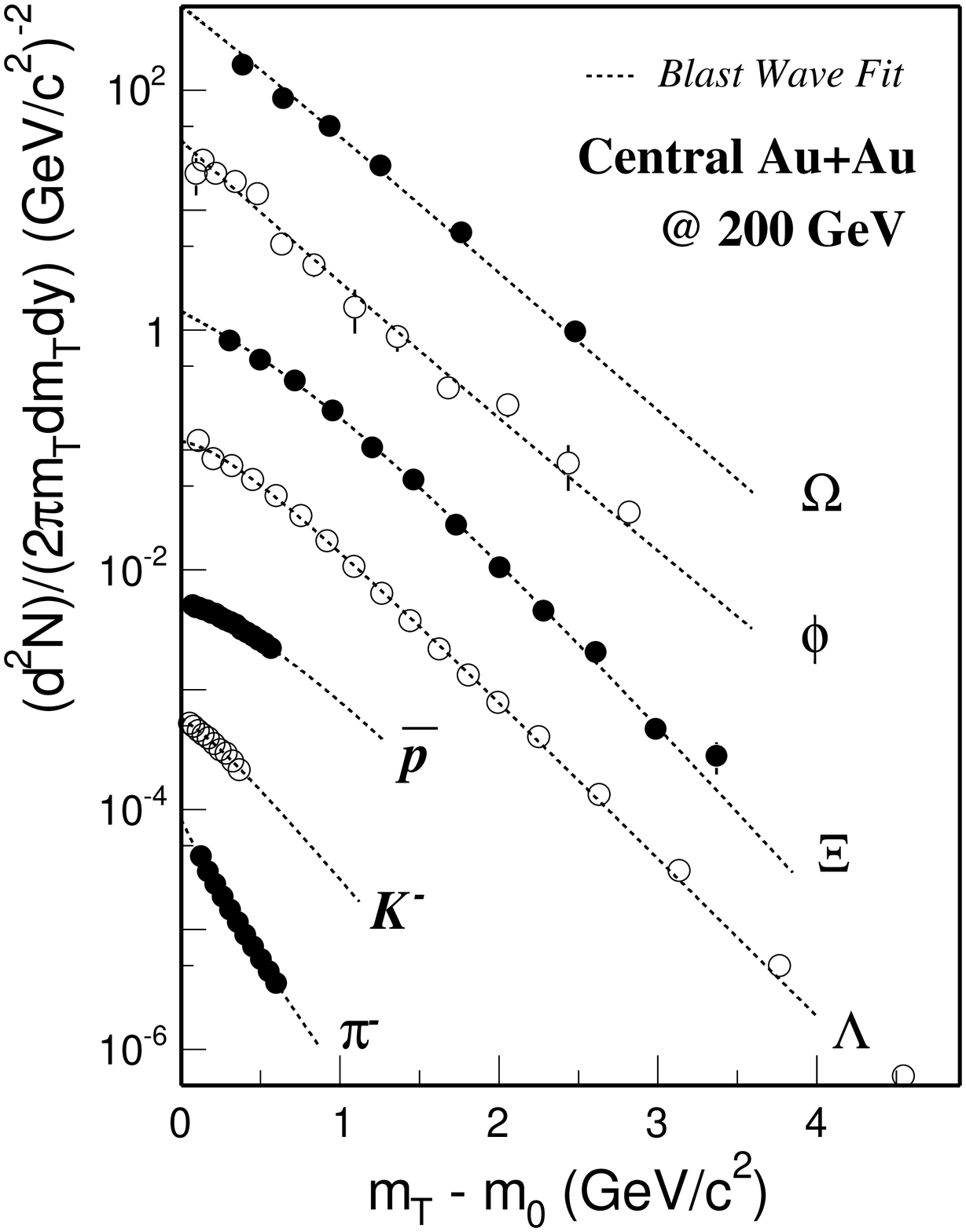}}
\caption[Blast-wave fit for PID spectra]{Identified particle
spectra in central Au+Au collisions at \sNN = 200 GeV and the
Blast Wave fit results. The BW fits were done for $\pi^-$, $K^-$,
$\bar{p}$ simultaneously and for other particles separately.}
\label{particlespecta} \ef

The plot shows the slopes of particle spectra changes for
different particles (masses), indicating the strong collectivity
of final state particles. The dashed lines depict the fit results
from the Blast Wave thermal model~\cite{blastwave}. In thermal
models, local thermal equilibrium is assumed and hence particles
spectra only depend on the mass of particles and the temperature
of system. "Blast wave" means that particles freeze out from the
system surface simultaneously when dense matter becomes dilute
enough. Under the assumption of simple cylindrical source and
boost invariance in rapidity, there are only two parameters to
describe the particles spectra: freeze-out temperature $T_{fo}$
and average transverse velocity $\langle\beta_{T}\rangle$.
Fig.~\ref{thermalResult} shows the fit results for different
particles. For the simultaneous fit to $\pi^-$, $K^-$, $\bar{p}$,
stronger and stronger collectivity is observed from peripheral to
central collisions. But the fit to $\Omega$ and $\phi$ in central
\AuAu collisions shows higher freeze-out temperature and lower
transverse velocity, indicating those particles leave the system
at the earlier stage than stable hadrons. This is not surprising
since those multi-strange baryons/mesons are expected to have much
smaller hadronic scattering cross sections and thus the
temperature from the fit to those particles may reflect the
chemical freeze-out temperature $T_{ch}$. This temperature is
close to the critical temperature $T_{c}$, meaning the temperature
of the system created in the collisions is greater than $T_{c}$
and hence the phase transition may take place at RHIC energy.

\bf \centering\mbox{
\includegraphics[width=0.60\textwidth]{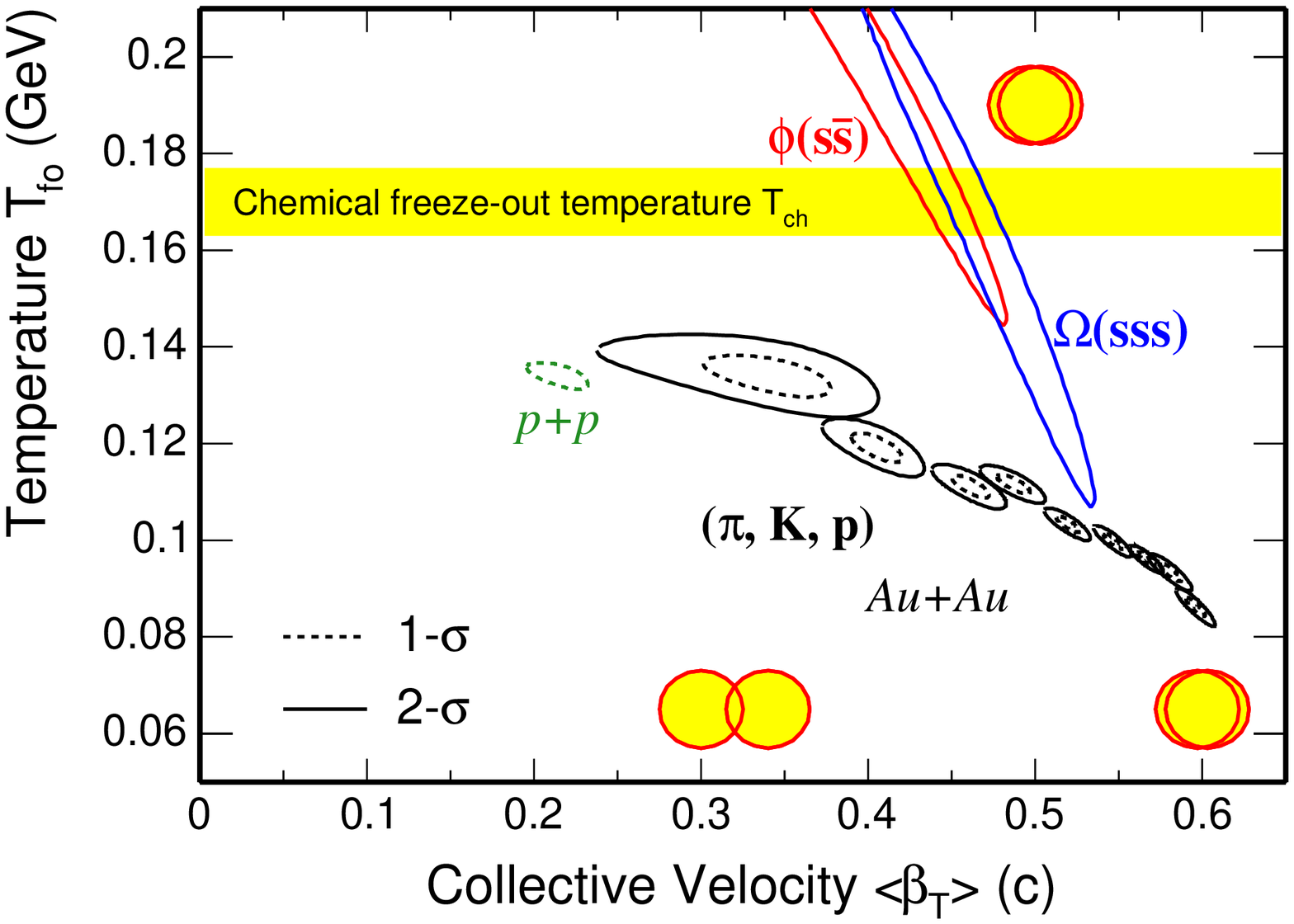}}
\caption[Blast-wave fit results]{Blast-wave parameters $T_{fo}$
vs. $\la\beta_T\ra$ contour plot from the simultaneous fits to
stable hadrons ($\pi,K,p$) spectra in \AuAu collisions and \pp
collisions and separate fits to multi-strange hadrons
$\phi(s\bar{s})$, $\Omega(sss)$ spectra in central \AuAu
collisions. The contours of $\pi,K,p$ fits in \AuAu are for
peripheral collisions in the left and central collisions in the
right. The contours of $\phi$ and $\Omega$ are for central \AuAu
collisions.} \label{thermalResult} \ef

\subsection{Transverse azimuthal anisotropy}
In non-central heavy ion collisions, the overlapping region of the
two nuclei will form an anisotropy in coordinate space. Because of
rescattering, the anisotropic pressure gradient will lead to an
anisotropy in momentum space. The dynamic expansion of the system
will wash out the coordinate-space-anisotropy, while the
momentum-space-anisotropy will saturate during the evolution of
the system~\cite{hydrointro}. The final state particle spectrum in
momentum space can be expanded into a Fourier series as
Eq.~\ref{floweq}. \begin{subequations}\label{floweq}
\begin{align} E\frac{d^{3}N}{dp^{3}}&=\frac{d^{2}N}{2\pi
p_{T}dp_{T}dy} (1+\sum_{n=1}^{\infty}2v_{n}cos[n(\phi-\Psi_{rp})])
\\ v_n &= \la cos[n(\phi-\Psi_{rp})]\ra \end{align}
\end{subequations} where $\Psi_{rp}$ denotes the direction of the
reaction plane. The first and second harmonic coefficients $v_1$,
$v_2$ are called directed and elliptic flow. Due to the
approximate elliptic shape of the overlapping region, the elliptic
flow $v_2$ is the largest harmonic observed in mid-rapidity.
Because of the quenching of coordinate-space-anisotropy, elliptic
flow can reveal early information about the system and because it
depends on rescattering, elliptic flow is sensitive to the degree
of thermalization of the system in the early stage.

\bf \centering\mbox{
\includegraphics[width=0.6\textwidth]{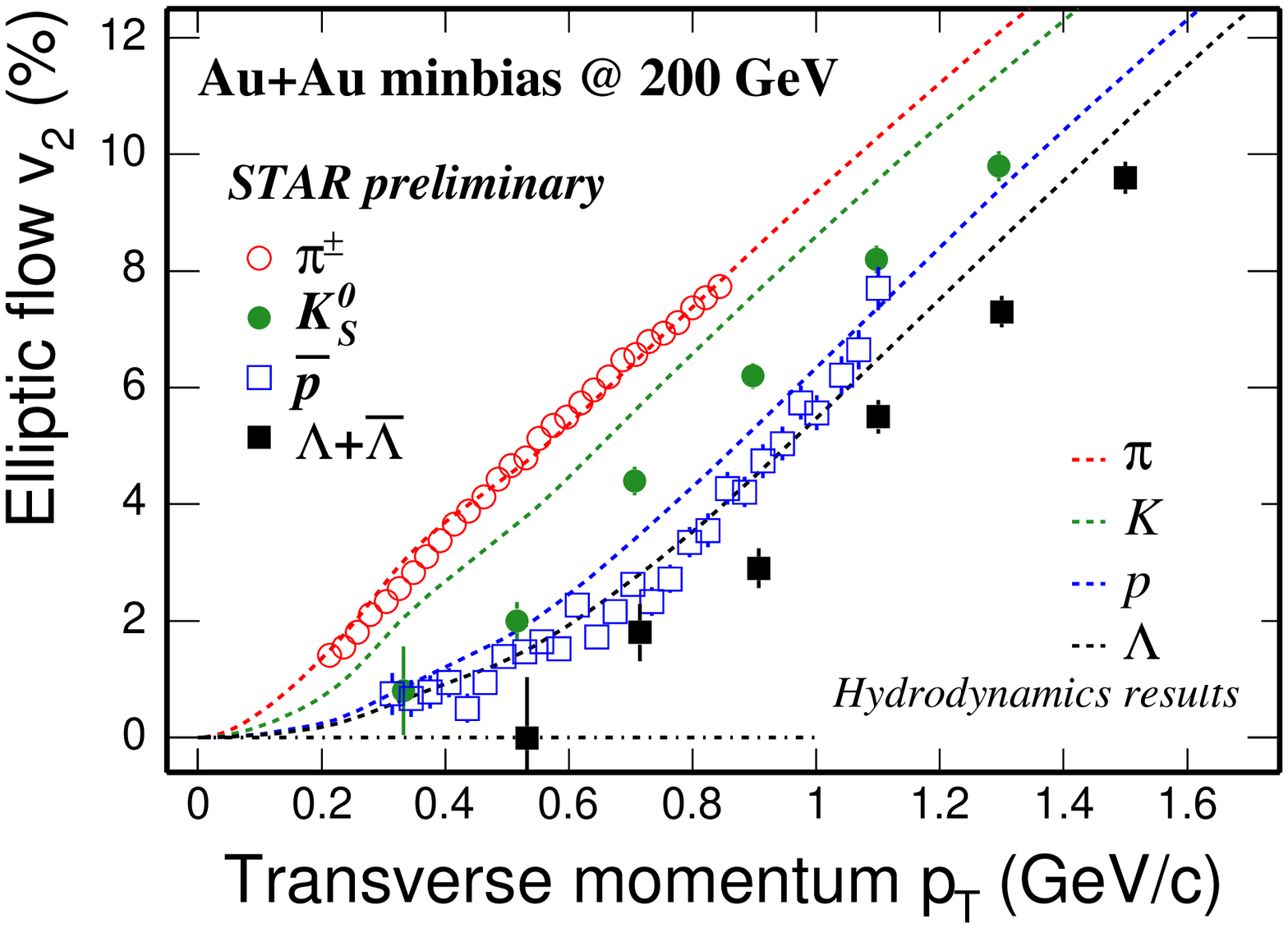}}
\caption[PID $v_2$ at low \pT]{Elliptic flow $v_2$ of identified
particles as a function of \pT at low \pT region compared with
hydrodynamic model predictions.} \label{v2lowpt} \ef

Fig.~\ref{v2lowpt} shows identified particle $v_2(p_T)$ and the
hydrodynamic model predictions for $p_T<2$
GeV/c~\cite{starwhitepaper}. In this low \pT region, $v_2$ has
larger values for lower mass particles. This mass ordering is
reasonably described by the hydrodynamic models, which assume
ideal relativistic fluid flow and negligible relaxation time
compared to the time scale of the equilibrated system. The
agreement implies early thermalization, {\em i.e.} strongly
interacting matter with a very short mean free path dominates the
early stages of the collisions.

\bf \centering\mbox{
\includegraphics[width=0.95\textwidth]{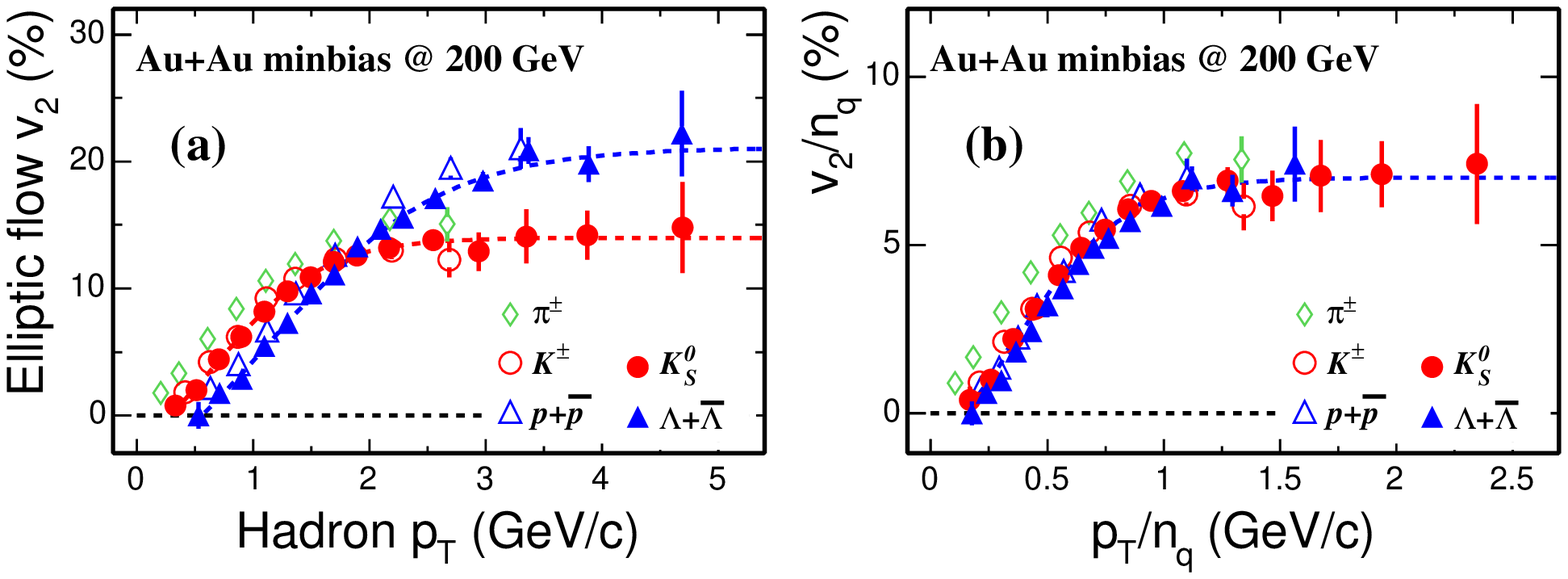}}
\caption[PID $v_2$ scaling]{Left: Identified particle $v_2$ up to
intermediate \pT. Right: After scaling the $v_2$ and $p_T$ with
number of constituent quarks $n_q$, all particles fall onto one
universal curve at $p_T/n_q>0.6$ GeV/c.} \label{v2highpt} \ef

Fig.~\ref{v2highpt} shows $v_2(p_T)$ in a larger \pT range for
different particles~\cite{phenixpikpv2,k0slambda}. $v_2$ for all
particles saturates above a certain \pT($\sim2-3$ GeV/c). On the
left panel, in addition, particles show two groups on the plots:
mesons and baryons. After scaling both $v_2$ and \pT with {\em the
Number of the Constituent Quarks} (NCQ) in the corresponding
hadron, all particles with $p_T/n_q>0.6$ GeV/c fall onto one
universal curve except pions (due to resonance decay
effect~\cite{kopiv2,minepiv2}), shown on the right panel. This
meson/baryon grouping phenomenon was also observed in the nuclear
modification factor $R_{CP}$ at intermediate
\pT($1.5<p_T$/(GeV/c)$<5$)~\cite{k0slambda,paulThesis}.
Coalescence models~\cite{coalKo,coalMolnar} which assume hadrons
are formed through coalescing of constituent quarks provide a
viable explanation for these observations. This indicates the flow
developed during a sub-hadronic (partonic) epoch, and offers a
strong evidence of deconfinement at RHIC.

\subsection{Heavy flavor production in HIC}
Heavy flavor hadrons are expected to be unique tools to probe the
early stage information in heavy ion collisions due to the
following features. The creation of heavy flavor quarks is
dominated by the initial gluon fusion processes and is negligible
in the pre-equilibrium stage~\cite{LinThesis}. Systematic studies
of charm production in \pp, and \pA collisions have been proposed
as a sensitive way to measure the PDF in the nucleon, and the
nuclear shadowing effect~\cite{LinGyulassy}. Due to their heavy
masses, energetic heavy flavor partons are expected to lose less
energy than light quarks from gluon radiations when traversing the
dense medium ("dead-cone" effect)~\cite{deadcone}. At RHIC
energies, due to the possible large production of charm quarks,
coalescence processes might not be negligible, especially for
closed charm production, indicating the standard \Jpsi suppression
scenario might be
invisible~\cite{luicjpsi,kinematicjpsi,mclerranjpsi,pbmjpsi}.
Recent studies propose that charm quark flow not only offers a
check of hydrodynamic models and/or coalescence models, but also
could imply the thermalization of the light quarks in the
system~\cite{minepiv2,jamiecharmflow,kocharmflow}.

Theoretically, there are lots of calculations of heavy quark
energy loss~\cite{heavyDMPRL,heavyDMNPA,heavyDMPLB}, nuclear
shadowing~\cite{LinGyulassy}, cronin
enhancement~\cite{vogtCronin}, charmonium
production~\cite{jpsiOctet} and charm quark
flow~\cite{linCharmflow,xuCharmflow} {\em etc.} for the collisions
at RHIC. However, we have very few experimental results to test or
prove these features so far. The first publication of charm
production at RHIC energy was a single-electron measurement by the
PHENIX collaboration at \sNN = 130 GeV in \AuAu
collisions~\cite{phenix130e}. They claimed that charm production
in heavy ion collisions obeys the number of binary collisions
scaling and the spectrum is consistent with PYTHIA~\cite{pythia}
(pQCD) calculations. Both direct reconstruction and indirect
semi-leptonic decay electron measurements (spectrum and elliptic
flow) at top RHIC energy offer us an opportunity to understand not
only pQCD, but also the early stage features of the matter created
in heavy ion collisions.

\subsection{What have we learned so far?}

So what is happening after heavy ions collide?
Fig.~\ref{timescale} is a cartoon showing the evolution of
excitation energy $E^*$ (represented as temperature $T$ for a
thermally equilibrated system) and the particles being created and
decoupling from the system. There is no absolute scale on the plot
because we don't know most of these variables yet.

\bf \centering\mbox{
\includegraphics[width=0.6\textwidth]{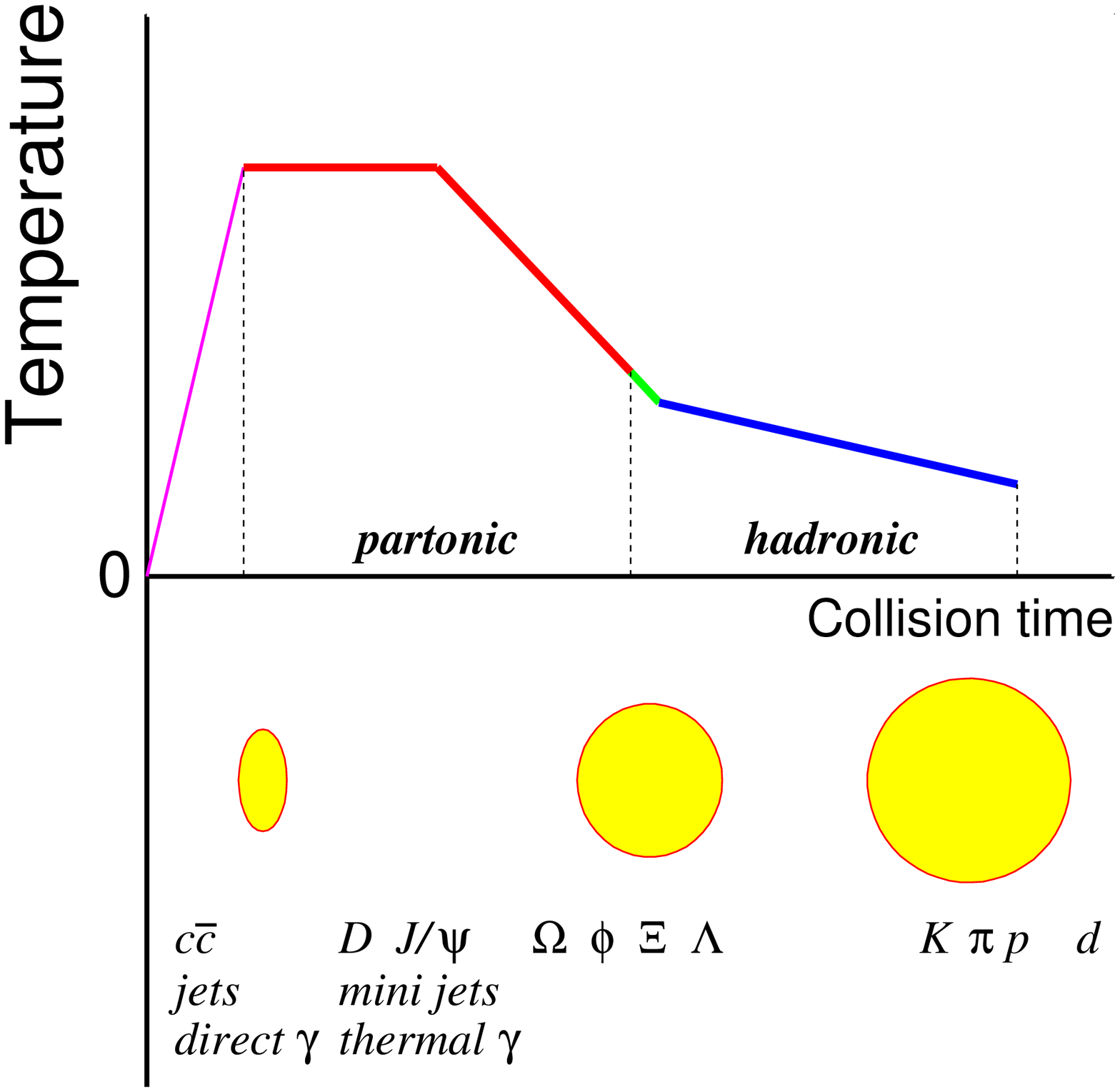}}
\caption[Time scale diagram]{A time scale plot to show the
evolution of the created matter. The several ellipses depict the
system shape during evolution. And creation and decoupling of
different identified particles are also shown on the plot.}
\label{timescale} \ef

At RHIC, after two heavy nuclei collide, plenty of energy ($\sim
70\%$ of the initial energy per participant~\cite{brahmsStop}) is
deposited into a compact overlap region suddenly. The initial
Bjorken energy density is well above the critical energy density
for a phase transition~\cite{phenixEt}. Those particles from hard
processes (such as jets and heavy quarks) as well as direct
photons ($qg\rar q\gamma$, $q\bar{q}\rar g\gamma$) are created in
these initial collisions. The overlap region is an ellipse in
coordinate space. Then rescatterrings between components in the
system make the system expand and approach thermal equilibrium.
Soft mini-jets ($p_T<\sim 2$ GeV/c) are created abundantly. The
anisotropy in coordinate space quenches due to the expansion, but
the anisotropy in momentum space saturates quickly during this
stage~\cite{hydrointro}. The system will sooner or later reach its
maximum temperature, and the matter at this stage is expected to
have the partonic {\em d.o.f.} equation of state. Fast partons
traversing through this hot dense matter will lose large amounts
of their energy (jet quenching). Meanwhile, particles (photons,
leptons, $g$, $u$, $d$, $s$, but very small fraction of heavy
quarks) with relative small masses are produced thermally in this
stage. With the continuous expansion of the system, the
temperature starts to drop, the matter becomes more and more
dilute and those partonic components begin to hadronize from the
system either through coalescence or fragmentation. In the stage
after hadronization, hadrons keep rescatterring between each other
and fractions between different hadrons are still varying. During
this stage, some hadrons with small scattering cross sections
start to freeze out from the system with fixed momentum such as
charm hadrons ($D$, $J/\psi$), multi-strange hadrons ($\Omega$,
$\phi$, $\Xi$). Then the system becomes dilute enough that there
is no inelastic scattering between different hadrons and the
fractions of different hadrons are stable. This point is called
chemical freeze-out. Some hadrons ($\pi$, $K$, $p(\bar{p})$) in
the system keep scattering elastically and finally all particles
freeze out from the system with each particle having stable
momentum. This is called kinetic freeze-out.

In heavy ion collisions, what we want to study ultimately is the
equilibrated matter with partonic {\em d.o.f.}. Multi-strange
hadrons, and more early freeze-out charm hadrons are ideal probes
to illustrate the properties of this matter created in collisions.
Based on the knowledge we have learned from three years of RHIC
runs, we need these probes to answer the questions: whether the
partonic matter is locally thermalized or not, and if yes, what
are its thermal and symmetry properties?

\chapter{Experimental Set-up}

\section{RHIC accelerator}

The Relativistic Heavy Ion Collider (RHIC) at Brookhaven National
Laboratory (BNL) is designed to accelerate and collide heavy ions
and polarized protons with high luminosity, allowing physicists to
explore the strong interaction through many extensive and
intensive measurements. It is the first facility to collide heavy
ion beams and the top center-of-mass collision energy is 200 GeV
per nucleon pair, which is about more than 10 times greater than
the highest energy reached at previous fixed target experiments.
The purpose of this extraordinary new accelerator is to seek out
and explore new high-energy forms of matter and thus continue the
centuries-old quest to understand the nature and origins of matter
at its most basic level. RHIC is also delivering polarized proton
beams up to center-of-mass energy 500 GeV to carry on vigorous
spin scientific programs.

Fig.~\ref{rhic} shows a diagram of the RHIC machine complex,
including a Van de Graaff facility, a linear proton accelerator,
the booster synchrotron, the Alternative Gradient Synchrotron
(AGS), and ultimately the RHIC synchrotron ring. For Au beam
operations, the Au ions with charge $Q=-1e$ are created using the
Pulsed Sputter Ion Source. Then they are accelerated through the
Van de Graaff facility and a series of stripping foils, and the Au
ions at the exit are with a kinetic energy of 1 MeV/nucleon and a
net charge of $Q=+32e$. The ions are then injected into the
booster synchrotron and accelerated to an energy of 95
MeV/nucleon. After the Au ions leave the booster, they are further
stripped to $Q=+77e$ and transferred into the AGS, where they are
accelerated to 8.86 GeV/nucleon and sorted into four final
bunches. Finally, the ions are injected into RHIC and stripped to
the bare charge state of $Q=+79e$ during the transfer. For \pp
operations, protons are injected from the 200 MeV Linac into the
booster, followed by acceleration in the AGS and injection into
RHIC.

\bf \centering\mbox{
\includegraphics[width=0.9\textwidth]{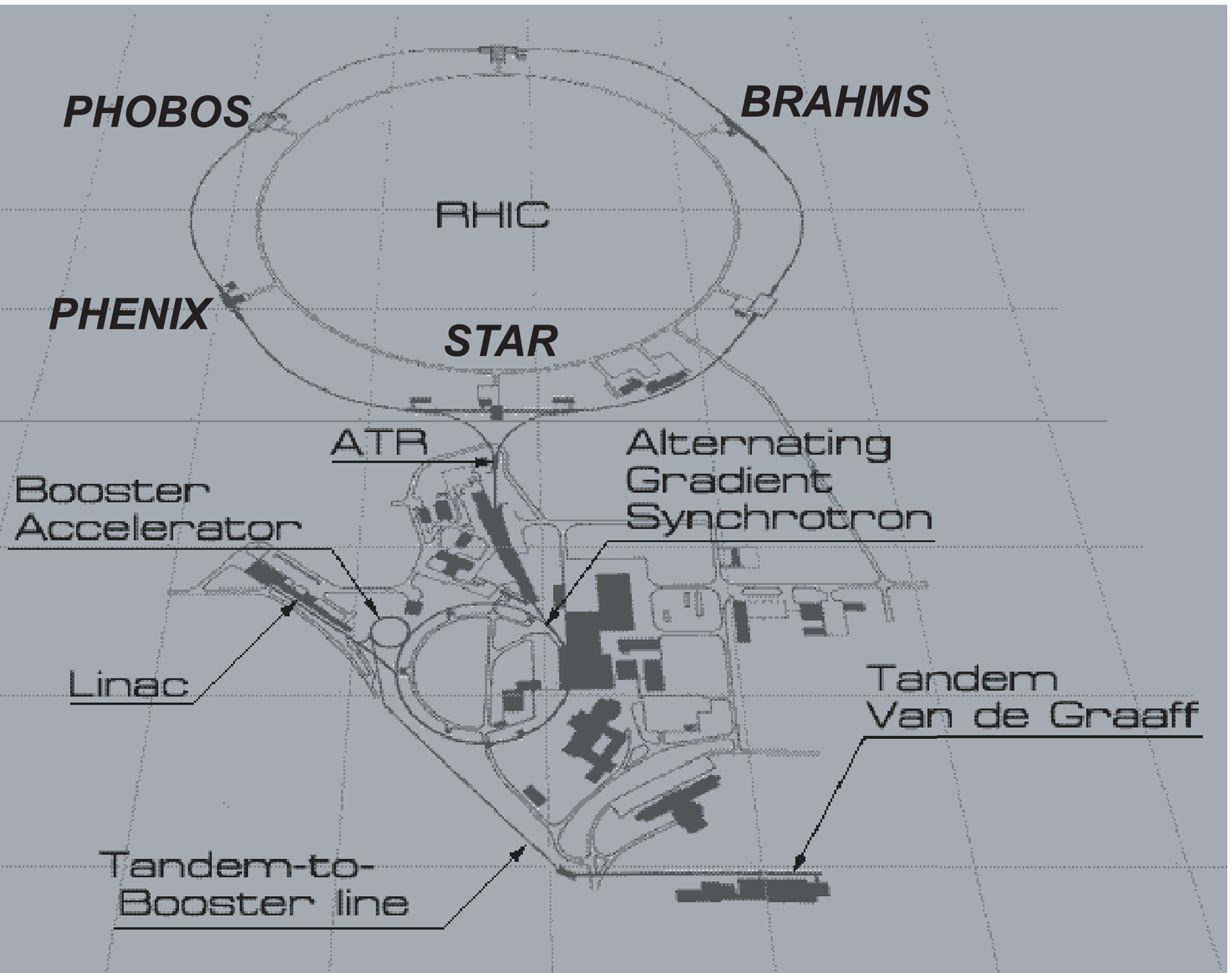}}
\caption[The RHIC complex]{Schematic of the RHIC complex. RHIC's
two 3.8-kilometer rings collide relativistic heavy ions and
polarized protons at six intersection points.} \label{rhic} \ef

RHIC consists of two concentric super-conducting storage rings
that are called blue and yellow rings, respectively. Each ring has
its own dependent set of bending and focusing magnets as well as
ratio frequency cavities, but both share a common horizontal plane
in the tunnel. The rings have six interaction points, and 4 of
them are equipped with detectors. They are two large experiments
STAR (6 o'clock), PHENIX (8 o'clock) and two small ones PHOBOS (10
o'clock) and BRAHMS (2 o'clock), respectively.

To date, RHIC has been run in \pp, \dAu, \AuAu and Cu + Cu
configurations.

\section{STAR detector}

The Solenoidal Tracker at RHIC (STAR) is a specially designed
detector to track thousands of particles simultaneously produced
by each ion collision at RHIC. It has an azimuthal symmetric
acceptance and covers large range around mid-rapidity. STAR
consists of several subsystems and a main tracker - the {\em Time
Projection Chamber} (TPC) located in a homogenous solenoidal
analyzing magnet.

\bf \centering\mbox{
\includegraphics[width=0.8\textwidth]{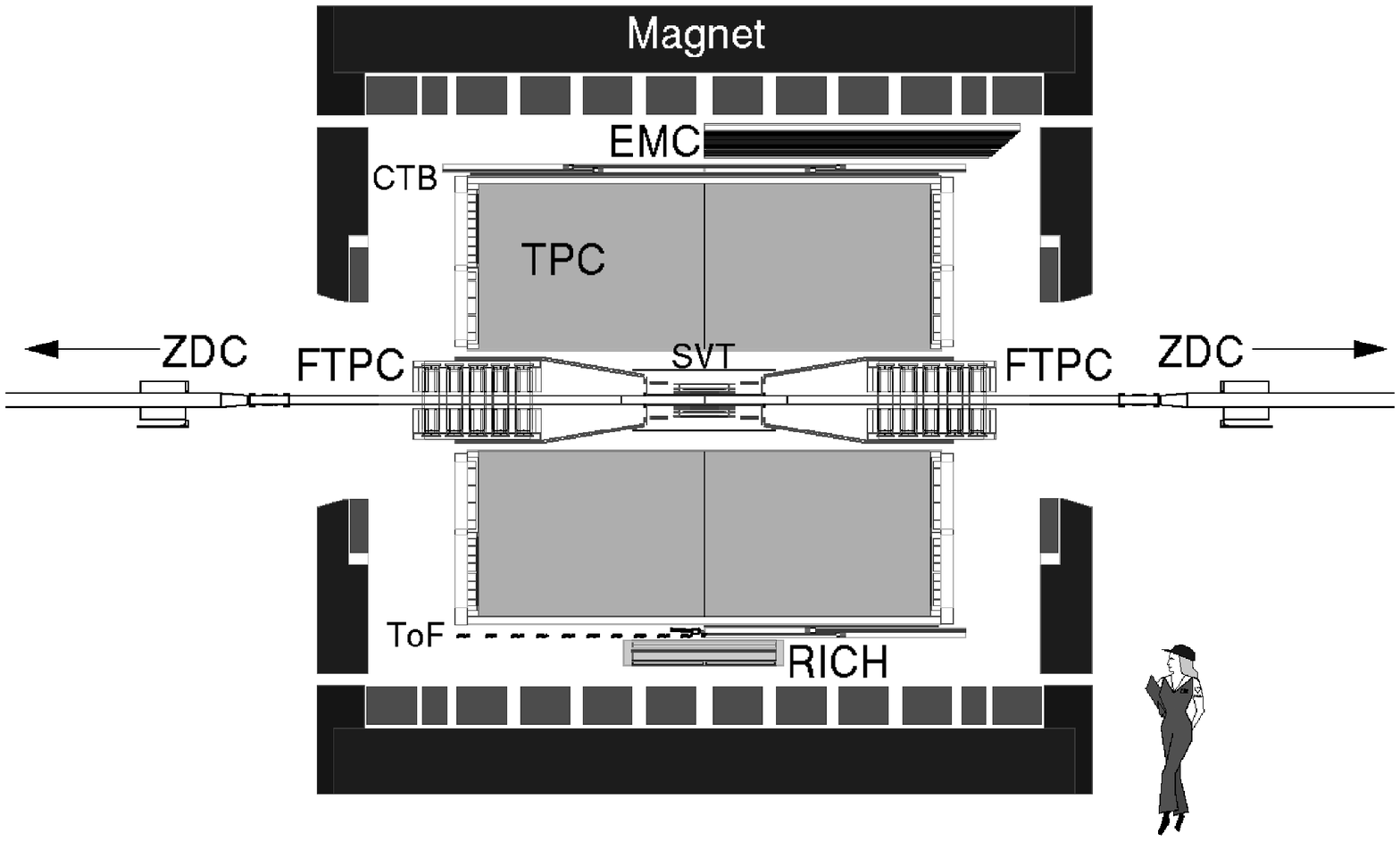}}
\caption[Cutaway view of STAR]{Cutaway view of the STAR detector.
It includes the partial installed ElectroMagnetic Calorimeter
(EMC) and two prototypes Time-of-Flight (TOF) detectors.}
\label{star} \ef

Fig.~\ref{star} shows the cutaway view of the STAR detector. The
main tracker - TPC covers the $|\eta|<1.5$ and $2\pi$ in azimuth.
The details of TPC detector will be discussed in the next section.
There are inner detectors {\em Silicon Vertex Tracker} (SVT) and
{\em Silicon Strip Detector} (SSD) close to the beam pipe, which
provides additional high precision space points on track so that
it improves the position resolution and allows us to reconstruct
the secondary vertex of weak decay particles. There are two {\em
Forward TPC} (FTPC) detectors covering $2.8<|\eta|<3.8$ to track
particles at forward and backward rapidity. One prototype tray of
{\em Time-Of-Flight} (TOF) detector using scintillator materials
(TOFp) was installed since Run II and another prototype tray of
TOF detector using {\em Multigap Resistive Plate Chamber} (MRPC)
technology (TOFr) was installed since Run III. Each replaces one
{\em Central Trigger Barrel} (CTB) tray (120 in total) surrounding
TPC. They are used to test the performance of upcoming full barrel
TOF detector upgrade which is expected to extend the PID
capability of STAR greatly. Part of barrel {\em ElectronMagnetic
Calorimeter} (EMC) was also installed since Run II. They are used
to measure the electromagnetic probes - electrons and photons.

There are some main trigger detectors: {\em Zero Degree
Calorimeter} (ZDC), CTB and {\em Beam-Beam Counters} (BBC). Two
ZDCs locates on each side $\sim 18$ m away from the collision
points. Each is centered at $0^\text{o}$ and covers $\sim 2.5$
mrad. The ZDCs are hadronic calorimeters to detect the outgoing
neutrons. They are put beyond the dipole magnets which bend away
the charged fragments. The ZDC signals are used for monitoring the
heavy ion beam luminosity and for the experiments triggers. The
CTB is a collection of scintillating tiles covering the whole
barrel ektexine of the TPC. The CTB will be mostly used to select
central triggered events in heavy ion collisions by measuring the
occupancy of those CTB slats. 
The BBC subsystem covers $3.3<|\eta|<5.0$, measuring the
"beam-jets" at high rapidity from {\em Non-Singly Diffractive}
(NSD) inelastic \pp interactions. It consists of two disk shaped
scintillating detectors, with one placed at each endcap of the TPC
(3.5 m from TPC center). Each BBC disk is composed of
scintillating tiles that are arranged in a hexagonal closest
packing. The \pp NSD trigger sums the output of all tiles on each
BBC and requires a coincidence of both BBC's firing above noise
threshold within a time window. Some other detectors are used for
special triggers, {\em e.g.} {\em pseudo Vertex Position
Detectors} (pVPDs) are used for TOF triggered events (this will be
discussed in the following section), and EMC is used to trigger on
high \pT particle events {\em etc}.


The STAR magnet is cylindrical in design with a length of 6.85 m
and has inner and outer diameters of 5.27 m and 7.32 m,
respectively. It generates a field along the length of the
cylinder having a maximum of $|B_z|=0.5$ T. It allows the tracking
detectors to measure the helical trajectory of charged particles
to get their momenta. To date, the STAR magnet has been run in
full field, reversed full field and half filed configurations.


\section{Main tracker - TPC}

TPC is the ``heart'' of the STAR detector~\cite{tpctech}.
Consisting of a 4.2 m long cylinder with 4.0 m in diameter, it is
the largest single TPC in the world. The cylinder is concentric
with the beam pipe, and the inner and outer radii of the active
volume are 0.5 m and 2.0 m, respectively. It can measure charged
particles within momentum $0.15<p_T$/(GeV/c)$<30$ (0.075 GeV/c low
limit for 0.25 T). The TPC covers the full region of azimuth
($0<\phi<2\pi$) and covers the pseodurapidity range of $|\eta|<2$
for inner radius and $|\eta|<1$ for outer radius. Fig.~\ref{tpc}
shows a cutaway view of the structure of the TPC.

\bf \centering\mbox{
\includegraphics[width=0.8\textwidth]{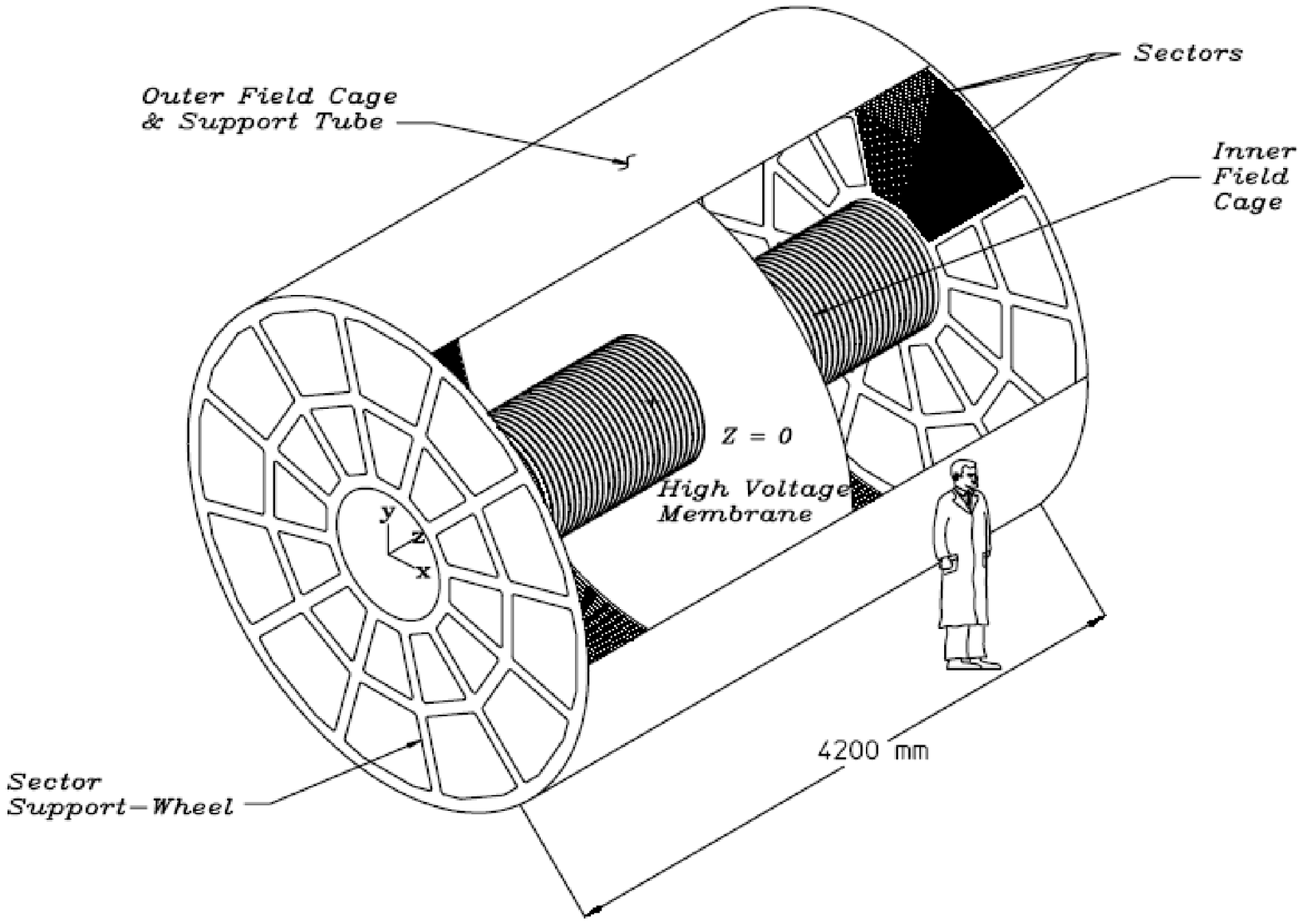}}
\caption[Cutaway view of the TPC detector]{Cutaway view of the TPC
detector at STAR.} \label{tpc} \ef

The TPC is divided into two parts by the central membrane. It is
typically held at 28 kV high voltage. A chain of 183 resistors and
equipotential rings along the inner and outer field cage create a
uniform drift filed ($\sim 135$ V/cm) from the central membrane to
the ground planes where anode wires and pad planes are organized
into 12 sectors for each sub-volume of the TPC. The working gas of
the TPC is two gas mixture $-$ P10 (Ar $90\%$ + CH$_4$ $10\%$)
regulated at 2 mbar above the atmospheric pressure. The electron
drift velocity in P10 is relatively fast, $\sim 5.45$ cm/$\mu$s at
$130$ V/cm drift field. The gas mixture must satisfy multiple
requirements and the gas gains are $~\sim 3770$ and $\sim 1230$
for the inner and outer sectors working at normal anode voltages
(1170 V for inner and 1390 V for outer), respectively. Each
readout plane is instrumented with a thin {\em Multi-Wire
Proportional Chamber} (MWPC) together with a pad chamber readout.
Each pad plane is also divided into inner and outer sub-sectors,
while the inner sub-sector is designed to handle high track
density near collision vertex. 136,608 readout pads provide
$(x,y)$ coordinate information, while $z$ coordinate is provided
by 512 time buckets and the drift velocity. Typical resolution is
$\sim 0.5-1.0$ mm.

When charged particles traverse the TPC, they liberate the
electrons from the TPC gas due to the ionization energy loss
($dE/dx$). These electrons are drifted towards the end cap planes
of the TPC. There the signal induced on a readout pad is amplified
and integrated by a circuit containing a pre-amplifier and a
shaper. Then it is digitalized and then transmitted over a set of
optical fibers to STAR {\em Data AcQuisition system} (DAQ).

The TPC reconstruction process begins by the 3D coordinate space
points finding. This step results in a collection of points
reported in global Cartesian coordinates. The {\em Timing
Projection chamber Tracker} (TPT) algorithm is then used to
reconstruct tracks by helical trajectory fit. The resulted track
collection from the TPC is combined with any other available
tracking detector reconstruction results and then refit by
application of a Kalman filter routine $-$ a complete and robust
statistical treatment. The primary collision vertex is then
reconstructed from these global tracks and a refit on these tracks
with the {\em distance of closest approach} ($dca$) less the 3 cm
is preformed by a constrained Kalman fit that forces the track to
originate from the primary vertex. The primary vertex resolution
is $\sim 350$ $\mu$m with more than 1000 tracks. The refit results
are stored as primary tracks collection in the container. The
reconstruction efficiency including the detector acceptance for
primary tracks depends on the particle type, track quality cuts,
\pT, track multiplicity {\em etc}. The typical value for the
primary pions with $N_{fit}>24$ and $|\eta|<0.7$, $dca<3.0$ cm is
approximate constant at $p_T>0.4$ GeV/c: $>\sim 90\%$ for \AuAu
peripheral collisions and $\sim 80\%$ for central collisions,
respectively.

The TPC can also identify particles by the $dE/dx$ of charged
particles traversing the TPC gas. The mean rate of $dE/dx$ is
given by the Bethe-Bloch equation~\ref{dEdxBB}~\cite{PDG}:

\be -\frac{dE}{dx} =
Kz^2\frac{Z}{A}\frac{1}{\beta^2}\left[\frac{1}{2}\ln\frac{2m_ec^2\beta^2\gamma^2T_{max}}{I^2}-\beta^2-\frac{\delta}{2}\right]
\label{dEdxBB} \ee

The meaning of each symbol can be referred to~\cite{PDG}.
Different types of particles (different rest masses) with the same
momentum have different kinematic variables $\beta$ ($\gamma$),
which may result in distinguishable $dE/dx$. The typical
resolution of $dE/dx$ in \AuAu collisions is $\sim 8\%$, which
makes the $\pi$/$K$ separation up to $p\sim 0.7$ GeV/c and
proton/meson separation up to $p\sim 1.1$ GeV/c.

A new recent technique was developed to identify high momentum
($p>3$ GeV/c) pions and protons in the relativistic rising region
of $dE/dx$~\cite{xzbDPF} benefiting from the advantage that the
mean rates of $dE/dx$ for different particles have a visible
separation in the relativistic rising region ($\sim 2\sigma$
separation for pions and protons). Due to large acceptance of the
TPC, using the topology of their weak decay in the TPC, the
$K_{S}^{0}$, $\Lambda$($\bar{\Lambda}$) {\em etc.} can be
identified across \pT region $0.3<p_T$/(GeV/c)$<7.0$ (upper edge
limited by statistics). Resonances ($K^{*}$, $\phi$, $\Delta$ {\em
etc.}) can be reconstructed through the event mixing
technique~\cite{KstarPRC}. Fig.~\ref{STARPID} shows the PID
capabilities up to date with the TPC. In addition, the TOF PID
capability is also shown on the plot which will be discussed in
the next section and the analysis part as well.


\bf \centering\mbox{
\includegraphics[width=0.9\textwidth]{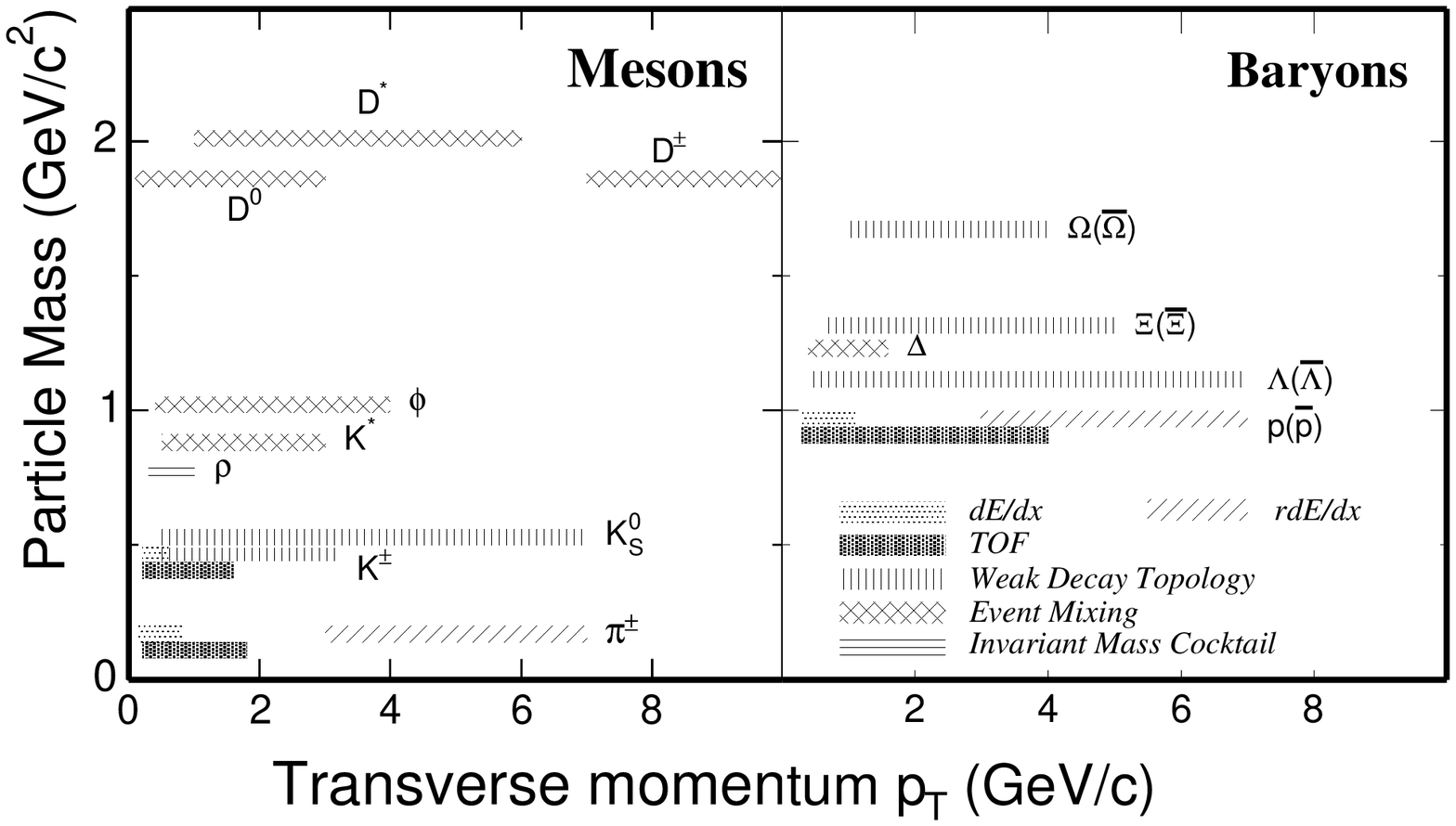}}
\caption[PID capability of STAR]{$p_T$ reach of particle
identification capability with STAR detectors for Run II and Run
III configurations. The upper edges of $rdE/dx$, weak decay
topology, event mixing are limited by statistics.} \label{STARPID}
\ef

\section{Prototype TOF detector}

STAR has proposed the full barrel {\em Time-Of-Flight} (TOF)
detector upgrade based on the {\em Multigap Resistive Plate
Chamber} (MRPC) technology in the coming future. The TOFp detector
(a prototype based on scintillator technology) was installed since
Run II~\cite{tofpBill}. It replaced one of CTB trays, covering
$-1<\eta<0$, and $\pi/30$ in azimuth. It contains 41 scintillator
slats with the signal read out by {\em Photo Multiplier Tubes}
(PMTs). The resolution of TOFp is $\sim 85$ ps in \AuAu
collisions. However, due to the significant higher cost by the
PMTs, this design will not be used in the full TOF upgrade.

In Run III and Run IV, new prototypes of TOF detector based on
MRPC (TOFr) were installed. Each also replaced one CTB tray,
covering $-1<\eta<0$ and $\pi/30$ in azimuth too. In Run III, 28
MRPC modules were installed in the tray and 12 of them were
equipped with electronics, corresponding to $\sim 0.3\%$ of the
TPC acceptance~\cite{tofpikp}. In Run IV, 24 modules were
installed in a new tray and the tray was put in the same position
in STAR as Run III (but slightly global z position shift), but
only 12 modules were equipped with valid electronics, which means
the acceptance in Run IV was roughly similar to that in Run III.

Two pVPDs were installed as well since Run II to provide a
starting time for TOF detectors, each staying 5.4 m away from the
TPC center along the beam line~\cite{tofpBill}. Each pVPD consists
of three detecting element tubes covering $\sim 19\%$ of the total
solid angle in $4.43<|\eta|<4.94$. Due to different
multiplicities, the effective timing resolution of total starting
time is 25 ps, 85 ps and 140 ps for 200 GeV \AuAu, \dAu and \pp
collisions, respectively.


\bf \centering\mbox{
\includegraphics[width=0.8\textwidth]{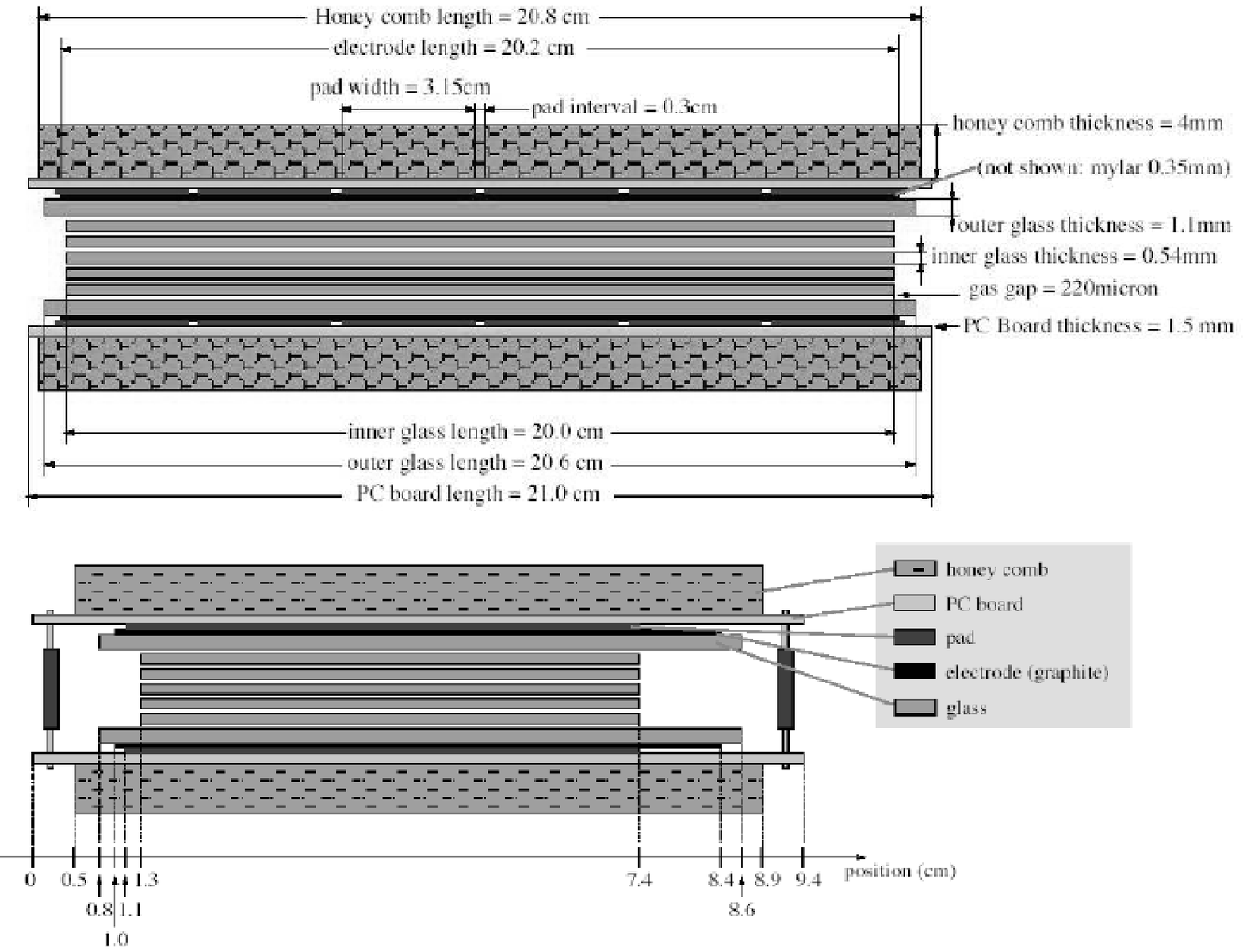}}
\caption[Two-side view of a MRPC module]{Two-side view of a MRPC
module~\cite{mrpctest}.} \label{mrpc} \ef

\begin{table}[hbtp]
\caption[TOF system performance]{TOF system performance in Run III
and Run IV} \label{tofCali} \vskip 0.1 in
\centering\begin{tabular}{c|c|c|c|c|c} \hline \hline
\multicolumn{3}{c|}{}    &  \multicolumn{3}{c}{Timing Resolution (ps)}\\
\cline{4-6} \multicolumn{3}{c|}{} & pVPDs & TOFr system & TOFp system \\
\hline
Run   & \multicolumn{2}{c|}{\dAu $@$ 200 GeV} & 85 & 120 (85)$^\dagger$ & 100-140 \\
\cline{2-6} III  & \multicolumn{2}{c|}{\pp $@$ 200 GeV} & 140 & 160 (85) & N/A \\
\hline
     & \multicolumn{2}{c|}{\AuAu $@$ 62.4 GeV} & 55 & 105 (89) & 110 (95)
     \\ \cline{2-6} Run & \AuAu & FF/RFF, w/o E pVPD$^\ddagger$ & 40 & 95
     (86) & 96 (87) \\ \cline{3-6}
     IV & $@$ & FF/RFF & 27 & 86 (82) & 92 (88) \\ \cline{3-6}
        & 200 GeV & HF$^{\star}$ & 20 & 82 (80) & 85 (83) \\ \hline \hline
\end{tabular}
\centering\mbox{$^\dagger$ numbers in the brackets are those for
intrinsic TOFr/TOFp detector.} \centering\mbox{$^\ddagger$ east
pVPD is not available for those runs from 5022001 to 5036000.}
\centering\mbox{$^\star$ HF - half field; FF - full field; RFF -
reversed full field.}

\end{table}

\bf \centering\mbox{
\includegraphics[width=0.6\textwidth]{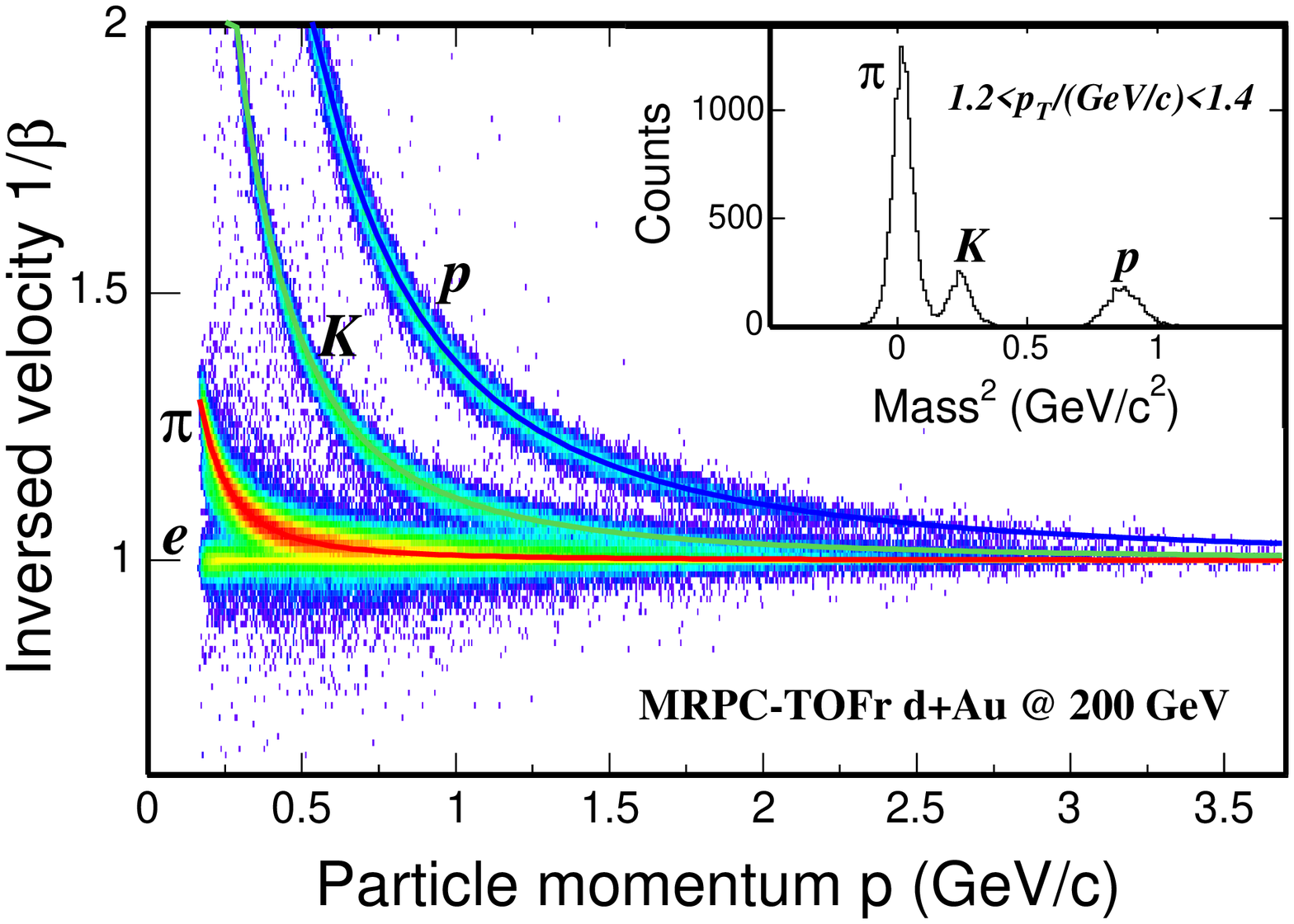}}
\caption[TOFr hadron PID]{$1/\beta$ vs. momentum ($p$) from 200
GeV \dAu collisions. Separations between pions and kaons, protons
and mesons are achieved up to $p_T\sim 1.6$ and $\sim 3.0$ GeV/c,
respectively. The insert plot shows $m^2=p^2(1/\beta^2-1)$ for
$1.2<p_T<1.4$ Gev/c.} \label{tofrPID} \ef

MRPC technology was first developed by the CERN ALICE group.
Fig.~\ref{mrpc} shows the two side views (long edge view on top
and short edge view on bottom) of an MRPC module appropriate for
STAR~\cite{mrpctest}. An MRPC basically consists a stack of
resistive plates with a series of uniform gas gaps. It works in
avalanch mode. Electrodes are applied to the outer surface of the
outer plates. With a strong electric field applied on, the
internal plates are left electrically floating and they will keep
the correct voltage due to the flow of electrons and ions created
in avalanches. There are six read-out strips on each module in
this design. The first beam test for 6-gap MRPCs at CERN PS-T10
facility with $p_{lab} = 7$ GeV/c pions beam resulted in a $\sim
65$ ps timing resolution with more than $~95\%$ detecting
efficiency and the module is capable of working at high event rate
(500 Hz/cm$^2$)~\cite{mrpctest}. These modules were then assembled
in a prototype TOF tray and tested in the AGS radiation area.
Similar resolution was obtained. In RHIC Run III and Run IV, the
MRPC modules in TOFr trays installed in the STAR detector were
applied on the high voltage of 14 kV and with the working gas of
$95\%$ freon and $5\%$ isobutane. The charged particle detecting
efficiency is $>95\%$ at high voltage plateau.

TOF system calibrations include the start time calibration from
pVPDs and TOFr/TOFp flight time calibration. The main sources need
to be considered are global time offset due to different
electronics delays, the correlation between the amplitude and the
timing signals, the correlation between the hit position and the
timing signals {\em etc}. Detailed calibrations on TOF systems can
be found in ~\cite{LijuanThesis,tofpikp} (TOFr) and
~\cite{tofpBill} (TOFp). Table.~\ref{tofCali} lists the calibrated
timing resolution results for TOFr and TOFp system in Run III and
Run IV (Run IV 200 GeV \AuAu results are from test sample so far).
The results show that the intrinsic timing resolution of TOFr was
$\sim 85$ ps and this performance was quite stable in two-year
runs. Fig.~\ref{tofrPID} shows the hadron PID capability of TOFr
system in 200 GeV \dAu collisions~\cite{tofpikp}. The performance
of TOFr satisfied STAR TOF system upgrade requirements.

\chapter{Single electron transverse momentum distributions}
Single lepton measurement has been proposed as a feasible and
effective way to extract heavy flavor production since long time
ago. Since the decay kinematics of heavy flavor hadrons to leptons
is well known, the transverse momentum ($p_T$) distribution of
single lepton can reveal that of heavy flavor hadrons. And
furthermore, many topics may be carried on, such as total cross
section, heavy quark fragmentation, heavy quark energy loss in
medium in A + A collisions, {\em etc}. In this chapter, analysis
details of single electron \pT distributions from 200 GeV \dAu,
\pp collisions and 62.4 GeV \AuAu collisions will be presented.

\section{Single electrons from \dAu and \pp collisions at \sNN = 200 GeV}

\subsection{Data sets and Trigger}
In RHIC Run III, STAR detector has accumulated several data sets
from \dAu and \pp collisions at \sNN = 200 GeV. Besides the
{\em minimum bias} triggered data from \dAu collisions and the
{\em Non Singly Diffractive} (NSD) collisions from \pp collisions,
special triggers were set up for TOFr detector data accumulation
in both \dAu and \pp collisions since TOFr detector has a relative
very small acceptance. Table~\ref{dataset} lists all the data sets
under the selections used in this analysis from Run III.

\begin{table}[hbt]
\caption[Data sets list]{Data sets from Run III used in this analysis}
\label{dataset}
\vskip 0.1 in
\centering\begin{tabular}{c|c|c|c} \hline \hline
Collision System & Trigger & Vertex Z Selection & Data Sample Size \\
\hline \dAu & minbias & $|V_{Z}|<30$ cm & $6.42$ M \\ \cline{2-4}
     & dAuTOF  & $|V_{Z}|<50$ cm & $1.89$ M \\ \hline
\pp  & NSD     & $|V_{Z}|<30$ cm & $4.35$ M \\ \cline{2-4}
     & ppFPDTOFu & $|V_{Z}|<50$ cm & $1.08$ M \\
     \hline \hline
\end{tabular}
\end{table}

The \dAu minimum bias trigger required at least one beam-rapidity
neutron in the ZDC in the Au beam outgoing direction, which is
assigned negative pseudorapidity ($\eta$)~\cite{dAuhighpt}. The
trigger accepts $95\pm3\%$ of the \dAu hadronic cross section. The
NSD \pp events are triggered on the coincidence of two BBCs, which
are angular scintillator detectors covering
$3.3<|\eta|<5.0$~\cite{auauhighpt}. The NSD cross section was
measured to $30.0\pm3.5$ mb by a {\em van der Meer} scan and
PYTHIA simulation of the BBC acceptance~\cite{tofpikp}. The TOF
trigger set up in Run III was to select events with a valid pVPD
coincidence and at least one TOFr hit (out of 72). The trigger
enhancement factors in comparison to minimum bias trigger are
$\sim$10 in \dAu and $\sim$40 in \pp collisions.

Centrality definition in \dAu collisions was based on the charged
particle multiplicity in $-3.8<\eta<-2.8$, measured by the FTPC in
the Au beam direction~\cite{dAuhighpt}. But in this analysis, due
to the limited statistics, the analysis was done on minimum bias
\dAu events, not on each specified centralities.

\subsection{Electron identification and hadron contamination}
TPC is the main detector in STAR for tracking and identifying
charged particles. With TPC only, electron identification is
complicated because the electron band crosses the hadrons bands. A
previous analysis on conversion electron was done for Run II
\AuAu data using pure topological method, however, electron
identification was limited up to $p_T\sim0.8$
GeV/c~\cite{IanPRC,IanThesis}.

In addition to its hadron identification capabilities, TOFr
detector can be used to identify electrons in combination with the
$dE/dx$ in the TPC, shown in Fig.~\ref{ePID}. The top panel of
left plot shows the 2-D scattering plot of $dE/dx$ vs. particle
momentum ($p$) for the charged particles with good TOFr matched
hits from \dAu collisions~\cite{LijuanThesis}. The selection
criteria are listed in Table ~\ref{electroncut}.
If additional particle velocity ($\beta$) cut $|1/\beta-1|<0.03$
is applied on, this plot is shown in the bottom panel. Slow
hadrons were eliminated and electrons band, then, can be separated
from those fast hadrons band more readily. With combination of
$dE/dx$ in the sTPC and $\beta$ in TOFr, electrons can be identified
starting from $p\sim0.15$ GeV, while the upper limit in momentum is
constrained by the statistics in this analysis.

\bf \bmn[c]{0.6\textwidth} \centering\mbox{
\includegraphics[width=0.95\textwidth]{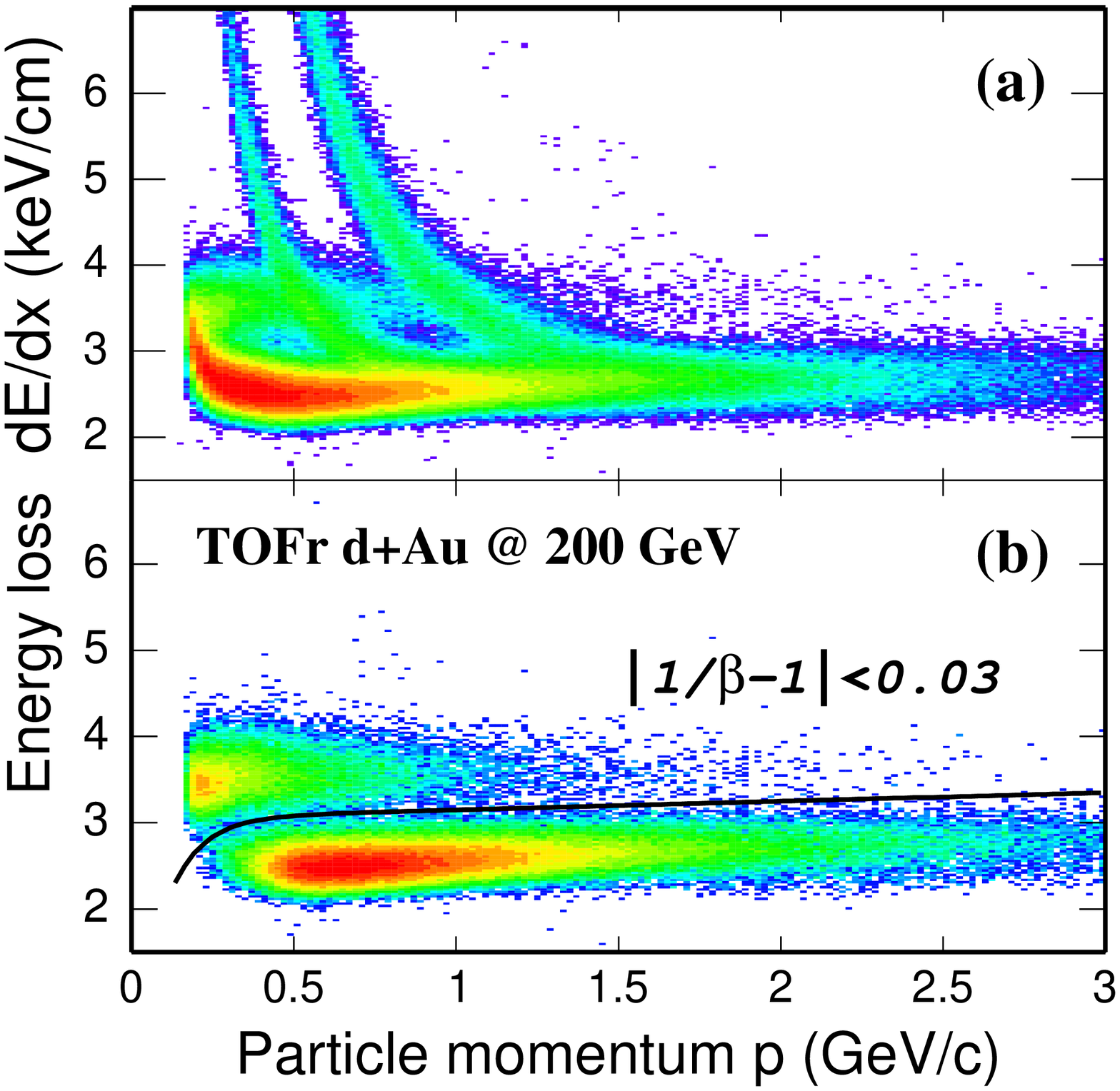}}
\emn \bmn[c]{0.4\textwidth} \centering\mbox{
\includegraphics[width=0.95\textwidth]{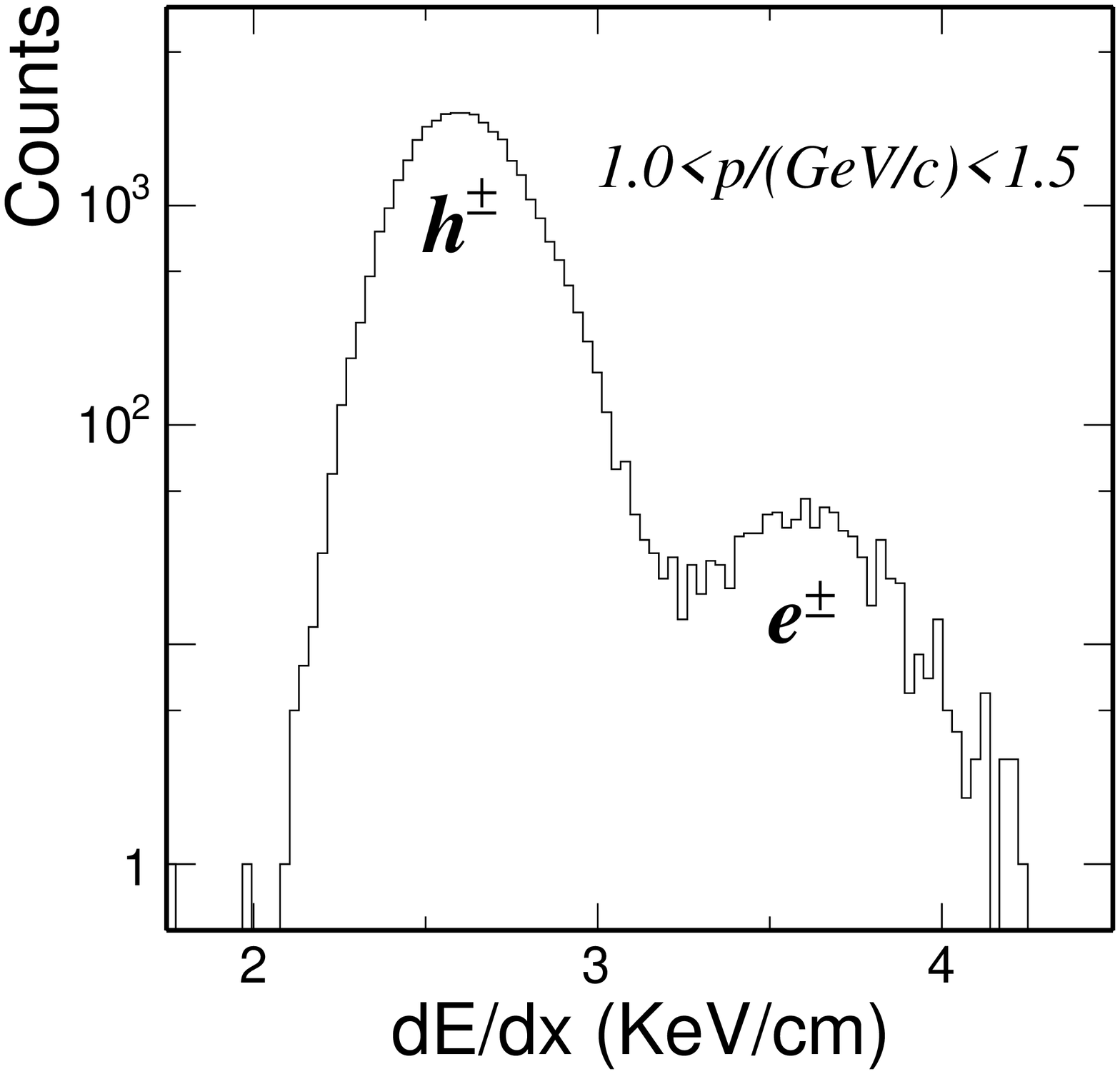}}
\emn \caption[Electron PID with TOF and $dE/dx$]
{Left: Electron identification by combining $\beta$ from
TOFr and $dE/dx$ from TPC. With the cut of $|1/\beta-1|<0.03$, electron
band can be separated from the hadrons readily, shown in the bottom panel.
Right: The $dE/dx$ projection plot in $1.0<p_T$/(GeV/c)$<1.5$.}
\label{ePID} \ef

Electrons were selected under the following cut: \be
dE/dx(p)>2.4+0.65\times(1-e^{-(p-0.15)/0.1})+0.1\times p
\label{tofCut} \ee where $p$ is in the unit of GeV and $dE/dx$ is
in keV/cm. Hadron contamination under this cut was studied by
fitting to the $dE/dx$ distributions in each \pT bin. The
2-Gaussian function cannot describe the shoulder region of fast
hadron peak very well in lower \pT region, a function of
exponential+gaussian was used in the fit. At \pT$\sim2-3$ GeV,
statistics cannot enable us to distinguish the difference of these
two fitting methods, so 2-Gaussian fit was performed in this \pT
region. Fig.~\ref{dEdxFitTof} shows the fitting results in several
typical \pT bins from \dAu collisions. The arrows denote the cut
from Eq.~\ref{tofCut}. Hadron contamination ratio was estimated
from these fits, shown in Fig.~\ref{hadronCom}, and the efficiency
for electrons was corrected for in the final spectra and will be
discussed in the sections later.

\bf \centering \bmn[b]{0.48\textwidth} \centering
\includegraphics[width=1.0\textwidth]{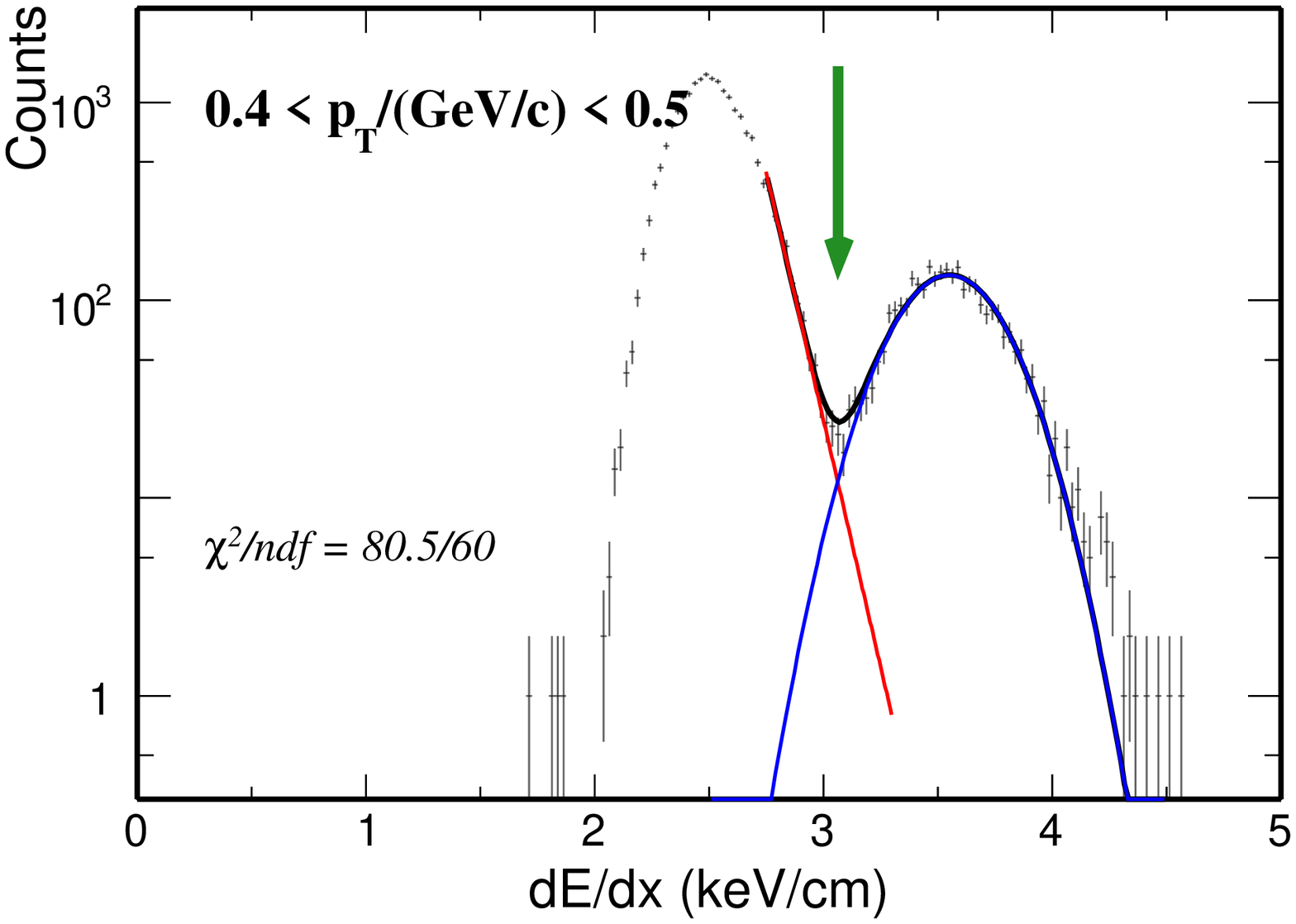}
\emn%
\bmn[b]{0.48\textwidth} \centering
\includegraphics[width=1.0\textwidth]{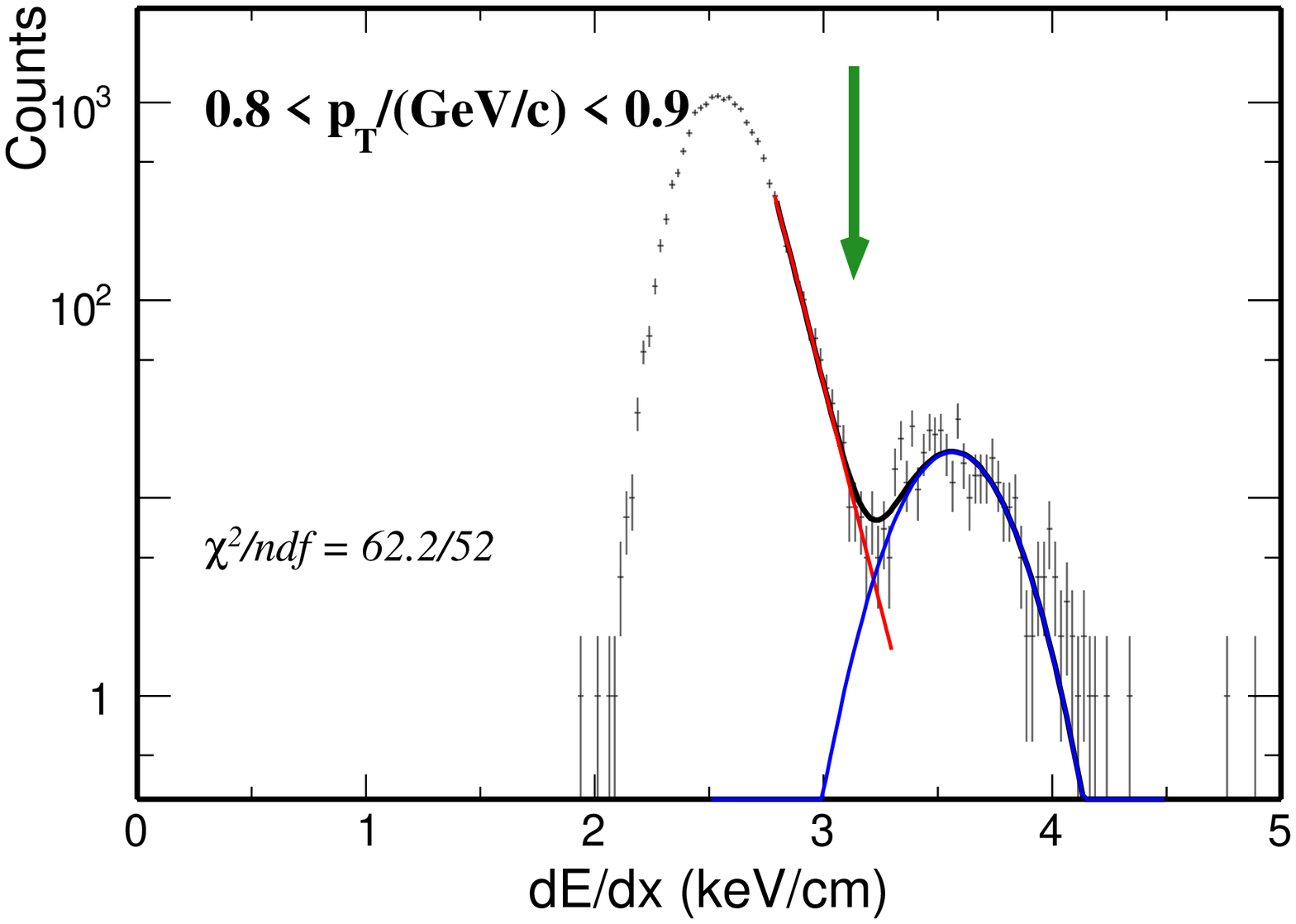}
\emn\\[10pt]
\bmn[b]{0.48\textwidth} \centering
\includegraphics[width=1.0\textwidth]{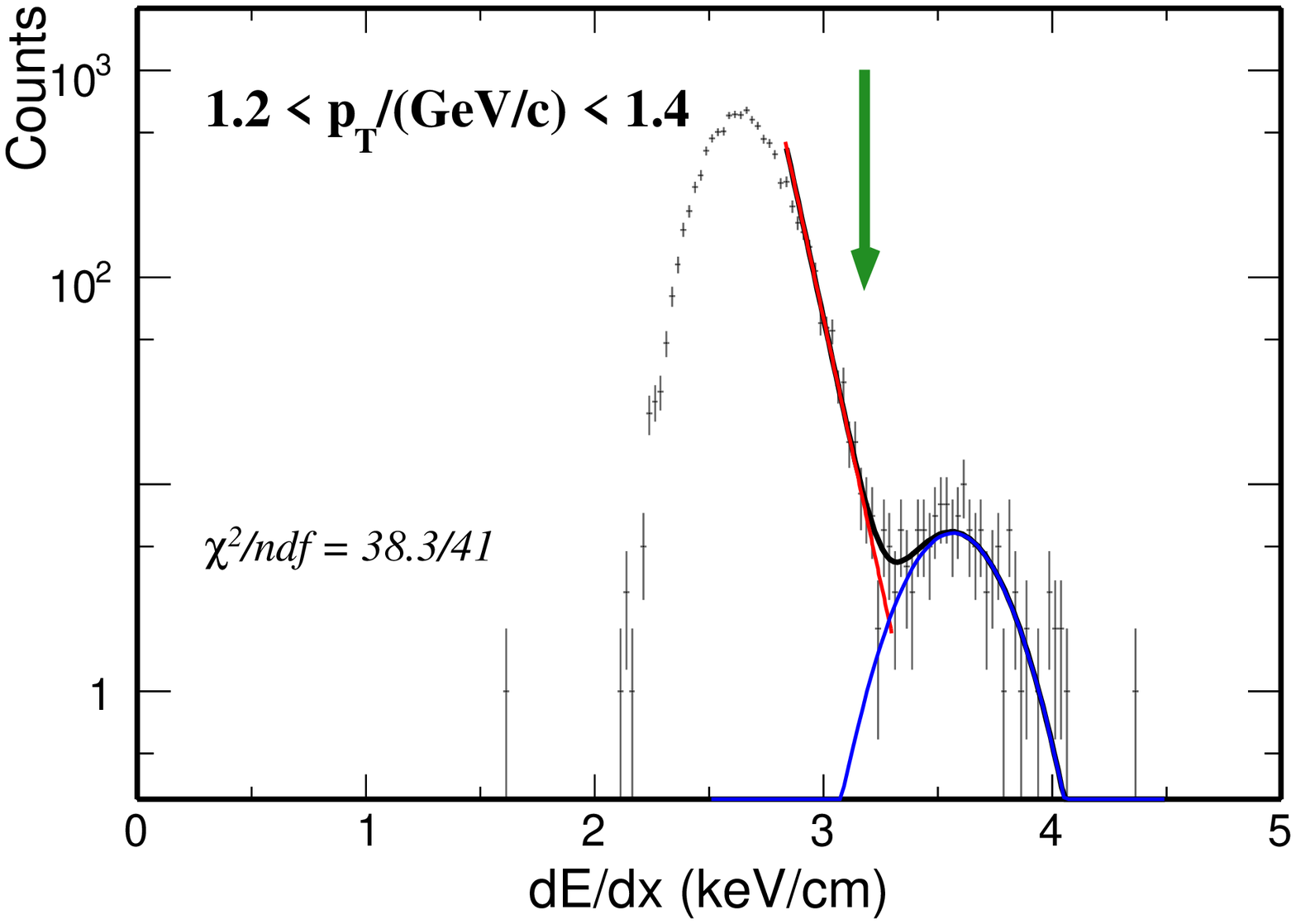}
\emn%
\bmn[b]{0.48\textwidth} \centering
\includegraphics[width=1.0\textwidth]{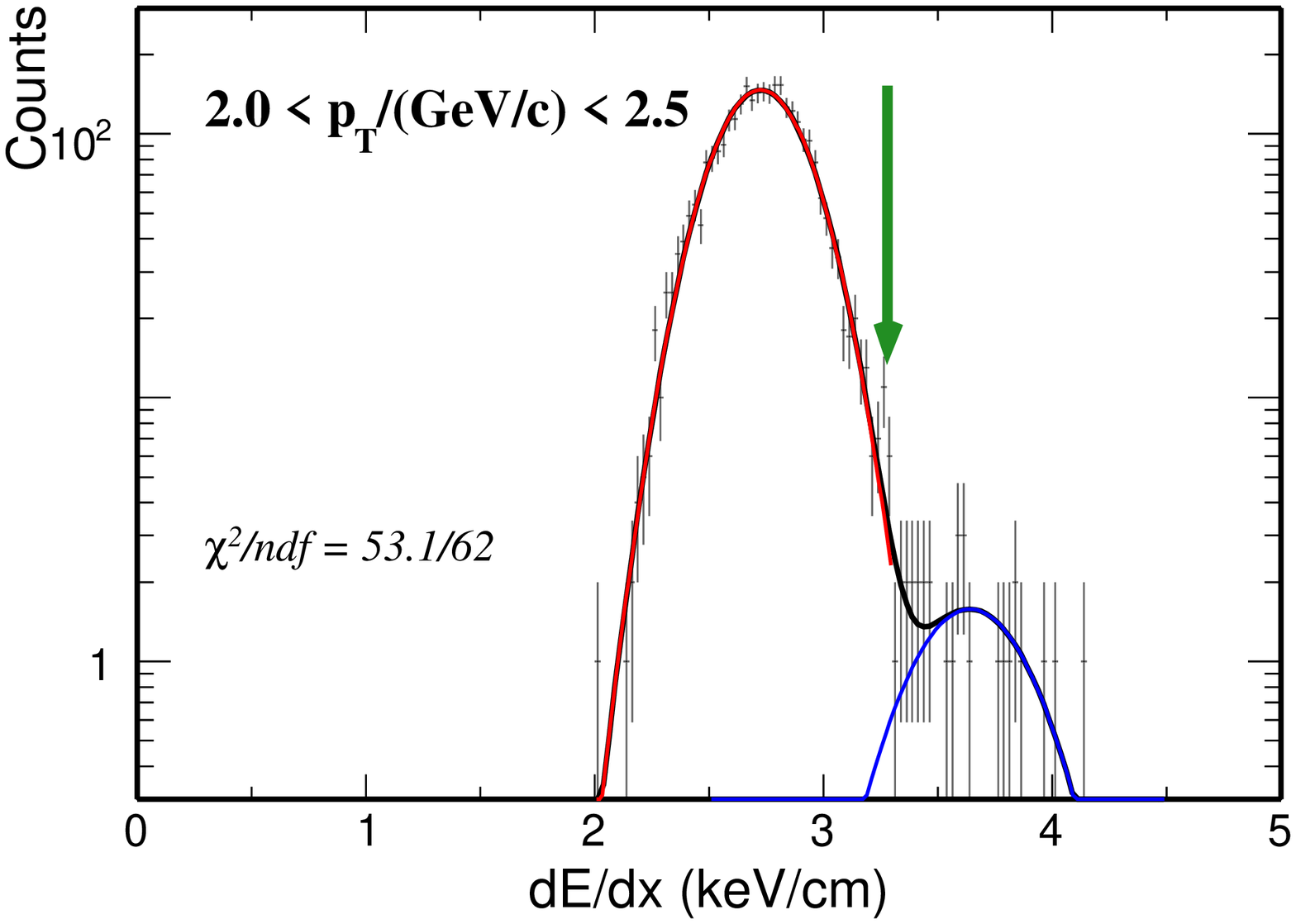}
\emn%
\caption[$dE/dx$ fit with TOF PID cut]
{$dE/dx$ projections in several \pT bins with the combination of
TOFr PID cut. The electron yields were extracted from the fit. The
green arrows denote the $dE/dx$ cut to select
electrons in each \pT bin.} \label{dEdxFitTof} \ef

At higher \pT ($2-4$ GeV/c), electrons could be identified
directly in the TPC since hadrons have lower $dE/dx$ due to the
relativistic rise of the $dE/dx$ for electrons. Positrons are more
difficult to be identified using $dE/dx$ alone because of the
large background from the deuteron band. Additional track quality
cuts were used in this selection: number of fit points is required
to be at least 25 and the number of $dE/dx$ points is required to
be at least 16. Fig.~\ref{dEdxLogFit} shows the logarithm of
$dE/dx$ distributions in each \pT bin. A 3-gaussian function
($e+\pi+K/p$) fit to the distribution was performed based on the
assumptions that all the particles have the same $dE/dx$
resolution and one gaussian function can describe kaons and
protons. The results are shown in Fig.~\ref{dEdxLogFit}. Electrons
with $dE/dx$ greater than the peak mean value were selected.

\bf \centering \bmn[b]{0.48\textwidth} \centering
\includegraphics[width=1.0\textwidth]{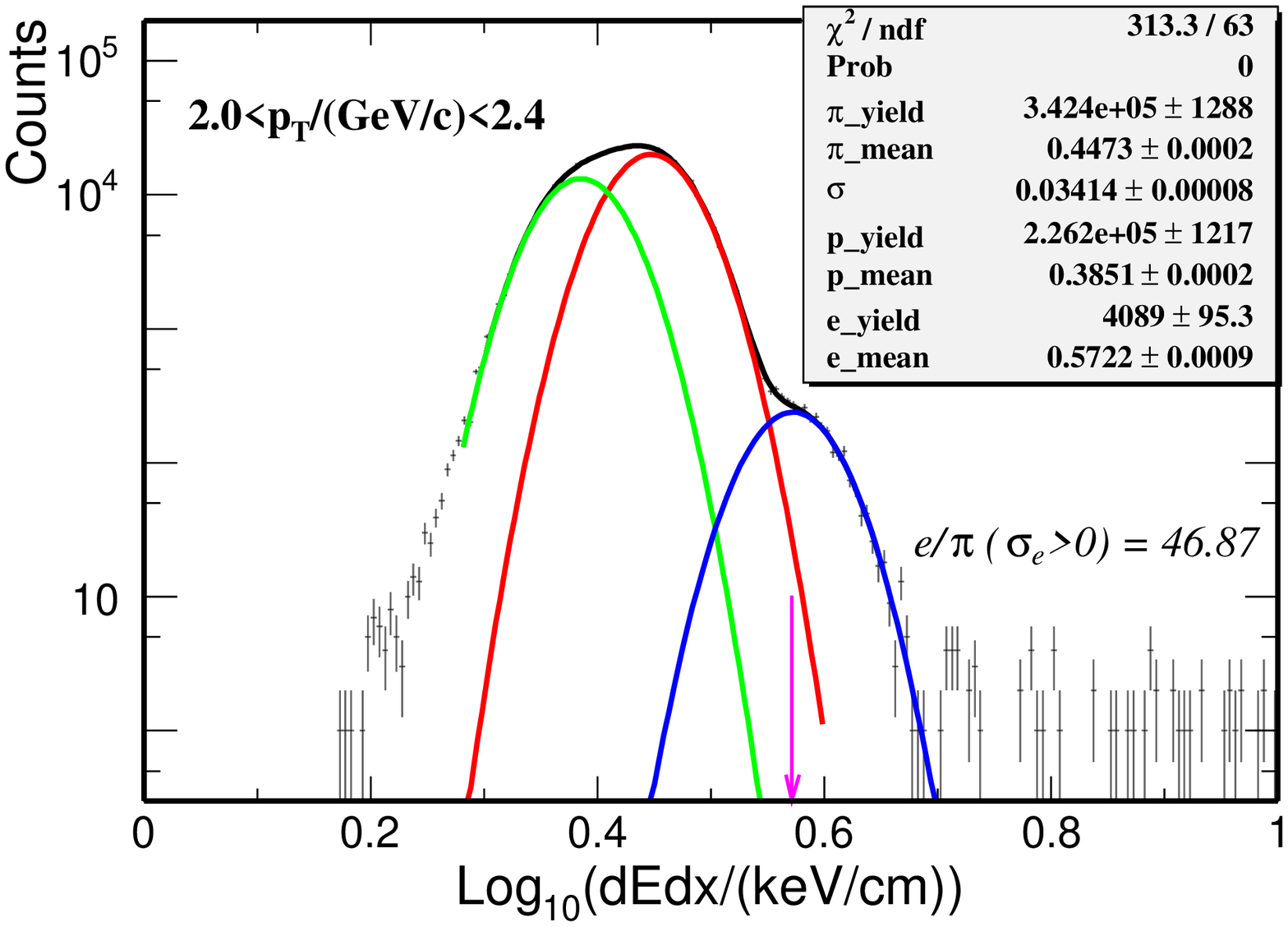}
\emn%
\bmn[b]{0.48\textwidth} \centering
\includegraphics[width=1.0\textwidth]{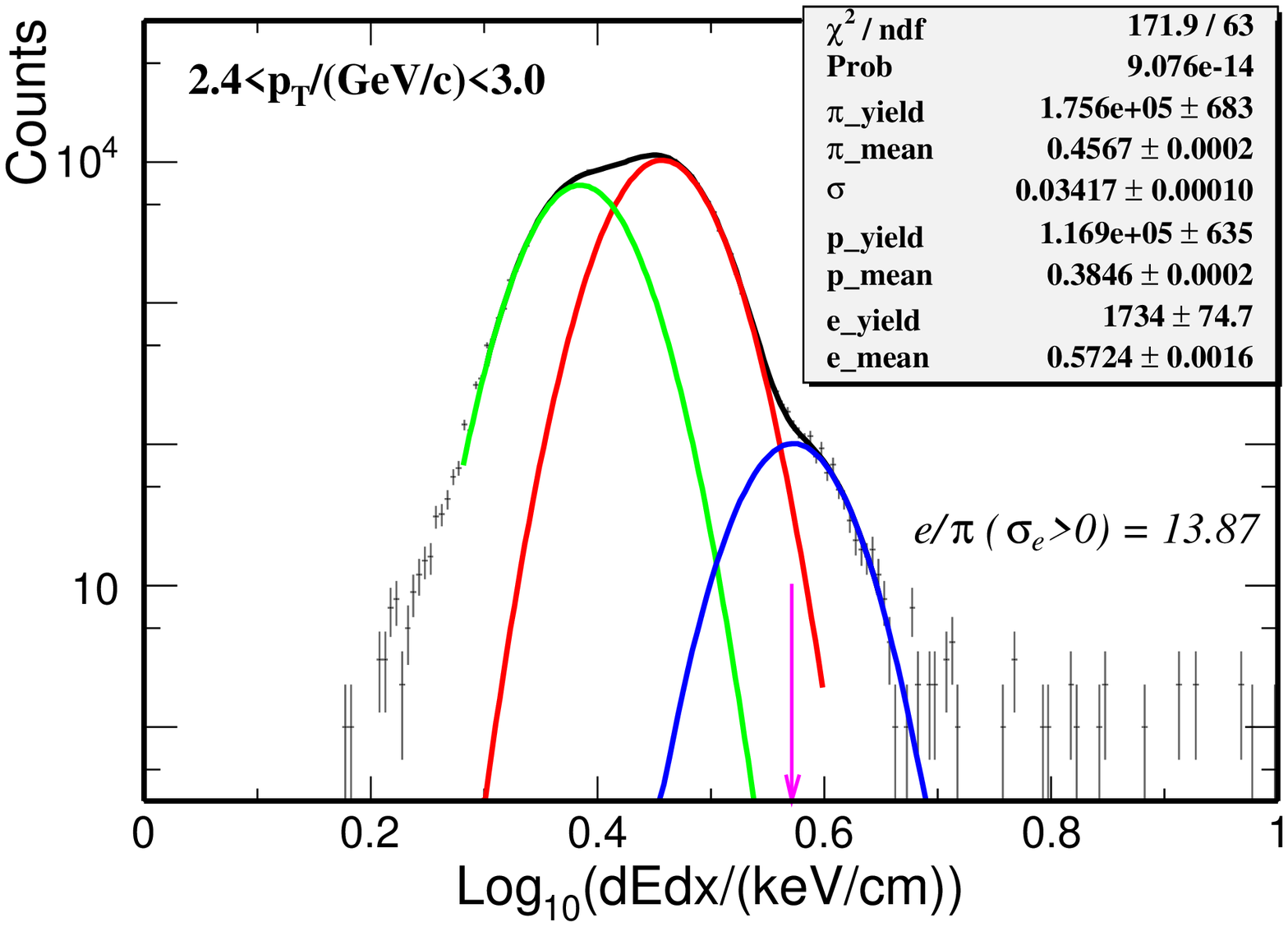}
\emn\\[10pt]
\centering \bmn[b]{0.48\textwidth} \centering
\includegraphics[width=1.0\textwidth]{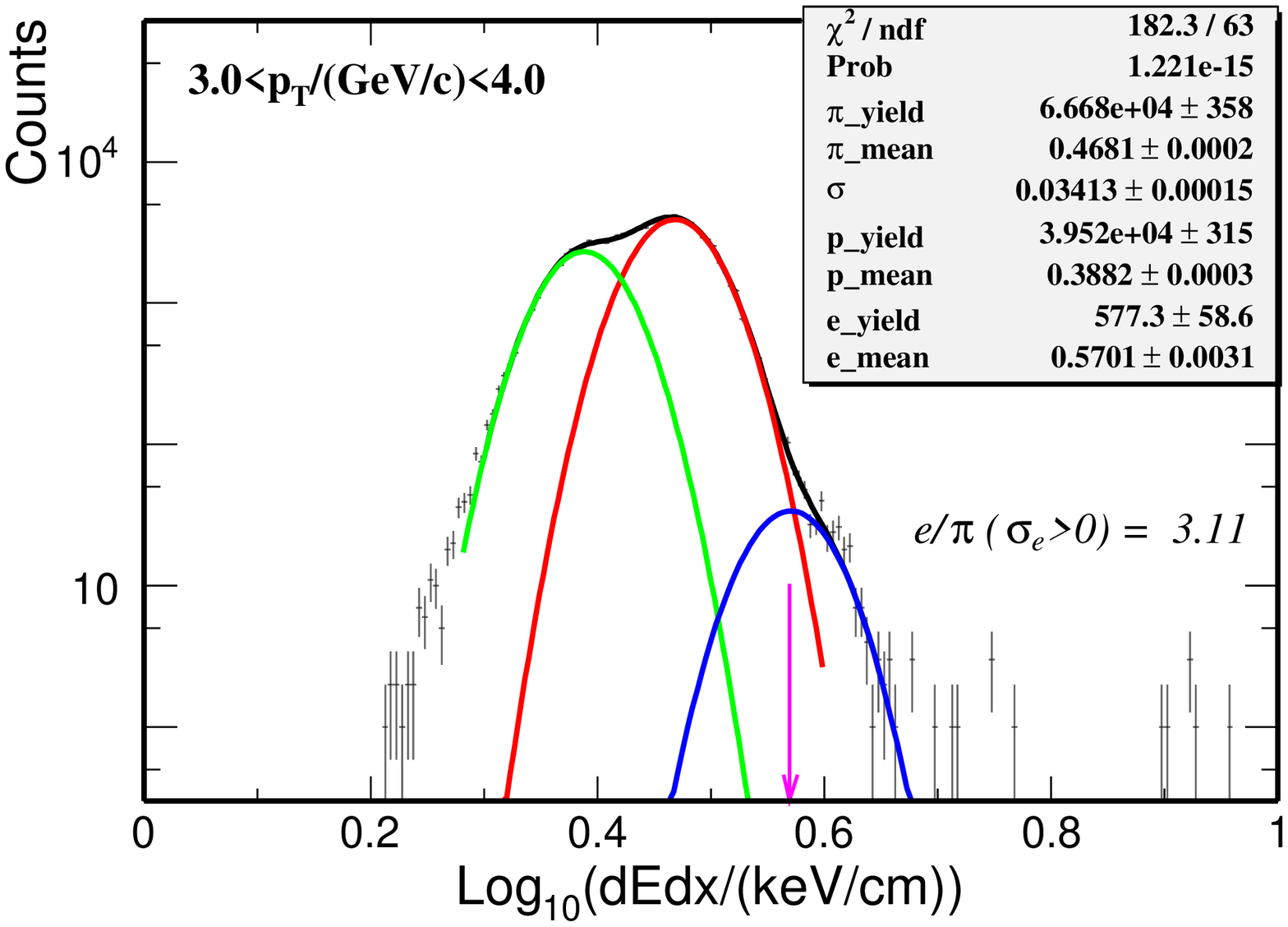}
\emn%
\caption[$dE/dx$ fit at higher \pT]
{Logarithm of $dE/dx$ distributions for
negative charged particles in three \pT bins between $2-4$ GeV/c.
A 3-Gaussian fit was used to extracted the electron yield and hadron
contamination. The pink arrows denote the electrons selection cut.}
\label{dEdxLogFit} \ef

Hadron contamination was extracted from the fits in these two
methods, respectively, shown in Fig.~\ref{hadronCom} for both \dAu
and \pp collisions. 
The hadron contamination not only need to
be subtracted statistically in the final yield, but also should be
considered carefully on the track-by-track analysis later on. This
will be discussed in detail in the next section.

\bf \centering\mbox{
\includegraphics[width=0.48\textwidth]{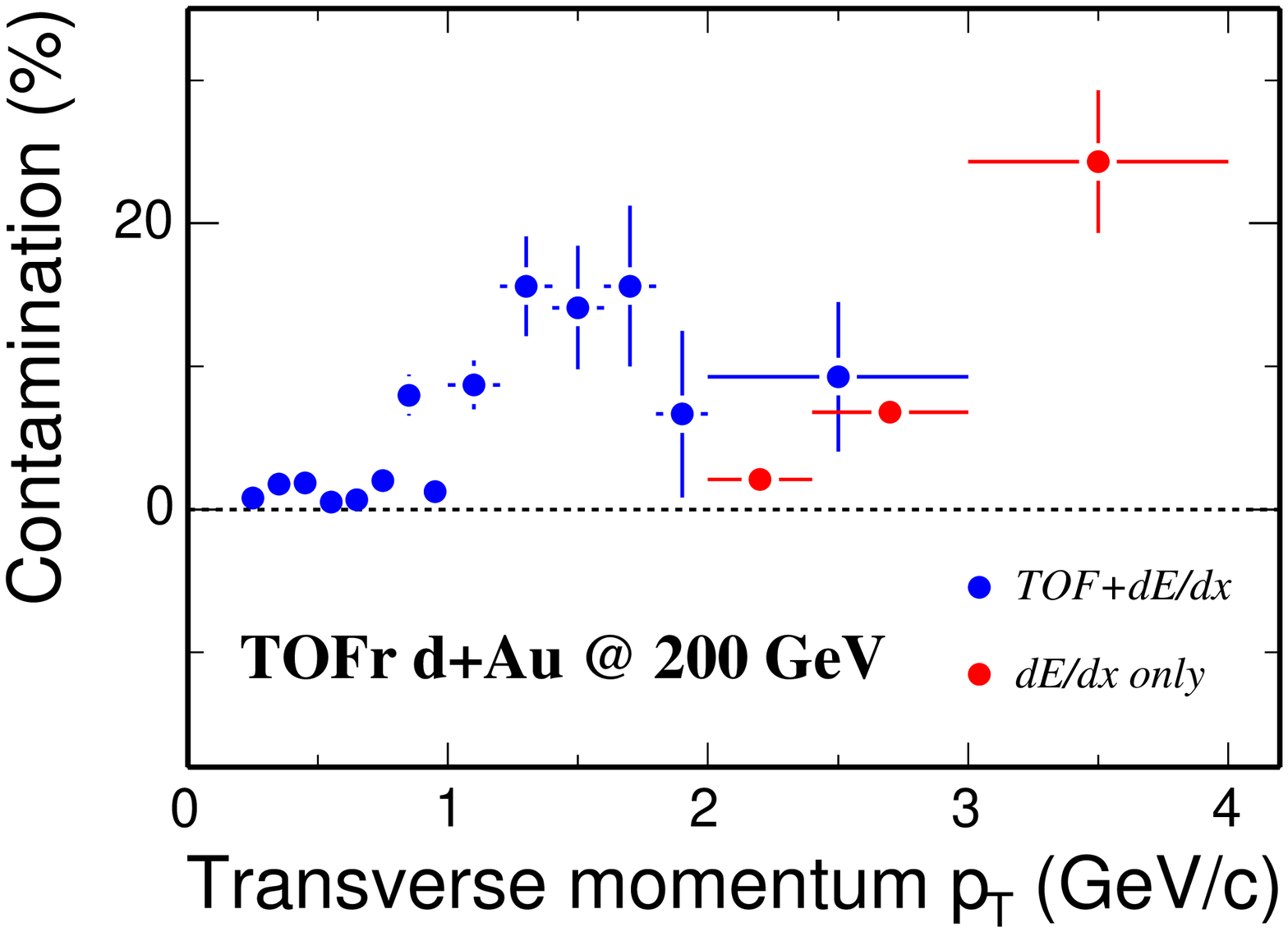}
\includegraphics[width=0.48\textwidth]{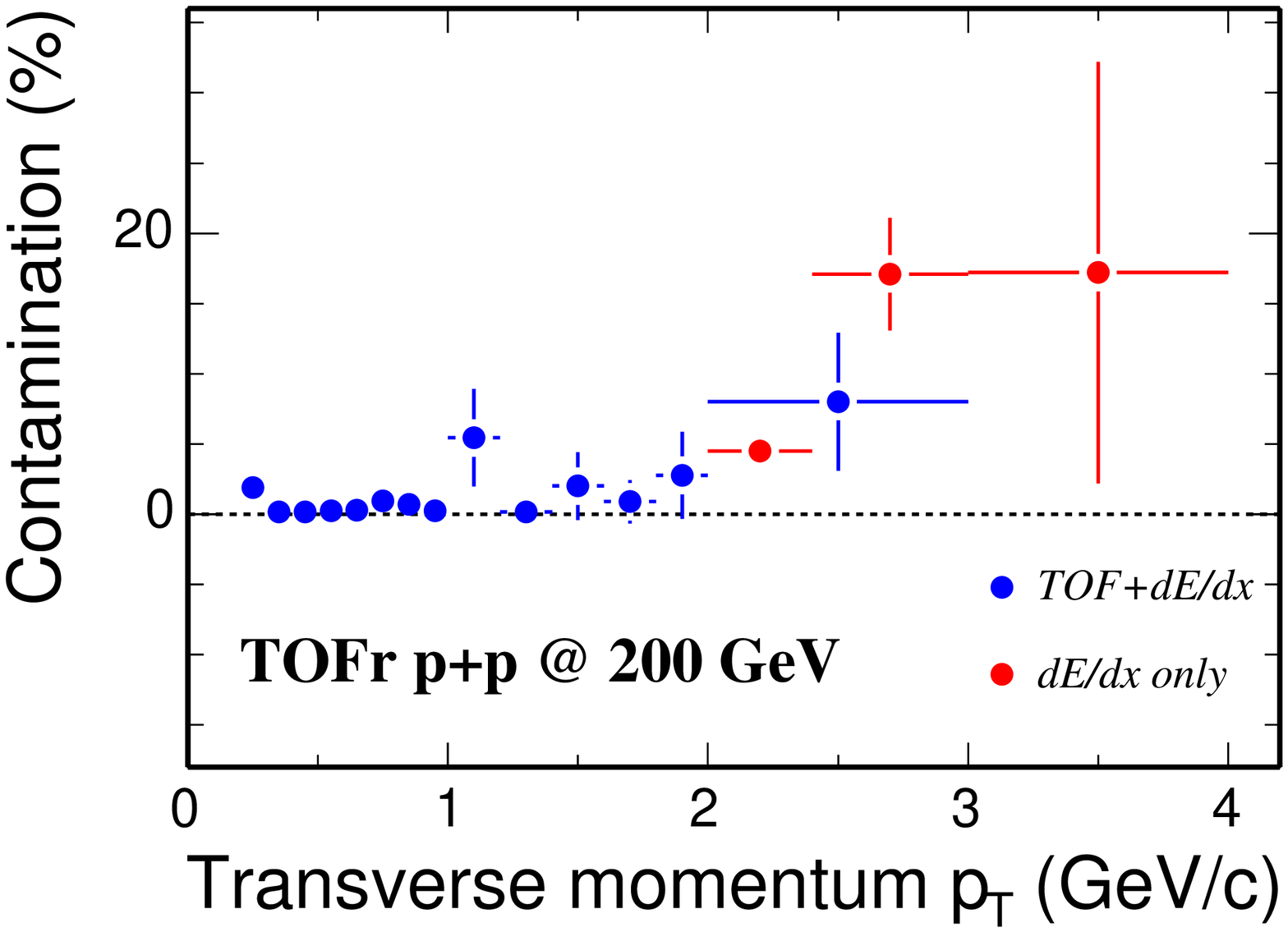}}
\caption[Hadron contamination] {Hadron contamination fractions as
a function of \pT from two electron PID methods in \dAu(left) and
\pp(right) collisions.} \label{hadronCom} \ef

Low hadron contamination and high electron efficiency are two
competiting aspects. We performed the cut scanning to optimize the
electron selection. Fig. ~\ref{effcon} shows the electron
efficiency vs. hadron contamination. Since we are limited for the
statistics currently, optimization results in the selection shown
in Eq.~\ref{tofCut} with $\sim90\%$ electron efficiency and $
\sim10\%$ hadron contamination. Fig.~\ref{ehratio} shows the \pT
dependent ratio of electrons to the total hadrons. It shows
$\sim1\%$ ratio at $p_T>1$ GeV/c. Combining the $\sim10\%$
contamination ratio under the selection shown in
Fig.~\ref{hadronCom}, the estimated hadron rejection power under
such selection is $\sim10^3$ at $p_T>1$ GeV/c and $\sim10^4-10^5$
at $p_T\sim 0.5$ GeV/c.

\bf \centering \bmn[c]{0.45\textwidth} \centering\mbox{
\includegraphics[width=1.0\textwidth]{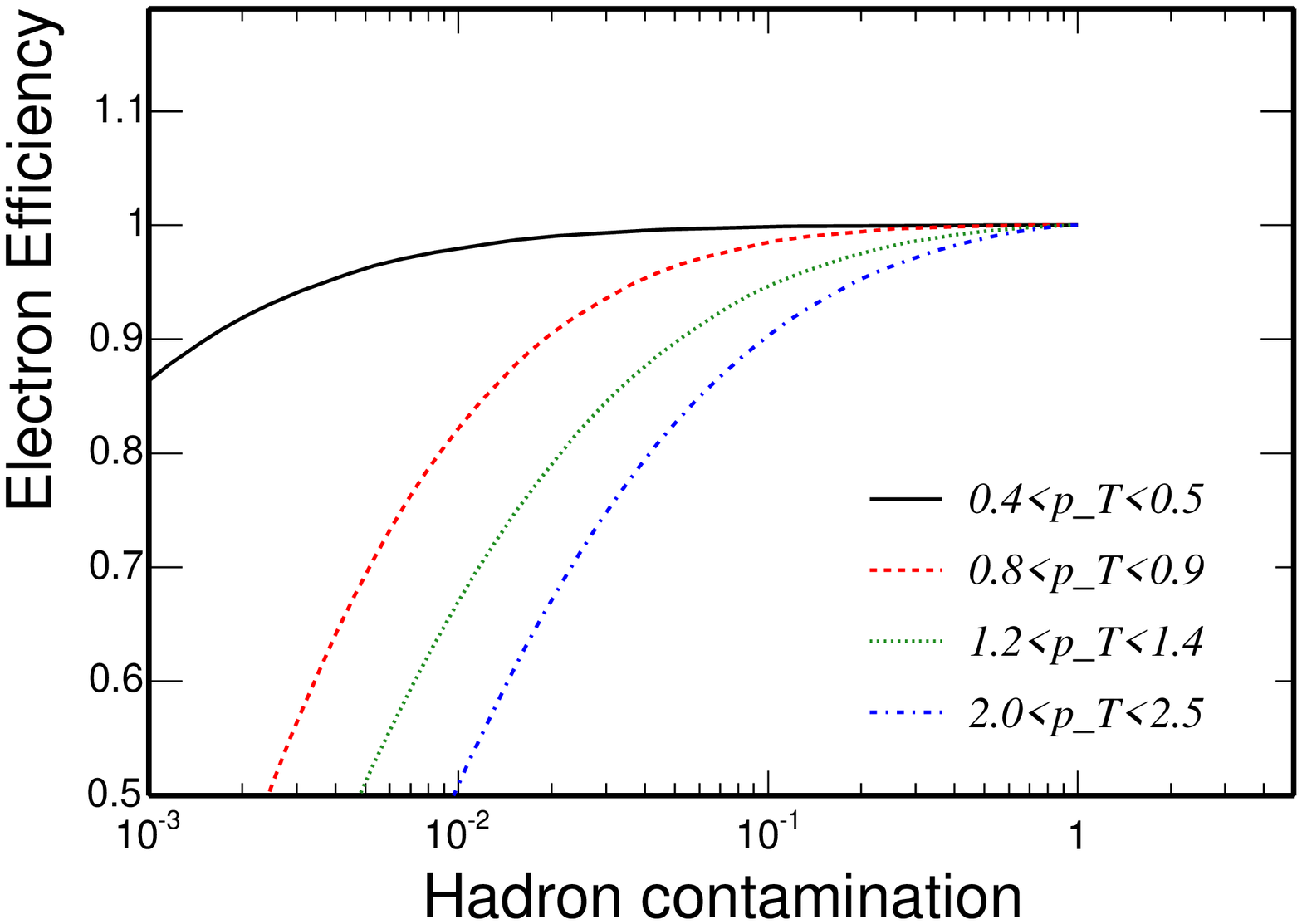}}
\caption[Efficiency vs. contamination] {Electron efficiency vs.
hadron contamination fraction by varying the $dE/dx$ cut to
selection electrons for 4 \pT bins.} \label{effcon} \emn \hskip
0.5 in \bmn[c]{0.45\textwidth} \centering\mbox{
\includegraphics[width=1.0\textwidth]{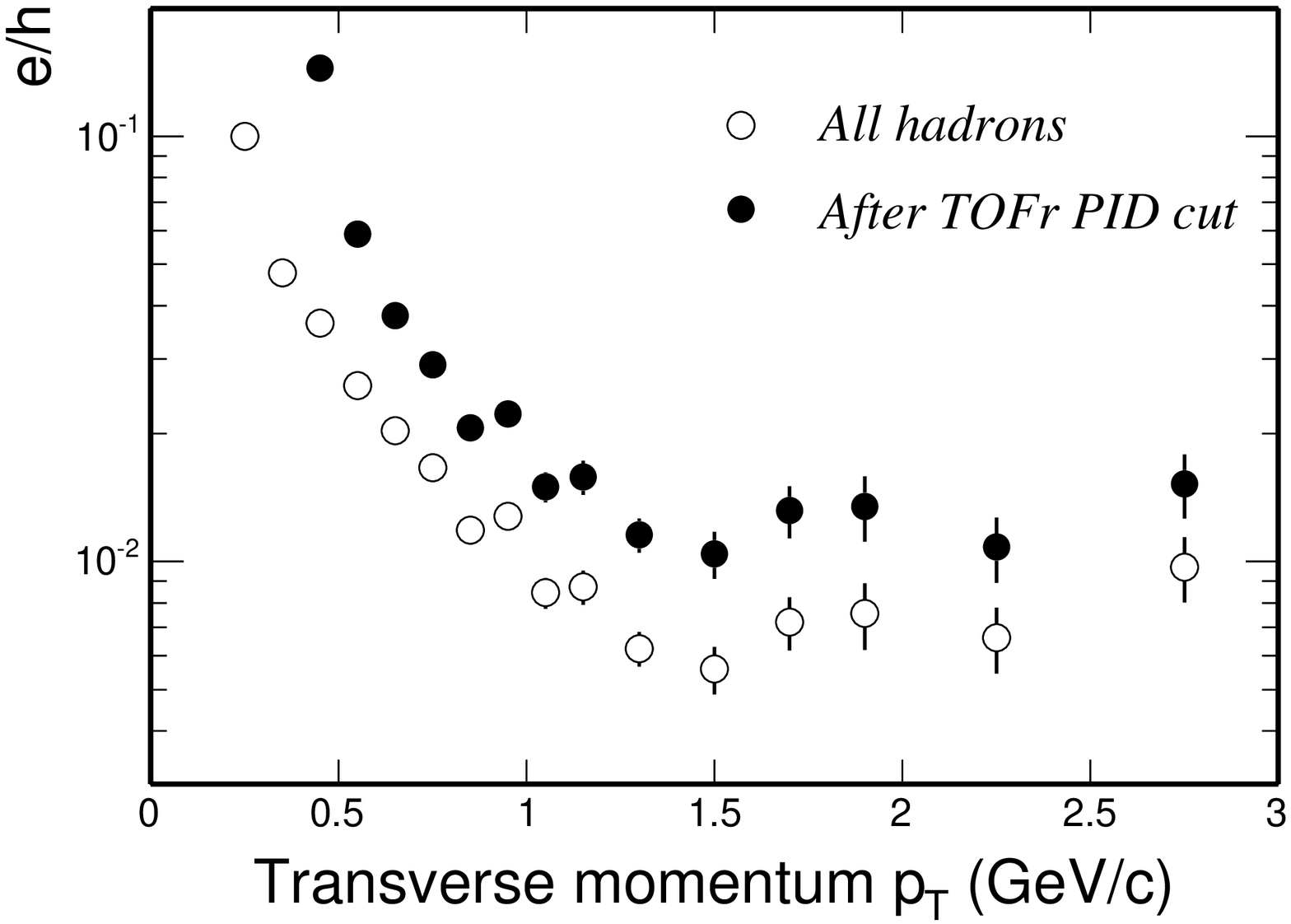}}
\caption[Electron to hadron ratio]
{The ratio of electron yield to total hadron yield.
By combining Fig.~\ref{hadronCom}, the hadron rejection
power under the current selection (Eq.~\ref{tofCut}) was estimated
to be $\sim10^3$ at $p_T>1$ GeV/c.} \label{ehratio}
\emn
\ef

Raw yields of inclusive electrons were then calculated from the
fits to $dE/dx$ distributions at each \pT bin.
Table.~\ref{electroncut} lists the event-wise and track-wise cuts
used in the electron selection.

\begin{table}[hbt]
\caption{Electron selection criteria}
\label{electroncut}\vskip 0.1 in
\centering\begin{tabular}{c|cc} \hline \hline
Method               & TOF+$dE/dx$       & $dE/dx$ \\ \hline
triggerword          & \multicolumn{2}{c}{2300 for \dAu}  \\
                     & \multicolumn{2}{c}{1300 for \pp} \\
$|VertexZ|<$         & 50 cm             & 30 cm \\ \hline
primary track ?      &   Yes             &  Yes \\
nFitPts $\geqslant$  &  15               &   25  \\
nFitPts/nMax $>$       &   0.52          &  N/A \\
ndEdxPts $\geqslant$ & N/A               &  16 \\
rapidity             & (-0.5, 0)         & (-0.5, 0) \\
$\beta$ from TOFr    & $|1/\beta-1|<0.03$ & N/A \\
TOFr hit quality     & $10<ADC<300$      & N/A \\
                     & $-2.7<z_{local}/cm<3.4$ & N/A \\
     \hline \hline
\end{tabular}
\end{table}

\subsection{Acceptance, Efficiency and Trigger bias}
Raw yields need to be corrected for the detector acceptance,
reconstruction and selection efficiency {\em etc.}, which were
done through {\em Monte Carlo} (MC) simulations. Full detector
simulations and/or embedding study show a $\sim90\%$
reconstruction efficiency for $e^+/e^-$ tracks, shown in
Fig.~\ref{efftpc}. As for the tracks hitting TOFr detector, there
are two ways to determine the acceptance/efficiency: the first one
is divided into two steps, TPC reconstruction efficiency and the
matching efficiency from TPC to TOFr from real data; the other one
is get the total efficiency (including acceptance) from embedding
data. Detailed matching efficiency study is referred to
~\cite{LijuanThesis}. Due to the lower statistics of electrons,
matching efficiencies of $\pi^{\pm}$ got from real data was used
for electrons and a correction factor ~90\% was used to account
for the loss (scattering and decay) of pions flying from TPC to
TOFr. In the later method, \dAu HIJING events with an additional
MC $e^+/e^-$ track within TOFr acceptance were embedded into real
zero-biased events, and the total efficiency was extracted from
the embedding data directly. Fig.~\ref{effcom} shows the total
efficiency (including acceptance) correction factor from two
methods in \dAu collisions. The efficiency correction in \pp
collisions was used as the same number as those from \dAu results
because of the similar multiplicities.

\bf
\centering
\bmn[c]{0.45\textwidth}
\centering\mbox{
\includegraphics[width=1.0\textwidth]{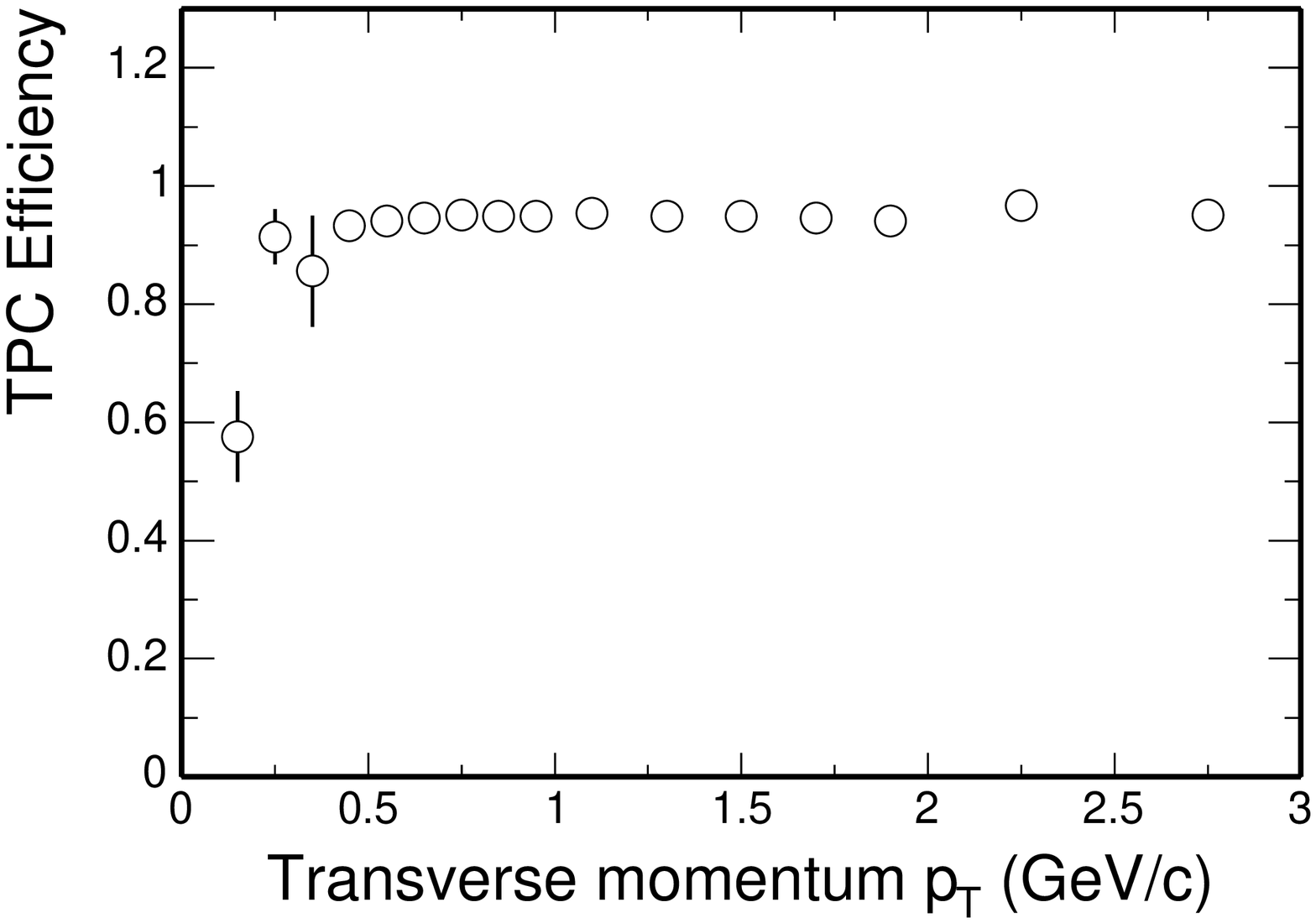}}
\caption[TPC efficiency for electrons]
{TPC track reconstruction efficiency for electrons from
embedding data.} \label{efftpc}
\emn
\hskip 0.5 in
\bmn[c]{0.45\textwidth}
\centering\mbox{
\includegraphics[width=1.0\textwidth]{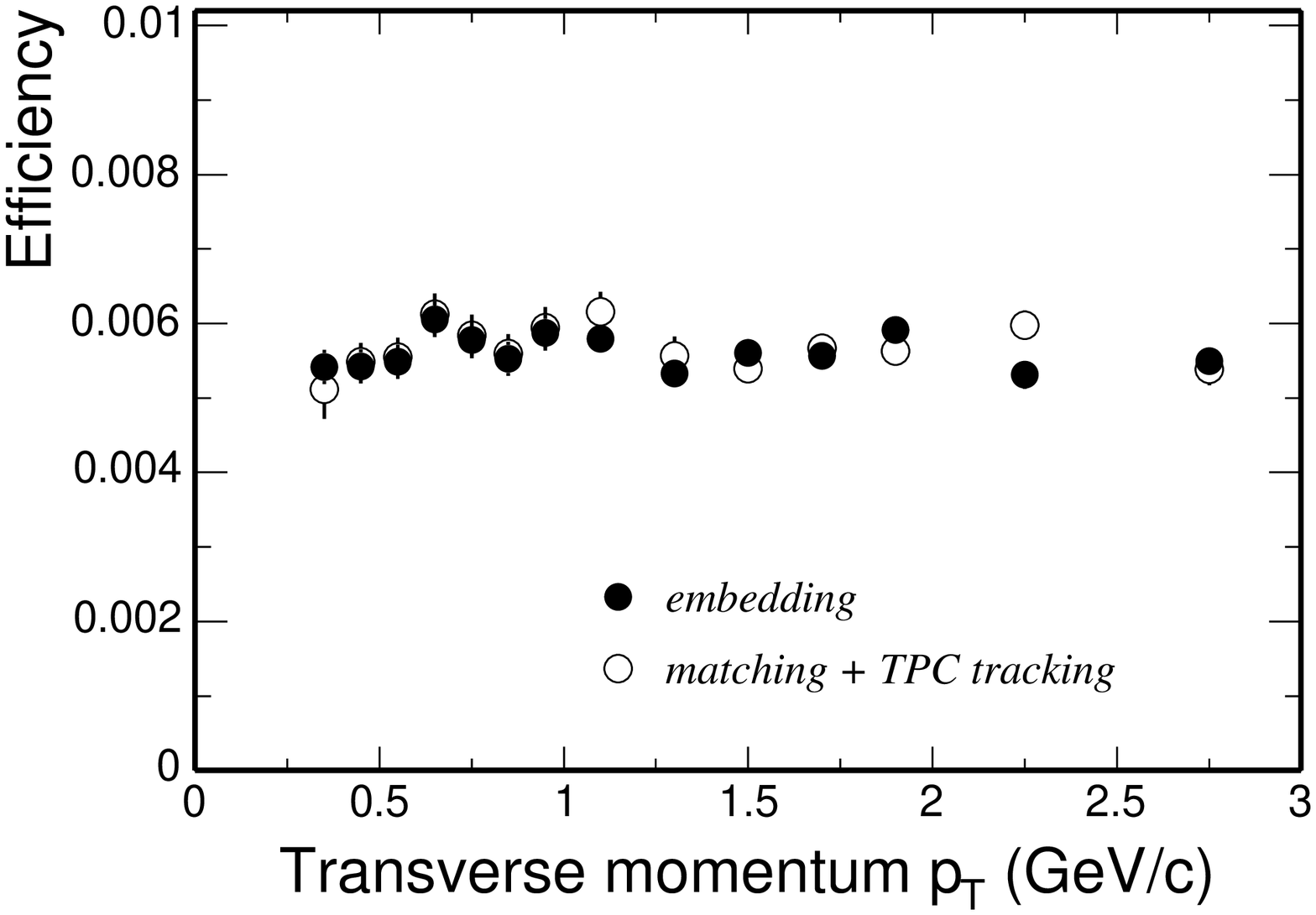}}
\caption[Total efficiency for electrons]
{Acceptance included efficiency for TOFr + $dE/dx$
electrons selection from two methods.} \label{effcom}
\emn
\ef

The additional TOFr PID cut $|1/\beta-1|<0.03$ used for slow
hadrons rejection will naturally lead to some loss of the
electrons. A simple fast simulation based on the TOFr timing
resolution (120 ps in \dAu and 160 ps in \pp), charged particle
phase space ($p_T$,$\eta$) distributions was used to determine
this correction. The result shows $\sim95\%$ efficiency in \dAu
and $\sim87\%$ efficiency in \pp under this cut and this
correction was applied in addition to the others above.

As for the TOF triggered data sets, the bias study of this trigger
is mandatory. Detailed investigation was introduced in
~\cite{LijuanThesis}. The results show no significant \pT
dependence for the spectrum at $p_T>$ 0.3 GeV/c according to the
TOFr trigger and the charged particle multiplicity bias correction
was done for both \dAu and \pp data sets.

To improve statistics, the $\eta$ cut $-0.5<\eta<0$ was removed.
This leads to a $\sim44\%$ increase for the TOFr acceptance and no
\pT dependence was found. Then the yields will be corrected to a
definite $\eta$ range for calculation according to this factor.
The final results from loosing $\eta$ cut are consistent within
statistical errors with those with $\eta$ cut applied. In the
following text, we will remove this cut in the discussion.

\subsection{Photonic background contribution}
There are several sources that can contribute to the final single
electron spectra in STAR environment:
\begin{itemize}
\item Photon (from $\pi^{0}$, $\eta$ {\em etc.}) conversions in inner detectors
\item $\pi^{0}$, $\eta$ {\em etc.} scalar meson Dalitz decay
\item $\rho$, $\omega$, $\phi$ vector meson di-electron decay and/or Dalitz decay
\item $K^{\pm}$ decay $-$ $K_{e3}$
\item heavy flavor ($c$, $b$) hadron semi-leptonic decay
\item other sources (Dell-Yan, heavy quarkonium decay, thermal electrons,
direct photon conversion {\em etc.})
\end{itemize}
In this analysis, the first four sources are considered to be the
background $-$ photonic background, which need to be subtracted from
the inclusive yield. As for the last item,
theoretical predictions~\cite{pythia,thermalE,directPhoton} show
no significant contributions with large uncertainties. This part of
contribution is neglected in this analysis. The heavy flavor hadron decay
is considered to be the only signal.

From the previous measurement~\cite{phenix130e} and the estimation
from the STAR environment, electrons from photon conversion and
$\pi^{0}$ Dalitz decay dominate the total yield, especially at low
\pT. At \pT $>2$ GeV/c, the signal may become visible due to the
different shape between the total spectrum and background
spectrum. In PHENIX collaboration single electron analysis, they
used {\em cocktail} method to subtract each contribution one by
one~\cite{phenix130e}. However, the \pT distribution and the total
yield of each source particles were used as assumed input in their
analysis, which may introduce large uncertainties. In STAR, due to
the large acceptance of TPC, background from photon conversion
{\em etc.} may be reconstructed experimentally. The pioneering
analysis using topological method were done in
~\cite{IanPRC,IanThesis}. In the analysis we report here, since we
have already tagged one electron/positron track, we don't want to
use the whole topological reconstruction. Instead, we try to use
the kinematical features of photon conversion and $\pi^0$ Dalitz
processes to justify whether the tagged electron/positron is from
photonic background or not. Firstly, use TOFr or $dE/dx$ at high
\pT to tag one electron/ positron track from the primary collision
vertex, then loop all other opposite-sign global tracks, whose
helixes have a {\em distance of closest approach} ($dca$) to the
tagged tracks' less than 3 cm, to find the other partner track in
the TPC. Because the TPC acceptance is large, the pair
reconstruction efficiency is reasonable.

To study the kinematics of the photon conversion and $\pi^0$
Dalitz decay processes, simulations of HIJING~\cite{hijing} for
\dAu system and PYTHIA~\cite{pythia} for \pp system plus full
detector description GEANT were investigated in detail. The data
sets used are 1.38 M \dAu HIJING (v1.35 + GCALOR) minimum bias
events and 1.2 M \pp PYTHIA (v6.203 + GCALOR) events. The tagged
electrons were required with the same cut as those for data, but
all good TPC tracks were selected to increase the statistics. Fig.
~\ref{invmassangle} shows the invariant mass square and opening
angle distribution of electron pairs from different sources.

\bf
\centering\mbox{
\includegraphics[width=0.48\textwidth]{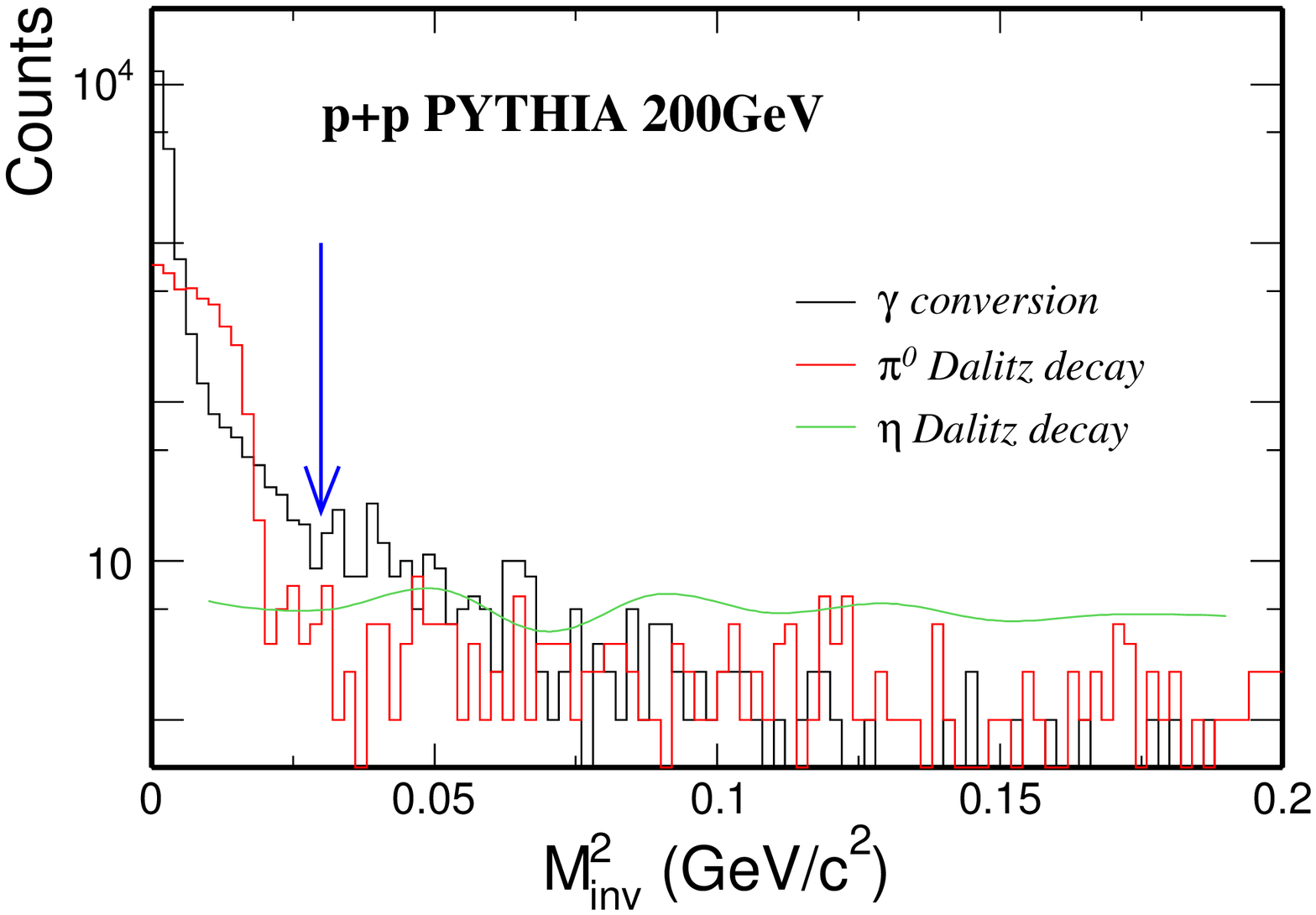}
\includegraphics[width=0.48\textwidth]{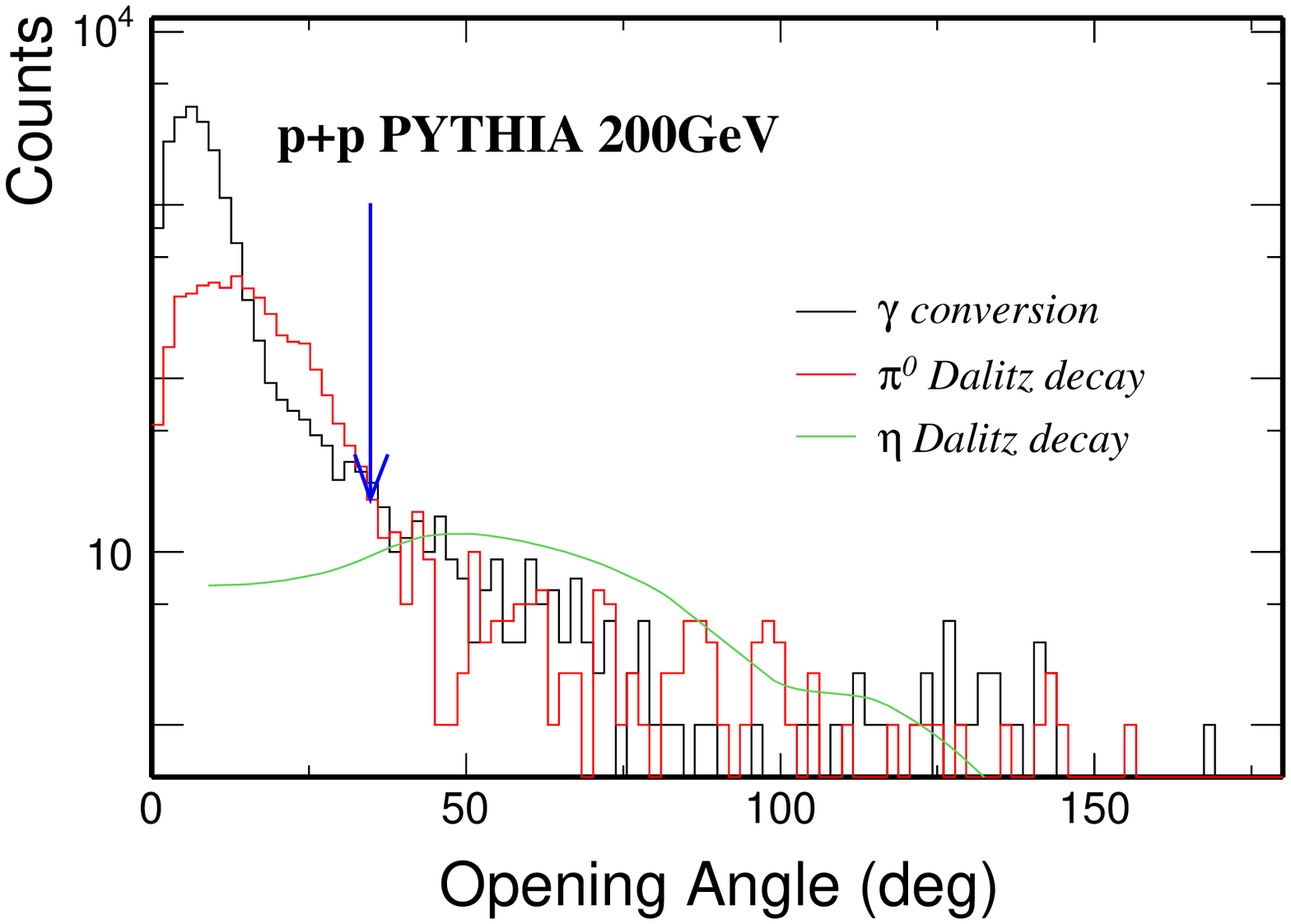}}
\caption[$M_{e^+e^-}^2$ and opening angle distributions]
{Invariant mass square and opening angle distributions of electron
pairs from several sources in simulation. The blue arrows depict
the cut thresholds in this analysis. } \label{invmassangle} \ef

As expected, the results show those background mostly has small
invariant mass square and/or small opening angle. This offers us a
good opportunity to estimate the single electron background with
almost topology-blind method. Those tagged electrons/positrons
with the invariant mass square or opening angle below certain
thresholds will be considered to be background from photon
conversion and/or $\pi^0$ Dalitz decay, and be rejected in the
analysis.

To understand the background reconstruction capability in our
environment, simulation study were done on the same data sets.
Fig.~\ref{bkgdeff} shows the reconstruction efficiency of photon
conversion and $\pi^0$ Dalitz decay in the TPC. The simulation
shows an almost constant efficiency $\sim60\%$ at \pT above 1
GeV/c. From the $\pi^0$ embedding data, in which MC $\pi^0$ tracks
through full detector response simulation were embedded into real
zero-bias events , the similar result within errors was also
obtained.

\bf
\centering
\bmn[c]{0.45\textwidth}
\centering\mbox{
\includegraphics[width=1.0\textwidth]{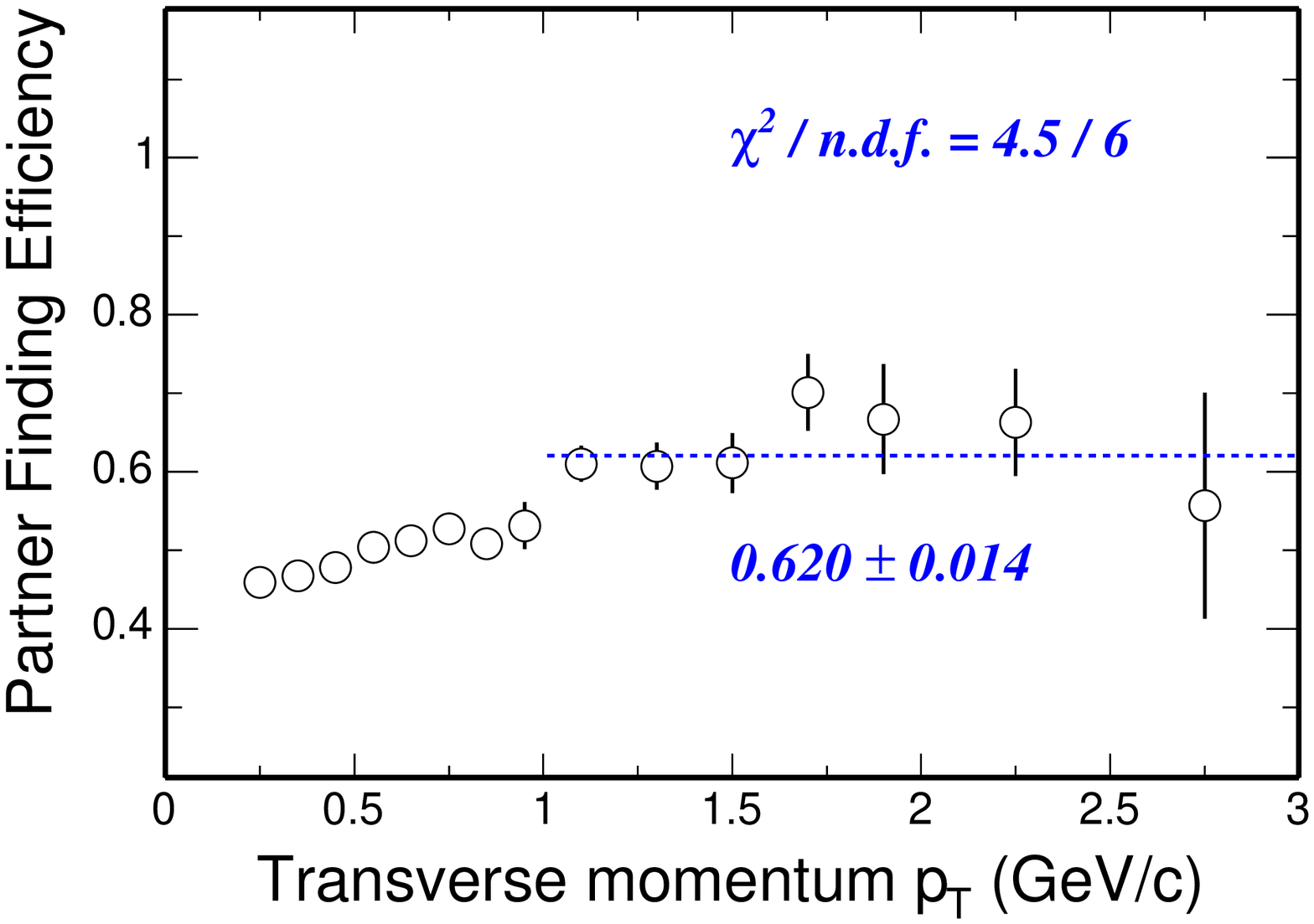}}
\caption[Conversion efficiency]
{Photon conversion and $\pi^0$ Dalitz decay background reconstruction
efficiency from simulations.} \label{bkgdeff}
\emn
\hskip 0.5 in
\bmn[c]{0.45\textwidth}
\centering\mbox{
\includegraphics[width=1.0\textwidth]{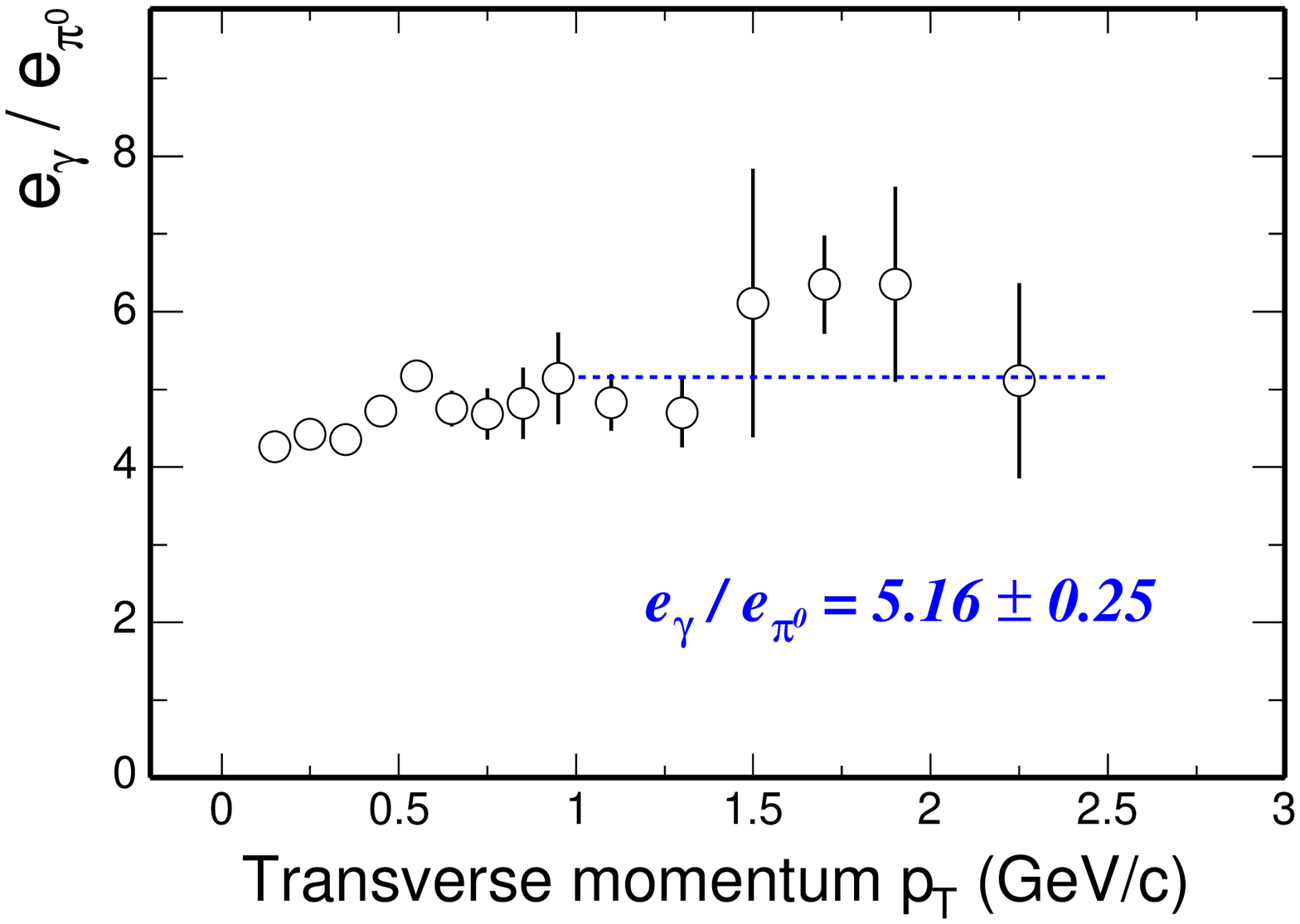}}
\caption[Ratio of conversion to Dalitz]
{The ratio of single electrons from photon conversion to those from
$\pi^0$ Dalitz decay from simulations.} \label{gamma2pi0}
\emn \ef

By applying this efficiency to the spectrum after the kinematic
cuts, one can extract the estimated total background from the main
sources: photon conversion and $\pi^0$ Dalitz decay. Other
contributions as well as the fractions of those could not be
extracted from data directly, and we determined them from
simulations. Fig.~\ref{gamma2pi0} shows the ratio of electrons
from photon conversion to those from $\pi^0$ Dalitz decay. The
electron spectrum from $\pi^0$ Dalitz was then used as a reference
for those contributions from other light meson decays.
Fig.~\ref{bkgdDis} shows the spectra of electrons from different
light hadron decays and the contribution of each source w.r.t
$\pi^0$ contribution from \dAu HIJING simulations. From this , we
found that the total contribution of photon conversion and $\pi^0$
Dalitz decay constitutes $\sim95\%$, which were measured
experimentally. Table~\ref{bkgdfrac} shows the fractions of each
electron background sources to the total background, averaging
over $1<p_T$/(GeV/c) $<3$. And the upper panel of
Fig.~\ref{electronBkgd} shows the inclusive electron spectra and
the photonic background contributions in \dAu and \pp collisions.
Clear excess of electrons at \pT $>1$ GeV/c is shown and the
bottom panels of that plot shows the ratio of inclusive electrons
to the total background electrons. The grey bands depict a ~20\%
systematic uncertainty of the total background.

\bf
\centering\mbox {
\includegraphics[width=0.60\textwidth]{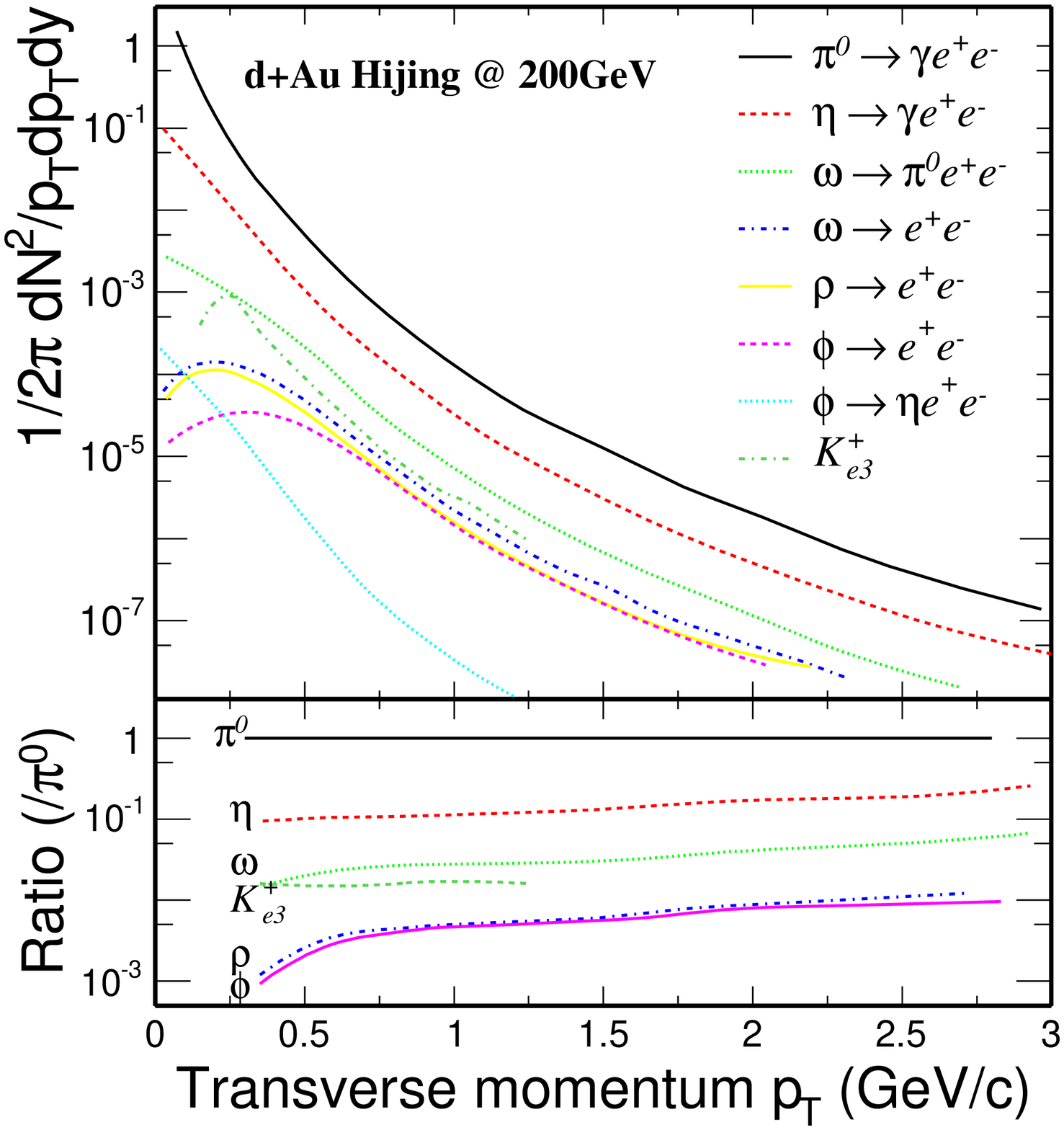}}
\caption[Other background fractions] {The single electron spectra
from various source contributions from simulations. The bottom
panel shows the relative fractions to $\pi^0$ Dalitz decay, which
were used in the estimation of other source contributions in data
analysis.} \label{bkgdDis} \ef

\bf
\centering\mbox {
\includegraphics[width=0.86\textwidth]{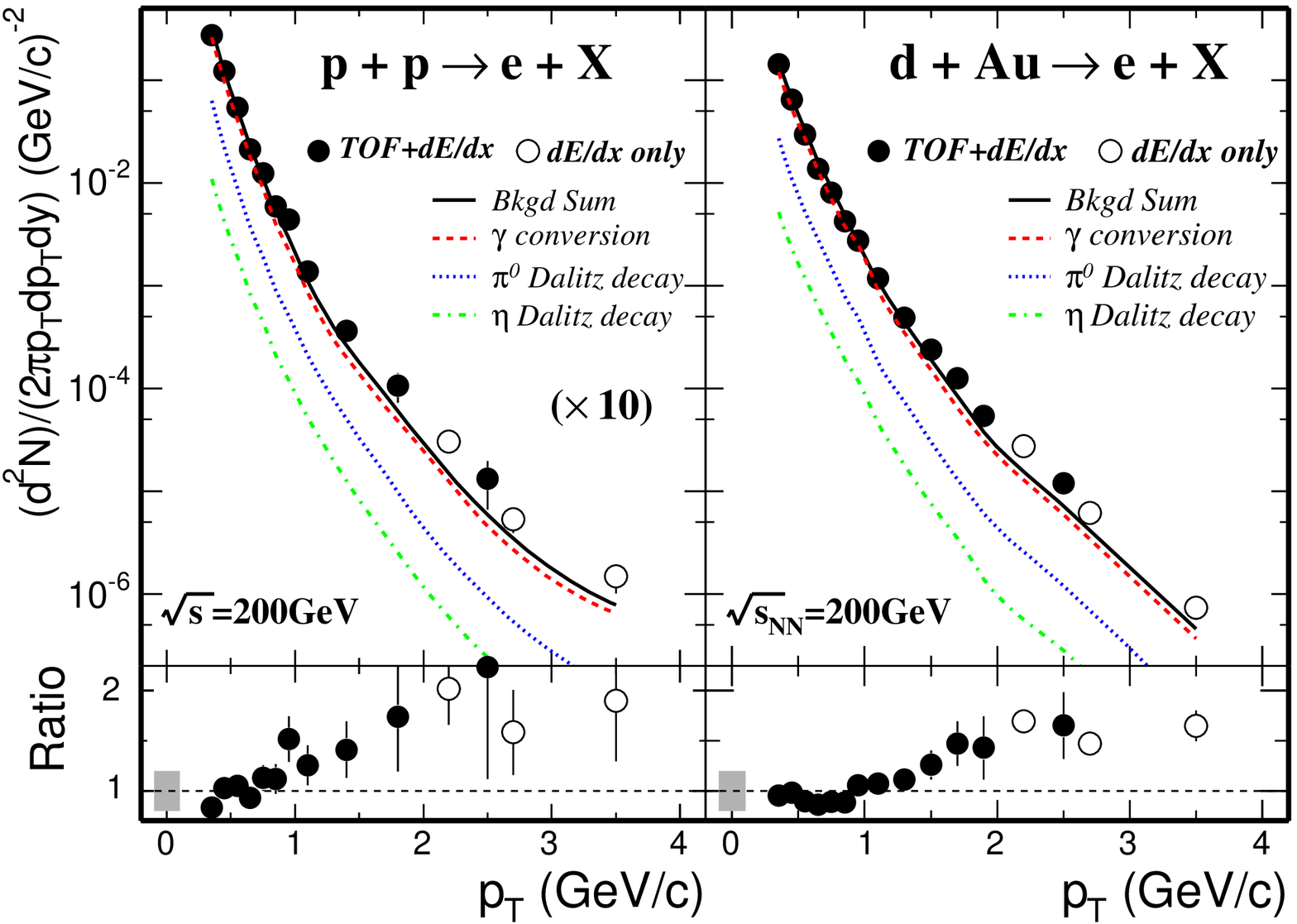}}
\caption[Electron spectra]{Upper panels: Electron distributions
from p+p (left) and d+Au (right) collisions. Solid and open
symbols depict electrons/positrons (($e^{+}$+$e^{-}$)/2)
identified via a combination of TOFr and $dE/dx$, and electrons
($e^{-}$) identified via $dE/dx$ alone. The measured total
photonic backgrounds are shown as solid lines. Dashed lines depict
the various contributing sources. The fractions were derived from
simulations. Bottom panels: the ratio of inclusive electrons to
the total backgrounds. The gray band represents the systematic
uncertainty in each panel.} \label{electronBkgd} \ef

\begin{table}[htbp]
\caption[Fractions of background]
{Photonic sources contributions to single electron spectrum}
\label{bkgdfrac}\vskip 0.1 in
\centering\begin{tabular}{c|c} \hline \hline
Source & fraction (\%) \\
\hline $\gamma$ conversion & 80 $\pm$ 7 \\
       $\pi^0$ Dalitz & 15 $\pm$ 2 \\
       $\eta$ & 3.2 \\
       $\omega$ & 0.99 \\
       $\phi$ & 0.22 \\
       $\rho$ & 0.19 \\
       $K_{e3}$ & 0.21 \\
     \hline \hline
\end{tabular}
\end{table}

\subsection{Signal extraction}
One simple way to extract the signal so far is to just subtract
the estimated total background from the total inclusive electron
spectrum. In practice, to reduce the statistically subtracting
errors, we subtracted the background from the left spectrum after
the kinematical cuts applied. Besides the real background one can
reject under such cuts, some signals may also be cut out due to
the random combination with the hadron tracks (mostly pions),
which need to be compensated after subtracting those background
from photon conversion and $\pi^0$ Dalitz decay. Hadrons
contaminating in the electron selection will be considered as
following the random rejection power behavior under the above
kinematical cuts. Fig.~\ref{randomRejection} shows the random
rejection fractions as a function of \pT from the MC studies.

\bf
\centering\mbox{
\includegraphics[width=0.6\textwidth]{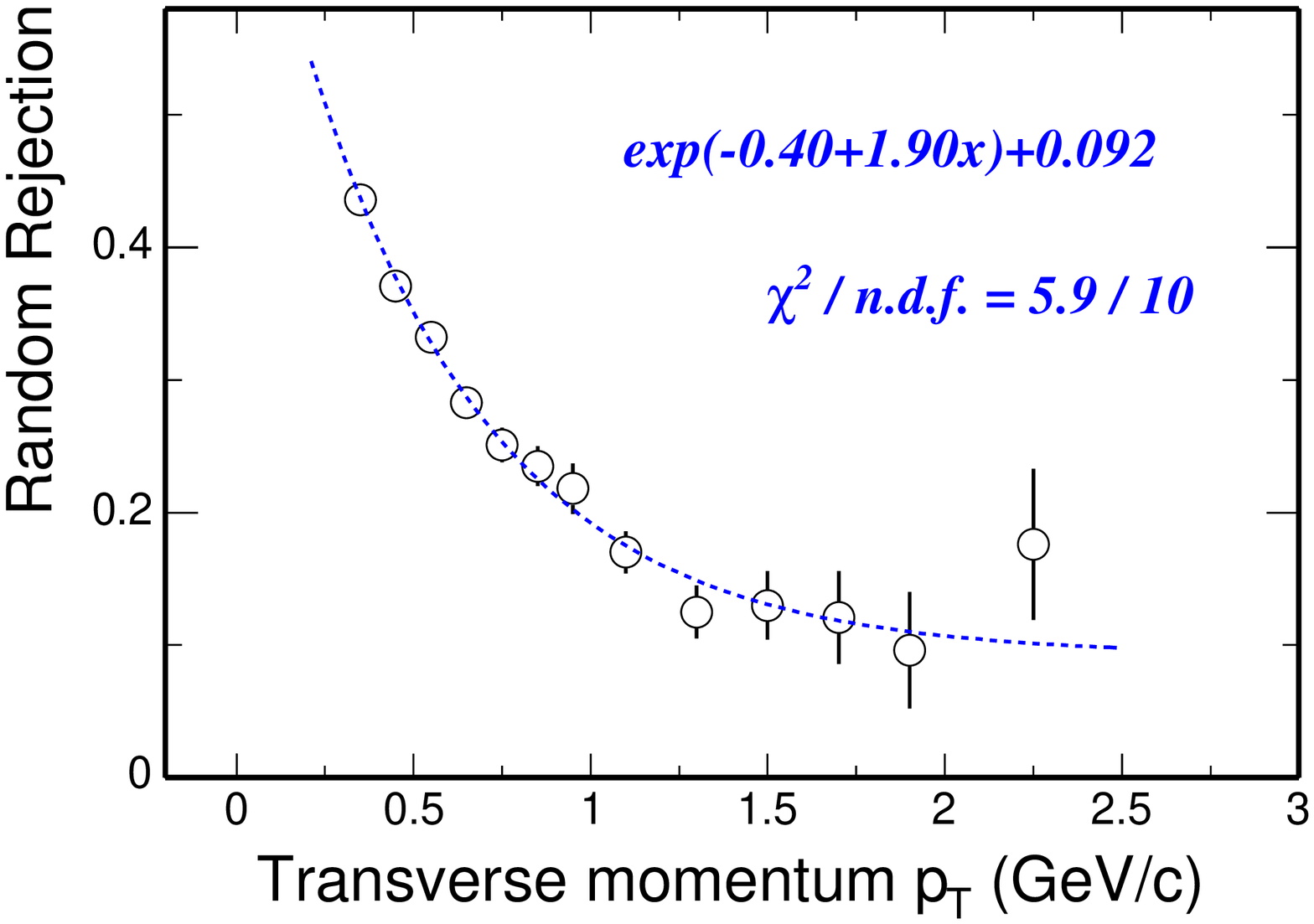}}
\caption[Random rejection fractions]{The electron fraction
rejected by random particles as a function of \pT.}
\label{randomRejection} \ef

The signal fraction was then extracted from the following equations:
\begin{subequations}\label{grp}
\begin{align}
eff\times B + ran\times ( S + H ) &= cut \label{first}\\
( 1 - eff )\times B + ( 1 - ran )\times ( S + H ) &= sur \label{second}\\
( S + B ) / H &= r \label{third}
\end{align}
\end{subequations}

where $B$, $S$, $H$ represent background electrons, signal
electrons and contaminating hadrons, respectively. $eff$ denotes
the efficiency of the background reconstruction, and $ran$ denotes
the random rejection efficiency for signals and hadrons. $cut$ and
$sur$ represent the yields being cut out and surviving cuts in
data, respectively. The electrons to hadrons ratio (related to
hadron contaminations) $r$ has already been obtained from previous
fit to $dE/dx$ distributions. From \eqref{grp}, we have \be S =
\frac{( eff + ran/r )\times sur - ( 1 - eff + ( 1 - ran )/r
)\times cut} {( eff - ran )\times ( 1 + 1/r )} \label{singaleq}
\ee when hadron contamination is negligible ( $r\gg1$ ), the above
formula can be written as \be S = \frac{ eff\times sur - ( 1 - eff
)\times cut }{ eff - ran } \ee Signal yields were extracted in
each \pT bin.

The bin-by-bin systematic errors are dominated by the
uncertainties of the background reconstruction efficiency from MC
simulations and the hadron contamination. The errors were then
calculated by varying the efficiency to $1\sigma$ uncertainty away
and changing the $dE/dx$ cut in the electron selection for
different hadron contamination levels. Table~\ref{datapoints} list
the non-photonic electron spectra data points and the errors.

\begin{table}[hbt]
\caption[Non-photonic electron spectra]{Non-photonic electron
spectra, the numbers represent yields $\pm$ statistical errors
$\pm$ systematical errors.} \label{datapoints}\vskip 0.1 in
\centering\begin{tabular}{c|c|c|c} \hline \hline
  &          & $p_T$ (GeV/c) &  data points\\ \hline \hline
  &          & $1.0-1.2$ & $(1.30 \pm 0.71 \pm 0.83) \times 10^{-4}$ \\
  &          & $1.2-1.4$ & $(5.52 \pm 3.66 \pm 2.41) \times 10^{-5}$ \\
  & TOFr +   & $1.4-1.6$ & $(6.42 \pm 2.43 \pm 3.40) \times 10^{-5}$ \\
\dAu  & $dE/dx$  & $1.6-1.8$ & $(3.88 \pm 1.61 \pm 0.88) \times 10^{-5}$ \\
minbias  &          & $1.8-2.0$ & $(1.55 \pm 0.85 \pm 0.26) \times 10^{-5}$ \\
  &          & $2.0-3.0$ & $(4.82 \pm 1.99 \pm 0.47) \times 10^{-6}$ \\ \cline{2-4}
  &          & $2.0-2.4$ & $(7.53 \pm 1.12 \pm 2.41) \times 10^{-6}$ \\
  & $dE/dx$  & $2.4-3.0$ & $(1.19 \pm 0.41 \pm 0.36) \times 10^{-6}$ \\
  &          & $3.0-4.0$ & $(2.45 \pm 1.22 \pm 1.72) \times 10^{-7}$ \\ \hline \hline

  &          & $1.0-1.2$ & $(2.23 \pm 1.36 \pm 1.43) \times 10^{-5}$ \\
  & TOFr +   & $1.2-1.6$ & $(7.02 \pm 4.51 \pm 3.30) \times 10^{-6}$ \\
  & $dE/dx$  & $1.6-2.0$ & $(3.59 \pm 2.05 \pm 0.79) \times 10^{-6}$ \\
\pp  &          & $2.0-3.0$ & $(7.16 \pm 4.57 \pm 0.70) \times
10^{-7}$ \\ \cline{2-4}
NSD  &          & $2.0-2.4$ & $(1.10 \pm 0.30 \pm 0.36) \times 10^{-6}$ \\
  & $dE/dx$  & $2.4-3.0$ & $(1.89 \pm 0.85 \pm 0.68) \times 10^{-7}$ \\
  &          & $3.0-4.0$ & $(4.11 \pm 3.02 \pm 1.68) \times 10^{-8}$ \\ \hline \hline

\end{tabular}
\end{table}

\subsection{Background subtraction check}
The photon conversions from the inner detectors are dominated by
the inner SVT detector and its supporting materials. These
materials are mainly within $|z|<\sim50$ cm ~\cite{IanThesis}. As
a double check the background subtraction, we made the vertex z
position distributions of electron tracks normalized to the event
vertex z position distributions, shown as solid symbols in
Fig.~\ref{VzCheck} upper panel. The "M" structure reveals the
conversion material positions. The open symbols in the same panel
shows the same distributions for the cut out electrons, which are
considered as conversion background. Both distributions show
similar structure. If we take into account the background
efficiency estimated from the discussion above, and subtract them
from the total distribution, the result is depicted in the bottom
panel in that figure. We cannot see significant material effect.
The flat distribution demonstrates that the background subtracting
method is reasonable.

\bf
\centering\mbox{
\includegraphics[width=0.6\textwidth]{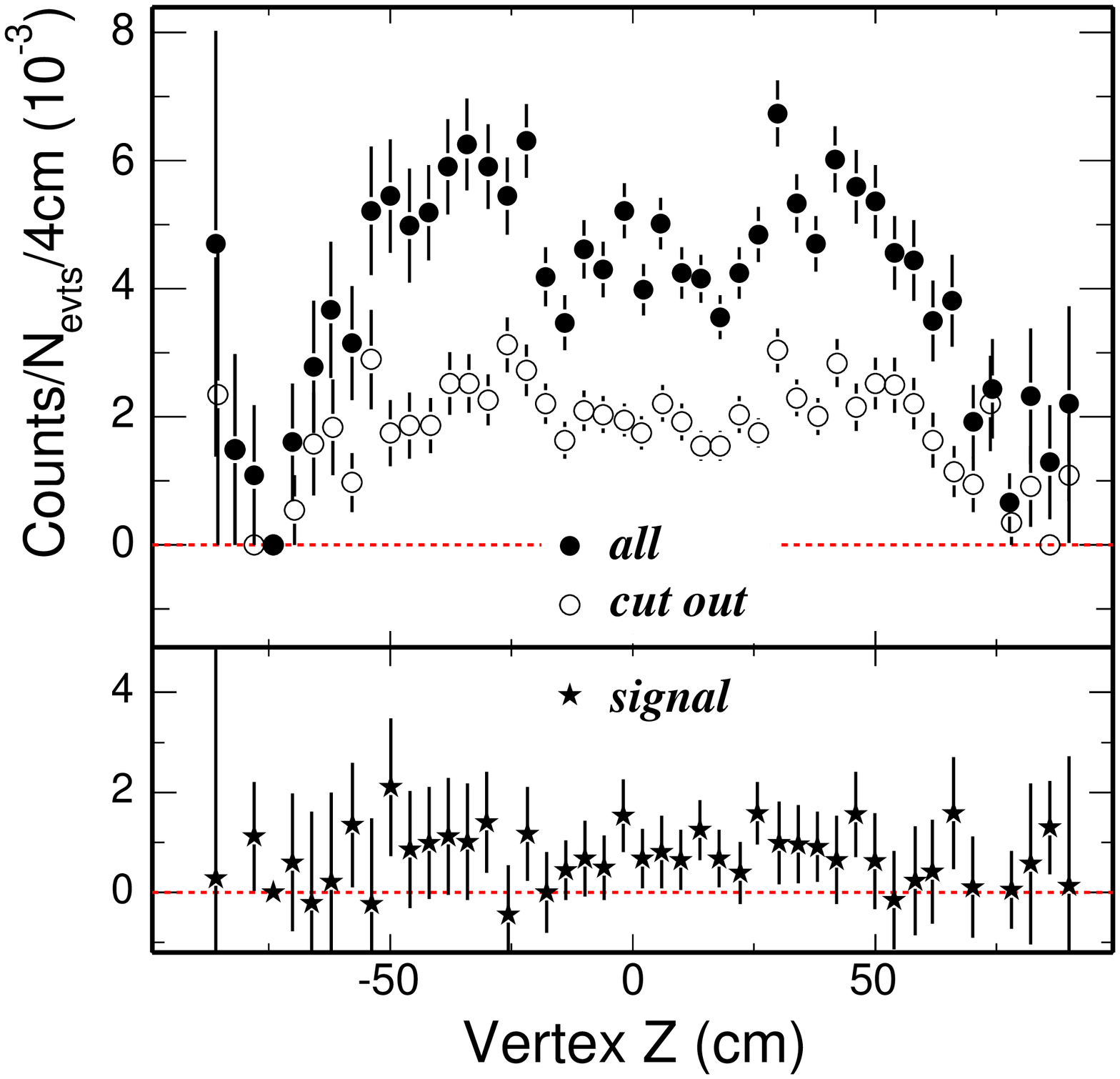}}
\caption[Vertex Z check]{Normalized vertex z position
distributions of electron tracks for total, cut-out background and
expected signal.} \label{VzCheck} \ef

Another important check is the electron/positron symmetry.
Fig.~\ref{ePlusMinus} shows the non-photonic yield of electrons
and positron separately. Within statistical errors, they are
consistent with each other. This indicates the possible $\delta$
electrons , which are knocked out from atoms in material by
charged particles, does not contribute significantly in the final
spectrum.

\bf
\centering\mbox{
\includegraphics[width=0.6\textwidth]{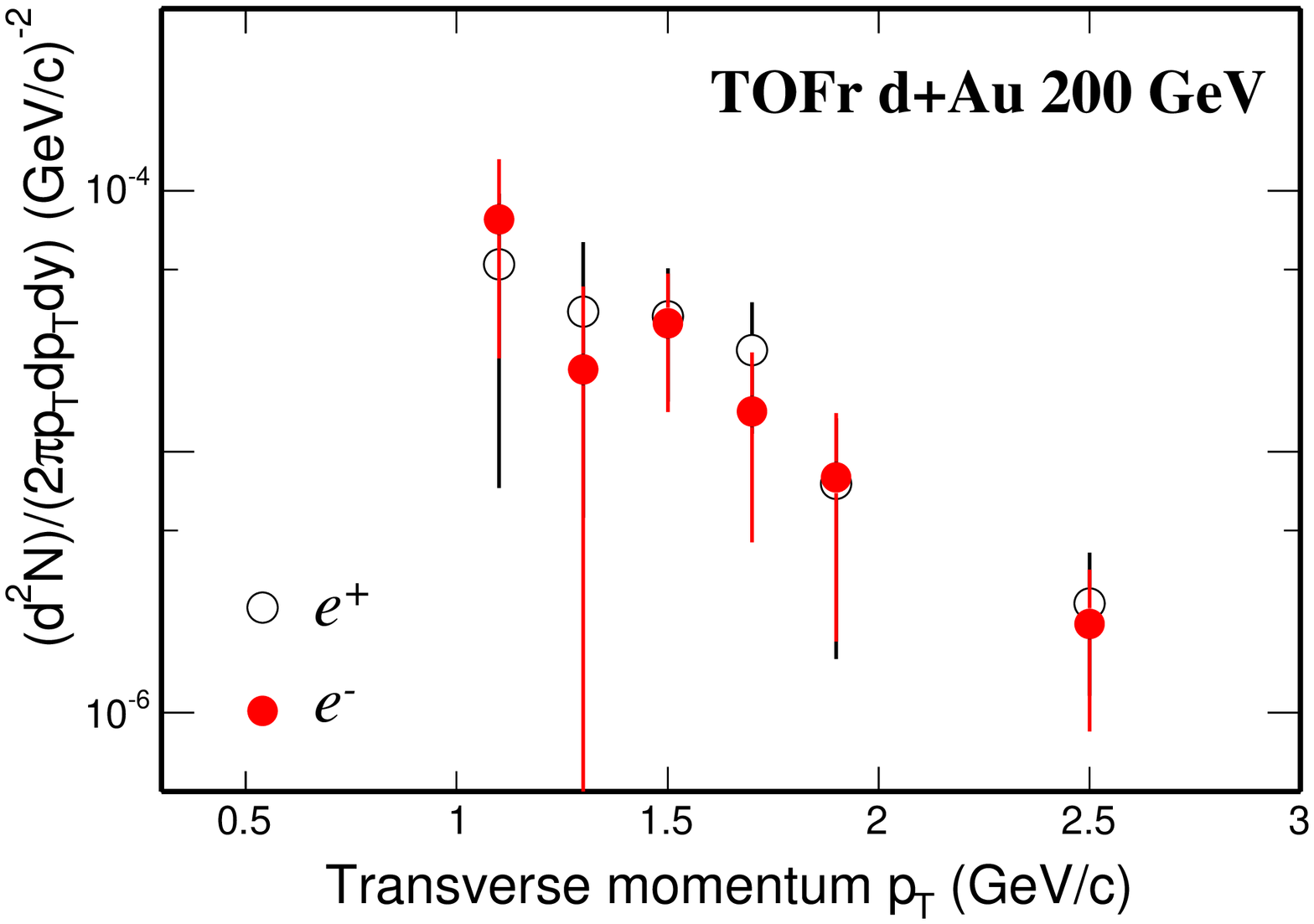}}
\caption[Separate $e^+$ and $e^-$ spectra]{Non-photonic electrons
and positrons spectra from \dAu collisions.} \label{ePlusMinus}
\ef

\subsection{Electron $dN/dy$ and implications for total
charm production cross section}
Many previous measurements tried to extract the charm production
cross section from electron spectra only~\cite{UA2,phenix130e}.
However, this method is strongly model dependent for the charm
hadron \pT spectrum is unknown and as well as the fractions
between different charm hadrons. In this analysis, the electron
spectrum only covers $1-4$ GeV/c, which is corresponding to $\sim
2-6$ GeV/c in the charm hadron spectrum. The direct extraction
from electron spectrum will have large systematic uncertainties.

On the other hand, STAR collaboration has measured charm mesons
through hadronic decay channels in \dAu
collisions~\cite{LijuanQM04,AnQM04}. The low \pT
$D^{0}$($\overline{D^0}$) (we use $D^{0}$ implying
$(D^{0}+\overline{D^0})/2$ in the following text) from channel
$D^{0}$($\overline{D^0}$)$\rightarrow K^{\mp}\pi^{\pm}$ ({\em
Branching ratio} ($B.R.$) $=3.8\%$) was reconstructed through
event mixing method~\cite{KstarPRC}. In the following part, we
will perform a fit combining both $D^{0}$ and electron data points
in \dAu collisions.

Firstly, the decay kinematics of charm hadrons to electrons should
be fixed. Two different decay packages were used: one is the
GENBOD from CERNLIB~\cite{cernlib}, which does the multi-body
decay in phase space; the other one is the PY1ENT function from
PYTHIA~\cite{pythia}, which is believed to be a more reliable one
since other effects (such as spin {\em etc.}) have been taken into
account. We implement both for the systematic studies.
Fig.~\ref{DePtCorr} shows a 2D plot of the correlation between
$p_T$ of charm hadrons and decayed daughter electron $p_T$. The
result shows the parent charm hadrons have much higher $p_T$ in
comparison with the decayed electrons.

\bf
\centering\mbox{
\includegraphics[width=0.5\textwidth]{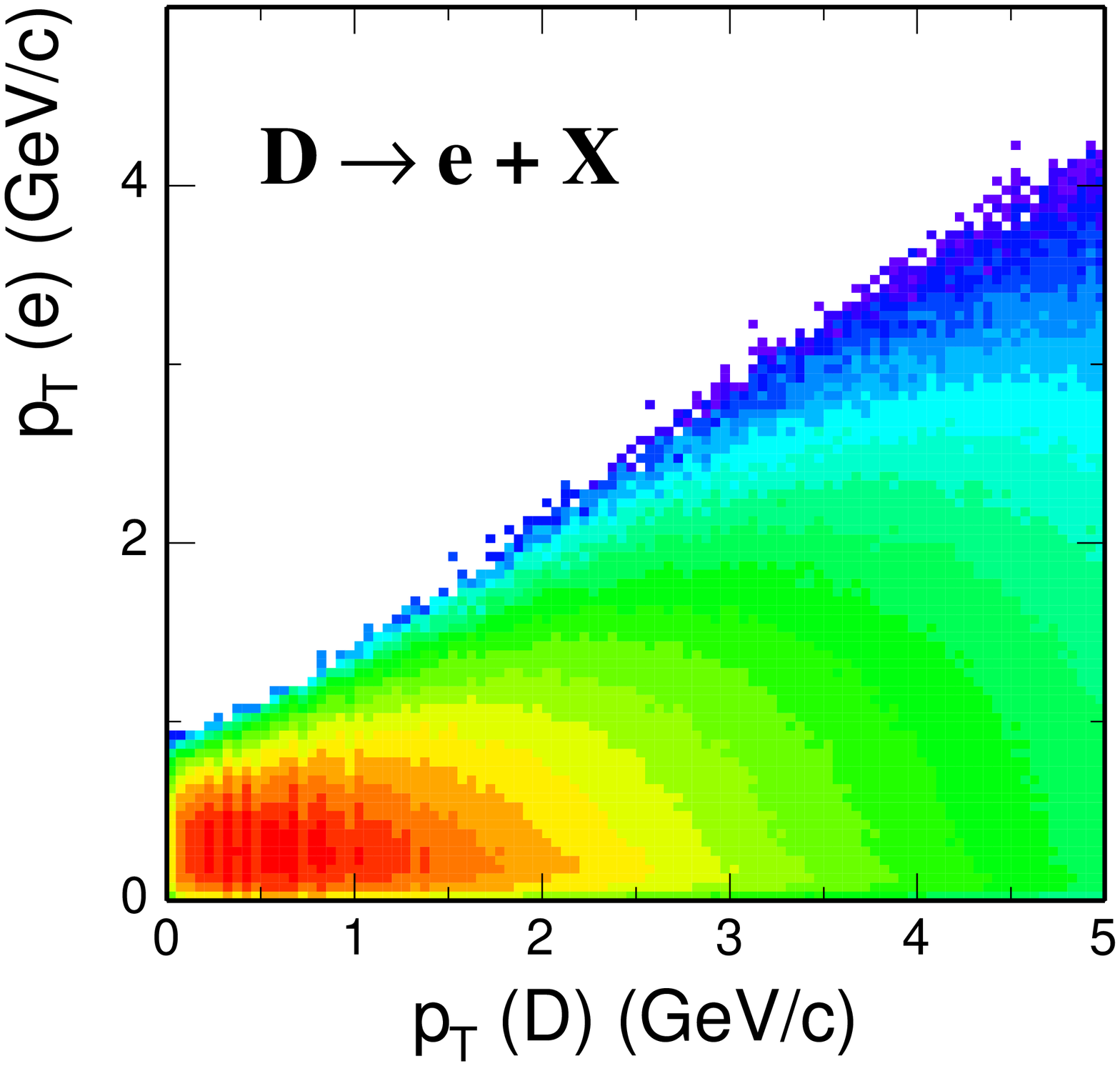}}
\caption[$p_T$ correlation of $D$ and $e$]{2D correlation plot of
$p_T$ for charm hadrons and decayed electrons. The $\la p_T\ra$ of
parent charm hadrons for decayed electrons at $p_T = 2-3$ GeV/c is
$\sim 4.8$ GeV/c.} \label{DePtCorr} \ef

The fractions of different charm hadrons are also needed for the
decay. Previous measurements as well as the model predictions show
almost consistent ratios in broad energies and different
collisions systems~\cite{AnQM04}. We quote the values from
$e^+e^-$ collisions at \s = 91 GeV from PDG~\cite{PDG}: $R\equiv
N_{D^0}/N_{c\bar{c}}= 0.54\pm0.05$. And $D^{+}/D^0\approx 0.4$
from STAR \dAu measurement. According to the large uncertainty of
the $B.R.$ of $D_{s}^{+}\rar e^{+} + X$ and $\Lambda_{c}^{+}\rar
e^{+} + X$, and both are similar to that of $D^0$, and the
distributions of decay daughter electrons have similar shapes, we
took $D_{s}^+$ and $\Lambda_{c}^{+}$ as $D^{0}$. Thus the final
input charm mesons are: 79\% $D^0$ and 21\% $D^+$. This ratio will
be tested for the systematic study later.

The combined fit was then performed under the following assumptions:
\begin{itemize}
\item charm meson spectrum follows a power law function up to $p_T\sim6$ GeV/c.
\item similar \pT spectrum shape between different charm hadrons.
\item the signal electrons are all from charm decays (bottom quark contribution
will be discussed in chapter 5).
\end{itemize}
The power law function is written as the following: \be
\frac{d^2N}{2\pi p_Tdp_Tdy} = \frac{2(n-1)(n-2)}{\pi(n-3)^2{\la
p_T\ra}^2}\frac{dN}{dy} \left( 1 + \frac{p_T}{\dfrac{n-3}{2}\la
p_T\ra} \right)^{-n} \label{powerlaw} \ee There are only 3 free
parameters: $dN/dy$, $\la p_T\ra$ and the power $n$. A specified
charm hadron with a set of these parameters was input into the
generators, and the decayed electrons spectrum was obtained. A
3-dimensional scan on the ($dN/dy$, $\la p_T\ra$, $n$) ``plane"
was done to fit $D^{0}$ and electron data points simultaneously.
The point with the smallest $\chi^2$ value was set to be the fit
result. $\chi^2$ was calculated from the following equation. \be
\chi^2 = \sum_{D} \bigl(\frac{y_D-f_D}{\sigma_D}\bigr)^2 +
         \sum_{e} \bigl(\frac{y_e-f_e}{\sigma_e}\bigr)^2
\label{chi2} \ee where $y_D$, $y_e$ denote the measured yields of
$D^{0}$ and electrons and $\sigma_D$, $\sigma_e$ denote the
measured errors. $f_D$, $f_e$ denote the expected values from
input power law function for $D^0$ and electrons respectively. To
avoid the \pT position issue in large \pT bins, we used the yield
$dN$ instead of $dN/p_Tdp_T$ in each \pT bin.
Table~\ref{combinefit} lists all the fitting characters.

\begin{table}
\caption[Combined fit]{The Combined fit characters for $D^0$ and electrons.}
\label{combinefit}
\vskip 0.1 in
\centering
\begin{tabular}{c|c|c|c|c} \hline \hline
$par.$        & scan range   & step   & result          &  $\chi^2$/ndf \\ \hline
$dN/dy(D^0)$  & (0.02, 0.04) & 0.0004 & $0.029\pm0.004$ &     \\ \cline{1-4}
$\la p_T\ra$  & (0.9, 1.6)   & 0.05   & 1.15            &  18.8/9 \\ \cline{1-4}
$n$           & (7, 16)      & 0.5    & 11              & \\ \hline \hline
\end{tabular}
\end{table}

The error estimation was through the contour scan in the 3-D
"plane" with the $\chi^2 =\chi^2_{min}+1$. The error of $dN/dy$
was then obtained by projecting this 3-D contour into $dN/dy$
axis. Fig.~\ref{combineFitFig} shows the plots of corrected
non-photonic electrons spectra and the combining fit results for
$D^0$ and electrons spectra in \dAu collisions. The curve for \pp
collisions is by scaling down that of \dAu collisions by $\la
N_{bin}\ra = 7.5\pm0.4$.

\bf
\centering\mbox{
\includegraphics[width=0.6\textwidth]{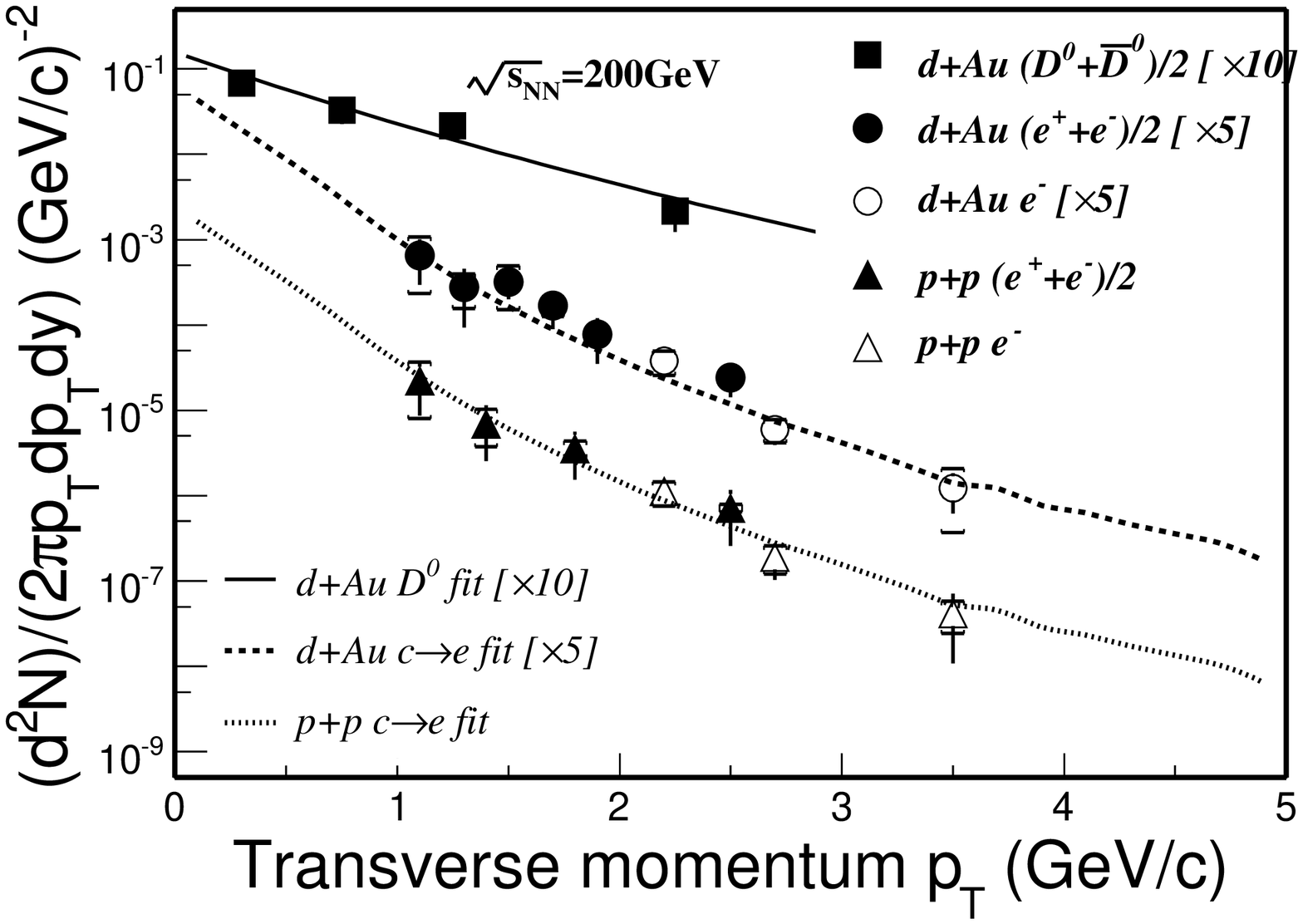}}
\caption[Combined fit]{Reconstructed $D^0$ (solid squares)
$p_{T}$ distributions from d+Au collisions and non-photonic electron
\pT distributions from p+p collisions (triangles) and d+Au
collisions (circles). Solid and dashed lines are the fit results
from both $D^{0}$ and electron spectra in d+Au collisions. The
dotted line is scaled down by a factor of $\la N_{bin}\ra=7.5\pm0.4$
from d+Au to p+p collisions. } \label{combineFitFig}
\ef

Once the $dN/dy$ was extracted, the total charm production cross
section per nucleon-nucleon interaction at RHIC energy can be
calculated from Eq.~\ref{Xsec}: \be \sigma_{c\bar{c}}^{NN} =
\frac{d\sigma_{c\bar{c}}^{NN}}{dy}\Bigr\rvert_{y=0}\times f =
\frac{dN_{D^0}}{dy}\rvert_{y=0}\times R
\times\frac{\sigma_{inel}^{pp}}{\la N_{bin}\ra}\times f
\label{Xsec} \ee In this equation, the factor $R$ is the $D^0$
fraction in total charm hadrons, as mentioned before, and $f$ is a
factor when extrapolating the $dN/dy$ at mid-rapidity to full
rapidity. This factor $f$ was extracted from PYTHIA model study
and we estimated a $\sim15\%$ systematic error on this factor from
different parameters in that model. The charm production cross
section per nucleon-nucleon collision at mid-rapidity and total
cross section at \s = 200 GeV is: \be
\frac{d\sigma_{c\bar{c}}^{NN}}{dy}\Bigr\rvert_{y=0}(\sqrt{s} =
200~GeV) = 0.30 \pm 0.04~mb \ee \be
\sigma_{c\bar{c}}^{NN}(\sqrt{s} = 200~GeV) = 1.42 \pm 0.20~mb \ee

\subsection{Systematic error study}
The systematic errors contributing to the final total charm cross
section were studied. One of the important sources is the
uncertainty of charm hadron \pT spectrum and the fractions of
different charm hadrons. The effect caused by the charm hadron \pT
spectrum was studied by changing the power law parameters to
$1\sigma$ away. The correlations between parameters were not taken
into account in the estimation, instead, we looped all the
combinations to find the largest deviation as the estimation
systematic uncertainty. The effect caused by the uncertain
fractions of different charm hadrons was studied by changing the
effective fraction in the decay $D^0:D^+=0.79:0.21$ to $0.90:0.10$
and $0.70:0.30$. For the charm decaying to electron kinematics,
the difference of two decay codes was take into the systematic
errors. Table~\ref{sysError} shows the contributions from possible
sources.

\begin{table}[htbp]
\caption[Systematic errors]{Systematic errors to the final charm
cross sections.} \label{sysError}\vskip 0.1 in
\centering\begin{tabular}{c|c} \hline \hline Source & Relative
contribution \\ \hline
charm hadron \pT spectrum & 13\% \\
fractions of charm hadrons & 10\% \\
decay codes & 12\% \\
$NFitHits$ $15\rar25$ & 14\% \\
rapidity distribution & 15\% \\
$\la N_{bin}\ra$ in \dAu & 5\% \\
Normalization & 10\% \\ \hline
Sum & 31\% \\ \hline \hline
\end{tabular}
\end{table}

\newpage

\section{Single electrons from \AuAu collisions at \sNN = 62.4 GeV}

The purpose of the data analysis of \dAu and \pp collisions is to
set up the baseline for heavy ion collisions. Since the large data
sample of \AuAu collisions at \sNN = 200 GeV is not available yet,
the smaller sample of \AuAu collisions at \sNN = 62.4 GeV has been
analyzed to gain understandings of the necessary techniques for
\AuAu collisions. Due to the significant increase of
multiplicities in \AuAu collisions, and the consequent decrease of
track qualities, the situation in \AuAu collisions becomes more
complicated. In this section, I would discuss the inclusive
electron spectrum from TOF + $dE/dx$ method and the photonic
background estimation. Since the charm yield is pretty low
compared to photonic contributions, we would not expect to extract
the signal with acceptable errors.

\subsection{Data set and electron PID}
RHIC has offered a relative short time \AuAu beams at \sNN = 62.4
GeV and STAR has accumulated $\sim15$ M events in total. With the
minimum bias trigger ($0-80\%$) and vertex z position selection,
the useful physics events number in this analysis is $\sim6.4$ M.


The calibration results for TOF detectors are $\sim110$ ps
resolutions for both TOFr and TOFp system, with $\sim55$ ps start
timing resolution included. The hadron PID capability was reported
in~\cite{swingHQ}. Similar as before, electrons can be identified
by combining TOF and $dE/dx$ in the TPC. Primary electrons were
selected under the criteria shown in Table.~\ref{eTrackAuAu}.

\begin{table}[hbt]
\caption{Electron selection criteria in \AuAu}
\label{eTrackAuAu}\vskip 0.1 in \centering\begin{tabular}{c|c}
\hline \hline Method & TOF+$dE/dx$       \\ \hline $|VertexZ|<$ &
30 cm
\\ \hline
primary track ?      &   Yes             \\
nFitPts $\geqslant$  &  25               \\
ndEdxPts $\geqslant$ & 15                \\
rapidity             & (-1.0, 0)         \\
$\chi^{2}/ndf$       & (0., 3.0)         \\
$\beta$ from TOF     & $|1/\beta-1|<0.03$\\
TOFr hit quality     & $30<ADC<300$      \\
                     & $-2.7<z_{local}$/cm$<3.4$ \\
                     & $|y_{local}-y_{C}|<1.9$ cm \\
TOFp hit quality     & $th_1<ADC<th_2$    \\
                     & $2.0<z_{local}$/cm$<18.0$ \\
                     & $0.4<y_{local}$/cm$<3.2$ \\
     \hline \hline
\end{tabular}
\end{table}

Fig.~\ref{ePIDAuAu} shows the $dE/dx$ vs. particle momentum after
a $\beta$ cut from TOF. Electrons band can be separated from
hadrons. The $dE/dx$ resolution in \AuAu collisions is worse than
that in \dAu and \pp due to much higher multiplicities, so the
separation in this plot is not as good as that in \dAu and \pp,
see Fig.~\ref{ePID}. We performed both two-gaussian function and
exponential+gaussian function fit to extract the electron raw
yields in each \pT bin.

\bf \centering\mbox{
\includegraphics[width=0.6\textwidth]{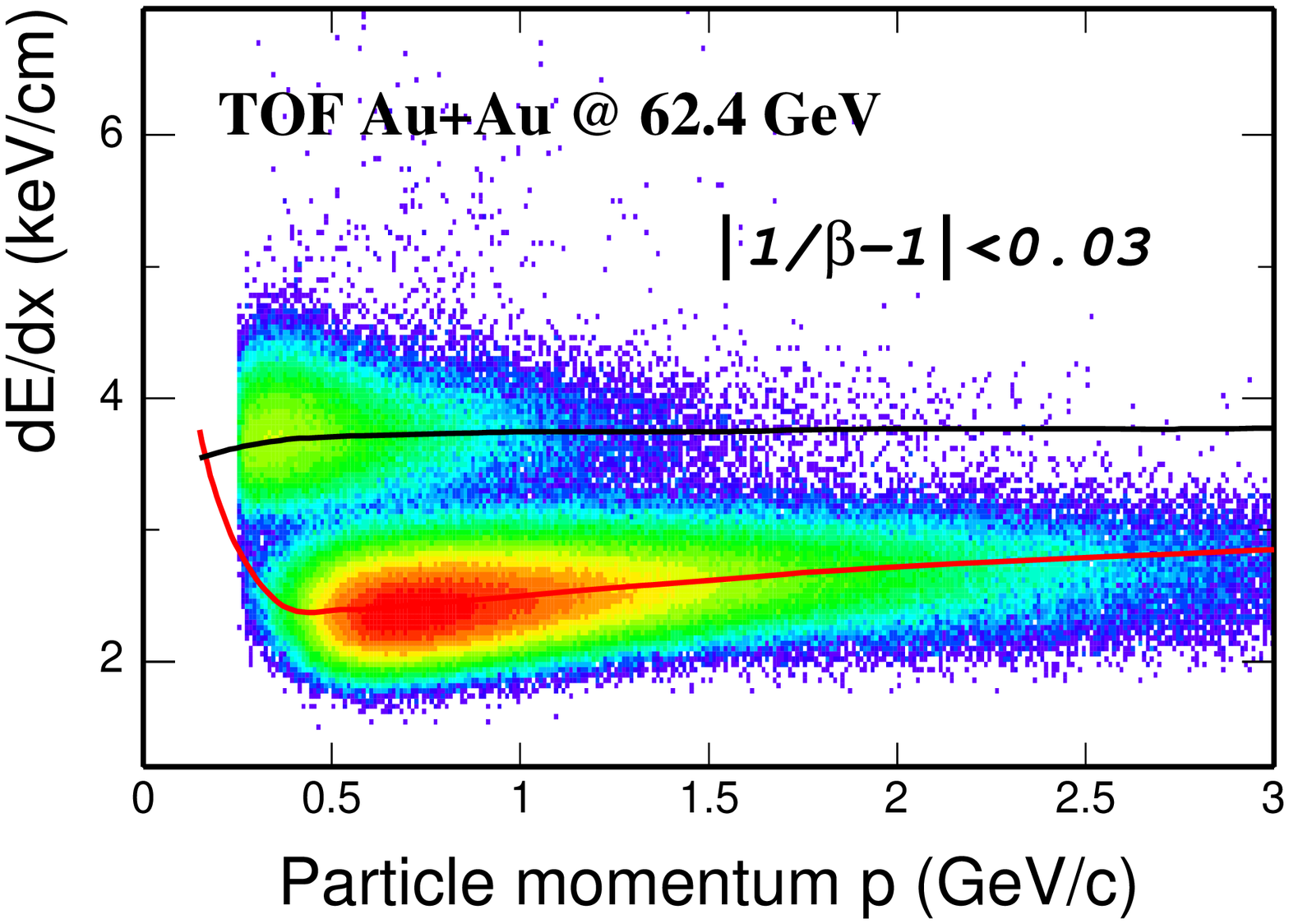}}
\caption[Electron PID in \AuAu 62.4 GeV]{$dE/dx$ vs. particle
momentum after a TOF $\beta$ cut ($|1/\beta-1|<0.03$)}
\label{ePIDAuAu} \ef

Hadron contamination becomes larger if we select electrons with
the same efficiency as before. We need to optimize the pure
electron sample selection for the background study later.
Fig.~\ref{hadComAuAu} shows the hadron contamination fractions
under different electron selections.

\bf \centering\mbox{
\includegraphics[width=0.6\textwidth]{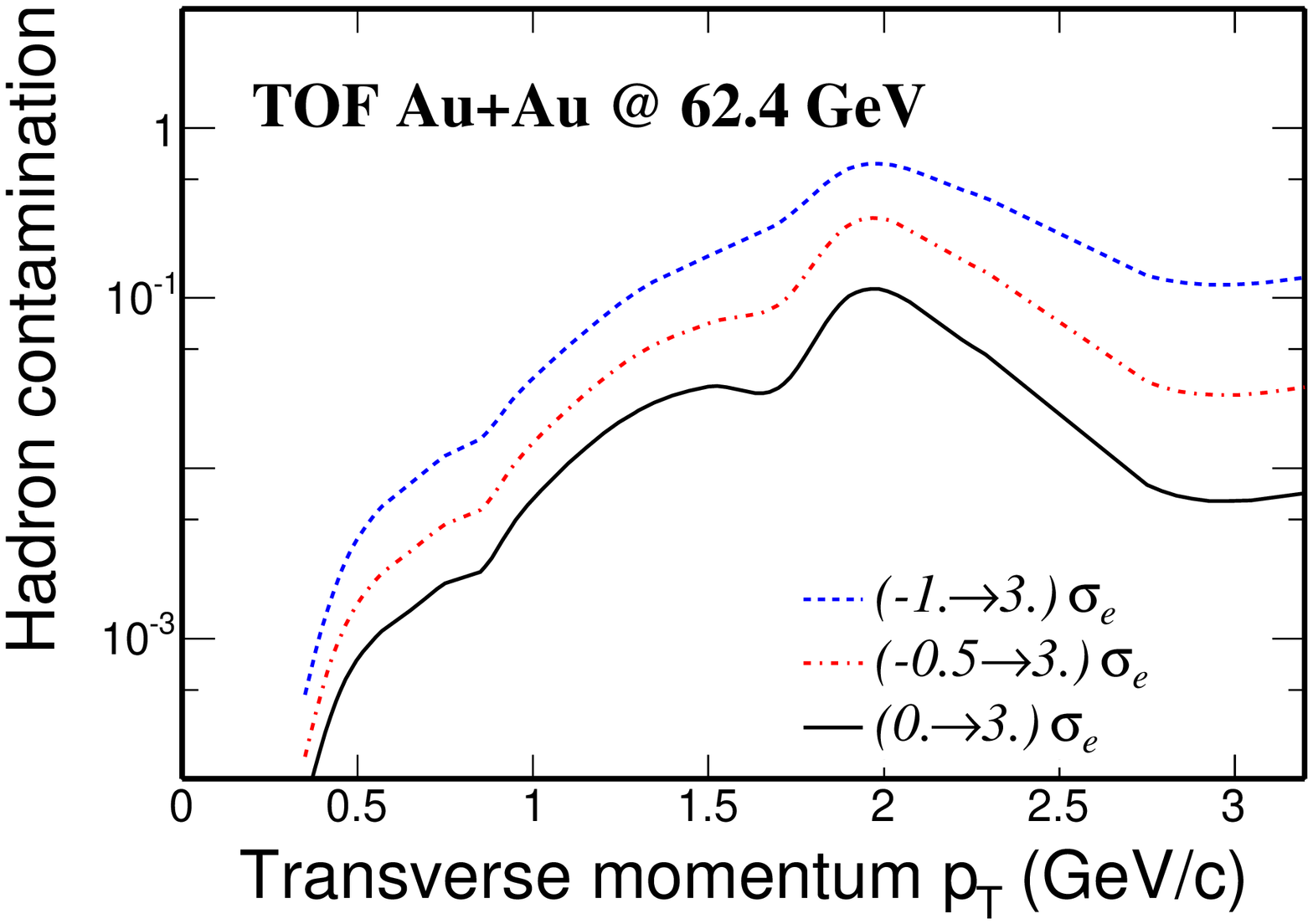}}
\caption[Hadron contamination in \AuAu 62.4 GeV]{Hadron
contamination fractions for different electron selections in \AuAu
62.4 GeV. Statistics are poor at $p_T>2$ GeV/c.}
\label{hadComAuAu} \ef

The fit results show that to have contamination under control
($<10\%$), we need the selection $\sigma_{e}>0$ with electron
efficiency $\sim50\%$.

\subsection{Non-photonic background estimation in \AuAu}
In \AuAu collision events, due to large multiplicity, the
PID-blinding partner track reconstruction will lead to a huge
combinatorial background. Even a critical $dE/dx$ selection is
applied, this combinatorial background is still significant.
Fig.~\ref{AuAuTOFsame} shows the electron pair candidate invariant
mass distribution. The tagged electron track was selected from TOF
with the cuts shown in Table.~\ref{eTrackAuAu} and additional
$0<\sigma_{e}<3$ (we will open it to $-1<\sigma_{e}<3$ later). The
partner track candidate was selected under the criteria shown in
Table.~\ref{ePartnerAuAu}.

\bf \centering\mbox{
\includegraphics[width=0.6\textwidth]{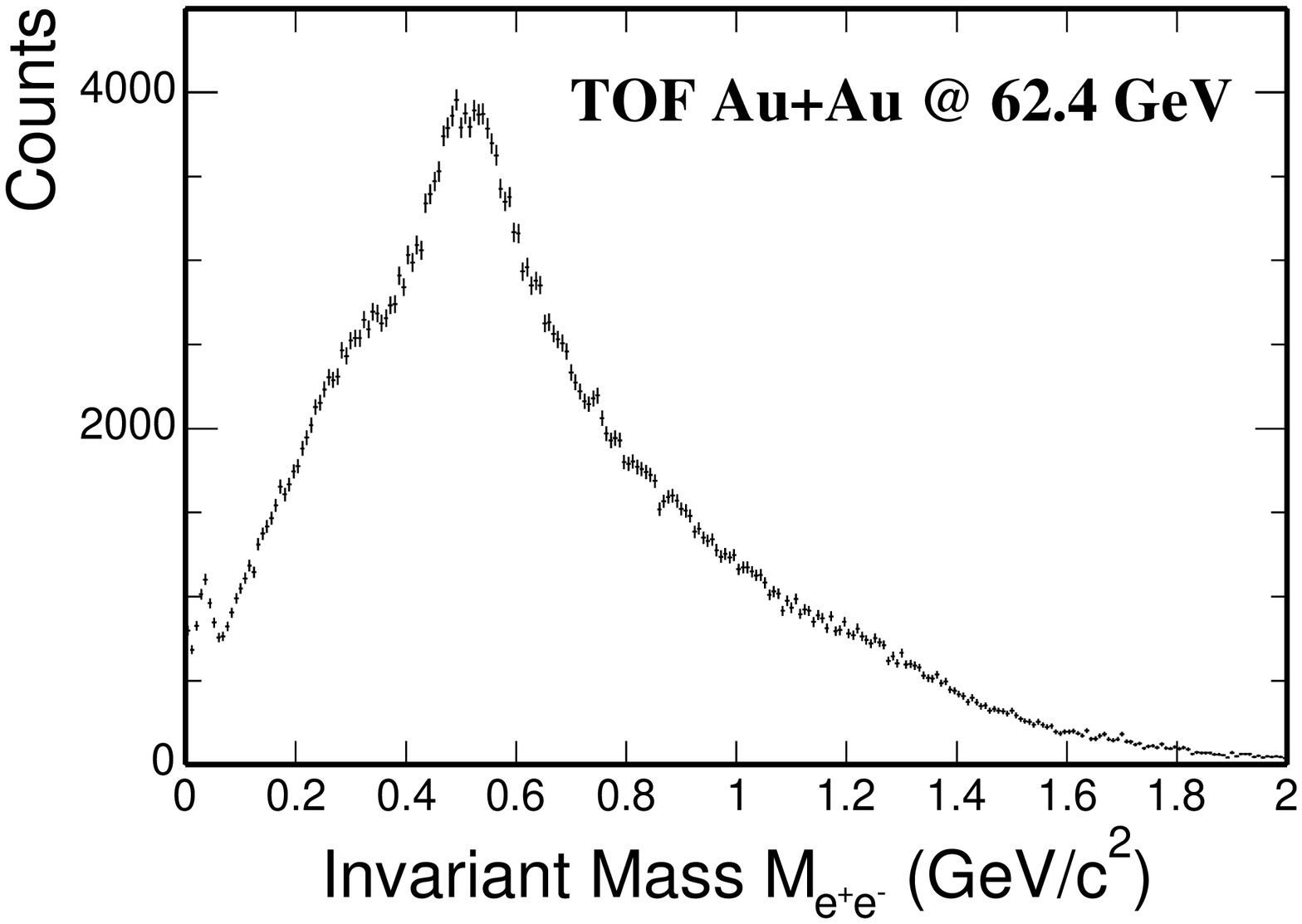}}
\caption[$M_{e^+e^-}$ in \AuAu 62.4 GeV]{Invariant mass of
electron candidates pair from the same event in \AuAu 62.4 GeV
collisions. The plot shows significant combinatorial contribution
near the conversion peak.} \label{AuAuTOFsame} \ef

\begin{table}[hbt]
\caption{Partner candidate selection criteria in \AuAu}
\label{ePartnerAuAu}\vskip 0.1 in \centering\begin{tabular}{c|c}
\hline \hline charge &  opposite to tagged track \\
primary/global ?     &   global             \\
nFitPts $\geqslant$  &  15               \\
nFitPts/nMax $>$     &  0.52          \\
$\chi^{2}/ndf$       & (0., 3.0)         \\
$\sigma_{e}$         & (-1., 3.0)$^\dagger$ \\ \hline $dca$ of
$e^{+}$,$e^{-}$ & (0.0, 3.0) cm \\
     \hline \hline
\end{tabular} \\
$^\dagger$ several different $\sigma_{e}$ cuts were tried.
\end{table}

\bf \centering \bmn[b]{0.48\textwidth} \centering
\includegraphics[width=1.0\textwidth]{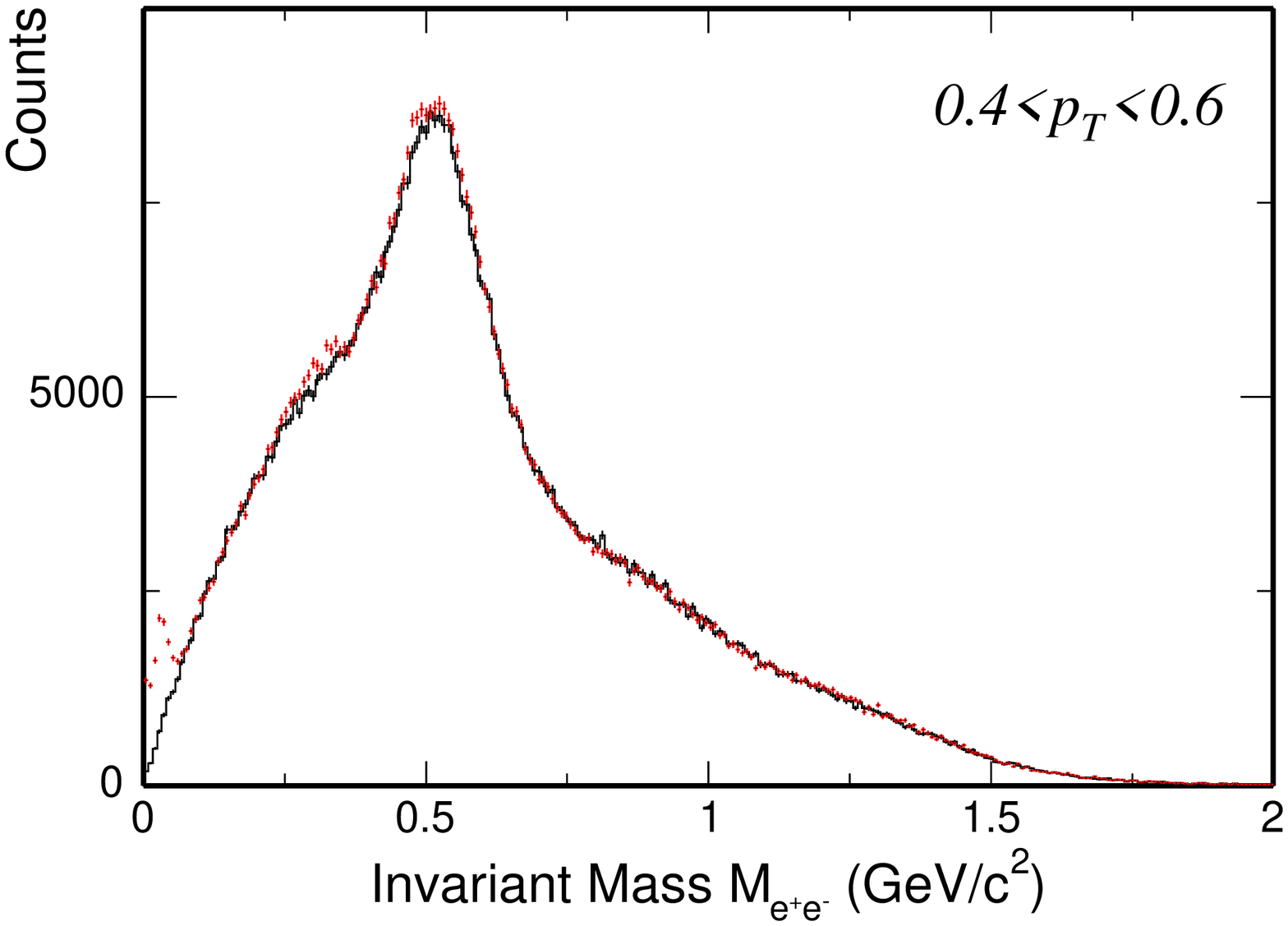}
\emn%
\bmn[b]{0.48\textwidth} \centering
\includegraphics[width=1.0\textwidth]{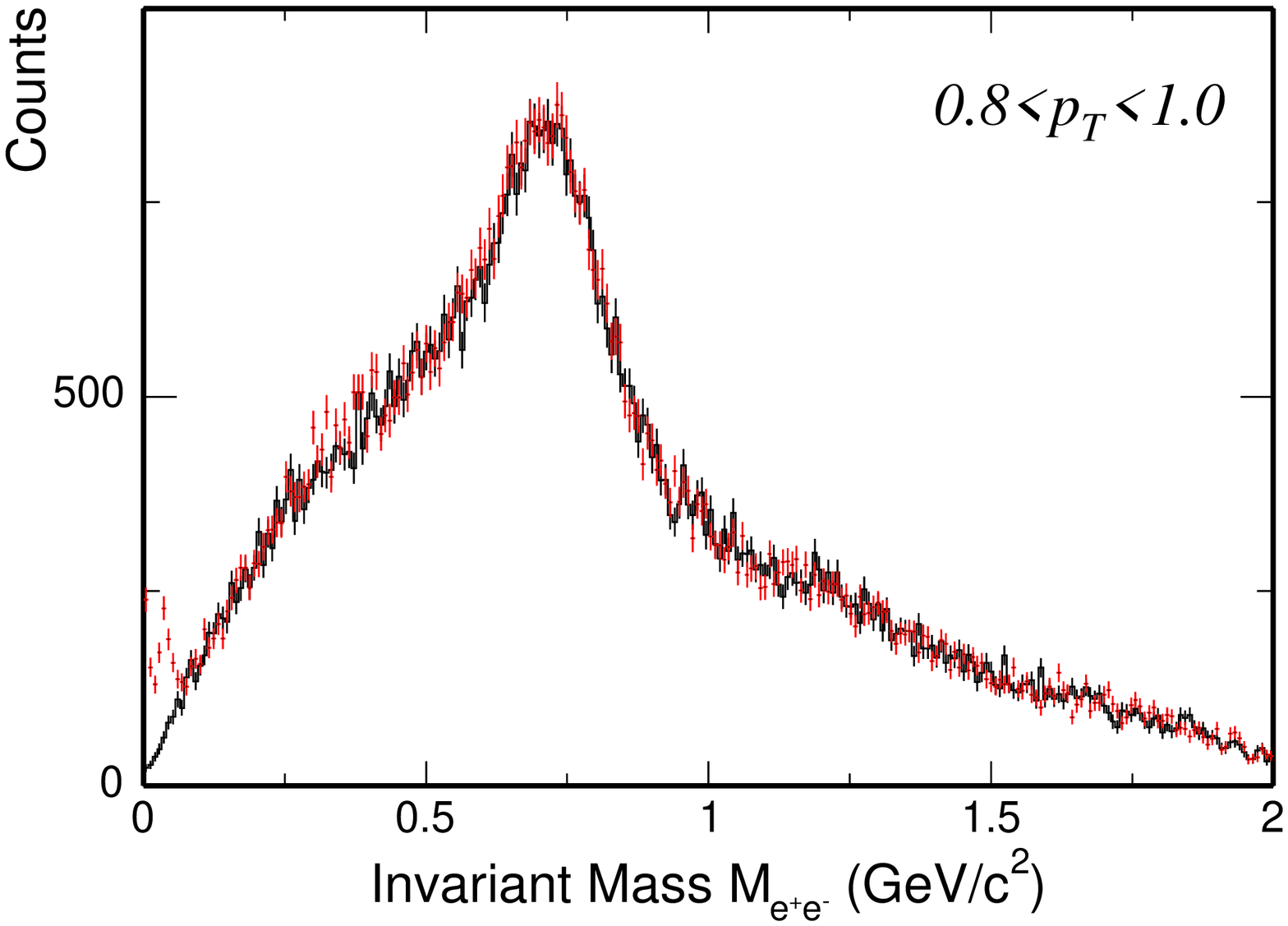}
\emn\\[10pt]
\bmn[b]{0.48\textwidth} \centering
\includegraphics[width=1.0\textwidth]{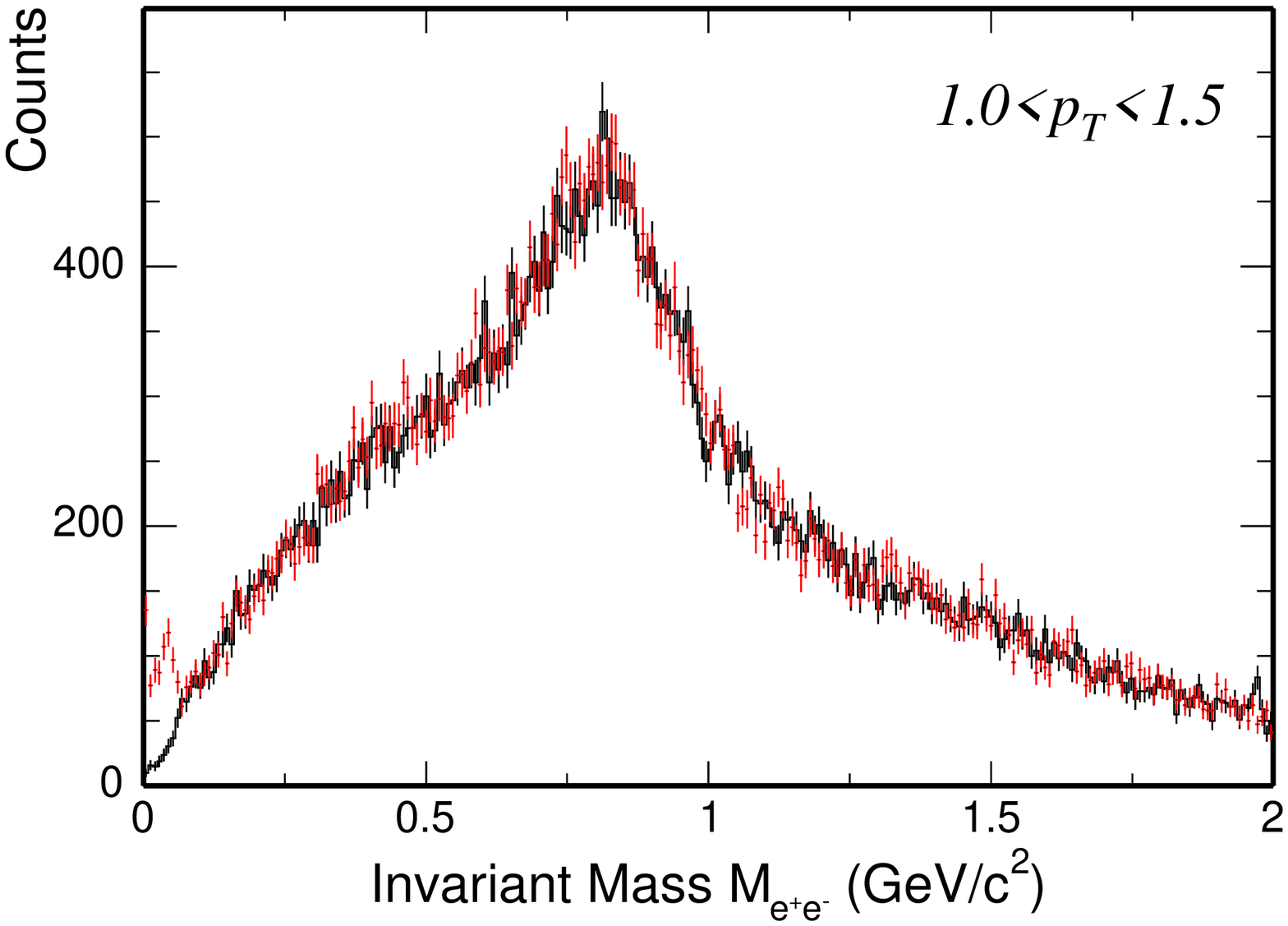}
\emn%
\bmn[b]{0.48\textwidth} \centering
\includegraphics[width=1.0\textwidth]{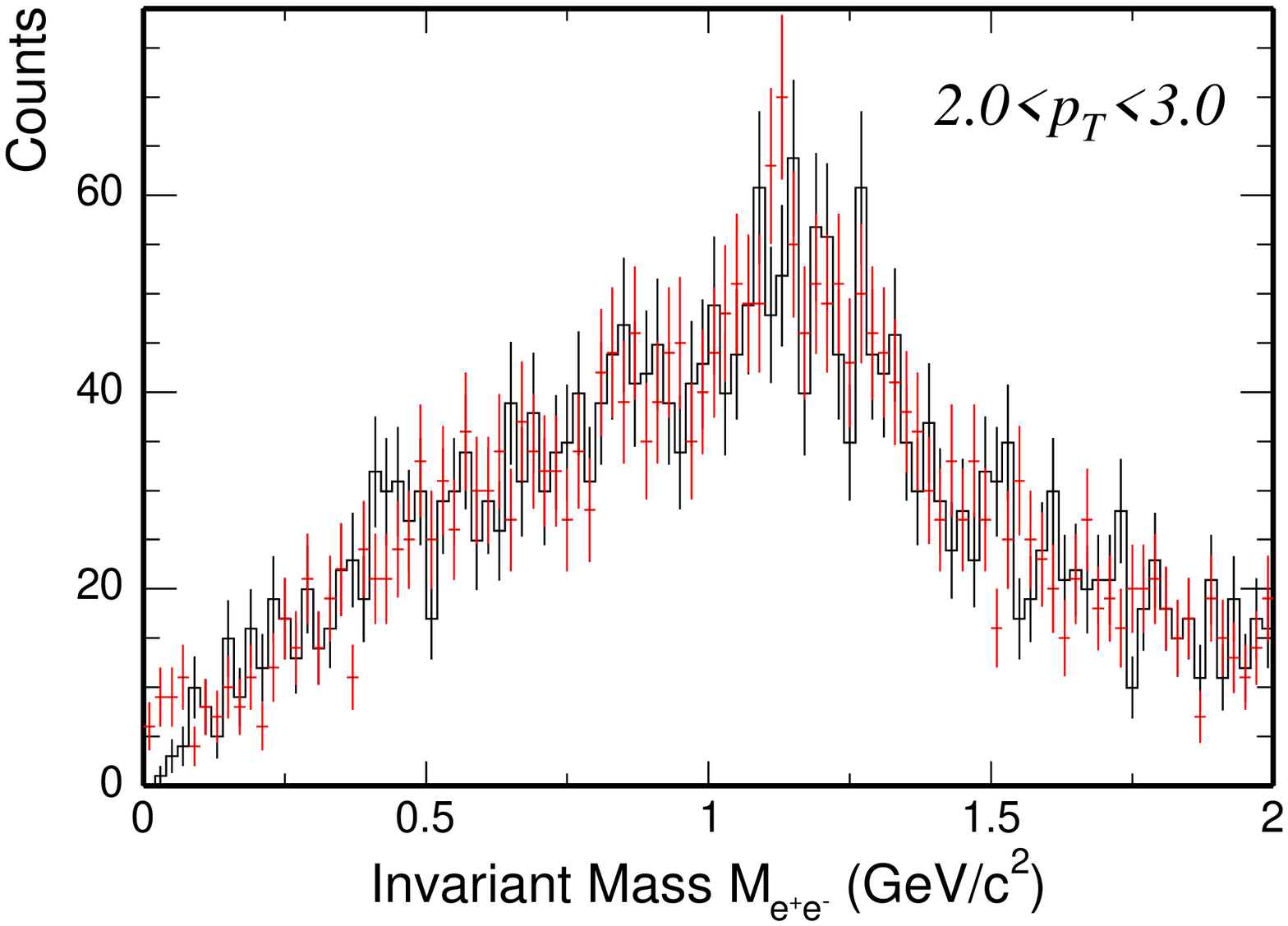}
\emn%
\caption[Background reconstruction in \AuAu 62.4 GeV]{Main
photonic background reconstruction in \AuAu collisions. The
crosses depict the real electron pair candidate invariant mass
distributions and the histograms represent the combinatorial
backgrounds. The photon conversion peak is clear in $\sim 0$ mass
region, while the $\pi^{0}$ Dalitz is hard to see.}
\label{eAuAubkgd} \ef

To reconstruct the photon conversion and $\pi^0$ Dalitz, we need
to understand the combinatory background shape in
Fig.~\ref{AuAuTOFsame}. In this step, the photon conversion and
other correlations will be considered as signals, and
combinatorial background was produced by rotating the partner
track momentum $\overrightarrow{p}\rightarrow
-\overrightarrow{p}$, then do the normalization of the mass
spectrum in the region $0.8<M_{e^+e^-}$/(GeV/c$^2$)$<2.0$.
Fig.~\ref{eAuAubkgd} shows this results in several \pT bins. Here
\pT is the \pT of tagged electrons. The plots show the
combinatorial background was very well reproduced. In the first
plot, there are two small bumps at $M_{e^{+}e^{-}}\sim$ 0.3
GeV/c$^{2}$ and 0.5 GeV/c$^{2}$. These are due to the
misidentified pions which come from $K_{S}^{0}$ and $\rho, \omega$
decays, respectively. This has been confirmed through MC decay
simulations.

The photon conversion peak is clearly seen near zero mass region,
and the offset from 0 is due to the opening angle
resolution~\cite{IanThesis}. The $\pi^0$ Dalitz contribution is
not visible in this case, possibly because the Dalitz contribution
is much smaller and distribution is much broader compared to
conversion processes. We subtracted the combinatorial background
from the real distribution, and integrated the remaining
distribution from $0-0.15$ GeV/c$^{2}$ to get the reconstructed
main photonic background raw yield. In this case, we assumed that
both photon conversion and $\pi^{0}$ Dalitz decays were
reconstructed.

The background raw yield need to be corrected for the
reconstruction efficiency, as we did it for \dAu and \pp
collisions. This efficiency was calculated from \AuAu 62.4 GeV
HIJING events plus full detector MC simulations. After
$|V_{Z}|<30$ cm cut, $\sim53$ K events were used in the
calculation. We took all TPC electron tracks without any $dE/dx$
and TOF hit cut to improve the statistics. The procedure is the
same as we did before. Fig.~\ref{bkgdeff62} shows the background
reconstruction efficiency in \AuAu 62.4 GeV compared with \dAu
results. Due to the higher multiplicities, the relative lower
reconstruction efficiency does make sense.

\bf \centering\mbox{
\includegraphics[width=0.6\textwidth]{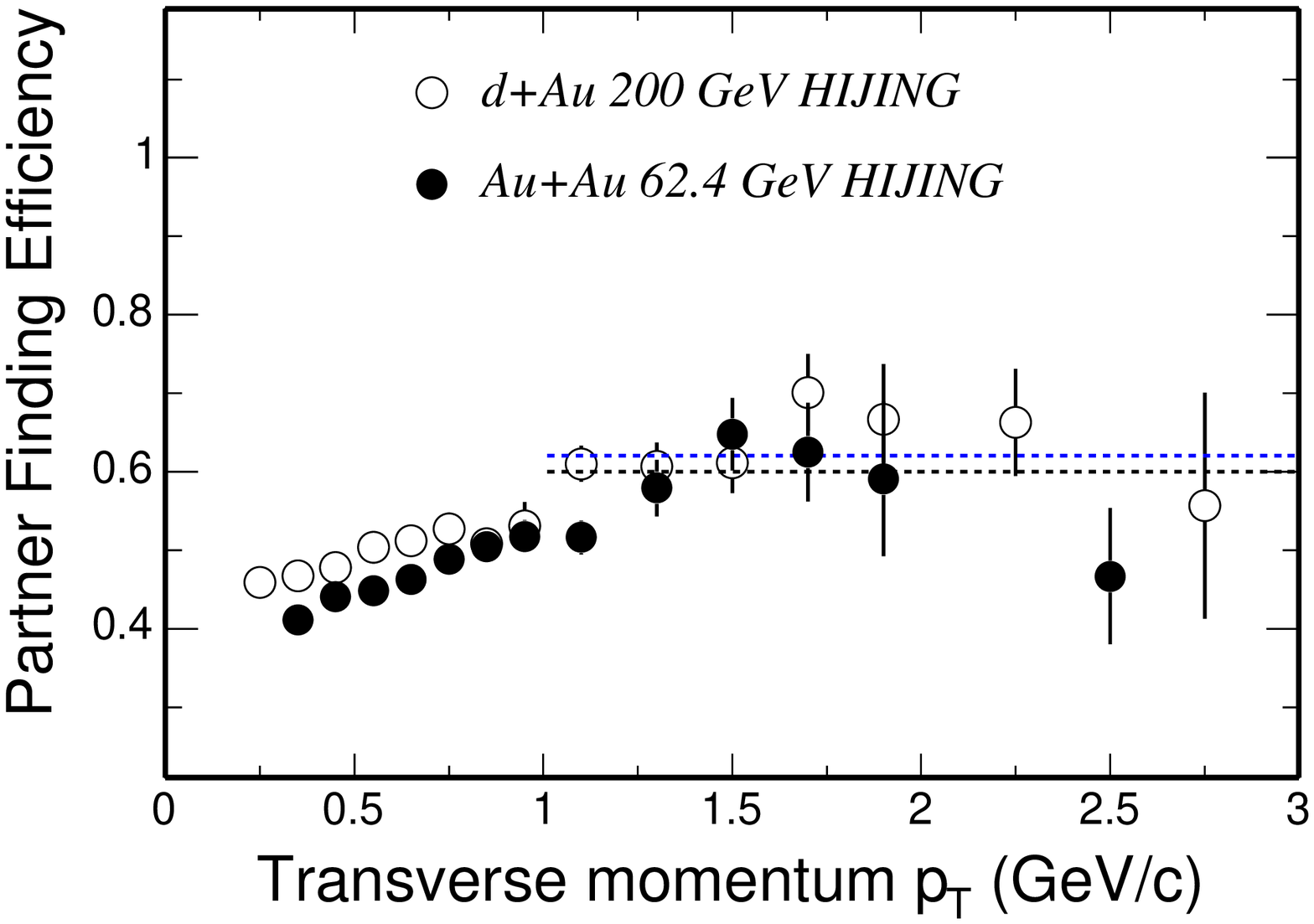}}
\caption[Background efficiency in \AuAu 62.4 GeV]{Photonic
background reconstruction efficiency from \AuAu 62.4 GeV HIJING
simulations. Also shown on the plot is that from \dAu 200 GeV
HIJING simulations.} \label{bkgdeff62} \ef

\bf \centering\mbox{
\includegraphics[width=0.6\textwidth]{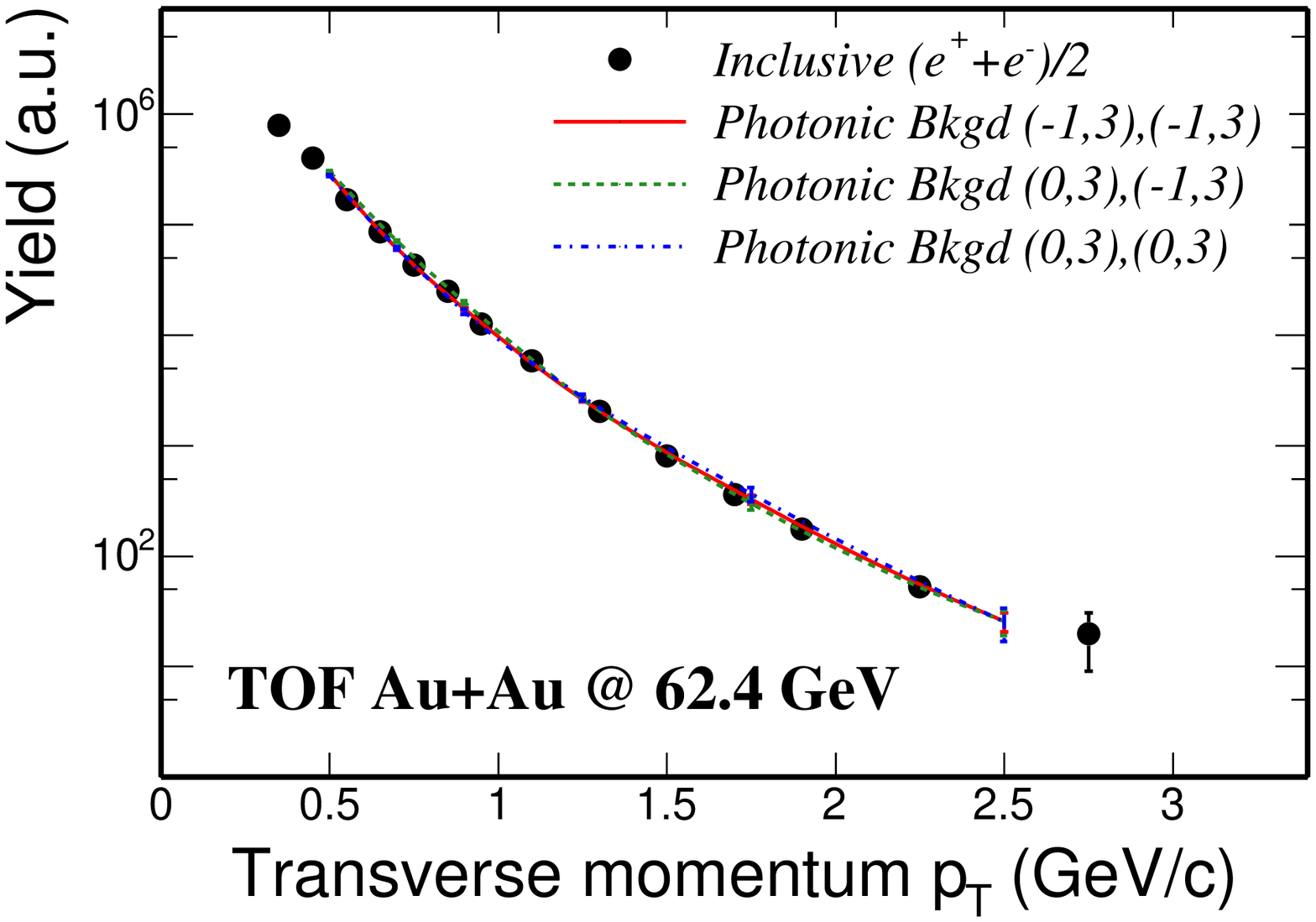}}
\caption[Photonic background contribution in \AuAu 62.4 GeV]{Raw
inclusive electron yield and the reconstruction efficiency
corrected photonic background contributions under different
electron/positron track selections. The numbers in the brackets on
the plot show the $\sigma_{e}$ cut for the tagged track and the
partner track, respectively.} \label{rawCom62} \ef

This efficiency was used to correct for the photonic background
raw yield obtained from above. In addition, a $\sim5\%$ fraction
of other photonic background (from \dAu results) and the
$\sigma_{e}$ selection efficiency were also included. Then we can
get the corrected background spectrum compared with the inclusive
spectrum, shown as Fig.~\ref{rawCom62}. All these were not
corrected for the single track efficiency and acceptance yet. The
reconstructed background matches the total inclusive electron
spectrum. This is not surprised since the charm yield is pretty
low at 62.4 GeV.

This improved method for \AuAu system working makes us confident
to extract the charm signal at $2-3$ GeV/c from the coming 30 M
minimum bias 200 GeV \AuAu data.

\subsection{Spectra from \AuAu collisions at \sNN =
62.4 GeV}

Single electron track efficiency and acceptance is needed to
correct the raw yield spectrum. The efficiency calculation is
similar to what has been introduced before. Here, the method
combining TPC tracking efficiency from MC simulation and TOF
matching efficiency from real data was used. Fig.~\ref{matcheff62}
shows the matching efficiency of pions from TPC to TOF (including
both TOFr and TOFp) within $-1<y<0$ window~\cite{swingHQ}.

\bf \centering\mbox{
\includegraphics[width=0.6\textwidth]{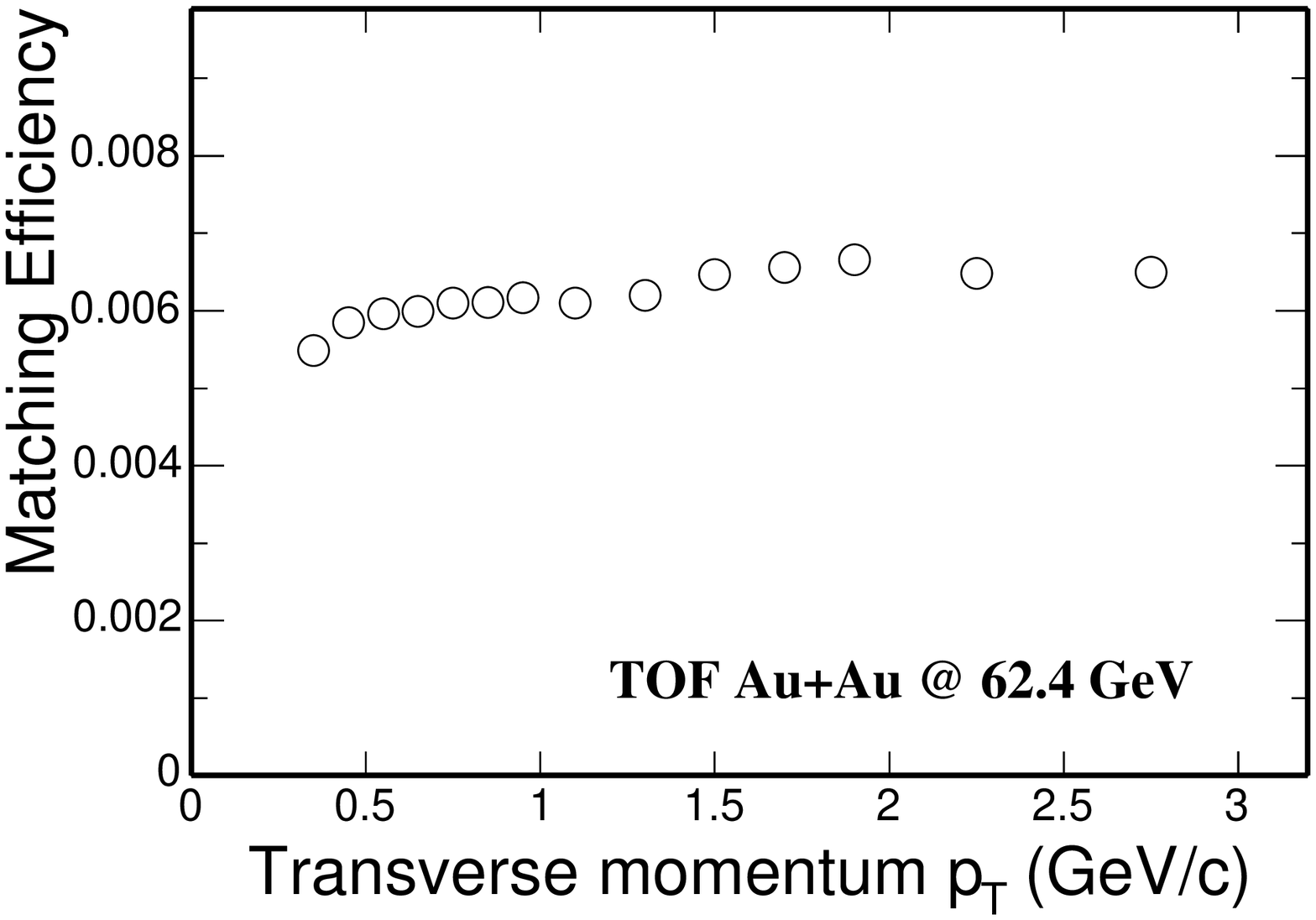}}
\caption[Matching efficiency in \AuAu 62.4 GeV]{Matching
efficiency (including detector response) from TPC to TOF
(including both TOFr and TOFp) of pions whin $-1<y<0$ window.}
\label{matcheff62} \ef

The TPC tracking efficiency from \dAu was lowered down by $5\%$ to
as an estimation for \AuAu 62.4 GeV collisions according to the
multiplicity difference in \dAu and \AuAu collisions. The TOF
velocity selection efficiency was estimated to be $\sim 95\%$.
With all these efficiency corrected, we can get the spectrum of
inclusive electrons, shown in Fig.~\ref{spectrum62}. Also shown in
the plot are inclusive charged pion spectrum from TOFr measurement
in \AuAu 62.4 GeV and a previous ISR non-photonic electron
spectrum from \pp 52.7 GeV collisions scaled by the $N_{bin}$ to
\AuAu collisions. This plot shows the expected non-photonic signal
is about an order of magnitude lower than the inclusive electron
spectrum. This is also consistent with the photonic background
reconstruction discussed in the previous section.

\bf \centering\mbox{
\includegraphics[width=0.6\textwidth]{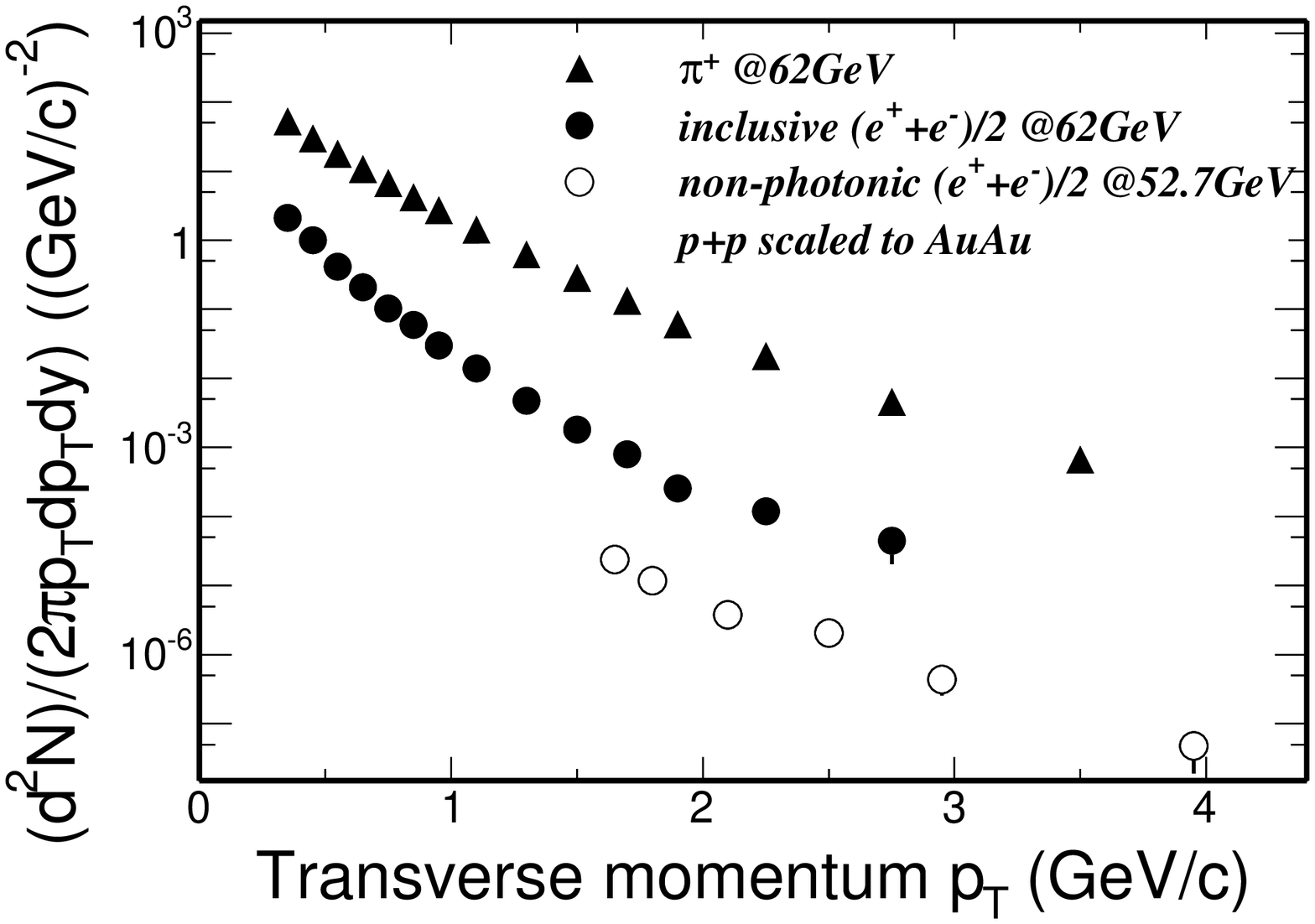}}
\caption[Electron spectrum in \AuAu 62.4 GeV]{Inclusive electron
spectrum in \AuAu 62.4 GeV compared with the charged pion spectrum
in the same system and the ISR non-photonic electron spectrum in
52.7 GeV \pp collisions scaled with $N_{bin}$ to \AuAu minimum
bias collisions.} \label{spectrum62} \ef

Although more than $\sim90\%$ of the electrons up to 3 GeV/c are
photonic background in \AuAu 62.4 GeV so that we cannot extract
the non-photonic signal with reasonable errors, this photonic
background reconstruction method works well and technically, it is
ready for the coming 200 GeV large data sample analysis to answer
some of the issues about the charm production in heavy ion
collisions.

\chapter{Single electron azimuthal anisotropy distributions}
The data set used in this analysis is \AuAu minimum bias triggered
($0-80\%$) events at \sNN = 62.4 GeV. This data sample is the same
as that described before. The anisotropic parameters were obtained
using the event plane analysis technique. The detailed analysis
method introduction can be found in~\cite{v2EPmethod}.

\section{Event plane and its resolution}

In heavy ion collisions, the event plane is reconstructed from the
detected final particle azimuths. The acceptance and efficiency of
the detectors in azimuth was corrected by compensating the azimuth
to a flat distribution with $\phi$ weights. Technically, the
$\phi$ weights were created for different days to deal with the
different situations in a long running period. Additional \pT
weights were also applied to improve the event plane resolution.
The second order harmonic event plane azimuth $\Psi_{2}$ can be
calculated from the $\overrightarrow{Q}$ vector, as
Eq.~\ref{eventP},\ref{Qvector}:

\be \Psi_2 = \left(\arctan\frac{Q_y}{Q_x}\right)/2, \hskip 1 in
0<\Psi_2<\pi \label{eventP} \ee \be \overrightarrow{Q}=(Q_x,
Q_y)=\left(\sum_{i}w_i\cdot \cos(2\phi_i), \hskip 0.2 in
\sum_{i}w_i\cdot \sin(2\phi_i)\right) \label{Qvector} \ee Here,
$w_i$ is the weight for each track included in the event plane
calculation, which includes both the $\phi$ weight and the \pT
weight. Fig.~\ref{phiWeight} shows a set of typical $\phi$ weight
distributions for a single day. The bumps at $\phi\sim 6$ in
"East" and "Fast East" plots are due to a bad sector on the east
side of the TPC. The tracks selected in the event plane
calculation should satisfy the criteria listed in
Table.~\ref{epTrack}.

\bf \centering\mbox{
\includegraphics[width=0.8\textwidth]{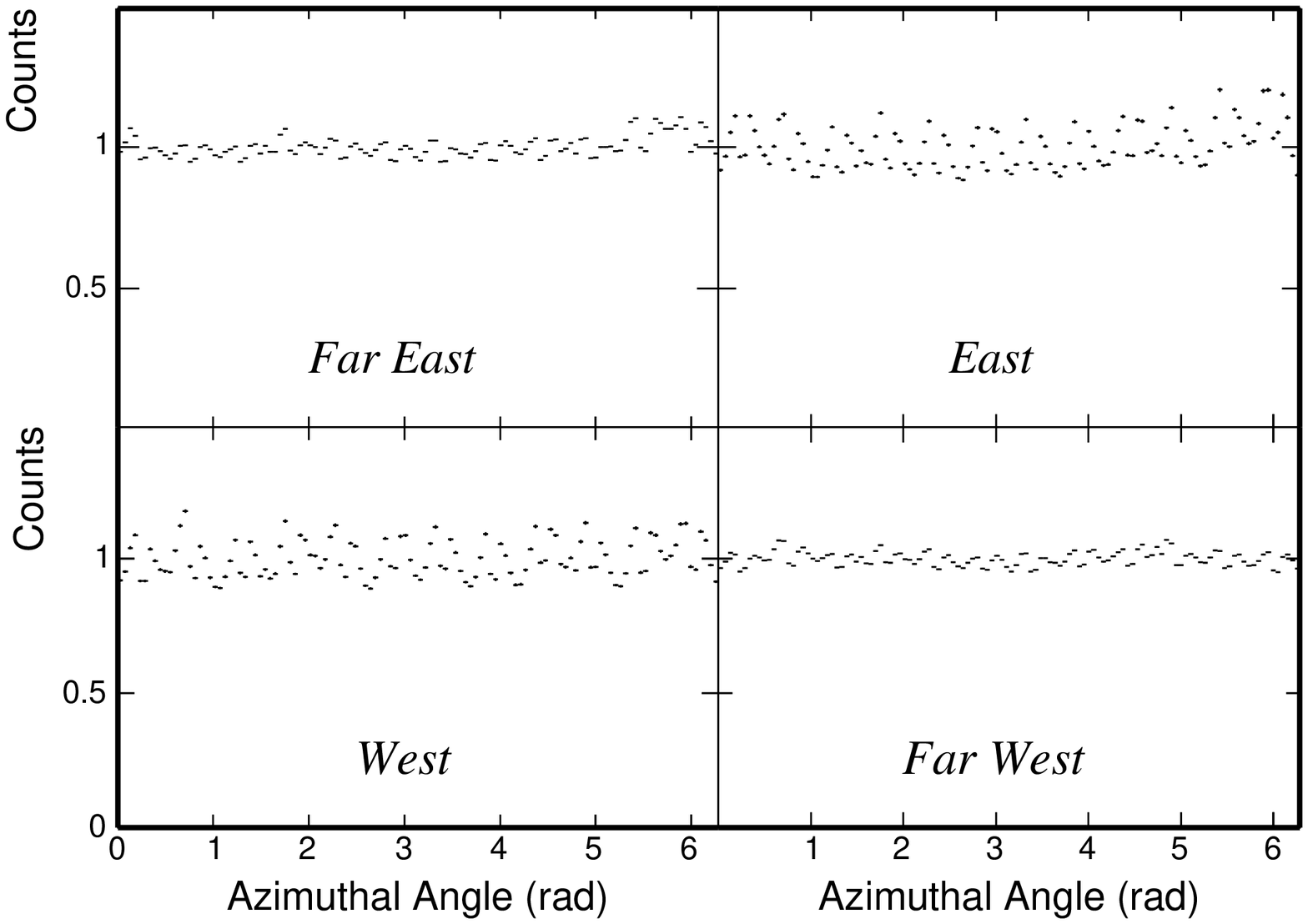}}
\caption[$\phi$ weight distribution]{$\phi$ weight used for a
single day for minimum bias 62.4 GeV \AuAu collisions. The bump
around $\sim6$ rad in $\phi$ in east and far east set is due to a
bad sector in east side of TPC.} \label{phiWeight} \ef

\begin{table}[hbt]
\caption{Track selection in event plane
calculation}\label{epTrack} \vskip 0.1 in
\centering\begin{tabular}{c|c} \hline \hline nFitPts & $\geq15$
\\ nFitPts/nMax & $>0.52$  \\   $p_{T}$ & (0.1, 4.0) \\
 $\eta$         & (-1.3, 1.3) \\   global $dca$ & (0.0, 3.0)
\\
     \hline \hline
\end{tabular}
\end{table}

Fig.~\ref{ePlane} shows the second order event plane azimuthal
angle distribution. This distribution was fit to a constant value
and the fit quality $\chi^2/ndf=126/119$ indicates a good event
plane reconstruction.

The resolution of the event plane was calculated using the
sub-event method~\cite{v2EPmethod}. Each event was divided into
two random sub-events with nearly equal multiplicity. The event
plane was reconstructed in each sub-event, denoted as
$\Psi_{2}^{a}$ and $\Psi_2^b$. Then the event plane resolution
$r=\la cos[2(\Psi_2-\Psi_{rp})]\ra$ can be calculated from Eq.(14)
and (11) from ~\cite{v2EPmethod}:

\be \la \cos[2(\Psi_2-\Psi_{rp})]\ra =
\frac{\sqrt{\pi}}{2\sqrt{2}}\chi_{2}^{}
\exp(-\chi_2^2/4)\times[I_0(\chi_2^2/4)+I_1(\chi_2^2/4)]
\label{evtResFunc} \ee \be \la \cos[2(\Psi_2^a-\Psi_{rp})]\ra =
\sqrt{\la \cos[2(\Psi_2^a-\Psi_2^b)]\ra} \label{subEvtResFunc} \ee
\be \chi_2^{}=v_2^{}/\sigma=v_2^{}\sqrt{2N} \label{chim} \ee

\bf \centering\mbox{
\includegraphics[width=0.6\textwidth]{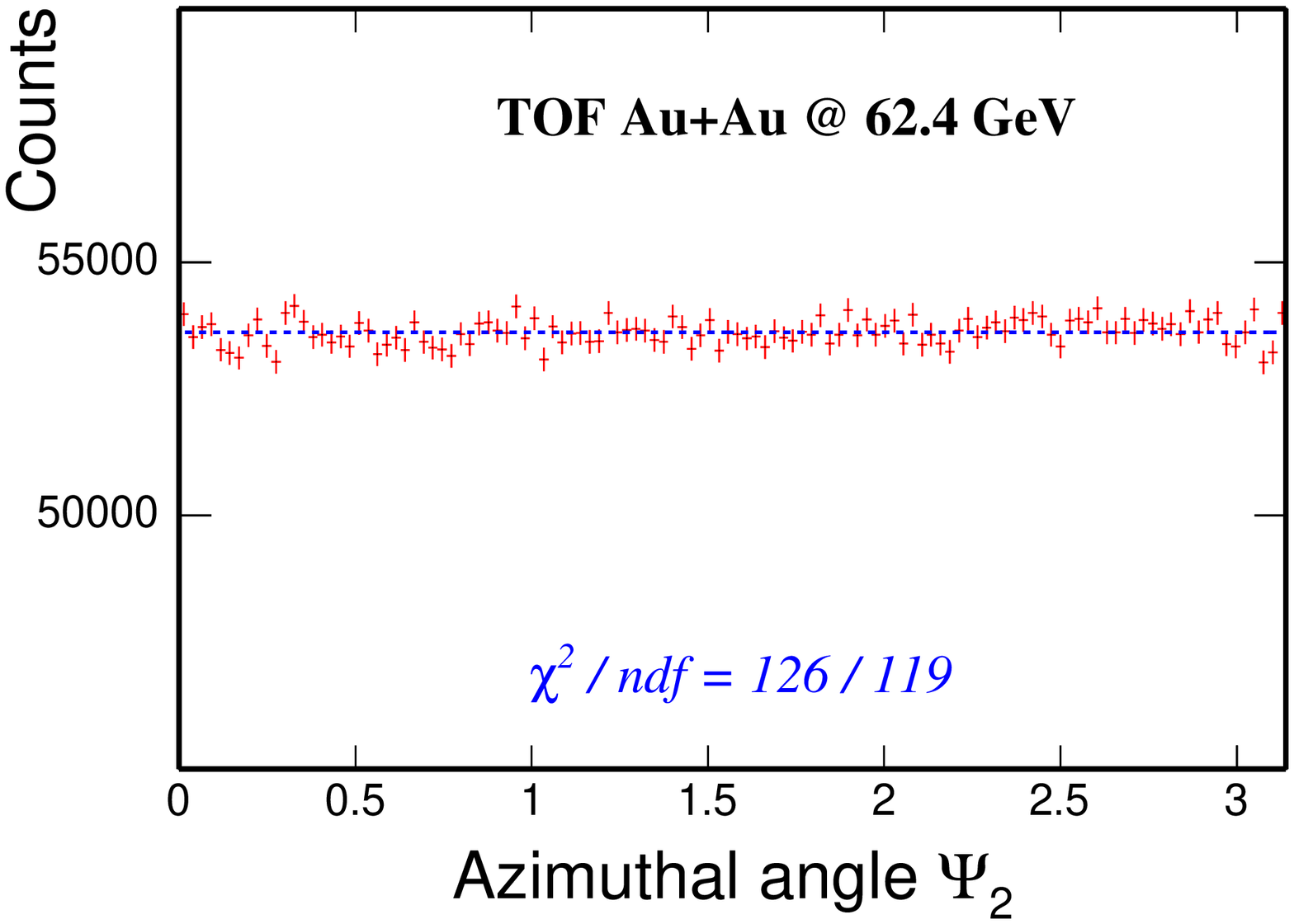}}
\caption[Event plane distribution]{The second harmonic event plane
azimuthal angle distribution. A constant fit with $\chi^2/ndf$
close to unity means a good event plane reconstruction.}
\label{ePlane} \ef

Firstly, we obtained the sub-event resolution $\la
\cos[2(\Psi_2^a-\Psi_{rp})]\ra$ from Eq.~\ref{subEvtResFunc}. Then
Eq.~\ref{evtResFunc} can be solved as an iterative routine to
extract the sub-event $\chi_2^a$. This variable is proportional to
$\sqrt{N}$ according to Eq.~\ref{chim}, so the total event
$\chi_2$ was obtained by $\chi_2^{}=\sqrt{2}\chi_2^a$. After
putting this $\chi_2^{}$ into Eq.~\ref{evtResFunc}, we calculated
the final full event resolution. The physical $v_2$ is calculated
as $v_{2} = v_{2}^{obs}/r$, where $v_2^{obs}$ is the observed
$v_2$ and $r$ is the event plane resolution. However,
experimentally, what we observe is $\langle v_{2}^{obs}\rangle$,
the averaged $v_2^{obs}$ over a data sample, such as the minimum
bias events. Then \be \langle v_2\rangle = \left\langle
\frac{v_2^{obs}}{r}\right\rangle \approx (but \neq) \frac{\langle
v_2^{obs}\rangle}{\langle r\rangle} \ee

This is not quite correct when we just divide the $\langle
v_2^{obs}\rangle$ by the event-wise averaged resolution $\langle
r\rangle$. Practically, we calculated a track-wise averaged
resolution by weighting each event with the number of observed
particles in the $v_2^{obs}$ calculation~\cite{v2obsres}.
Fig.~\ref{ePlaneRes} shows the event-wise averaged and track-wise
averaged event plane resolution for 9 centrality bins and minimum
bias ($0-80\%$) events. The final event plane resolution used for
correction is $68\%$ for minimum bias \AuAu 62.4 GeV collisions.

\bf \centering\mbox{
\includegraphics[width=0.6\textwidth]{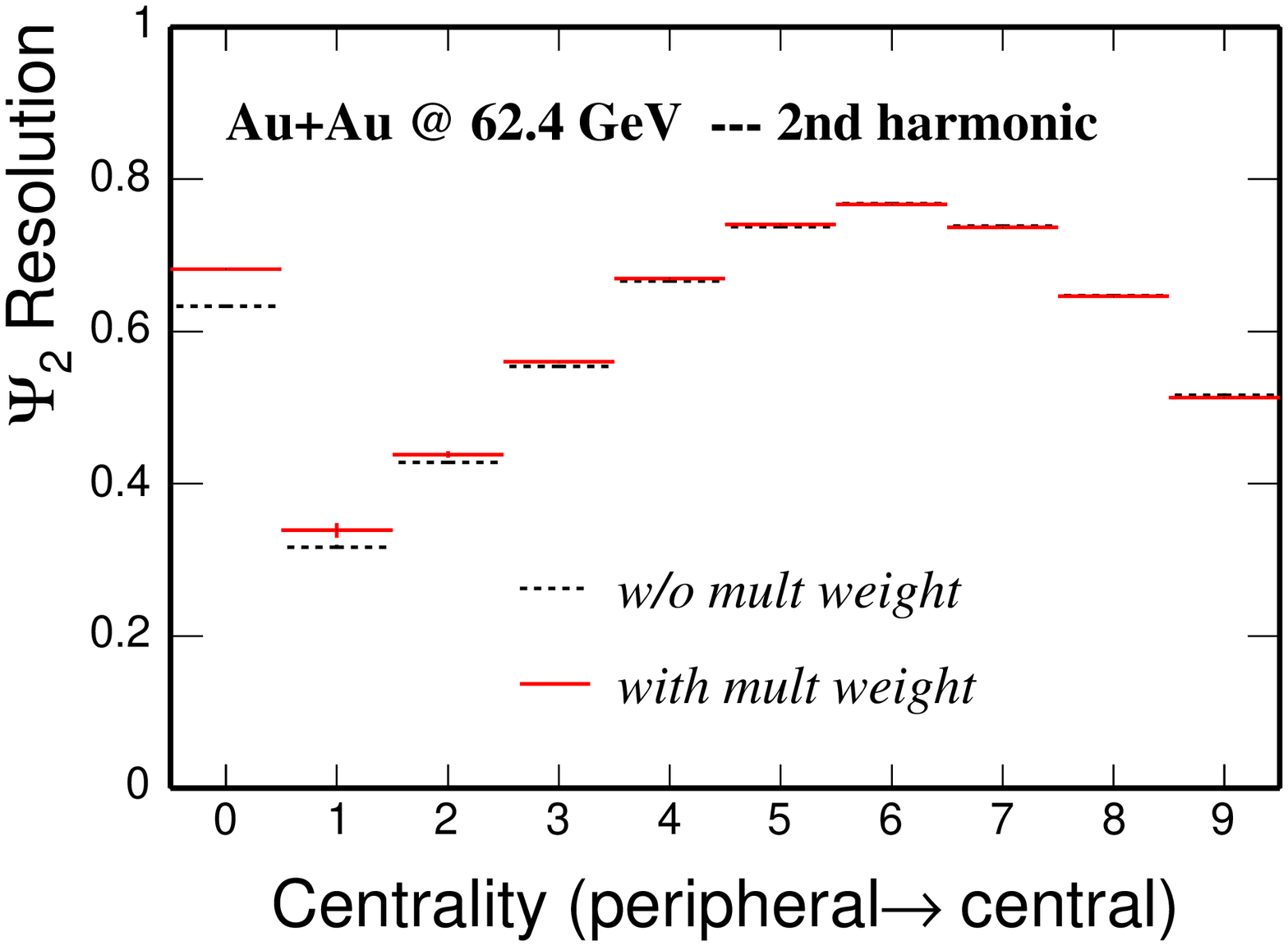}}
\caption[Event plane resolutions for all centrality bins]{Event
plane resolutions for all centrality bins (bin 1-9) and minimum
bias ($0-80\%$) events (bin 0). Black points show the resolution
with event weight only. Red points show the improved resolution
with multiplicity weight.} \label{ePlaneRes} \ef

\section{Elliptic flow of stable hadrons}

The TOF detectors were used for hadron identification in this
analysis. A multiple gaussian fit was applied to the $M^{2}$
distributions in each \pT bin. In those \pT bins where peaks start
to merge, a cut of more than $95\%$ purity was applied in the
particle selection. In the \pT region beyond the $\pi$ and $K$
separation with TOF only, $dE/dx$ was used additionally to help
identify pions while $K$ PID is limited by its low
yield~\cite{swingHQ}.

There is a certain correlation between the selected particle
$\phi_j$ and the event plane azimuth calculated including this
particle. This so-called auto correlation was removed by excluding
this selected particle in a new event plane azimuth calculation.
The new $\overrightarrow{Q_j}$ vector was constructed by:

\be \overrightarrow{Q_j} = (Q_{jx}, Q_{jy}) = \left(\sum_{i\neq
j}w_i\cdot \cos(2\phi_i), \hskip 0.2 in \sum_{i\neq j}w_i\cdot
\sin(2\phi_i)\right) \ee \be \Psi_{2j} =
\left(\tan^{-1}\frac{Q_{jy}}{Q_{jx}}\right)/2 \ee

Then the $v_{2}^{obs}$ in each \pT bin was calculated by \be
v_{2}^{obs} = \langle \cos[2(\phi_j - \Psi_{2j})]\rangle \ee

Fig.~\ref{v2pikp62} shows the measured resolution corrected $v_2$
of $\pi^{\pm}$, $K^{\pm}$ and $p$($\bar{p}$) from the TOF detector
for \AuAu 62.4 GeV minimum bias ($0-80\%$) collisions. The results
show: at low \pT region, the mass ordering is similar to what has
been observed in 200 GeV \AuAu collisions; and at intermediate \pT
( $>$ 2 GeV/c ) the proton $v_2$ overtakes pion $v_2$, which is
also similar to that in 200 GeV case. To check the NCQ scaling
that was observed in 200 GeV data, a function parametrization
Eq.~\ref{v2fitfun} was used to fit to the measured
$v_2$~\cite{minepiv2}.

\be v_2(p_T,n) = \frac{an}{1-\exp[-(p_T/n-b)/c]}-dn
\label{v2fitfun} \ee where $n$ is the number of constituent
quarks.

\bf \centering\mbox{
\includegraphics[width=0.6\textwidth]{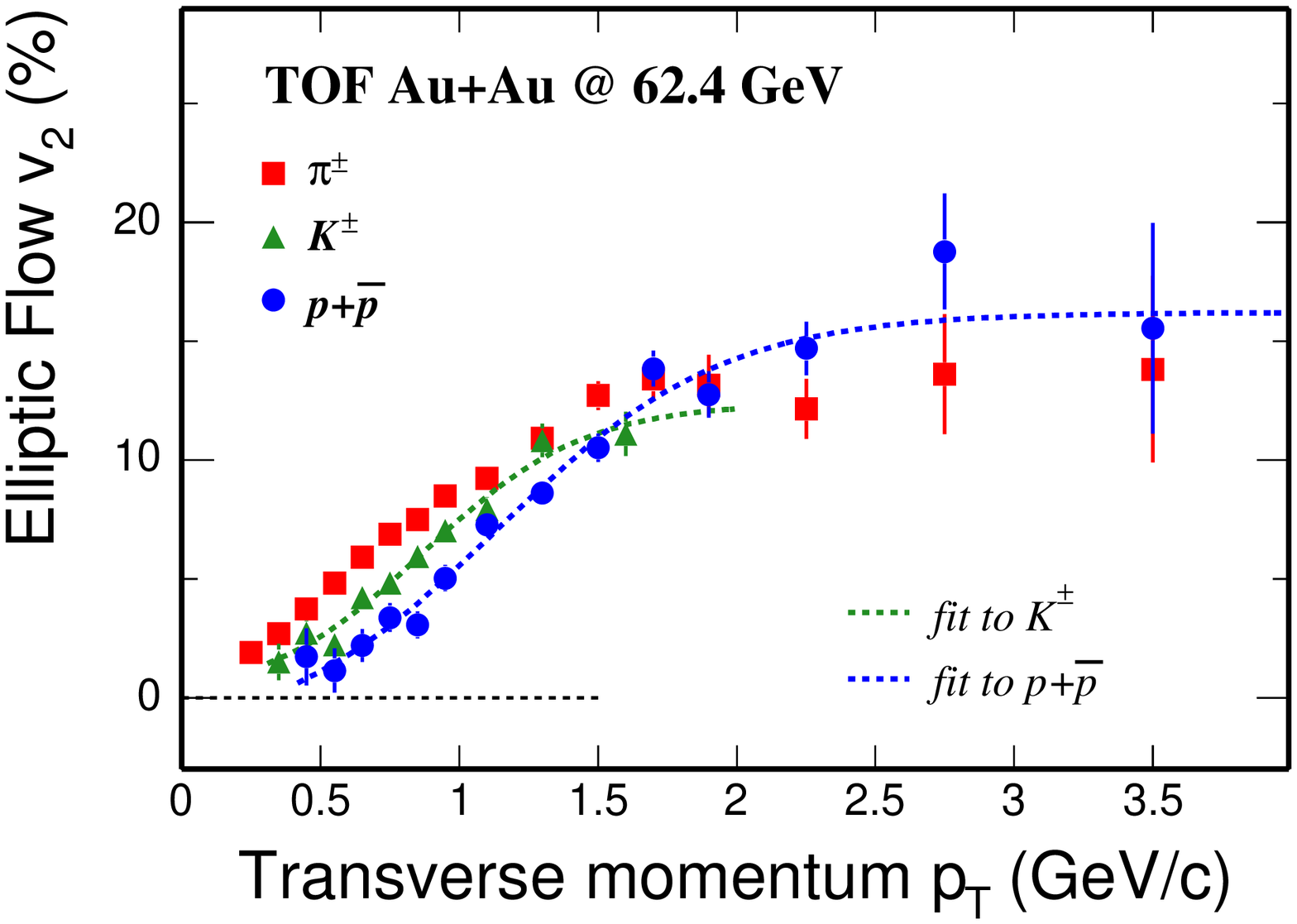}}
\caption[$v_{2}$ of $\pi^{\pm}$, $K^{\pm}$ and
$p$($\bar{p}$)]{Elliptic flow $v_2$ of $\pi^{\pm}$, $K^{\pm}$ and
$p$($\bar{p}$) from TOF detector at \AuAu 62.4 GeV minimum bias
($0-80\%$) collisions. Curves depict the parametrized function fit
from~\cite{minepiv2}.} \label{v2pikp62} \ef

\bf \centering\mbox{
\includegraphics[width=0.6\textwidth]{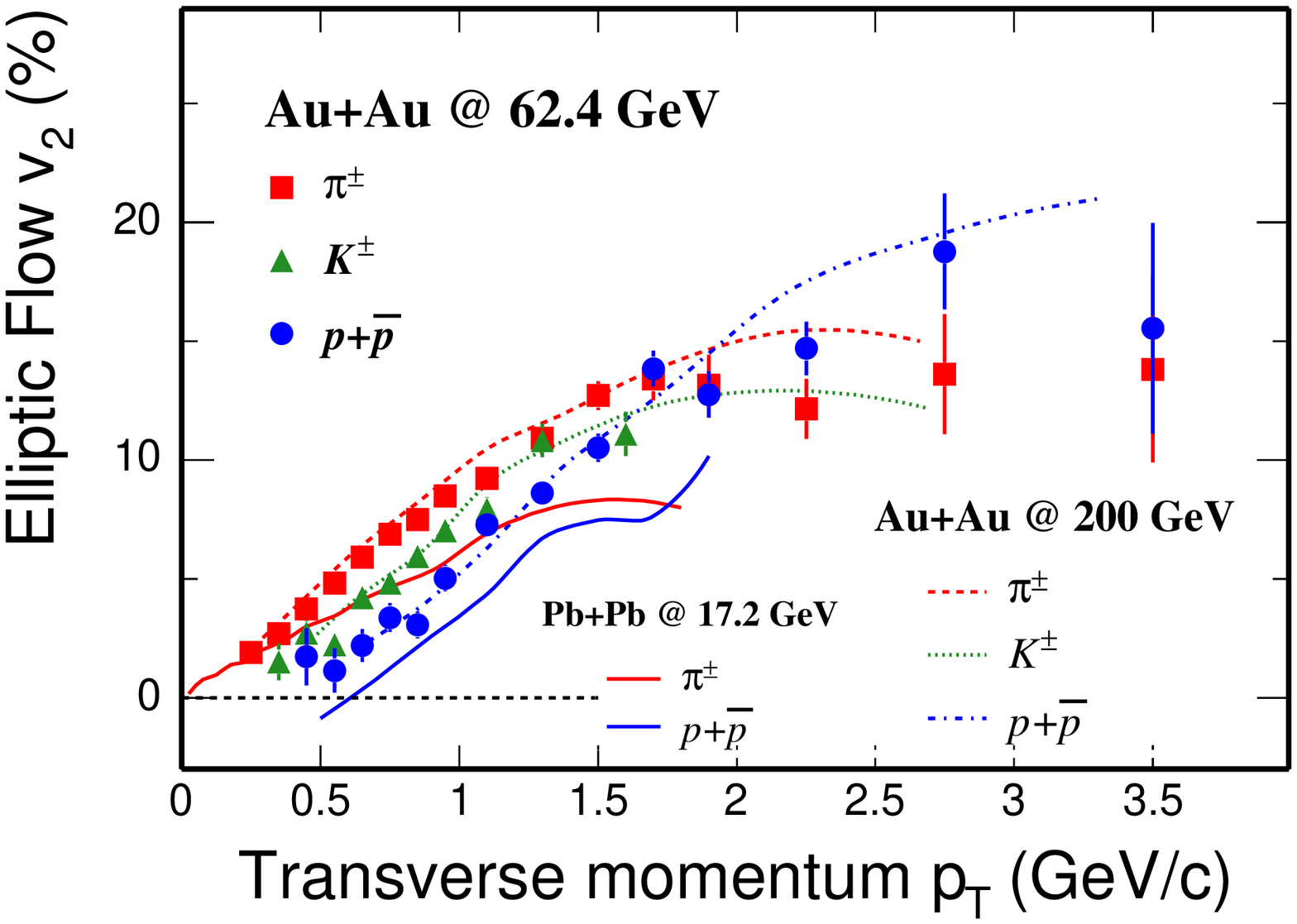}}
\caption[$v_{2}$ comparison with 200 GeV data]{Minimum bias \AuAu
62.4 GeV $v_{2}$ (data points) compared with those from 200 GeV
measurements in \AuAu collisions at RHIC (non-solid
lines)~\cite{phenixpikpv2} and 17.2 GeV measurements for pions and
protons (solid lines) in Pb + Pb collisions at
SPS~\cite{na49flow}.} \label{v2com200} \ef

The fit results are shown in Table.~\ref{v2fittable}. The
consistency of the fit parameters for $K^{\pm}$ and $p+\bar{p}$
indicates an agreement with NCQ scaling. This scaling in 62.4 GeV
was also further demonstrated by the measurements of $v_2$ of
$K_{S}^{0}$, $\Lambda$, $\Xi$ and $\Omega$~\cite{v2paper62}.

\begin{table}[hbt]
\caption{Parametrization for $v_2$ of $K^{\pm}$ and
$p+\bar{p}$}\label{v2fittable} \vskip 0.1 in
\centering\begin{tabular}{c|c|c} \hline \hline
     &  $K^{\pm}$  &   $p+\bar{p}$ \\ \hline
 a & $6.3 \pm 1.8$ &
$6.1 \pm 1.3$ \\ b & $0.44 \pm 0.04$ & $0.38 \pm 0.03$
\\
c & $0.14 \pm 0.06$ & $0.13 \pm 0.04$ \\
d & $0.05 \pm 1.05$  & $0.66 \pm 0.92$ \\
     \hline \hline
\end{tabular}
\end{table}

Comparisons of the identified particle $v_2$ from different energy
measurements are shown in Fig.~\ref{v2com200}. Those are
measurements from \sNN = 200 GeV \AuAu minimum bias ($0-92\%$)
collisions~\cite{phenixpikpv2} and \sNN = 17.2 GeV Pb + Pb minimum
bias collisions~\cite{na49flow}. The results show the $v_2$ from
62.4 GeV are very similar to those from 200 GeV measurement, while
they are significantly different from those in 17.2 GeV
measurement.

Charged pions can also be identified using the relativistic
$dE/dx$ (r$dE/dx$) method~\cite{xzbDPF} up to $\sim 7$ GeV/c
(limited by the statistics). A simple selection of
$0<\sigma_{\pi}<2$ was used to select a pion sample and the
contamination from kaons and protons was estimated to be less than
$3\%$~\cite{xzbDPF}. Fig.~\ref{pionv2all} shows the pion $v_2$
measured over a large \pT region ( $0.2-7.0$ GeV/c ). The event
plane method shows a continuous flat region up to almost 7 GeV/c.
However, the non flow effect in the high \pT region becomes
important and even more significant in central and peripheral
collisions~\cite{nonFlow}. To answer the question of high \pT
identified particle $v_2$, we need other methods (such as
multiparticle cumulant methods {\em etc}~\cite{v2CorrPRL}) to
minimize the non flow effects.

\bf \centering\mbox{
\includegraphics[width=0.6\textwidth]{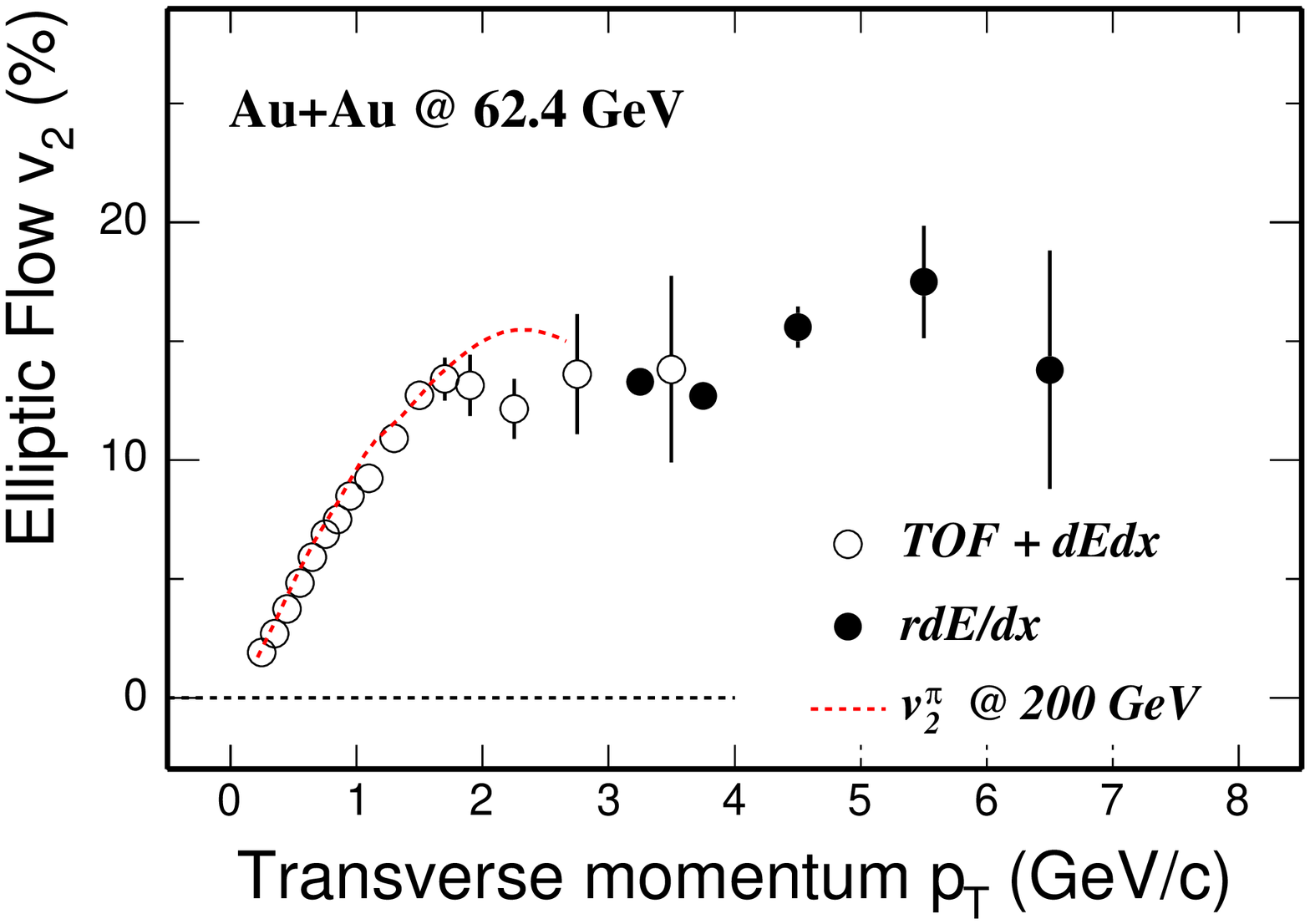}}
\caption[Pion $v_2$ in large \pT region]{Pion $v_2$ from TOF, TOF
+ $dEdx$ and relativistic $dE/dx$ (r$dE/dx$) measurements in \AuAu
62.4 GeV.} \label{pionv2all} \ef

\section{Elliptic flow of inclusive and photonic background
electrons}

The electron PID capability in \AuAu 62.4 GeV was already shown in
Fig.~\ref{ePIDAuAu}. With the selection of $0<\sigma_{e}<3$, we
identify electrons with the purity more than $90\%$ in $1-3$
GeV/c. The inclusive electron $v_2$ was extracted using a method
similar to that used for $v_2$ of $\pi^{\pm}$, $K^{\pm}$ and
$p$($\bar{p}$). Fig.~\ref{epiv2} shows the results compared with
$v_2$ of $\pi^{\pm}$. Since the dominant source of the electrons
are photon conversion where the photons are from two photon
$\pi^{0}$ decay processes, and $\pi^{0}$ Dalitz decays, the
electrons at lower \pT will carry the anisotropy of those parent
pions at higher $p_T$. This will be discussed in the next chapter.

\bf \centering\mbox{
\includegraphics[width=0.6\textwidth]{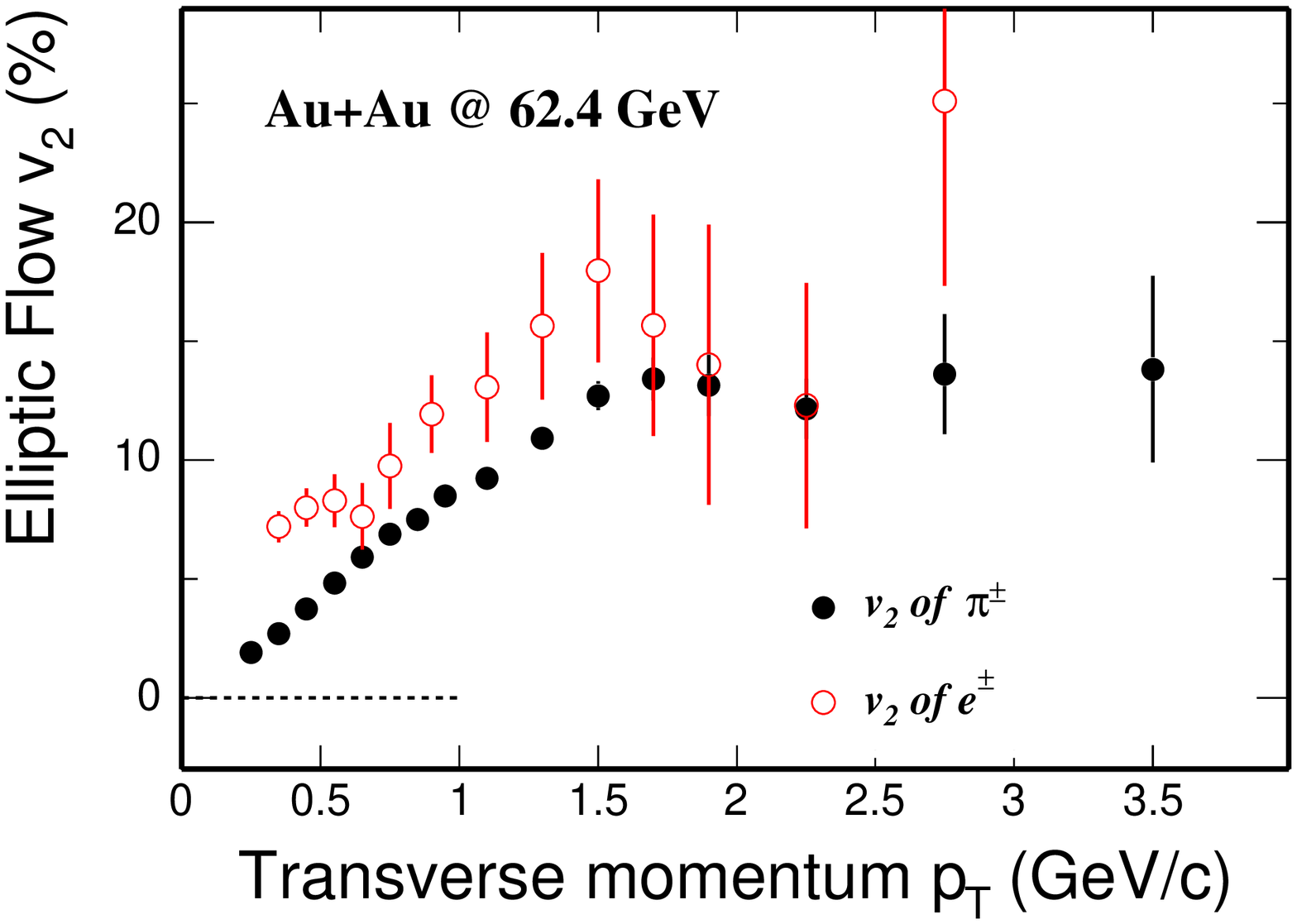}}
\caption[$v_2$ of inclusive electrons and pions]{$v_2$ of
inclusive electrons and pions from TOF detector in \AuAu 62.4 GeV
minimum bias ($0-80\%$) collisions.} \label{epiv2} \ef

Photonic electron $v_2$ is needed to extract the interesting
non-photonic electron $v_2$. We used the event mixing technique to
reconstruct the main photonic background from the TPC only. The
tagged and partner electron candidates are selected using the same
selection criteria shown in Table.~\ref{eTrackAuAu} and
~\ref{ePartnerAuAu}, but we used all TPC candidates instead of TOF
tracks only. Both tagged and partner track are required to have
$0<\sigma_{e}<3$ to reject hadrons. The event buffer used for
mixing is divided into a $10\times 10\times 9$ lattice in
($V_{z}$, Mult, $\Psi_2$) with lattice size of $6$ cm $\times$ 60
$\times$ 20$^{\text{o}}$. Two 3-D ($M_{e^+e^-}$, \pT,
$\Delta\phi=\phi-\Psi_2$) histograms for the same event and
mixed-event were stored. The additional dimension in $\Psi_2$ for
event mixing is to control the residual background in different
$\Delta\phi$ bins. The histograms were then projected onto the
$M_{e^+e^-}$ axis for each ($p_T$, $\Delta\phi$) bin. The mixed
event distribution, which is expected to describe the
combinatorial contribution, was normalized to the same event
distribution in the high $M_{e^+e^-}$ region, where there are no
significant decay correlations. The normalized mixed event
distribution was then subtracted from the same event distribution.
The photonic electrons were extracted for every $\Delta\phi$ bin
in each \pT bin. The $\Delta\phi$ distribution was fit to the
following function to get the observed $v_2^{obs}$.

\be \frac{dN}{d(\Delta\phi)}\propto 1 +
2v_2^{obs}\cos(2\Delta\phi) \label{dNdphi} \ee

Fig.~\ref{bkgdflow} shows the $M_{e^+e^-}$ distribution of the
electron candidate pairs from the same event (crosses) and the
mixed event (histograms) in different $\Delta\phi$ bin for tagged
electrons $2.0<p_T$/(GeV/c)$<2.5$. The combinatorial background
shows oscillation in $\Delta\phi$, which is expected from the
anisotropy from final state hadrons. After mixed combinatorial
background subtracted, the photonic source electron anisotropy is
shown in Fig.~\ref{bkgdflow_sub}. Since the photon conversion
($\pi^0$ Dalitz continuum may hide in) distribution is not a
gaussian-like peak, it is hard to do a signal+background fit as we
did for other resonances to extract the signal yields. We need to
understand the residual background well. So in the yield
extraction in each $\Delta\phi$ bin, we tried several residual
background estimations to extract the systematic errors: (i) no
residual. (ii) use a constant fit in
$0.15<M_{e^+e^-}$/(GeV/c$^2$)$<0.25$ and take it as a residual.
(iii) the same as (ii), but take the averaged value through all
$\Delta\phi$ bins as a common residual. Fig.~\ref{bkgdv2fit} shows
the yields from each $\Delta\phi$ bin using method (iii).
Eq.~\ref{dNdphi} was used to fit the distribution and the result
is $v_2^{obs} = (13.4\pm4.4)\%$. After resolution correction, we
can get the $v_2$ of electrons from photonic sources.
Fig.~\ref{epiv2all} shows the photonic electron $v_2$ as a
function of $p_T$, compared with the inclusive electron $v_2$ and
pion $v_2$. The photonic electron $v_2$ is consistent with that of
inclusive electrons, which is expected from the spectra analysis
in the previous chapter.

In general, the $v_2$ of inclusive electrons is the sum of
photonic (B) and non-photonic (S) electron $v_2$, weighted by each
yield fraction $f_{B}$, $f_{S}$. \be v_2^{tot} = f_{B}\times
v_2^{B} + f_{S}\times v_2^{S} \ee

\bf \centering\mbox{
\includegraphics[width=0.8\textwidth]{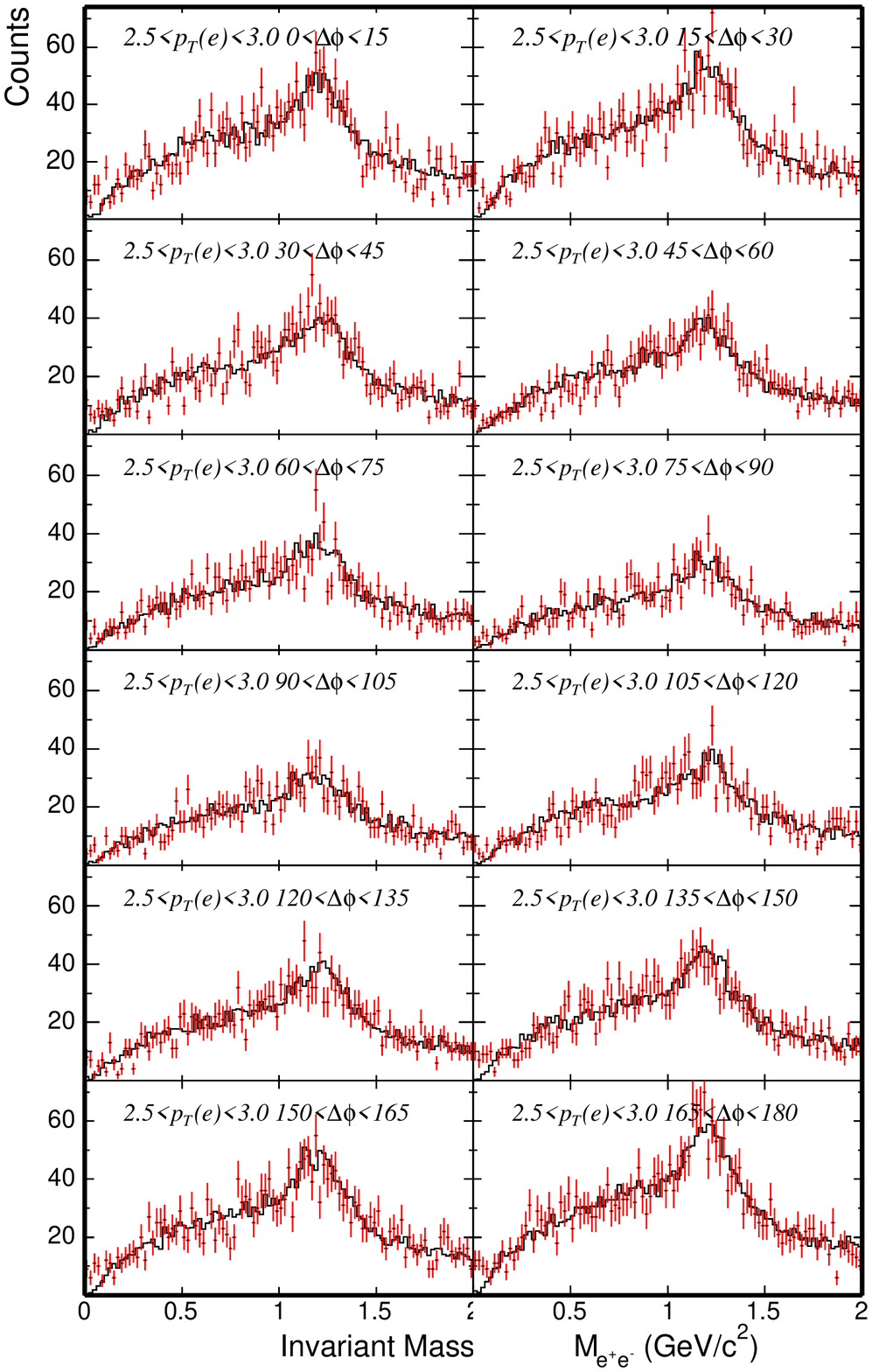}}
\caption[$M_{e^{+}e^{-}}$ distribution in every $\Delta\phi$
bin]{Invariant mass $M_{e^{+}e^{-}}$ distributions of electron
pair candidates and the combinatorial background from rotating
method in each $\Delta\phi$ bin for tagged electron
$2.0<p_T$/(GeV/c)$<2.5$.} \label{bkgdflow} \ef

\bf \centering\mbox{
\includegraphics[width=0.8\textwidth]{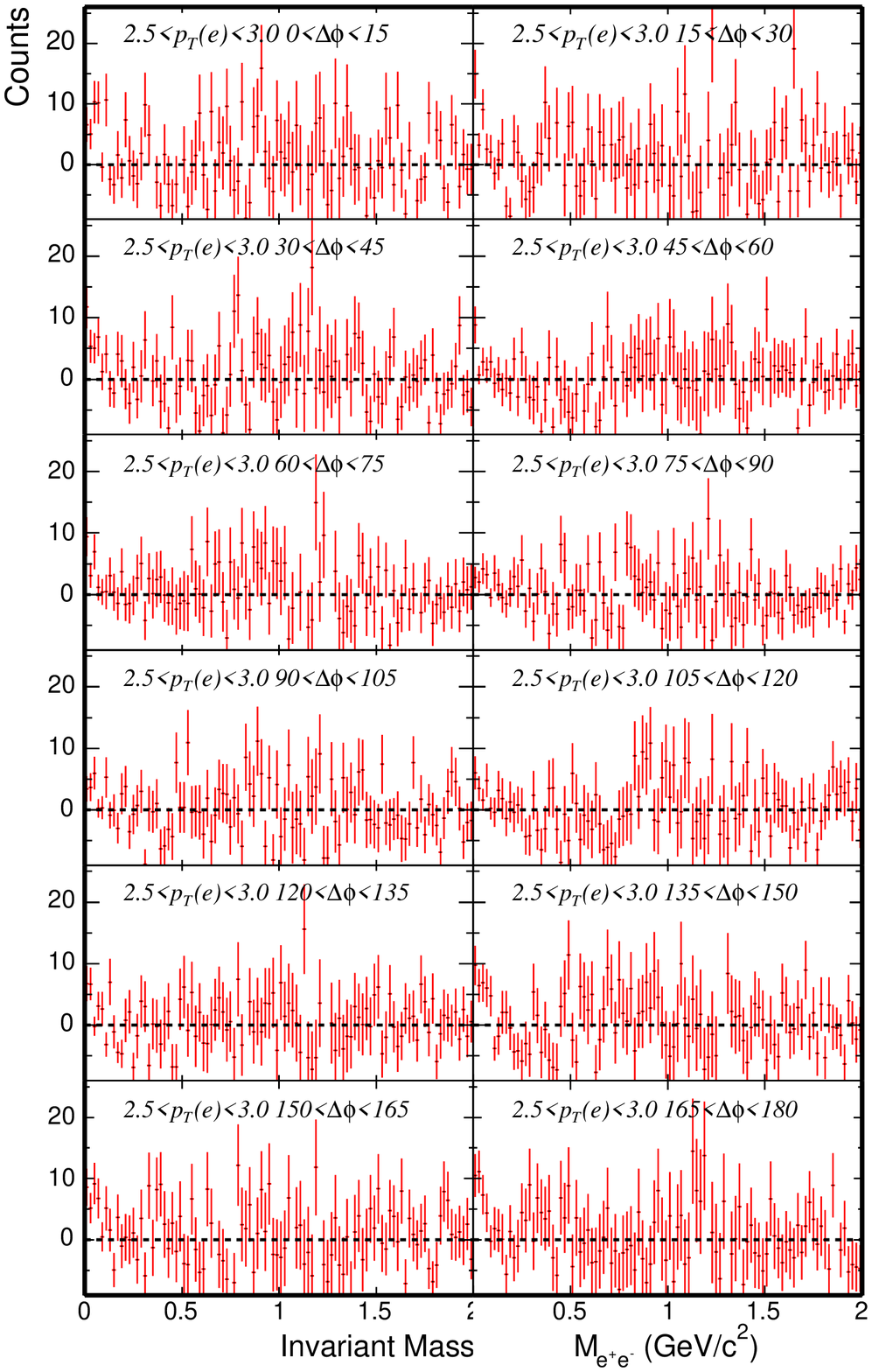}}
\caption[Subtracted $M_{e^{+}e^{-}}$ distribution in every
$\Delta\phi$ bin]{Invariant mass $M_{e^{+}e^{-}}$ distribution of
combinatorial background subtracted electron pair candidates in
each $\Delta\phi$.} \label{bkgdflow_sub} \ef

\bf \centering\mbox{
\includegraphics[width=0.6\textwidth]{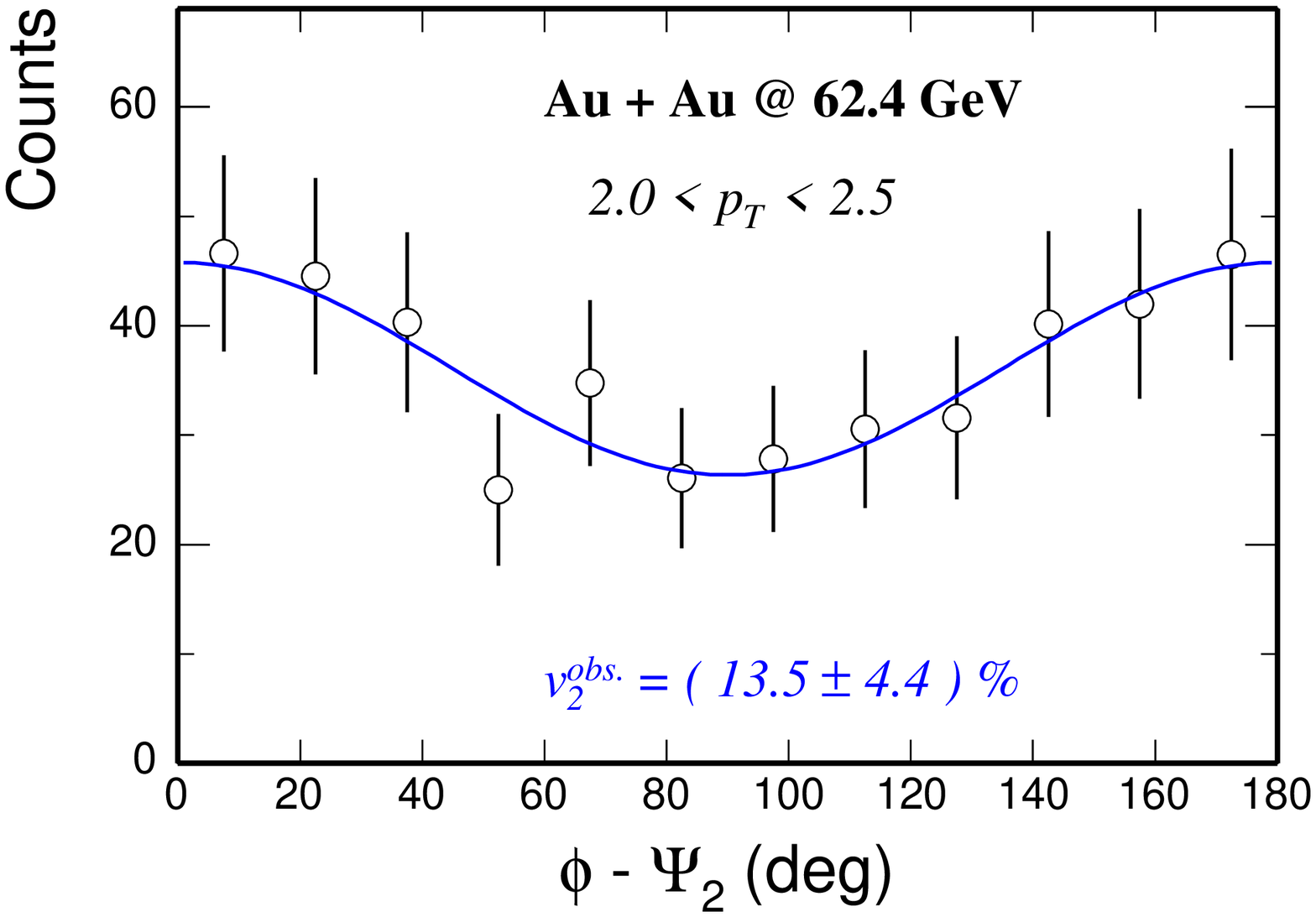}}
\caption[$\Delta\phi$ distribution of photonic
electrons]{$\Delta\phi$ distribution of photonic electrons in
$2.0<p_{T}$/(GeV/c)$<2.5$. The distribution was fit to
$[0]\times(1+2v_{2}^{obs.}\cos(2\Delta\phi))$ to extract the
observed $v_{2}^{obs.}$.} \label{bkgdv2fit} \ef

\bf \centering\mbox{
\includegraphics[width=0.6\textwidth]{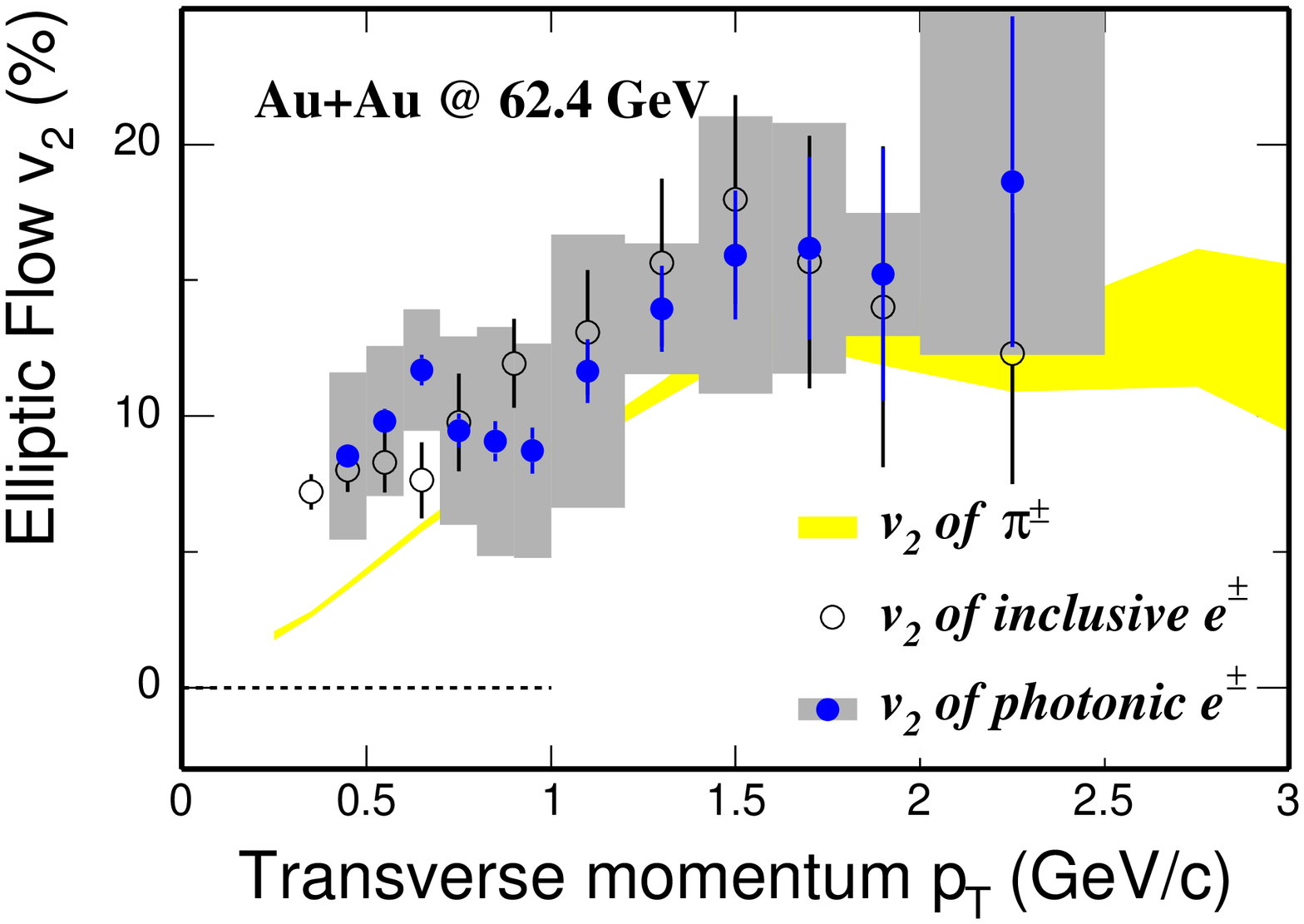}}
\caption[Electron $v_{2}$ result]{Inclusive electrons $v_2$ (from
TOF trays) and photonic electrons $v_2$ (from the TPC) compared
with charged $\pi^{\pm}$ $v_2$. The grey bands in each \pT bin for
photonic electrons $v_2$ depict the systematic uncertainties from
the combinatorial background estimation.} \label{epiv2all} \ef

So the $v_2^{S}$, the elliptic flow of non-photonic electrons, can
be calculated as \be v_2^{S} = \frac{v_2^{tot}-f_{B}\times
v_2^{B}}{f_{S}} \ee

From above formula, the ratio of S/B is needed in this analysis.
So the spectra analysis is necessary for the $v_2$ of non-photonic
electrons.

A recent technique was developed to deal with such signal $v_2$
extraction from background~\cite{v2obsres}. We can get the $v_2$
of inclusive electrons in different $M_{e^+e^-}$ bins $-$
$v_2(M)$. And if we know the signal to background ration $r(M)$,
then we can use two free parameters $v_2^{S}$ and $v_2^{B}$ to fit
the $v_2(M)$. This statistical fit method looks promising in the
$v_2$ calculation for V$_0$ particles, and $\pi^{\pm}$,
$p$,$\bar{p}$ in $dE/dx$ relativistical rising region. So this
method is feasible for the extraction of photonic electrons $v_2$
out of the mixture with combinatorial background, and also that of
non-photonic electrons $v_2$ out of inclusive electrons.

\chapter{Discussion}

\section{Open charm production in high energy collisions}

\subsection{Total charm cross section}
As we discussed in the introduction chapter, the heavy quark total
cross section measurement offers a powerful test for pQCD
calculations. Since plenty of measurements were made at low
energies while few were made at \s $>$ 100 GeV, theoretical
calculations are often tuned to match the low energy data points
and then extrapolated to high energies. Because the parameters in
the calculations (scales, heavy quark mass {\em etc.}) are not
understood well yet in all energies, the predictions at high
energies differ significantly. Fig.~\ref{XsecEnergy} shows almost
all the total charm cross section measurements made so far
compared with several theoretical calculations.

\bf \centering\mbox{
\includegraphics[width=0.6\textwidth]{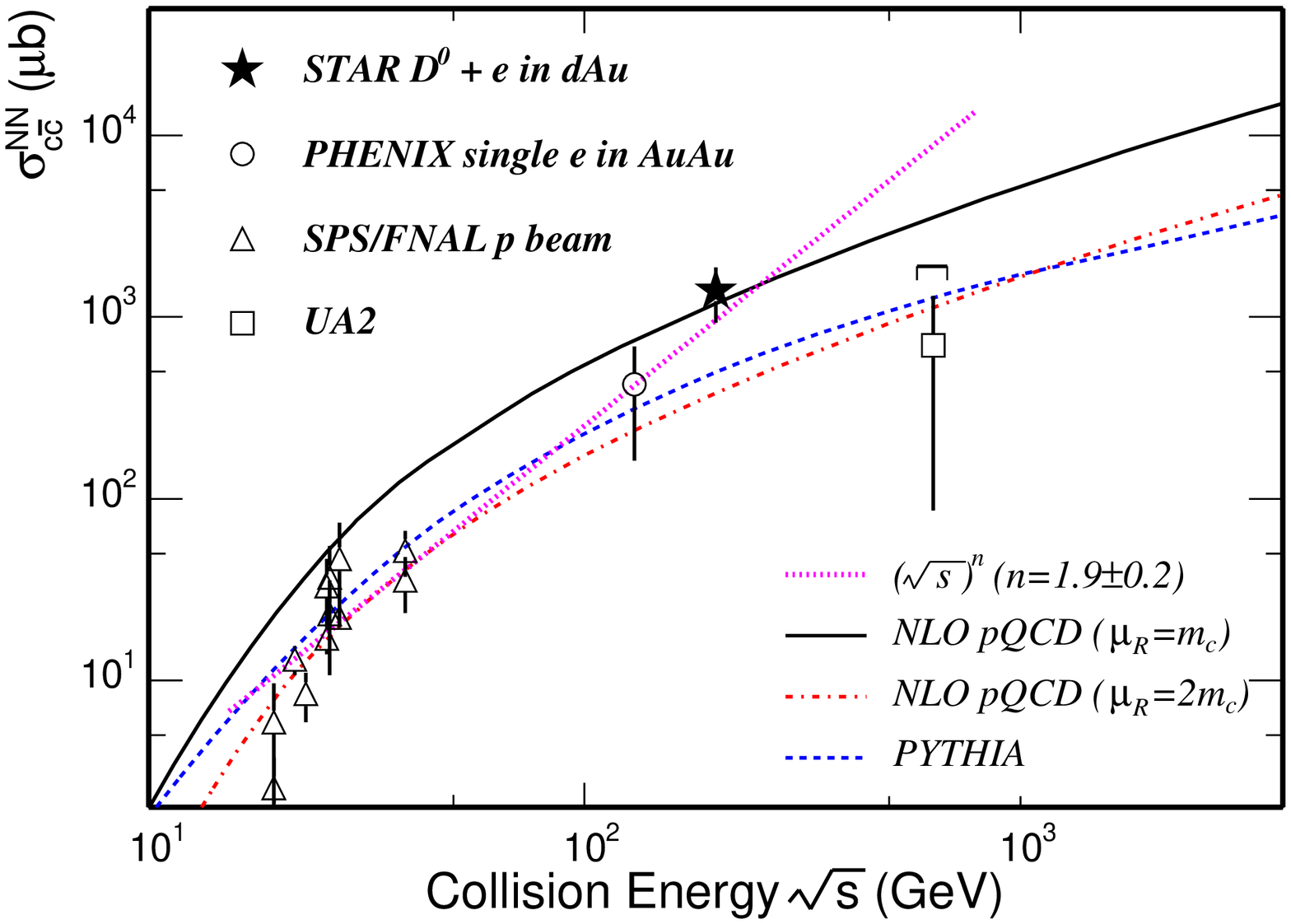}}
\caption[Total charm cross section]{Total $c\bar{c}$ cross section
per nucleon-nucleon collision vs.~the collision energy (\s). The
dashed line depicts a PYTHIA calculation with a specific set of
parameters~\cite{pythia}. The solid and dot-dashed lines depict
two NLO pQCD calculations with MRST HO, $m_{c}$ = 1.2 GeV/$c^{2}$,
$\mu_{F}=2m_{c}$ and specified $\mu_{R}$ shown on the
plot~\cite{vogtXsec}. The dotted line (in pink) depicts a power
law fit to the data points, with power $n\sim2$.}
\label{XsecEnergy} \ef

The low energy (\s $<$100 GeV) data points were taken from the
review paper~\cite{charmreview} (which summarized all the charm
measurements before 1987) and from several experiments after 1988
( refer to~\cite{phenix130e} for data points)\footnote{Detailed
data points selections are listed in Appendix C.}. A factor of 1.5
was applied on all these data points to include the contribution
of $D_{s}^{\pm}$ and $\Lambda_{c}^{\pm}$~\cite{heavyReview}. We do
not include total cross sections from those references that were
extrapolated from high $x_F$ and/or had extremely low efficiency
from correlation measurements.(with extrapolation factor of $\gg
10$). The PHENIX and UA2 experiments extracted the cross section
from the single electron spectrum in 130 GeV \AuAu and 630 GeV
\ppbar collisions, respectively.

The lines depict different theoretical calculations and model
fits. PYTHIA, a model based on pQCD calculations, is often used
for predictions. The dashed line depicts a typical pQCD
calculation with the only change of PDF from the default PYTHIA
(CTEQ5L$\rar$CTEQ5M1). The result underpredicts the measurements
at RHIC. Although the default PYTHIA takes initial radiation into
account, high order processes seem to still not be well predicted.
These include initial and final radiation, gluon splitting, and
production through parton showers {\em etc.}. Recent analysis for
CDF energy open beauty measurements showed these processes
contribute a large fraction in the heavy flavor creation at
Tevatron~\cite{cdfHeavy}. This discrepancy indicates these
processes may also play an important role at RHIC energy.

The two curves labeled ``NLO pQCD'' depict the full NLO pQCD
calculations from~\cite{vogtXsec,vogtPrivate}. ``Full'' means for
both PDFs and differential cross sections. The red dot-dashed line
shows a calculation tuned to the low energy data points. This
calculation underpredicts our measurements at 200 GeV as well. The
scale parameters used are $m_{c}$ = 1.2 GeV/$c^{2}$,
$\mu_{F}=\mu_{R}=2m_{c}$. When the NLO pQCD calculation is carried
out with $\mu_R=m_c$ instead of $2m_c$, it seems to reproduce the
measurement at 200 GeV, but misses other data points. An even
smaller scale ($\mu_{R}=m_{c}/2$) was tried, but it overshoots our
measurement by almost an order of magnitude. The K-factors in pQCD
calculations also differ significantly when changing the
scales~\cite{vogtXsec,vogtPrivate}. As we know, when the scales go
to $\sim600$ MeV, which is close to $\Lambda_{QCD}$, the
feasibility of pQCD is doubtful. The discrepancy indicates these
scales can be energy dependent.

On the other hand, recent $\pi^0$ spectrum measurements at both
mid-rapidity~\cite{phenixpi0} and forward rapidity~\cite{starpi0}
at RHIC seem to offer clear evidence that pQCD can reproduce the
$\pi^0$ spectrum very well. This agreement even reaches as low as
$p_T\sim1$ GeV/c. Because of the large $Q^2$, one might naturally
think the agreement for heavy flavor production should be better.
However, our measurement indicates a negative result, which means
there can be some differences in the calculations between light
flavor hadrons and heavy flavor hadrons.

We fit a power law function to the data points in
Fig.~\ref{XsecEnergy} and the result is depicted as a pink dotted
line with the power $n\sim1.9\pm0.2$. This power law dependence
has already been proposed in ~\cite{cosmic} and they gave the
power $n\sim 1.6$. The power law dependence of charged hadron
production ($n\sim 0.3$) has been observed in A + A collisions
from AGS to RHIC~\cite{pbmjpsi} and also predicted in the
saturation approach~\cite{saturation}. Also for pion production, a
linear scaling as a function of
$F\equiv(\sqrt{s_{NN}}-2m_{N})^{3/4}/\sqrt{s_{NN}^{1/4}}\approx\sqrt{\sqrt{s_{NN}}}$
was observed~\cite{na49pi} in central A + A collisions, which is
equivalent to a power law dependence of $\sqrt{s_{NN}}$ with
$n\sim 0.5$. For the kaons that have strange valence quarks,
similar scaling seems to hold from AGS to RHIC, but with a larger
power value $n\sim 0.6-1$. Hence, this power scaling for charm
quark production up to RHIC is not surprising. One interesting
thing is that heavier quark production has a larger value of
power, which may be caused by a threshold effect in the
production: once the production channels are opened, or above and
near production threshold, the total cross section distribution on
different flavors is likely to favor heavier quarks.

Fig.~\ref{dsigmady} shows $d\sigma/dy$ compared to theoretical
predictions with smaller systematic uncertainties. This figure
shows clearly that the theoretical calculations underpredict out
measurement.

\bf \centering\mbox{
\includegraphics[width=0.6\textwidth]{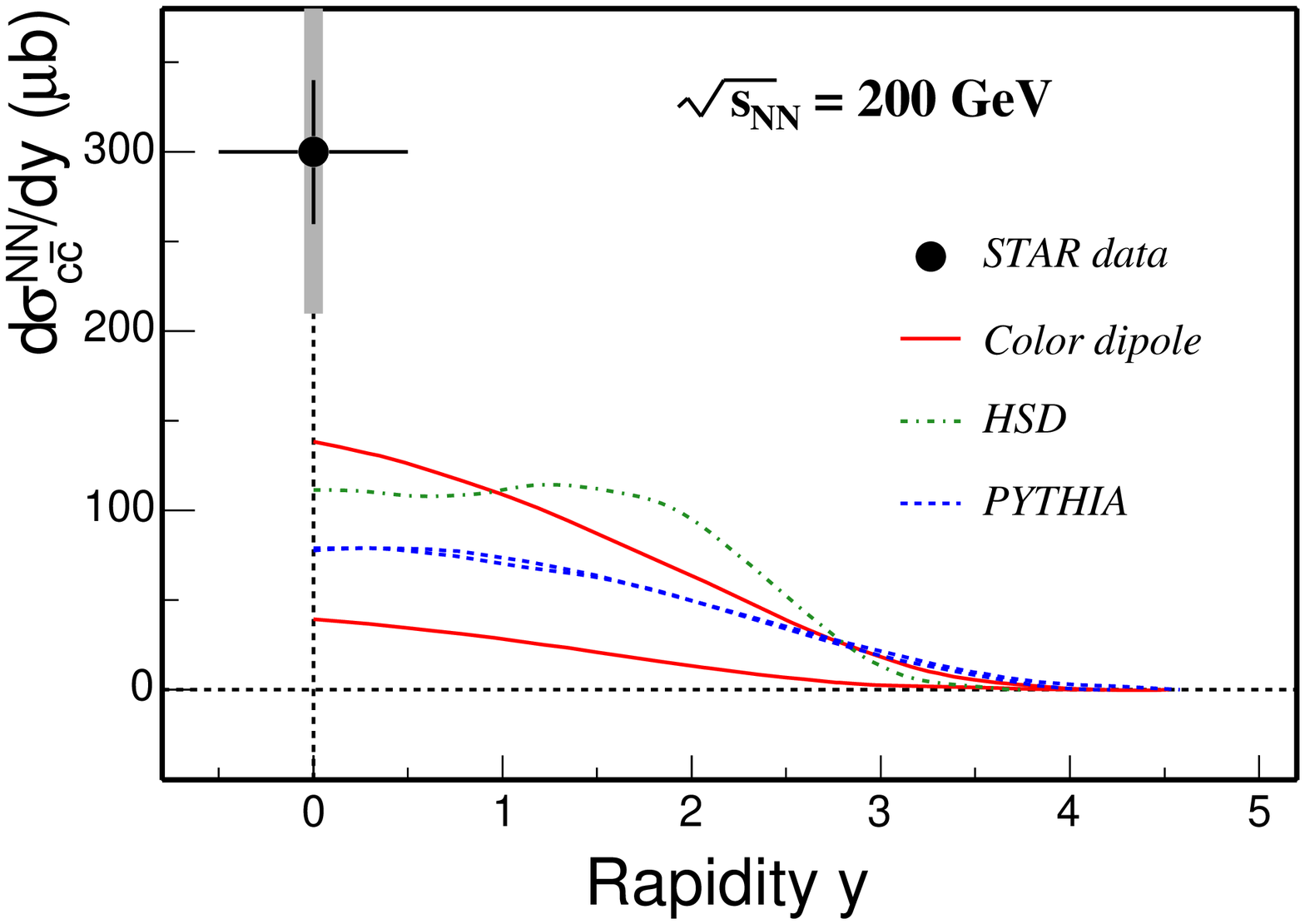}}
\caption[Charm differential cross section at
mid-rapidity]{$d\sigma/dy$ of charm quarks from STAR measurement
compared with different theoretical
predictions~\cite{vogtPrivate,raufeisenXsec,HSD}.}
\label{dsigmady} \ef

\subsection{Spectrum comparison}
Open charm hadrons, including $D^0$($K^-\pi^+$), $D^{\star
+}$($D^0\pi^+$), $D^+$($K^-\pi^+\pi^+$) and $D^0$($K^-\pi^+\rho$),
were reconstructed directly in \dAu collisions (the measurements
were statistically limited in \pp). A consistency check between
the open charm spectrum and its expected semi-leptonic decay
electron spectrum were made. The PYTHIA function for particle
decay was used for generating the electron spectrum. The input
open charm spectrum was fit to a power law function, and the
parameters are~\cite{AnQM04}: $dN/dy=0.0265\pm0.0035$, $\langle
p_{T}\rangle=1.32\pm0.08$, and $n=8.3\pm1.2$.
Fig.~\ref{spectraCom} shows the electron spectra compared with the
expected contribution from charm quark semi-leptonic decays. The
bands include the uncertainties from those parameters. The
comparison demonstrates a consistency between the two
measurements.

\bf \centering\mbox{
\includegraphics[width=0.6\textwidth]{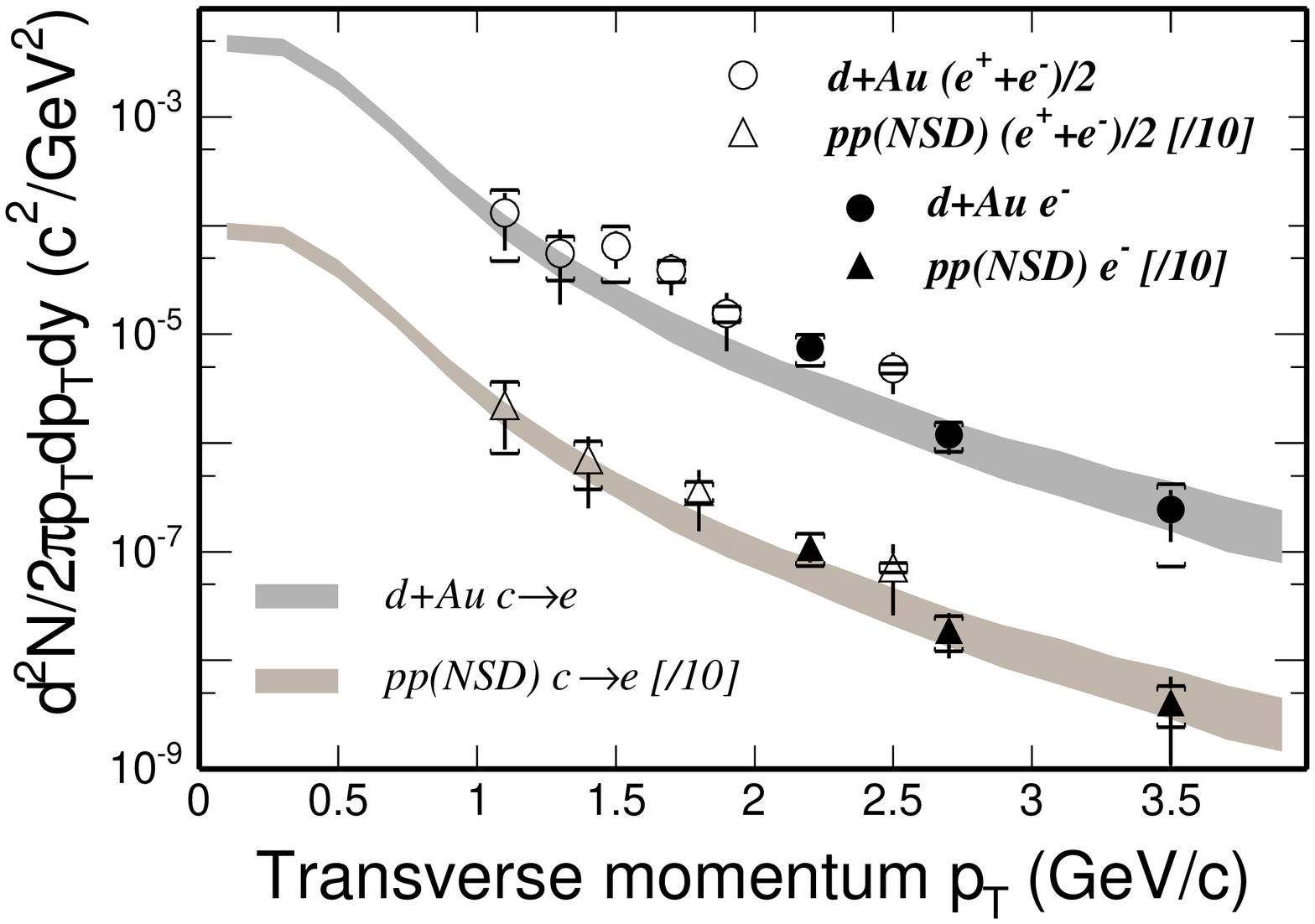}}
\caption[Spectra consistency check]{Non-photonic electron spectra
compared with the expected ones from the semi-leptonic decays of
measurement open charm spectrum.} \label{spectraCom} \ef

There are already several electron spectra from charm semileptonic
decay measurements. In Fig.~\ref{spectraAllEnergy} they are shown.

\bf \centering\mbox{
\includegraphics[width=0.6\textwidth]{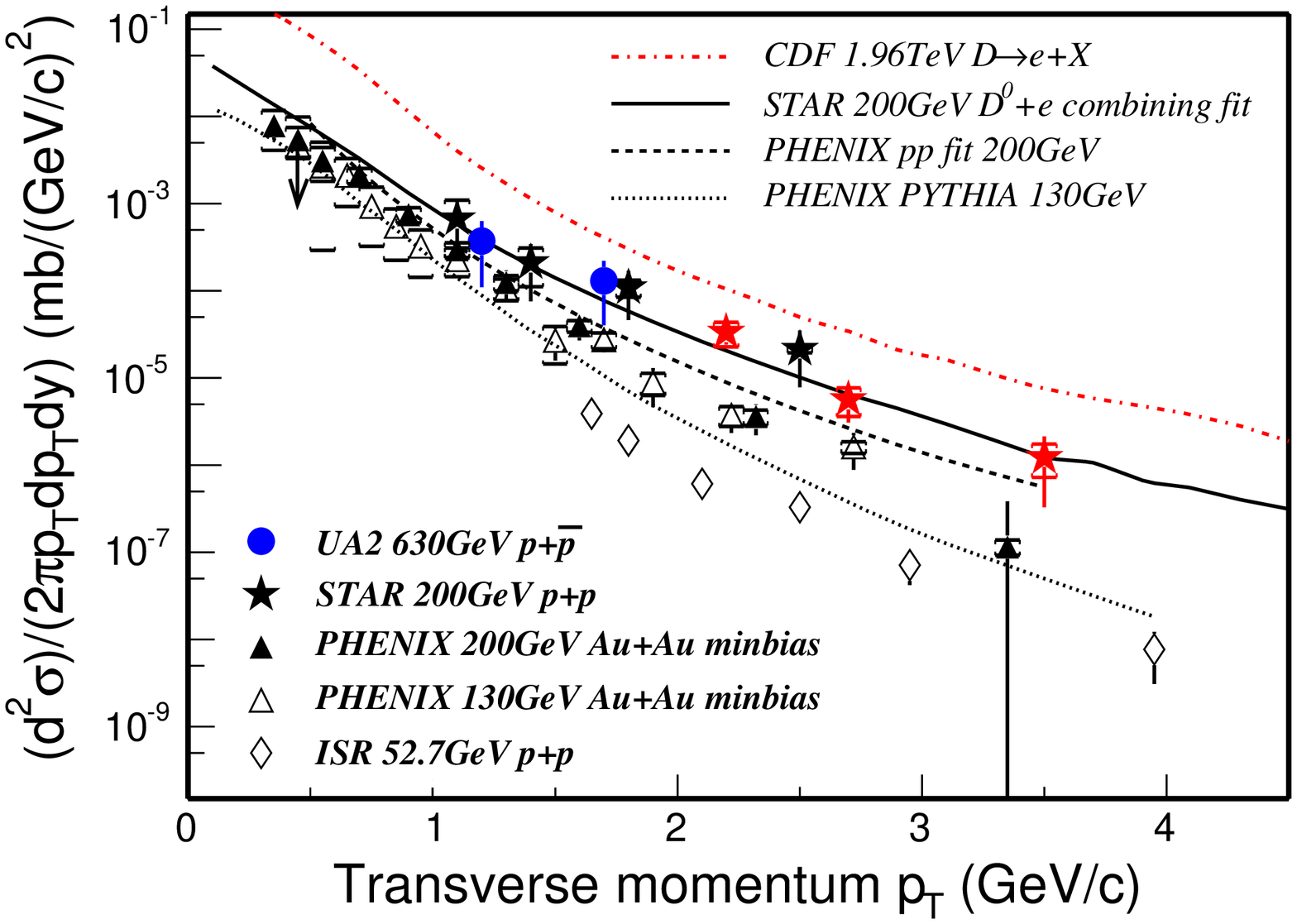}}
\caption[Electron spectra from different energies]{Non-photonic
electron spectra from several collision energies and collision
systems. Data points are taken
from~\cite{ISRe,phenix130e,phenix200e,UA2}. Curves are expected
spectra from models or measured open charm spectrum.}
\label{spectraAllEnergy} \ef

The comparison shows a continuous increase in the yields with the
increase of beam energies. However, from RHIC 130 GeV to UA2 630
GeV, the errors on the data points conceal the detailed structure.
Not only the total yield, but also the shape of the spectra seem
to become harder with the increase of
collision energy. 

\bf \centering\mbox{
\includegraphics[width=0.6\textwidth]{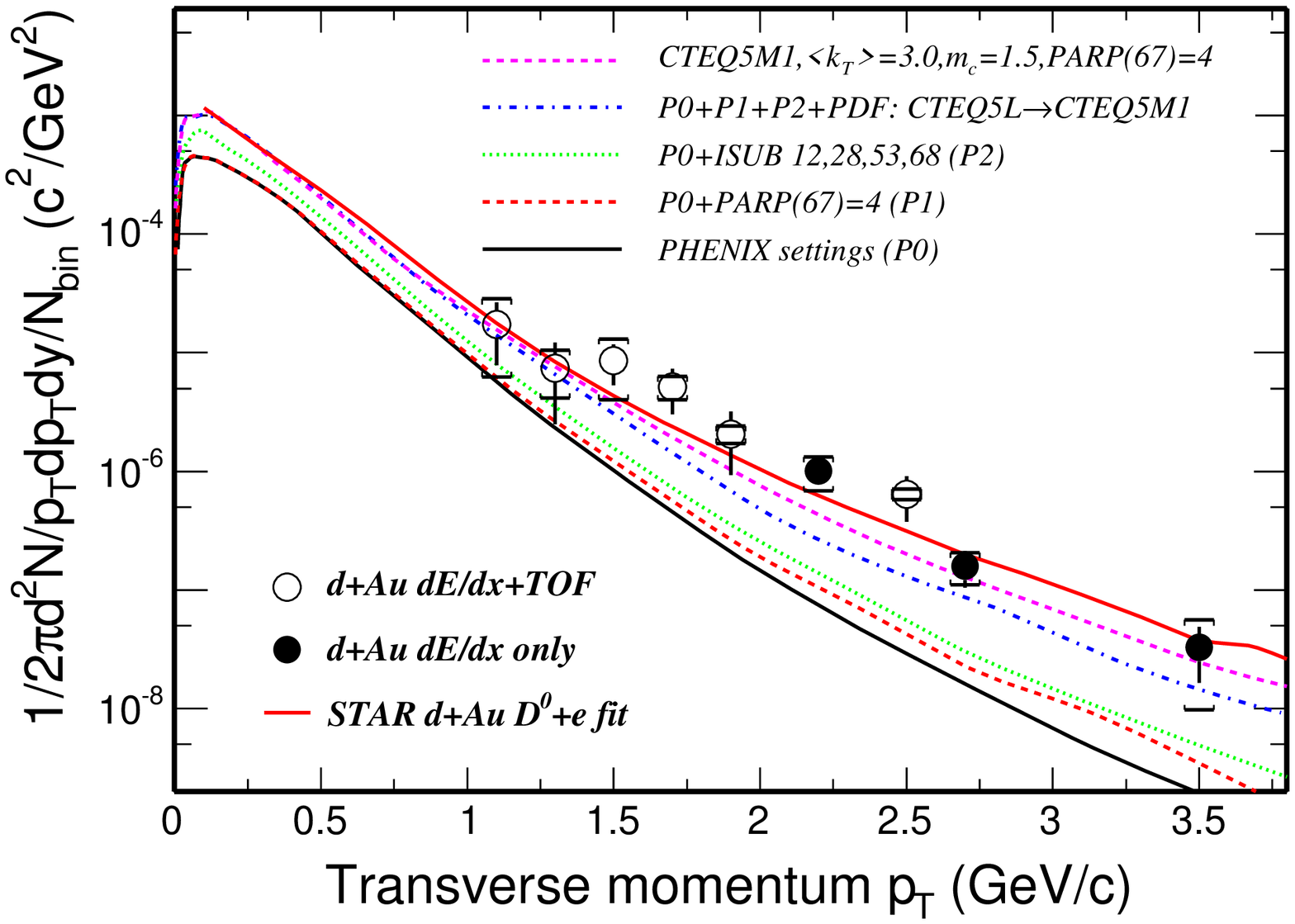}}
\caption[Comparison with PYTHIA]{Measured non-photonic electron
spectrum compared with PYTHIA calculations with different
parameter settings at 200 GeV.} \label{pythiaCom} \ef

Fig.~\ref{pythiaCom} shows a comparison of our measurement to
PYTHIA calculations using different parameters. The PYTHIA
calculation from the PHENIX 130 GeV \AuAu paper~\cite{phenix130e}
was used as a starting line (black solid curve). This calculation
deviates significantly from the result of \dAu measurement at \sNN
= 200 GeV: it shows a lower yield and a steeper \pT spectrum.
Several parameters in PYTHIA were tuned to try to match our
measurement: we increased the abundance of parton showers, we
included higher order processes and gluon splitting processes, we
changed the PDFs, we increased the initial $\langle k_{T}\rangle$
and we changed the charm quark mass. Although one of the
calculations (red solid line) seems to match the spectrum, the
large initial $\langle k_{T}\rangle$ requirement is questionable.
Theoretical predictions~\cite{kTbroadening} as well as
measurements show a much smaller $\langle k_{T}\rangle$ is needed
for the 200 GeV \dAu system. In general, PYTHIA results seem to be
hard to match to our measured spectrum.

In the above comparisons, we didn't change anything about the
charm quark fragmentation function, which is a Peterson function
in the default PYTHIA~\cite{pythia}. The bare charm quark spectrum
from PYTHIA seems to match the open charm measurement, which means
a very hard fragmentation function (almost a $\delta$ function) is
needed for charm quarks~\cite{AnQM04} at RHIC energy. This may
indicate that charm quark coalescence processes become competitive
with fragmentation processes in the 200 GeV \dAu system. A recent
study shows this final state effect may play an important role
through soft and shower recombination in \dAu
collisions~\cite{recom,LijuanThesis}.

\subsection{Bottom contribution}
In the above discussions, the bottom contribution to the
non-photonic electron spectrum was neglected. Since there is no
bottom measurement around RHIC energy, this can only be studied
through models. We used the default PYTHIA as a first estimation.
Fig.~\ref{bcCom} shows the results from this estimation where we
used the scale $\sigma_{c\bar{c}}:\sigma_{b\bar{b}}=400:1$ as the
normalization~\cite{vogtXsec,raufeisenXsec}. At \pT above 1 GeV/c,
all the electrons from bottom are from direct semi-leptonic
decays. This contribution is $\sim25-40\%$ for $2<p_T$/(GeV/c)$<3$
and $\sim60-100\%$ for $3<p_T$/(GeV/c)$<4$ to the charm
contribution. The corresponding bottom contribution to the total
non-photonic electrons is $\sim20-30\%$ for $2<p_T$/(GeV/c)$<3$
and $\sim40-50\%$ for $3<p_T$/(GeV/c)$<4$. If we subtract this
estimated bottom contribution from our spectra, the change in the
total cross section from a combined fit is $~\sim0.09$ mb, within
the current statistical error of our result. This fraction is
strongly model dependent however, so we didn't include the bottom
contribution in the final calculation.

\bf \centering\mbox{
\includegraphics[width=0.6\textwidth]{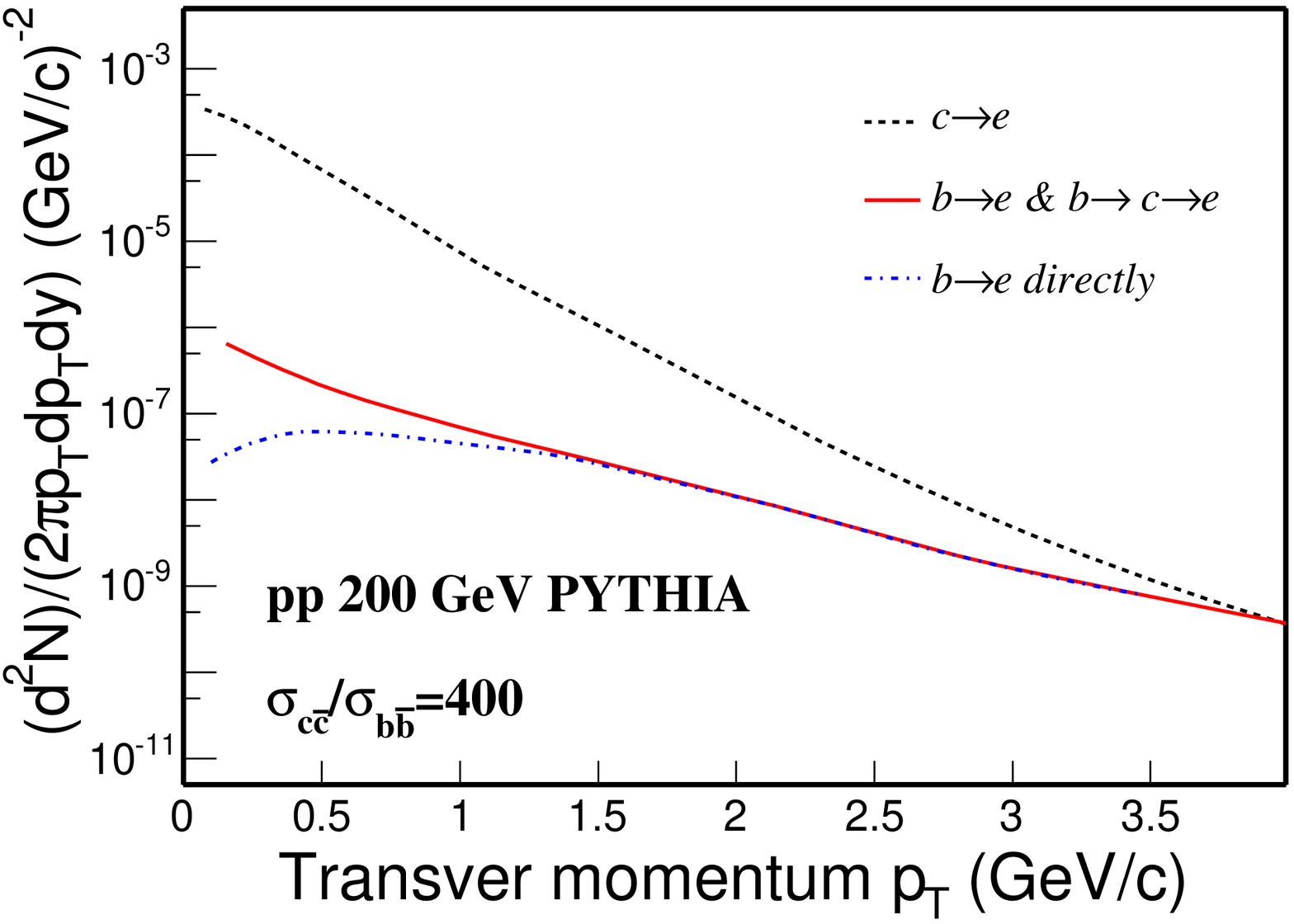}}
\caption[Charm and bottom contributions]{Comparison between the
contributions from charm quarks and bottom quarks from the default
PYTHIA.} \label{bcCom} \ef

We also tried several of the parameter settings from
Fig.~\ref{bcCom} to estimate the bottom contribution. Almost all
these results show the bottom contribution will begin to overcome
the charm contribution around \pT $\sim3-4$ GeV/c. However, in
these calculations, the heavy quark fragmentation functions have
not been modified. As we discussed above, there is a hint that
heavy quark may have a hard fragmentation function. In this case
the difference of the fragmentation between charm quarks and
bottom quarks may be much smaller than the default Peterson
settings in PYTHIA. A recent study shows that under such
assumptions, the bottom contribution on the electron spectrum does
not overcome the charm contribution until $p_T\sim10$
GeV/c~\cite{Xiaoyan}. Precise measurements of heavy flavor spectra
are needed to address these open issues.

\subsection{Cronin effect of charm hadrons in \dAu}
The well-known "Cronin effect" was named after the observation
that particle (differential) cross sections depend on the atomic
number ($A$) and the particle species from low energy \pA
collisions~\cite{cronin}. It is also characterized by the spectrum
broadening in the intermediate \pT in \pA collisions compared to
\pp collisions. The initial projectile partons' multi-scattering
with target partons has been proposed as an explanation for this
effect~\cite{multiparton,kTbroadening}. However, a recent
measurement on the pseudorapidity asymmetry of charged hadron
spectra shows a contrary result with these model predictions.
Another approach through the final state recombination of soft and
shower partons seems to reproduce the identified particle spectra
in \dAu collisions~\cite{LijuanThesis,recom}. Measuring the Cronin
effect of charm hadrons offers an additional opportunity to study
the initial $\langle k_{T}\rangle$ broadening and coalescence of
the final state charm quarks. Even though we don't have enough
statistics to measure the charm hadron spectra in both \dAu and
\pp collisions, estimates of $R_{dAu}^{e}$, the nuclear
modification factor of non-photonic electrons can give us some
hints about these processes for charm hadrons.

\bf \centering\mbox{
\includegraphics[width=0.6\textwidth]{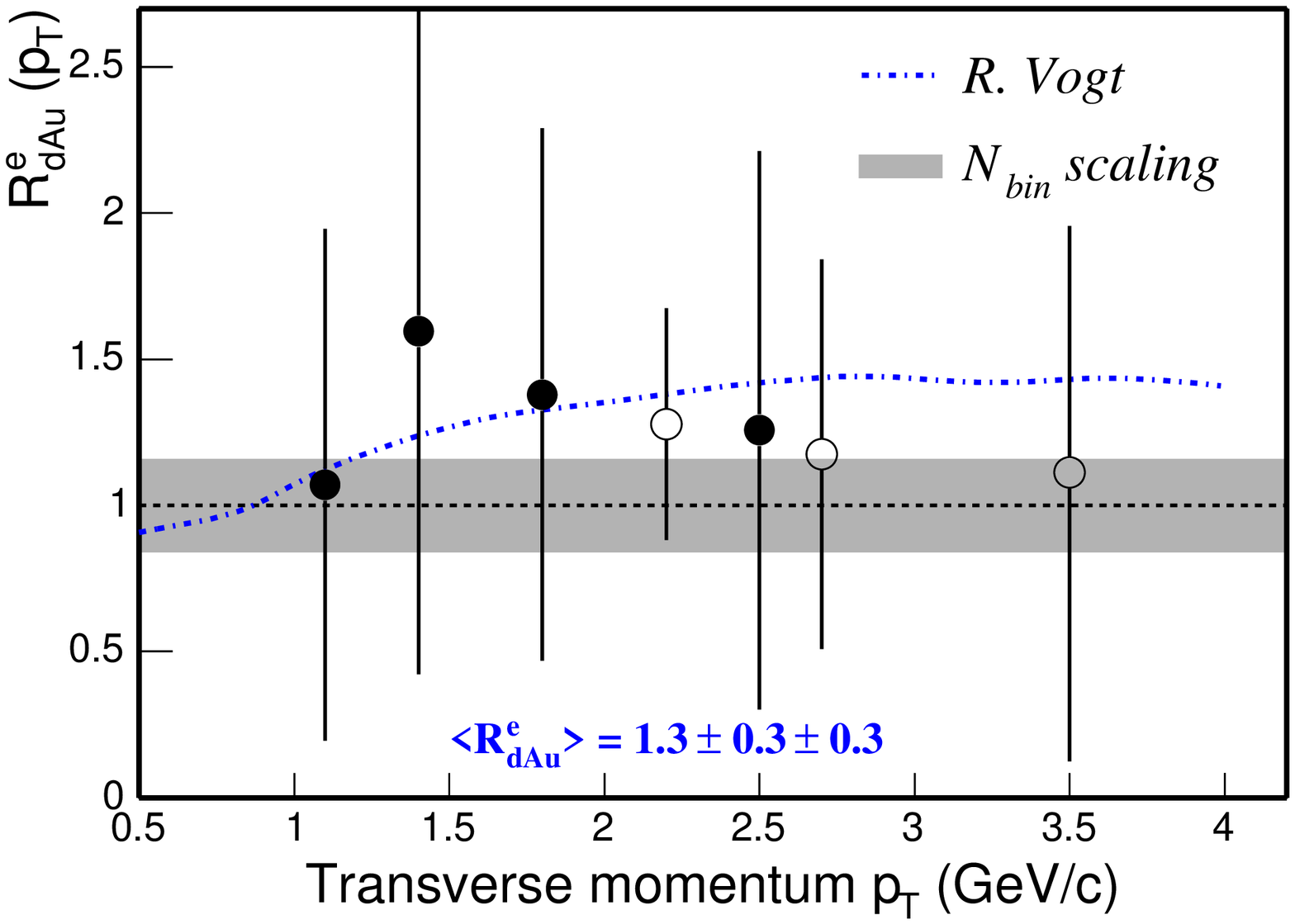}}
\caption[$R_{dAu}^{e}$ plot]{Cronin effect of non-photonic
electrons in \dAu collisions, compared with the theoretical
prediction of charm hadrons.} \label{RdAuE} \ef

Fig.~\ref{RdAuE} shows the $R_{dAu}^{e}$ for $1<p_{T}$/(GeV/c)$<4$
compared with a theoretical prediction curve based on the initial
multi parton scattering picture. We used the measured charm hadron
spectrum in \dAu collisions and generated that in \pp collisions
according to the $R_{dAu}^{D}$ from~\cite{vogtCronin}. Then we
obtained the decayed electron spectra from both and calculated the
predicted $R_{dAu}^{e}$. Due to poor statistics, the current
measurement cannot give definite conclusions. The averaged
$R_{dAu}^{e}$ in $1<p_{T}$/(GeV/c)$<4$ is $1.3\pm0.3\pm0.3$, which
is consistent with $N_{bin}$ scaling within present uncertainties.

\section{Closed charm production}

\subsection{$J/\psi$ production in \pp collisions}

The heavy quarkonia production mechanism in hadron hadron
collisions is not well understood. Understanding this production
mechanism in \pp collisions is needed to search for \Jpsi
suppression as a signature of QGP formation in central \AuAu
collisions at RHIC energy. Recent theoretical
progress~\cite{jpsiOctet} and the measurement from PHENIX
collaboration~\cite{phenixjpsipp} strongly support the {\em Color
Octet Model} (COM) for \Jpsi production in \pp collisions at \s =
200 GeV. Fig.~\ref{JpsiXsec} shows the measured \Jpsi production
cross sections compared with the total charm quark pair production
cross sections in different energies. The ratio of
$\sigma_{J/\psi}/\sigma_{c\bar{c}}$ is on the order of $10^{-2}$
and shows a slight decrease from low energies up to RHIC energy.

\bf \centering\mbox{
\includegraphics[width=0.6\textwidth]{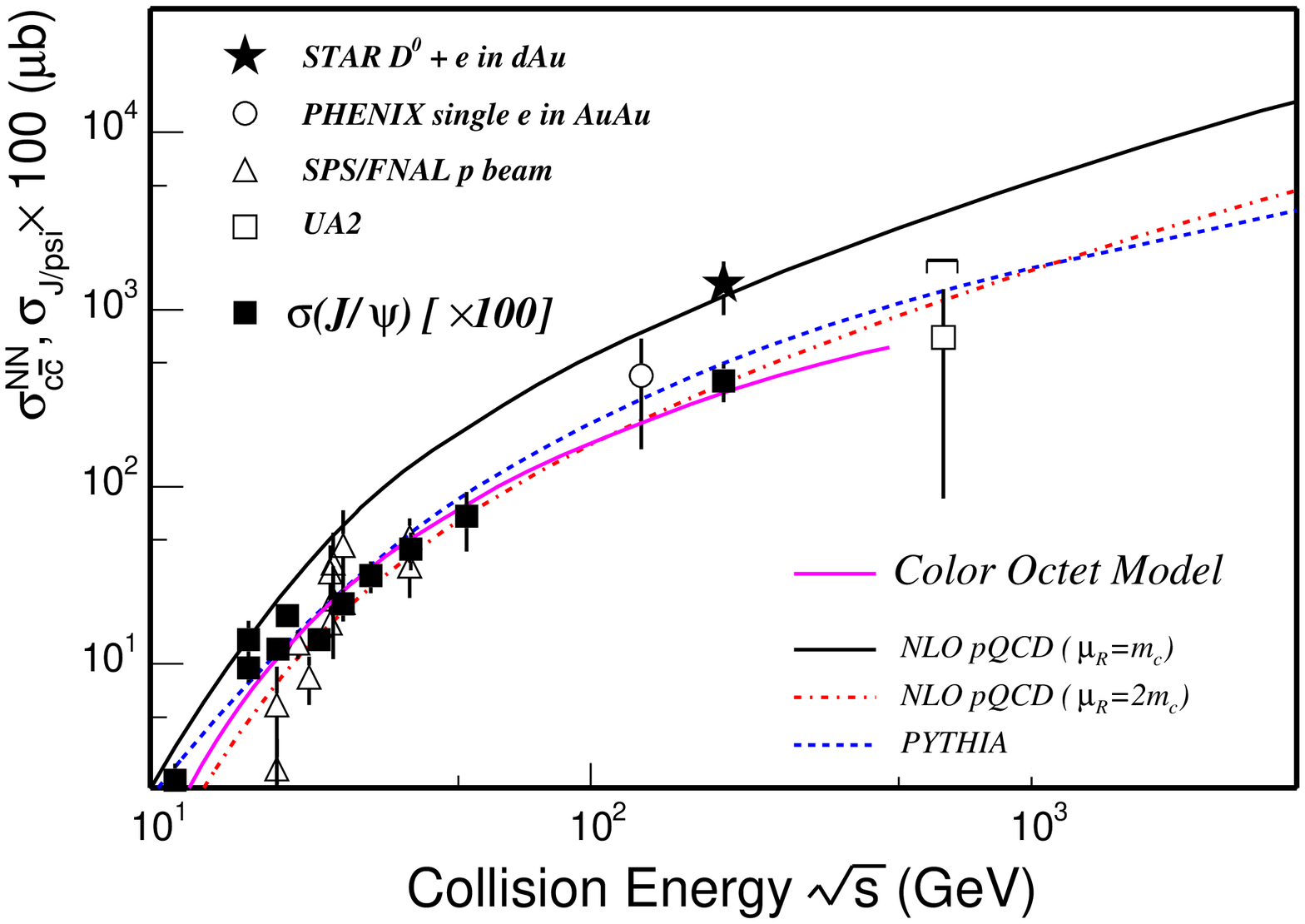}}
\caption[$\sigma_{J/\psi}$ compared with
$\sigma_{c\bar{c}}$]{$\sigma_{J/\psi}$ as a function of collision
energy in comparison to the total charm cross section. \Jpsi
production mechanism is proved to validate the COM model at RHIC
energy.} \label{JpsiXsec} \ef

\subsection{$J/\psi$ production in heavy ion collisions}

In the hot dense QGP matter, the color screening effect is
predicted to prevent heavy quark pairs from forming quarkonium
bound states, and thus leads to a decrease of the ratio of closed
charm (bottom) to open charm (bottom)~\cite{jpsisuppression}. This
heavy quarkonium suppression has been proposed as a signature of
QGP formation. In low energy SPS 17.2 GeV Pb + Pb collisions, the
NA50 experiment reported a suppression of heavy quarkonium
suppression relative to normal nuclear absorption. This was
interpreted as an anomalous suppression due to the dissociation of
$c\bar{c}$ pairs interacting with co-movers~\cite{na50}.

In high energy heavy ion collisions, since there may be multiple
$c\bar{c}$ pairs produced in a single collision, charmonium
production through coalescence of charm quarks may become
important: $D+\bar{D}\rightarrow J/\psi+X$. Since $c\bar{c}$ pairs
are mostly produced from initial scatterings~\cite{LinThesis}, the
partition of these charm quarks between open and closed ones is
deduced in statistical coalescence models~\cite{pbmjpsi} from
Eq.~\ref{ccbar} .
\begin{subequations}\label{ccbar}
\begin{align}
N_{c\bar{c}}^{Tot} = \frac{1}{2}N_{oc} &+ N_{c\bar{c}} \\
N_{oc} &\propto g_{c}n_{oc} \label{first}\\
N_{c\bar{c}} &\propto g_{c}^{2}n_{c\bar{c}} \label{second}
\end{align}
\end{subequations}
In these models, the total charm yield is needed as an input to
calculate $g_{c}$. The models often use $300-400$ $\mu$b for
$\sigma_{c\bar{c}}$ at RHIC energy, which is a typical value from
NLO pQCD calculations~\cite{manganoXsec,vogtXsec,raufeisenXsec}.
With this input, the statistical coalescence model, without any
additional absorption effects, seems to agree with the \Jpsi
measurement in \AuAu collisions within
errors~\cite{phenixjpsiauau}. However, if our measurement
$\sigma_{c\bar{c}}^{NN} = 1.4$ mb is used as the input, assuming
the statistical coalescence models work in central \AuAu
collisions at RHIC and ignoring the absorption effect, one would
expect at least a factor of 3 increase above the number of binary
scaling in the \Jpsi yield in central \AuAu collisions. The upper
limit from PHENIX measurement on \Jpsi production in central \AuAu
collisions seems to invalidate this expectation.

Additional improvements for the nucleon absorption effects on
\Jpsi were taken into account in~\cite{luicjpsi}. But in central
collisions, the production through coalescence can still make up a
significant portion of the total production. Fig.~\ref{JpsiModel}
shows the predictions from different calculations.

\bf \centering\mbox{
\includegraphics[width=0.6\textwidth]{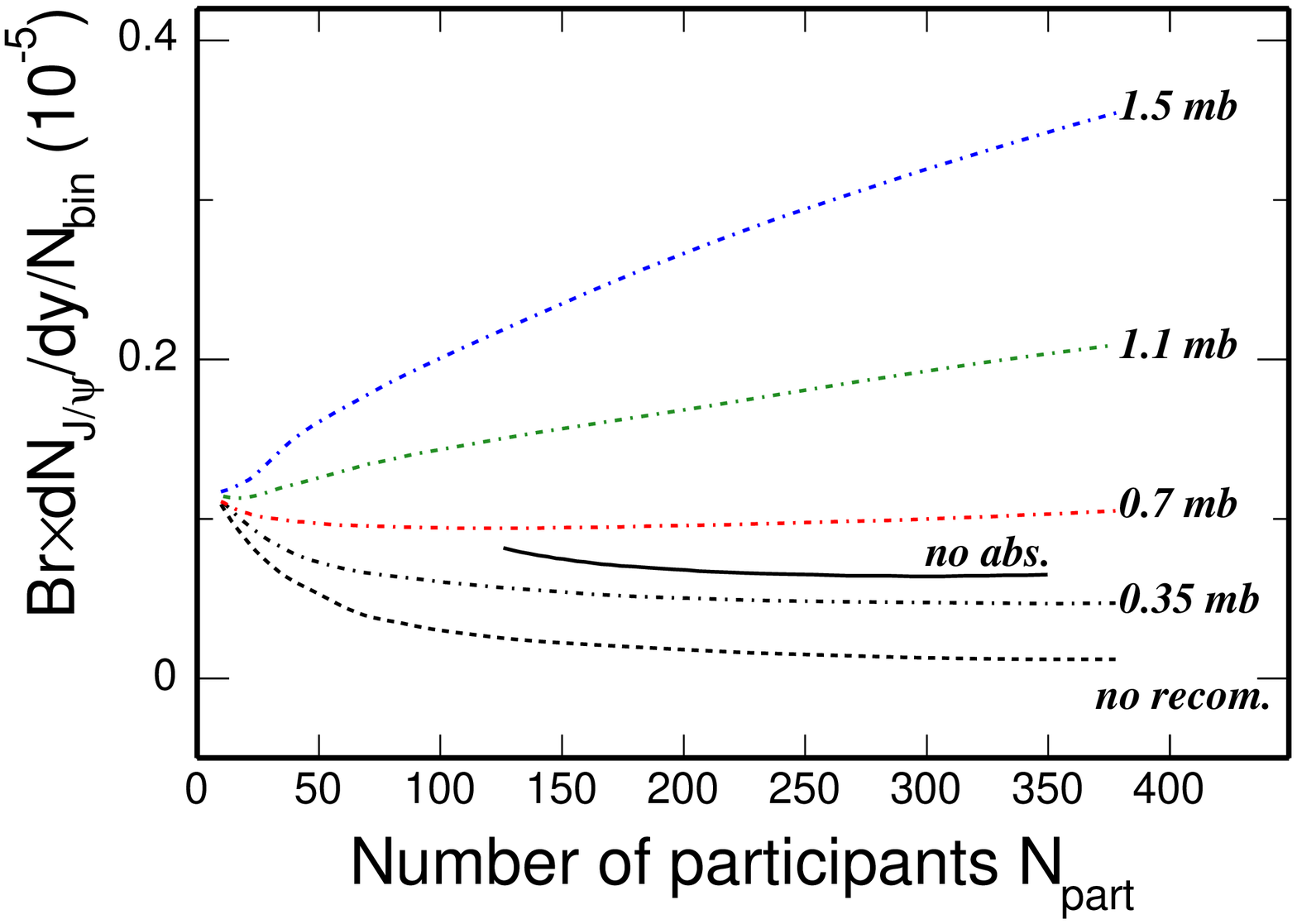}}
\caption[\Jpsi predictions from models]{\Jpsi yields from models
as a function of number of participants. The black dashed curve
depicts a calculation only includeing \Jpsi absorption effects,
and no recombination effects from~\cite{luicPrivate} with
$\sigma_{c\bar{c}}=0.35$ mb. The black solid curve depict a
calculation from statistical coalescence model and there is no
absorption effect in it~\cite{pbmjpsi} with
$\sigma_{c\bar{c}}=0.39$ mb. The dot-dashed lines depict several
calculations including both absorption and recombination effects
with different cross sections as input~\cite{luicPrivate}.}
\label{JpsiModel} \ef

From Fig.~\ref{JpsiModel}, because of the dominant contribution
from charm quark recombination, the expected \Jpsi yield in
central \AuAu collisions inferred from our \dAu result is still
significantly higher than expectations from number of binary
scaling. Even when nuclear effects in the \dAu system are taken
into account, which means $\sigma_{c\bar{c}}\sim 1.0$ mb is used
as the input, this prediction is still above number of binary
scaling. In the QGP, due to color screening, the \Jpsi yield will
be suppressed comparing to $N_{bin}$ scaling. Since most of the
suppression will occur at low \pT, the $\la p_T\ra$ of \Jpsi will
be efficiently increased. On the other hand, if the QGP is formed,
partons will reach kinetic equilibrium~\cite{pbmjpsi} and charm
quark coalescence will be important for \Jpsi production, the
\Jpsi yield will be ``enhanced" in this case. Furthermore, the
\Jpsi from coalescence production will have a larger $\la p_T\ra$
than those from direct production. Therefore, both enhancements in
the \Jpsi yield and $\la p_T\ra$ could be a good signature for QGP
production at RHIC. The combination of the suppression and
possible coalescence enhancement will, in principle, further
enforce this argument. However, the energy loss of initially
produced heavy quarks traversing the medium might complicate this
picture. Detailed analysis of \Jpsi yields as well as transverse
momentum distributions are necessary in order to distangle the
problem.

\section{Elliptic flow of charm quarks and thermalization}
In heavy ion collisions, the momentum-space azimuthal anisotropy
can reveal the early stage system information~\cite{v2early}. The
large $v_2$ measured for multi-strange hadrons and the NCQ scaling
of hadrons at intermediate \pT indicate a partonic level
collectivity has been developed in \AuAu collisions at RHIC.
However, partonic collectivity is necessary but not sufficient to
show that local thermal equilibrium has been established. Parton
thermalization is crucial to demonstrate the discovery of QGP in
heavy ion collisions at RHIC.

Charm quarks are believed to be mostly produced through initial
scatterings. Hence the study of elliptic flow of charm quarks may
give us some hints of properties of the produced matter in heavy
ion collisions. Charm quarks, because of their heavy masses, need
many more rescatterings to develop flow similar to light quarks
($u$, $d$, $s$). If charm quarks flow as much as light quarks,
this may tell us there are frequent rescatterings happening
between the light quarks and thus provide a strong clue to
illustrate the thermalization of the light ($u$, $d$, $s$)
quarks~\cite{minepiv2}.

\bf \centering\mbox{
\includegraphics[width=0.60\textwidth]{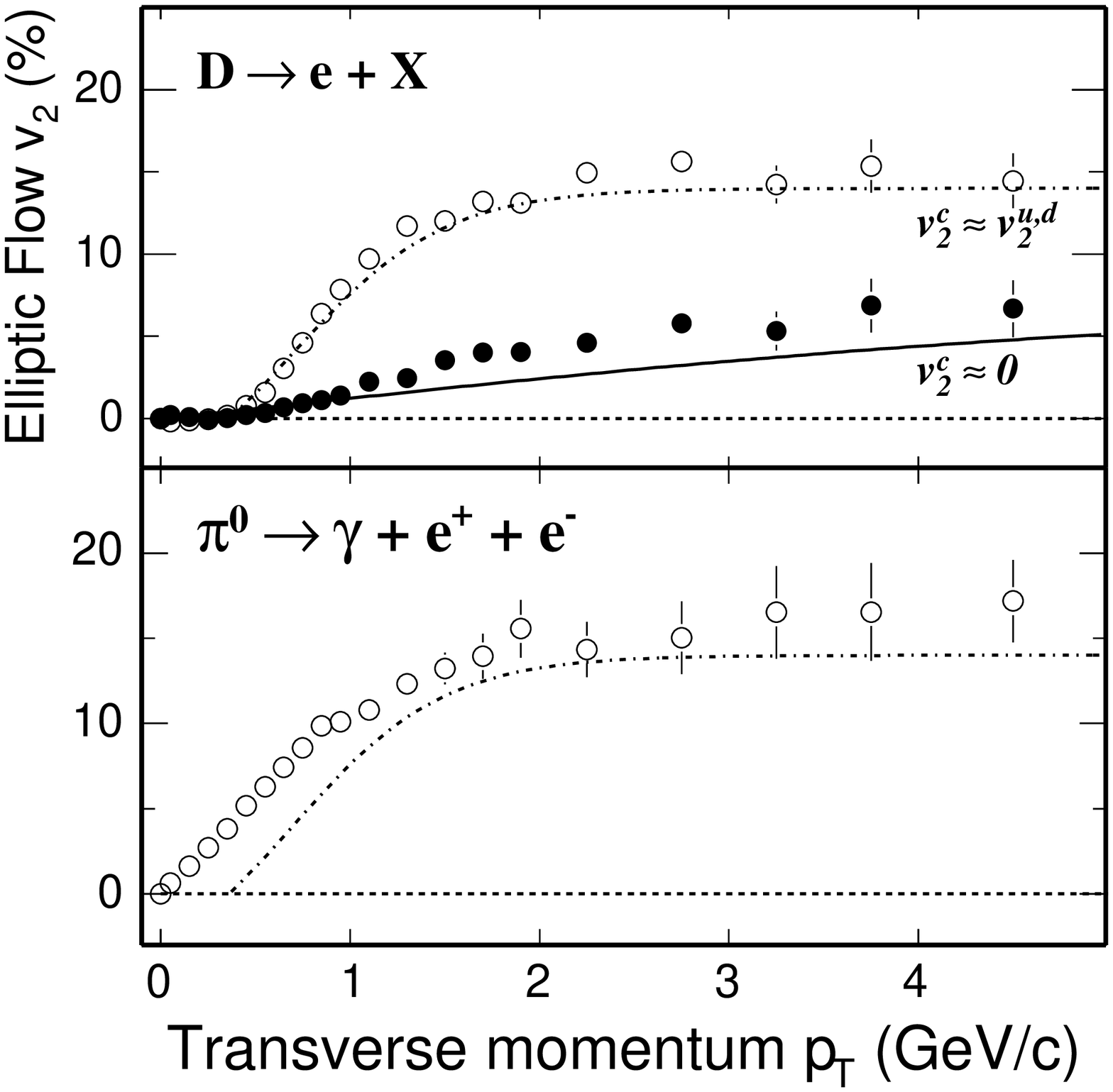}}
\caption[$v_{2}$ of electrons from $D$ and $\pi^{0}$]{$v_{2}$ of
electrons from the decays of charm hadrons and $\pi^{0}$. Lines
depict the input $D$ or $\pi^0$ $v_2$ while the points depict the
$v_2$ of decay electrons. Upper panel: the input D $v_2$ are
assumed as $v_{2}^{c}\approx v_{2}^{u,d}$ (dot-dashed line) and
$v_{2}^{c}\approx 0$ (solid line), respectively. Bottom panel: the
input $v_{2}(\pi^{0})$ (dot-dashed line) is from 200 GeV
measurement.} \label{Dv2} \ef

At present, the direct reconstruction of open charm hadrons uses a
mixed-event method. In this method, there are huge combinatorial
backgrounds under the signal peaks, which leads to large
uncertainties to extract $v_2$ from this method. A new way to
measure the charm quark $v_2$ is to measure the $v_2$ of their
semi-leptonic decayed electrons. Fig.~\ref{Dv2} upper panel shows
the input open charm hadron $v_{2}$ and decayed electron $v_{2}$
when charm quarks flow or not. About 50 M $D^{0}$s were input into
PYTHIA to obtain the decayed electrons. The shape of the $D^0$ \pT
spectrum is a power law function which we discussed before. The
results show that, at \pT above 2 GeV/c, the electrons will carry
almost all the saturated $v_2$ of open charm
hadrons~\cite{minepiv2} and the electron $v_2$ can distinguish
between $v_2^{c}\approx v_2^{u,d}$ and $v_2^{c}\approx 0$. This
demonstrates that measuring non-photonic electron $v_2$ for \pT
above 2 GeV/c is a feasible and efficient way to measure the charm
quark $v_2$~\cite{jamiecharmflow,minepiv2}.

Neutral pion decay is the dominant background source in the single
electron analysis. The two photon processes are followed by
conversion into electrons in the detector, and the rate of
conversion depends on detector materials. These background
electrons can be rejected by a topological
method~\cite{IanThesis}. The remaining dominant background source
is $\pi^{0}$ Dalitz decay, which can only be subtracted
statistically. The bottom panel of Fig.~\ref{Dv2} shows the
electron $v_2$ from $\pi^0$ Dalitz decays. The background electron
flow leads to large uncertainties in the statistical subtraction.

Electrons from heavy flavor decays begin to dominate the electron
spectrum above $p_T \sim 3$~GeV/c. With the knowledge of the pion
yield, D-meson yield, the pion $v_2$, and the electron $v_2$, it
is possible to extract the D-meson $v_2$. These measurements can
be made by both the PHENIX and STAR collaborations at RHIC. Direct
photon $v_2$ can also be measured with this method.


\chapter{Outlook}

\section{Detector upgrade proposals}

STAR has proposed two important sub-detector upgrades: a full
barrel Time-Of-Flight (TOF) detector~\cite{tofproposal} and a
Heavy Flavor Tracker (HFT)~\cite{hftproposal}.

The proposed full barrel TOF detector would surround the outer
edge of the TPC, and cover $-1<\eta<1$ and $\sim 2\pi$ in azimuth.
By using the recently developed technology - MRPC, the TOF system
can achieve the required timing resolution $<100$ ps and the
required particle detecting efficiency $>95\%$. This will
significantly improve the STAR PID capability: $\pi$/$K$ can be
separated to 1.8 GeV/c and $p$/meson can be separated up to $\sim
3$ GeV/c. The TOF detector allows us to greatly reduce the
integrated luminosity needed for key measurements (collectivity,
elliptic flow {\em etc.}), to significantly extend the \pT reach
of resonance measurements, and to study large and small scale
correlations and fluctuations {\em etc.}. Good performance of the
prototype TOF detector (TOFr) in Run III and Run IV strongly
offers us confidence about the full coverage barrel TOF detector.

The Heavy Flavor Tracker detector will be placed between the beam
pipe and the SVT detector. It will retain the ($p_T$, $\eta$)
coverage of the current TPC and have two layers of silicon pixel
detectors. The 10 $\mu m$ vertex resolution provided by the HFT
will allow us to reconstruct open charm and bottom hadrons
directly and make high precision measurements of \pT spectra,
particle ratios, azimuthal anisotropy {\em etc.}. With the help of
the SVT, it can be used to remove photon conversion electrons,
which will make the measurement of the low mass $e^+e^-$ spectrum
possible.

At RHIC, several key measurements such as jet quenching, bulk
collective motion, and partonic collectivity {\em etc.} have
provided strong hints for the discovery of QGP. However, at least
two remaining critical points need to be demonstrated: whether the
system is thermalized and whether the chiral symmetry is restored?
These two sub-detector upgrade will help us to understand these
questions and the two following measurements should be
investigated.

\section{Open charm measurements}
\bf \centering\mbox{
\includegraphics[width=0.6\textwidth]{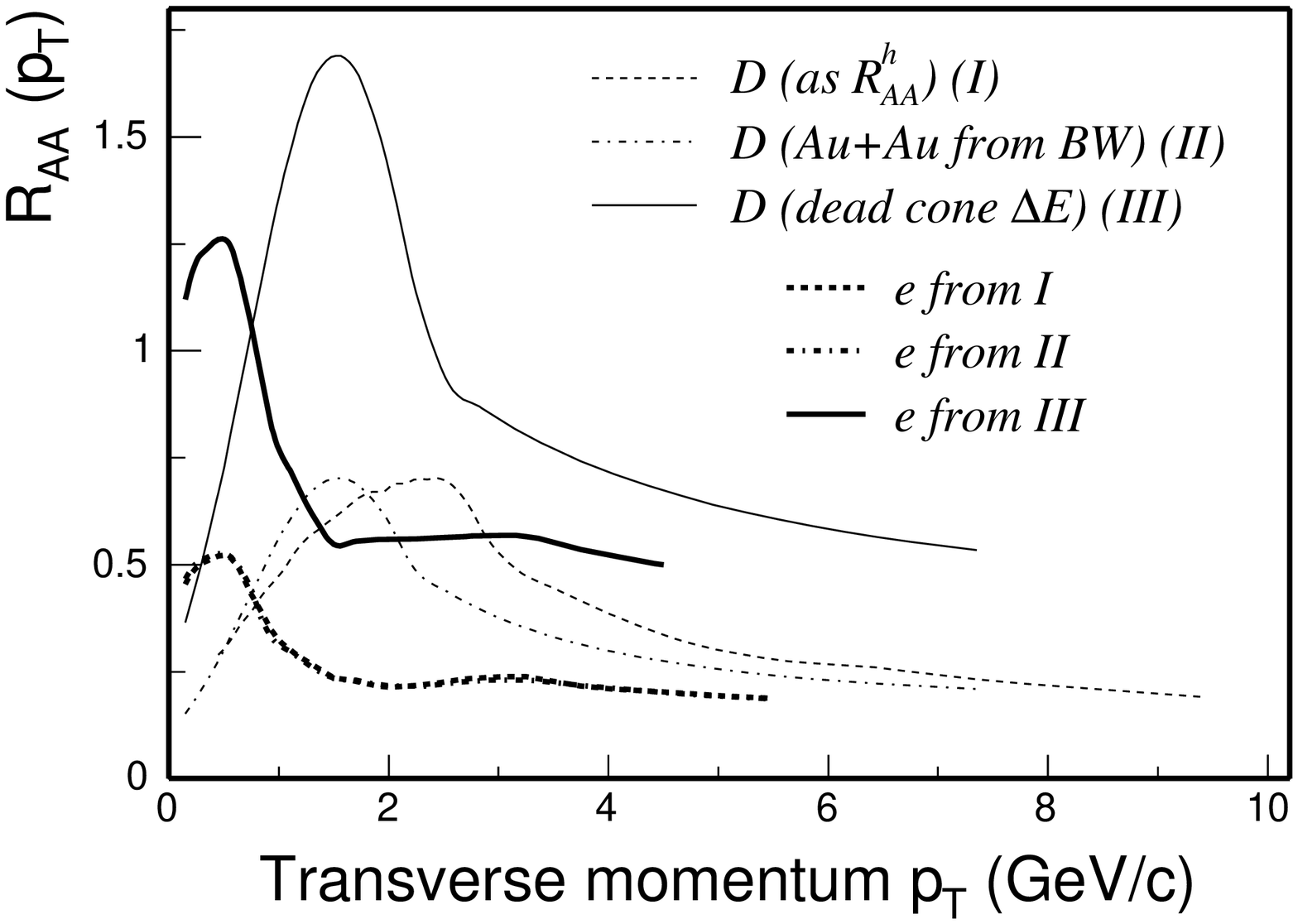}}
\caption[$R_{AA}$ of charm hadrons and decayed electrons]{$R_{AA}$
of charm hadrons (thin lines) and decayed electrons (thick lines)
for several assumed $D$ spectra in \AuAu central collisions. The
baseline used in \pp collisions is from STAR \dAu measurement. The
thin dashed line depicts the $R_{AA}$ of $D$ with the same
behavior as charged hadrons. The thin dot-dashed line is for $D$
with $T_{fo}=160$ MeV and $\la\beta_{T}\ra=0.4$ c blast wave
behavior in low \pT and with the same $R_{AA}$ as charged hadron
in high \pT. The thin solid line is for $D$ with the same BW
parameters as the dot-dashed line in low \pT, but the total yield
of $D$ is assumed to obey $N_{bin}$ scaling. While in high \pT,
the $R_{AA}$ is taken from dead-cone energy loss
calculation~\cite{gyulassyheavyloss}. Thick lines depict the
electron distributions from the corresponding $D$ distribution
with the same line style.} \label{RAAce} \ef

In the previous discussion, we proposed the measurement of charm
azimuthal anisotropy through its semi-leptonic decay channel.
However, the decay kinematics can smear the spectra differences.
So it is hard to study charm meson radial flow or hydrodynamic
behavior through its semi-leptonic decay channel. Fig.~\ref{RAAce}
shows $R_{AA}$ of charm hadrons and decayed electrons for several
assumed $D$ spectra in \AuAu central collisions. The difference
between the solid black line and the dashed black line for $D$ is
visible and it can be attributed to the smaller collectivity of
charm hadrons due to heavy masses. But this difference is washed
out by the decay kinematics and it is hard to tell the magnitude
of charm hadron collectivity from decayed electron spectra.

At high $p_T$, heavy quark may lose less energy due to the
``dead-cone'' effect~\cite{deadcone,gyulassyheavyloss}. A precise
charm-decayed electron spectrum measurement may give us some hints
from the suppression factor. But, the contribution to electrons
from bottom decays may become significant above 3 GeV/c and bottom
quarks are expected to lose much less energy. As such, the
electron spectrum suppression at high \pT is hard to interpret in
terms of charm hadrons and/or bottom hadrons separately. So the
direct charm spectrum measurement is crucial to answer the
hydrodynamic behavior and ``dead-cone'' effect of charm hadrons.
Together with those measurements for light hadrons, we can
understand the properties of the hot dense matter created in heavy
ion collisions.

Measured light hadron ($u$, $d$, $s$) yields and their ratios,
from AGS to RHIC energies, have been well described in statistical
models~\cite{PBMreview}. Precise measurement on charm hadron
yields and their ratios can help us understand the charm quark
chemistry and the kinematical equilibration property of charm
quarks in the medium. Meanwhile, since the charm quark yield is
quite high at RHIC energy, coalescence models predict
significantly different ratios of different charm
hadrons~\cite{charmRecom}. Furthermore, the standard $J/\psi$
suppression scenario may be modified when thermally produced
$J/\psi$ become significant. Precise measurements on these
sensitive probes: $D_{s}^{+}/D^{0}$, $D_{s}^{+}/D^{+}$ and
$J/\psi/D^0$ are needed to distinguish different
pictures~\cite{hftproposal}. Table.~\ref{Devts} shows the events
needed to observe 3$\sigma$ $D^0$ and $D_s^+$ signals with
different detector configurations. With the help of TOF and HFT
detectors, these precise measurements become feasible.

\begin{table}[hbt]
\caption[$N_{evt}$ to observe $3\sigma$ $D^0$ and $D_s^+$]{Number
of events needed to observe $3\sigma$ $D^0$ and $D_s^+$ signals in
\AuAu 200 GeV collisions with different detectors configurations.}
\label{Devts}\vskip 0.1 in \centering\begin{tabular}{c|c|c} \hline
\hline  &
$D^0$ & $D_s^+$ \\ \hline TPC+SVT & 12.6 M & 500 M ($K_{S}^0+K^+$) \\
\hline TPC+SVT+TOF & 2.6 M & 100 M \\ \hline
TPC+SVT+TOF+HFT & 5 K (Minuit) & 700 K (Minuit $\phi+\pi^+$) \\
\hline \hline
\end{tabular}
\end{table}

The development of elliptic flow requires frequent rescatterings
between components. Identified particle elliptic flow measurements
show strong evidence that a partonic level collectivity has been
established in \AuAu collisions at RHIC energy. However, it is
still not sufficient to prove that the system is thermalized.
Charm quarks are abundantly produced at RHIC energies. Due to
their large masses and expected smaller hadronic rescattering
cross sections, non-zero charm quark flow would provide a strong
indication that light quarks ($u$, $d$, $s$) are thermalized.
Although an electron $v_2$ measurement can provide an indirect
effective method, the technical systematics need to be understood
very well before we can draw conclusions. A precise direct
measurement of the charm hadron flow up to intermediate \pT can be
achieved with the upgraded detectors TOF and HFT. Simulations show
the statistical errors can reach $\sim 2-3\%$ at $p_T\sim 5$ GeV/c
in one year of \AuAu run.

\section{Low mass $e^+e^-$ spectrum} Electromagnetic probes have
great advantages to investigate the early stage information of the
system compared to hadronic probes because they don't have strong
interactions with the final state system. The di-electron
invariant mass $M_{e^+e^-}$ distribution has been proposed as a
unique tool to study the properties of vector mesons ($\rho$,
$\omega$, $\phi$, $J/\psi$ {\em etc}) in-medium effect during
heavy ion collisions. This is directly connected with the QGP
signature $-$ chiral symmetry restoration by measuring the mass
position change of these vector mesons. Once the broken chiral
symmetry is restored, $\langle \bar{\psi}\psi\rangle$ becomes 0
and the quark mass coupling with the Higgs field decreases to 0.
This is a robust proof of the discovery of QGP.

This measurement has been carried out in low energy Pb + Au
collisions at SPS. CERES/NA45 studied the $e^+e^-$ invariant mass
spectrum with projectile beams of 40A GeV and 158A GeV. They
observed an excess in the di-electron spectrum compared to light
hadron decays and attributed it to in-medium modifications of the
$\rho$ meson due to the restoration of approximate chiral
symmetry~\cite{na45}. However, the data points from SPS have large
errors both statistically and systematically, so we cannot draw a
strong conclusion from them so far. In the phase diagram shown in
Fig.~\ref{phasediagram}, the SPS machine sits near the crossover
boundary of the phase transition into QGP. While the $T_{ch}$ is
higher and $\mu_{B}$ is lower in RHIC, this signal should be
easier to observe. Since almost all other RHIC measurements
indicate the formation of QGP, a precise measurement of the low
mass $e^+e^-$ spectrum at RHIC energy is likely to be interesting.

This measurement needs both detector upgrade. By combining TOF and
$dE/dx$ in the TPC, electrons can be identified. This has been
demonstrated in Fig.~\ref{ePID} and ~\ref{ePIDAuAu}. Background
electrons from photon conversion and light hadron decays were
studied in the simulations. Fig.~\ref{epairbkgd} shows the
simulated electron pair invariant mass distributions for those
pairs from different sources in $\sim 1$ M PYTHIA + GEANT
simulation events. The background level in the interesting
$\omega$, $\rho$ mass region is shown in Table.\ref{epairbkgdtab}.

\bf \centering\mbox{
\includegraphics[width=0.6\textwidth]{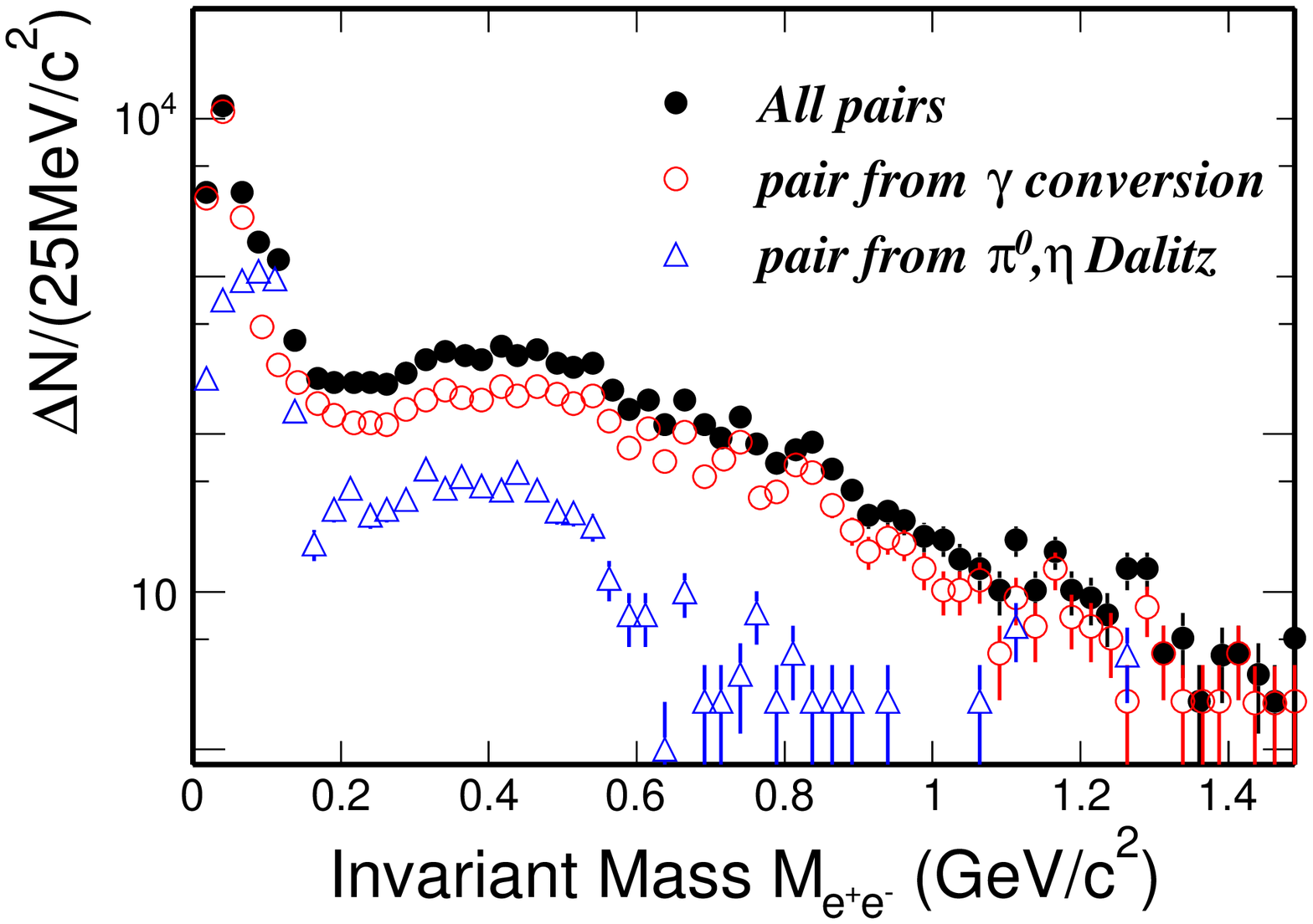}}
\caption[$M_{e^+e^-}$ of electron pairs from
simulation]{$M_{e^+e^-}$ of electron pairs from PYTHIA + GEANT
simulations for those from all sources (black solid circles),
photon conversion only (red open circles) and $\pi^0$,$\eta$
Dalitz decays (blue triangles).} \label{epairbkgd} \ef

\begin{table}[hbt]
\caption{Electron pair contribution from background}
\label{epairbkgdtab}\vskip 0.1 in \centering\begin{tabular}{c|c|c}
\hline \hline  & $\omega$ & $\phi$ \\ \hline Total bkgd &
$10^{-4}/(25$ MeV/c$^{2})$ & $2\times 10^{-5}/(25$ MeV/c$^{2})$
\\
\hline $\pi^0$ Dalitz & $5\times 10^{-6}/(25$ MeV/c$^{2})$ &
$5\times 10^{-7}/(25$ MeV/c$^{2})$
\\ \hline \hline
\end{tabular}
\end{table}

According to the material properties of the detectors and also
from simulations, photon conversions mostly happen in the SVT,
especially the first layer of the SVT (out of three). So requiring
charged particle hits in the SVT and HFT will significantly remove
the photon conversion electrons. Fig.~\ref{muVcon} shows the
electron \pT distribution from photon conversions with a
25-TPC-hit cut only (black), a 25-TPC-hit and a 2-SVT-hit cut
(red) and a 25-TPC-hit, a 2-SVT-hit cut and a 2-HFT-hit cut
(blue). With the help of SVT and HFT, $\sim 98\%$ of conversion
electrons can be removed. In this case, the $\pi^{0}$ will become
the dominant background source for primary electrons.

\bf \centering\mbox{
\includegraphics[width=0.6\textwidth]{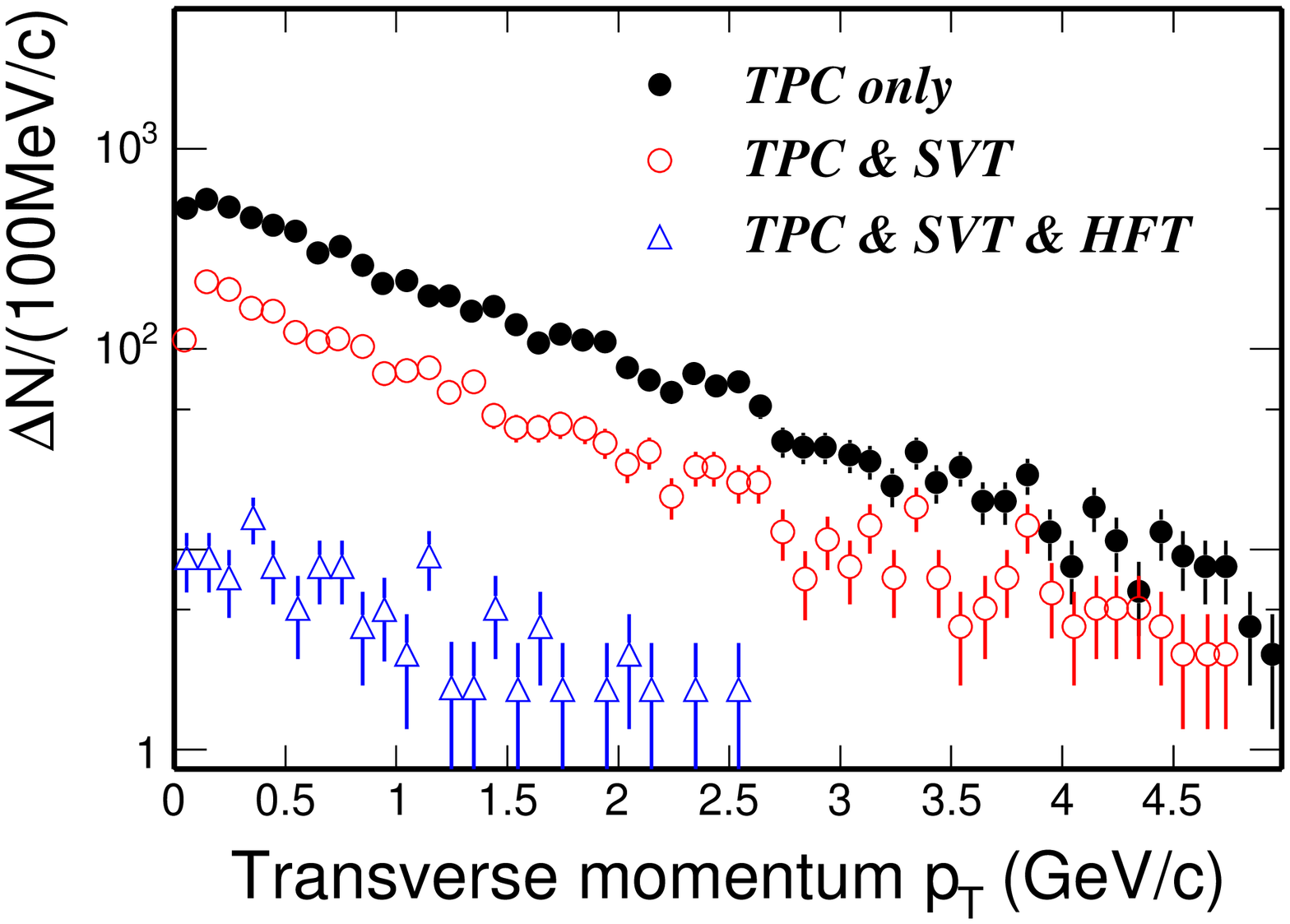}}
\caption[Conversion electrons rejection with HFT]{Electrons from
photon conversion \pT spectrum for a 25-TPC-hit and a 2-SVT-hit
cut (red) and a 25-TPC-hit, a 2-SVT-hit cut and a 2-HFT-hit cut
(blue). Only $\sim 2\%$ of total electrons from photon conversion
are left with combined TPC+SVT+HFT cut.} \label{muVcon} \ef

Furthermore, we can estimate the number of events needed to
observe the vector meson signals under different background
levels. From previous measurements and model simulations, we
assume that $dN^{\pi}/dy \approx 1$ in 200 GeV \pp collisions and
$dN^{\pi}/dy \approx 300$ in \AuAu collisions; $\omega/\pi\approx
0.15$, $\phi/\pi\approx 0.02$, $\sigma_{\omega}\approx 10$
MeV/c$^{2}$, $\sigma_{\phi}\approx 8$ MeV/c$^{2}$ from
simulations; The matching and electron PID efficiency with
TOF+$dE/dx$ method is $\sim 80\%$ and the matching efficiency of
the combined TPC+SVT+HFT is $\sim 60\%$. Then to observe $3\sigma$
signals for the $\omega$ and $\phi$, the number of events of \AuAu
collisions needed are shown in Table.~\ref{evtOmegaPhi}.

\begin{table}[hbt]
\caption[$N_{evt}$ to observe $3\sigma$ $\omega$ and
$\phi$]{Number of events needed to observe $3\sigma$ $\omega$ and
$\phi$ signals in \AuAu 200 GeV collisions with different
detectors configurations.} \label{evtOmegaPhi}\vskip 0.1 in
\centering\begin{tabular}{c|c|c} \hline \hline  &
$\omega$ & $\phi$ \\ \hline TPC+TOF & ~~~~~7 M~~~~~ & ~~~~~2 M~~~~~ \\
\hline TPC+TOF+SVT+HFT & ~~~~~800 K~~~~~ & ~~~~~150 K~~~~~ \\
\hline \hline
\end{tabular}
\end{table}

The measurements of vector mesons through electromagnetic decay
channels can shed light on whether chiral symmetry which is broken
in QCD vacuum is restored in the QGP, and furthermore indicate the
origin of mass in nature.

\appendix

\chapter{QCD Lagrangian}

\section{Notations}
We use the {\em natural units} throughout this thesis:
$c=\hbar=k_B=1$.

In the following part, we define the {\em contravariant vectors}
in for space-time coordinates and 4-momentum vector in the
Minkowski space. \be
x^{\mu}=(x^0,x^1,x^2,x^3)=(t,\boldsymbol{x})=(t,x,y,z) \ee \be
p^{\mu}=(p^0,p^1,p^2,p^3)=(E,\boldsymbol{p})=(E,\boldsymbol{p_T},p_z)=(E,p_x,p_y,p_z)
\ee The space-time {\em metric tensor} $g_{\mu\nu}$ is: \be
g_{\mu\nu}=g^{\mu\nu}=\begin{pmatrix} 1 & 0 & 0 & 0\\ 0 & -1 & 0 &
0 \\ 0 & 0 & -1 & 0 \\ 0 & 0 & 0 & -1 \end{pmatrix} \ee The {\em
covariant vector} is related to the contravariant vector through
$g_{\mu\nu}$ by \be x_{\mu}=g_{\mu\nu}x^{\nu} \ee and conversely,
\be x^{\mu}=g^{\mu\nu}x_{\nu} \ee In this thesis, we use the
notation that a repeated index implies a summation with respect to
that index, unless indicated otherwise. The scaler product of two
vectors $a$ and $b$ is defined as: \be a\cdot b\equiv
a^{\mu}b_{\mu}=g_{\mu\nu}a^{\mu}b^{\nu}=a^0b^0-\boldsymbol{a}\cdot\boldsymbol{b}
\ee

The gradient operator $\partial_{\mu}$ is the derivative with
respect to $x^{\mu}$: \be
\partial_{\mu}=\frac{\partial}{\partial x^{\mu}} \ee
The four-momentum operator $p^{\mu}$ in coordinate representation
is \be
p^{\mu}=i\partial^{mu}=ig^{\mu\nu}\partial_{\nu}=ig^{\mu\nu}\frac{\partial}{\partial
x^{\nu}}=(i\frac{\partial}{\partial
x^0},-i\frac{\partial}{\partial x^1},-i\frac{\partial}{\partial
x^2},-i\frac{\partial}{\partial x^3}) \ee It is convenient to work
with gamma matrices $\gamma^{\mu}$ later. They satisfy the
anticommutation relation \be \{\gamma^{\mu},\gamma^{\nu}\}\equiv
\gamma^{\mu}\gamma^{\nu}+\gamma^{\nu}\gamma^{\mu}=2g^{\mu\nu} \ee

\section{$SU(3)_{C}$ invariant QCD Lagrangian}
We denote $q_{f}^{\alpha}$ a quark field with color $\alpha$ and
flavor $f$ and adopt a vector notation in color space: $q_f\equiv
column(q_f^1,q_f^2,q_f^3)$. The free Lagrangian is written as: \be
\mathcal{L}_0=\sum_{f}\bar{q}_f(i\gamma^{\mu}\partial_{\mu}-m_f)q_f
\ee It is invariant under arbitrary global $SU(3)_C$
transformations in color space: \be q_f^{\alpha}\rar
(q_f^{\alpha})^{\prime}=U_{\beta}^{\alpha}q_f^{\beta}, \hskip 0.7
in UU^{\dag}=U^{\dag}U=1, \hskip 0.7 in \text{det}U=1 \ee The
$U(3)_C$ matrices can be written as \be
U=\exp\left(-ig_s\frac{\lambda^{a}}{2}\theta_a\right) \ee where
$\lambda^a$ ($a=1,2,...,8$) denote the generators of the
fundamental representation of the $SU(3)_C$ algebra, and
$\theta_a$ are arbitrary parameters. In $SU(3)_C$, the matrices
$\lambda^a$ correspond to the eight Gell-Mann matrices and they
satisfy the commutation relations \be [\lambda^a, \lambda^b]\equiv
\lambda^a\lambda^b-\lambda^b\lambda^a=2if^{abc}\lambda^c \ee with
$f^{abc}$ structure constants, which are real and totally
antisymmetric.

As in the QED case, to satisfy the invariance under {\em local}
$SU(3)_C$ transformations: $\theta_a=\theta_a(x)$, 8 independent
gauge bosons $G_a^{\mu}(x)$ - {\em gluons} are introduced. Define:
\be D^{\mu}q_f\equiv
\left[\partial^{\mu}-ig_s\frac{\lambda^a}{2}G_a^{\mu}(x)\right]q_f\equiv
[\partial^{\mu}-ig_sG^{\mu}(x)]q_f \ee To require $D^{\mu}q_f$ to
transform in exactly the same way as $q_f$, the transformation
properties of the gauge fields under an infinitesimal $SU(3)_C$
transformation: \be G^{\mu}_{a}\rar
(G^{\mu}_{a})^{\prime}=G^{\mu}_{a}-\partial^{\mu}(\delta\theta_a)+g_sf^{abc}\delta\theta_bG^{\mu}_{c}
\ee Unlike the QED case, the non-commutativity of the $SU(3)_C$
matrices gives rise to an additional term involving the gluon
fields themselves. To build a gauge invariant kinetic term for the
gluon fields, the field strengths were introduced: \be
\begin{split} G^{\mu\nu}(x)\equiv
\frac{i}{g_s}[D^{\mu},D^{\nu}]&=\partial^{\mu}G^{\nu}-\partial^{\nu}G^{\mu}-ig_s[G^{\mu},G^{\nu}]\equiv
\frac{\lambda^a}{2}G^{\mu\nu}_a(x) \\
G^{\mu\nu}_a(x)&=\partial^{\mu}G^{\nu}_a-\partial^{\nu}G^{\mu}_a+g_sf^{abc}G^{\mu}_bG^{\nu}_c
\end{split} \ee

The final $SU(3)_C$ invariant QCD Lagrangian is: \be
\mathcal{L}_{QCD}\equiv
-\frac{1}{4}G^{\mu\nu}_aG_{\mu\nu}^a+\sum_{f}\bar{q}_f(i\gamma^{\mu}D_{\mu}-m_f)q_f
\ee It is worthwhile to decompose the Lagrangian into different
pieces: \be
\begin{split}\mathcal{L}_{QCD}=&-\frac{1}{4}(\partial^{\mu}G^{\nu}_a-\partial^{\nu}G^{\mu}_a)
(\partial_{\mu}G_{\nu}^a-\partial_{\nu}G_{\mu}^a)+\sum_{f}\bar{q}^{\alpha}_{f}(i\gamma^{\mu}
\partial_{\mu}-m_f)q^{\alpha}_f \\
&+g_sG^{\mu}_a\sum_{f}\bar{q}^{\alpha}_f\gamma_{\mu}\left(\frac{\lambda^a}{2}\right)_{\alpha\beta}q^{\beta}_f
\\
&-\frac{g_s}{2}f^{abc}(\partial^{\mu}G^{\nu}_a-\partial^{\nu}G^{\mu}_a)G_{\mu}^bG_{\nu}^c
-\frac{g_s^2}{4}f^{abc}f_{ade}G^{\mu}_bG^{\nu}_cG^d_{\mu}G^e_{\nu}
\end{split} \ee
The first line contains the correct kinetic terms for different
fields. The second line shows the color interaction between quarks
and gluons. In the last line, owing to the non-abelian character
of the $SU(3)_C$ group, the $G^{\mu\nu}_aG_{\mu\nu}^a$ term
generates the cubic and quartic gluon self-interactions.

\section{Chiral symmetry and effective chiral Lagrangian}

In the absence of quark masses, the QCD Lagrangian can be written:
\be \mathcal{L}_{QCD}\equiv -\frac{1}{4}G^{\mu\nu}_aG_{\mu\nu}^a +
i\bar{q}_L\gamma^{\mu}D_{\mu}q_L +
i\bar{q}_R\gamma^{\mu}D_{\mu}q_R \ee. It is invariant under
independent {\em global} $G\equiv SU(N_f)_L\otimes SU(N_f)_R$
transformations of the left-and right-handed quarks in flavor
space: \be q_L\xrightarrow{G}g_Lq_L, \hskip 0.7 in
q_R\xrightarrow{G}g_Rq_R, \hskip 0.7 in g_{L,R}\in SU(N_f)_{L,R}
\ee

This chiral symmetry, which should be approximately good for light
quark sector ($u,d,s$), is however not seen in the hadronic
spectrum. Moreover, the octet of pseudoscalar mesons are much
lighter than all the other hadronic states. Hence, the ground
state of the theory (the vacuum) should not be symmetric under the
chiral group. In the Goldstone's theorem, an octet of pseudoscalar
massless bosons is introduced and the symmetry $SU(3)_L\otimes
SU(3)_R$ spontaneously breaks down and the quark condensate \be
v\equiv \la 0|\bar{u}u|0\ra = \la 0|\bar{d}d|0\ra = \la
0|\bar{s}s|0\ra < 0 \ee

The Goldstone nature of the pseudoscalar mesons implies strong
constraints on their interactions, which can be most easily
analyzed on the basis of an effective Lagrangian. The Goldstone
boson fields can be collected in a $3\times3$ unitary matrix
$U(\phi)$. \be \la 0|\bar{q}^j_Lq^i_R|0\ra \rar
\frac{v}{2}U^{ij}(\phi) \ee A convenient parametrization is given
by \be U(\phi)\equiv \exp(i\sqrt{2}\Phi/f) \ee \be \Phi(x)\equiv
\frac{\overrightarrow{\lambda}}{2}\overrightarrow{\phi}=\begin{pmatrix}
\dfrac{\pi^0}{\sqrt{2}}+\dfrac{\eta_8}{\sqrt{6}} & \pi^+ & K^+ \\
\pi^- & -\dfrac{\pi^0}{\sqrt{2}}+\dfrac{\eta_8}{\sqrt{6}} & K^0 \\
K^- & \bar{K}^0 & -\dfrac{2\eta_8}{\sqrt{6}} \end{pmatrix} \ee The
matrix $U(\phi)$ transforms linearly under the chiral group, but
the induced transformation on the Goldstone fields
$\overrightarrow{\phi}$ is highly non-linear.

We should write the most general Lagrangian involving the matrix
$U(\phi)$ and organize it in terms of increasing powers of the
momentum or, equivalently, in terms of an increasing number of
derivatives: \be \mathcal{L}_{eff}(U)=\sum_{n}\mathcal{L}_{2n} \ee
The terms with a minimum number of derivatives will dominate in
the low energy domain. To lowest order, the effective chiral
Lagrangian is uniquely given by the term \be \begin{split}
\mathcal{L}_2&=\frac{f^2}{4}\text{Tr}[\partial_{\mu}U^{\dagger}\partial^{\mu}U]
\\ &=\frac{1}{2}\text{Tr}[\partial_{\mu}\Phi\partial^{\mu}\Phi] +
\frac{1}{12f^2}\text{Tr}[(\Phi\overleftrightarrow{\partial}_{\mu}\Phi)
(\Phi\overleftrightarrow{\partial}_{\mu}\Phi)]+\mathcal{O}(\Phi^6/f^4)
\end{split}
\ee The non-linearity of the effective Lagrangian relates
amplitudes with different numbers of Goldstone bosons, allowing
for absolute predictions in terms of $f$.

Considering the quark masses, the corrections induced by the
non-zero masses are taken into account through the term \be
\begin{split}
\mathcal{L}_m&=\frac{|v|}{2}\text{Tr}[\mathcal{M}(U+U^{\dagger})],
\hskip 0.5 in \mathcal{M}\equiv \text{diag}(m_u, m_d, m_s) \\
\mathcal{L}_m&=|v|\left\{-\frac{1}{f^2}\text{Tr}[\mathcal{M}\Phi^2]
+ \frac{1}{6f^4}\text{Tr}[\mathcal{M}\Phi^4] +
\mathcal{O}(\Phi^6/f^6)\right\} \end{split} \ee

The relation between the physical meson masses and the quark
masses can be obtained rom the trace in the quadratic mass term.
Although the absolute values of the quark masses cannot be fixed
from this approach because of the factor $|v|/f$ in each mass, one
can obtain information about quark mass ratios after taking out
this common factor. One of the famous ratios advocated by Weinberg
is: \be m_u\colon m_d\colon m_s = 0.55\colon 1 \colon 20.3 \ee

\chapter{Kinematic variables}
Let us introduce several useful kinematic variables in
hadron-hadron interactions.
\begin{itemize}
\item Bjorken $x$, $x_{Bjorken}\equiv \dfrac{p_z(i)}{p_z(hadron)}$:
the longitudinal momentum
fraction carried by the parton $i$ to the total hadron momentum.
Usually denoted as $x$.
\item Feynman $x$, $x_{F}\equiv \dfrac{p_z^{\star}}{p_z^{\star}(max)}$:
the longitudinal momentum
fraction of final particle to the maximum momentum in the
center-of-mass system.
\item transverse mass $m_T\equiv \sqrt{p_T^2+m^2}$.
\item rapidity $y\equiv
\dfrac{1}{2}\ln\left(\dfrac{p_0+p_z}{p_0-p_z}\right)$. Then
$p_0=m_T\cosh y$, and $p_z=m_T\sinh y$.
\item pseudorapidity $\eta\equiv
\dfrac{1}{2}\ln\left(\dfrac{|\boldsymbol{p}|+p_z}{|\boldsymbol{p}|-p_z}\right)$.
Then $|\boldsymbol{p}|=p_T\cosh\eta$, and $p_z=p_T\sinh\eta$.
\end{itemize}

 We consider a two-parton interaction $1+2\rar
1^{\prime}+2^{\prime}+...$. Two partons are from the incoming
hadron beams of symmetric energy \s. Usually, the $z$ axis is
defined to parallel to the beam. At high energy ($\sqrt{s}\gg
m_h$), before interaction, the four-momentum of the two partons
are: \be
p^{\mu}_1=(x_1\dfrac{\sqrt{s}}{2},0,0,x_1\dfrac{\sqrt{s}}{2}),
\hskip 0.7 in
p^{\mu}_2=(x_2\dfrac{\sqrt{s}}{2},0,0,-x_2\dfrac{\sqrt{s}}{2}) \ee
The four-momentum of final state particle $j$ is: \be
p^{\mu\prime}_{j}=(p^{\prime}_{0j},\boldsymbol{p_T}^{\prime}_
j,p^{\prime}_{zj}) \ee Due the momentum conservation
\be\begin{split}
\frac{\sqrt{s}}{2}(x_1+x_2)&=\sum_{j}p^{\prime}_{0j} \\
\frac{\sqrt{s}}{2}(x_1-x_2)&=\sum_{j}p^{\prime}_{zj}
\end{split}\ee
Using $p^{\prime}_{0j}={p}^{\prime}_{Tj}\cosh y_j$ and
$p^{\prime}_{zj}={p}^{\prime}_{Tj}\sinh y_j$, we can obtain
\be\begin{split}
x_1&=\frac{1}{\sqrt{s}}\sum_{j}{p}^{\prime}_{Tj}e^{y_j} \\
x_2&=\frac{1}{\sqrt{s}}\sum_{j}{p}^{\prime}_{Tj}e^{-y_j}
\end{split} \ee
If we consider a two-particle final state, transverse momentum
conservation requires
$p^{\prime}_{T1}=p^{\prime}_{T2}=p_{T}$(neglect the symbol
$\prime$), at mid-rapidity where $y_{1,2}\sim 0$, we have: \be
x_1=x_2=x_T=\frac{2p_{T}}{\sqrt{s}} \ee In general cases, we often
use this treatment $x_{Bjorken}\approx x_T=2p_T/\sqrt{s}$.

Since the maximum fraction of the longitudinal momentum for a
certain parton is $\sqrt{s}/2$, the Feynman $x$ is easily to be
extracted: \be x_F=\frac{2p_z}{\sqrt{s}} \ee

\chapter{Low energy charm cross section data points selection}
There were a lot of charm cross section measurements at low energy
(\s $<$ 70 GeV). Table~\ref{gooddata} lists all the charm cross
section data points at low energies from journal publications. We
ignored those data points extrapolated from high $x_F$ and/or
extremely low efficiency from correlation measurements, with
extrapolation factor of $\gg 10$. The references for these data
points are listed as followings. The data values included in the
thesis are multiplied by 1.5 to account for the additional
$D_{s}^{\pm}$ and $\Lambda_c^{\pm}$
contribution~\cite{heavyReview}.

\begin{enumerate}
\item S.P.K. Tavernier, {\em Rep. Prog. Phys.} 50, 1439(1987) and
references therein.
\item NA32 collaboration, {\em Z. Phys. C} 39, 451(1988).
\item E769 collaboration, {\em Phys. Rev. Lett.} 77, 2388(1996).
\item NA16 collaboration, {\em Z. Phys. C} 40, 321(1988).
\item NA27 collaboration, {\em Phys. Lett. B} 135, 237(1984).
\item E743 collaboration, {\em Phys. Rev. Lett.} 61, 2185(1988).
\item E653 collaboration, {\em Phys. Lett. B} 263, 573(1991).
\end{enumerate}

\begin{table}[htbp]
\label{gooddata} \caption[Data points selection]{Overview of charm
cross section data points at low energies and the criteria to
include those in the thesis. Some experiments used
$d\sigma/dx_F=(1-|x_F|)^{-n}$ for extrapolation.} \vskip 0.1 in
\centering\begin{tabular}{c|c|c|c|c} \hline\hline Beam $p$ & Beam
& $\sigma_{c\bar{c}}$
($\mu$b) &  Comment & Included? \\
(GeV/c) & + Target & & & Ref. \\
\hline\hline
 200 & $p$ + Si &  $1.5\pm0.7$ & NA32,$D^0,D^+$,good &
 \checkmark ~~~[2]\\
 & & & acceptance (0-20\%) at $x_F>0$ & \\ \hline
 200 & $p$ + C$_3$F$_8$ & $3.9^{+2.5}_{-1.9}$ & NA25,Bubble chamber HOBC,
 1$\mu$ & \checkmark ~~~[1]\\
& & &  trigger, $E(\mu)>6$ GeV, good & \\
& & &  acceptance at $x_F>0$ (8\% efficiency) & \\ \hline 250 &
$p$ + Be,Cu, & $9.0\pm1.5$ & E769, $D^0,D^+$, good acceptance &
\checkmark ~~~[3]\\
& Al,W & &  at $x_F>0$ & \\ \hline
 280 & $n$ + He/Ne & $5.6\pm1.7$ & E630 streamer chamber, & \checkmark ~~~[1]\\
  & & & 1$\mu$ trigger, $E(\mu)>$6.5 GeV & \\ \hline
 350 & $p$ + Fe & $22\pm9$ & CIT-Stanford, prompt 1$\mu$ trigger, & \checkmark ~~~[1]\\
 & & & $E(\mu)>$20 GeV, acceptance 39\% & \\ \hline
 350 & $p$ + Fe & $11.3\pm2.0$ & FNAL-CCFRS, prompt 1$\mu$ trigger, & \checkmark ~~~[1]\\
 & & & $E(\mu)>$20 GeV, n=5, $x>$0.3 & \\ \hline
 360 & $p$ + C$_3$F$_8$ & $24.6^{+12.0}_{-8.3}$ & See 200GeV data
 & \checkmark ~~~[1]\\ \hline
 360 & \pp & $5.5^{+8.2}_{-4.6}$ & NA16, H$_2$ bubble chamber LEBC+EHS, & \checkmark ~~~[4]\\
 & & & good acceptance for $x_F>$0 & \\ \hline
 400 & \pp & $15.1\pm1.7$ & NA27, H2 bubble chamber & \checkmark ~~~[5]\\
 & & & good acceptance for $x_F>$0 & \\ \hline
 400 & $p$ + Fe & $31^{+29}_{-18}$ & CIT-Stanford, 1$\mu$, $E(\mu)>$20 GeV, & \checkmark ~~~[1]\\
 & & & $1.0<p_T<2.5$ GeV, 2.5\%, n$\sim$5 \\ \hline
 800 & \pp & $48^{+10}_{-8}$ & E743, LEBC-MPS, -0.1$<x_F<$0.5, n$\sim$8 & \checkmark ~~~[6]\\ \hline
 800 & \pp & $76^{+10}_{-19}$ & E653, -0.2$<x_F<$0.5, n$\sim$7 & \checkmark ~~~[7]\\ \hline
\hline
 400 & $p$ + Cu & $320^{+150}_{-100}$ & GGM $\nu$ beam dump, n=3, $\la x\ra=$0.8 &
 $\boldsymbol{\times}$ ~~~[1]\\ \hline
 400 & $p$ + Cu & $39\pm10$ & CERN 1st $\nu$ beam dump, CDHS, &
 $\boldsymbol{\times}$ ~~~[1]\\
 & & & $E(\mu)>20$ GeV, n=3-5, $\la x\ra=$0.8 & \\ \hline
 400 & $p$ + Cu & $17\pm4$ & CERN-BEBC, 2nd beam dump, & $\boldsymbol{\times}$ ~~~[1]\\
 & & & $E(\mu)>10$ GeV, n=4, $\la x\ra=$0.8 & \\ \hline
 400 & $p$ + Cu & $15\pm5$ & CERN-CHARM $\nu$ beam dump, & $\boldsymbol{\times}$ ~~~[1]\\
 & & & $E(\mu)>20$ GeV, n=4, $\la x\ra=$0.8 & \\ \hline
 400 & $p$ + W & $15.5\pm2.5$ & E613 FNAL $\nu$ beam dump, & $\boldsymbol{\times}$ ~~~[1]\\
 & & & $E(\mu)>20$ GeV, n=4, $\la x\ra=$0.45 & \\ \hline
 400 & $p$ + Cu & $\sim 17$ & WA66-BEBC, CERN 2nd $\nu$ beam dump, & $\boldsymbol{\times}$ ~~~[1]\\
 & & & $E(\mu)>10$ GeV, n$\sim$5 & \\ \hline
 400 & $p$ + Cu & $15.5\pm2.9$ & CHARM, CERN 2nd $\nu$ beam dump, n=5 &
 $\boldsymbol{\times}$ ~~~[1]
 \\ \hline
 400 & $p$ + Cu & $\sim 20$ & CDHS, CERN 2nd $\nu$ beam dump, n=5 &
 $\boldsymbol{\times}$ ~~~[1]
 \\ \hline
 \multicolumn{5}{c}{the above experiments cover high x, large extrapolation
 errors} \\ \hline \hline

 \end{tabular}
 \end{table}

\begin{table}[ht]
\mbox{continued} \vskip 0.1 in
\centering\begin{tabular}{c|c|c|c|c} \hline\hline Beam $p$ & Beam
& $\sigma_{c\bar{c}}$
($\mu$b) &  Comment & Included? \\
(GeV/c) & + Target & & & Ref. \\
\hline\hline
 400 & $p$ + Fe & $7\sim 20$ & CIT-Stanford, 2$\mu$+missing energy, & $\boldsymbol{\times}$ ~~~[1]\\
 & & & $p_T>$0.75 GeV, 0.1-0.4\% acceptance & \\ \hline
 400 & $p$ + W & $<10$ & $\mu+\mu$ mass & $\boldsymbol{\times}$ ~~~[1]\\
 \hline
\multicolumn{5}{c}{using correlation, low acceptance} \\ \hline
\hline
 \s & & & & \\ \hline
  63 & $p+p$ & $840\pm320$ & ISR, $$ trigger, $\Lambda_c(K\pi p)$ only, n$\sim$0 &
  $\boldsymbol{\times}$ ~~~[1] \\ \hline
  63 & $p+p$ & $150-450$ & ISR, $e$ trigger, $\Lambda_c(K\pi p)$ and $D(K\pi\pi)$ &
  $\boldsymbol{\times}$ ~~~[1] \\ \hline
  62 & $p+p$ & $650\pm222$ & ISR, CBF, $e$ trigger, n=3($D$),
  n=0($\Lambda_c$) &   $\boldsymbol{\times}$ ~~~[1]\\ \hline
  62 & $p+p$ & $129\pm75$ & ISR, CBF, new runs, $\Lambda_c(K\pi p)$ only &
  $\boldsymbol{\times}$ ~~~[1] \\ \hline
53-62 & $p+p$ & $1390\pm180$ & ISR, $\Lambda_c(K\pi p)$ only, &
$\boldsymbol{\times}$ ~~~[1] \\ & & &  n=0, $x>$0.75 & \\ \hline
\multicolumn{5}{c}{correlation, high $x_F$, large extrapolation
errors} \\ \hline \hline 53-62 & $p+p$ & $70\pm36$ & ISR, $e+\mu$
unlike-sign pairs, & $\boldsymbol{\times}$ ~~~[1] \\ & & &
0.0016\% efficiency, n$\sim$5 & \\ \hline 53-62 & $p+p$ &
$73\pm21$ & ISR, $ee$ pairs & $\boldsymbol{\times}$ ~~~[1] \\
\hline \multicolumn{5}{c}{correlation, low efficiency} \\ \hline
 \hline
\end{tabular}
\end{table}

\chapter{Presentations and publication list}
\vskip 0.1 in \Large{$\underline{\textbf{Presentations}}$}
\normalsize
\begin{itemize}
\item {\em Open charm production at RHIC - recent results from
STAR} \\
21st Winter Workshop on Nuclear Dynamics, Breckenridge, Colorado,
USA, 02/05/2005 - 02/12/2005.
\item {\em Elliptic flow of pion, kaon, proton from Au + Au collisions at 62.4 GeV} \\
2004 Fall Meeting of the Division of Nuclear Physics of APS,
Chicago, Illinois, USA, 10/27/2004 - 10/31/2004.
\item {\em Open Charm Yields in d + Au Collisions at 200 GeV}
(poster) \\
2004 Gordon Research Conference on Nuclear Chemistry, New London,
New Hampshire, USA, 06/13/2004 - 06/18/2004.
\item {\em Resonance decay effects on Anisotropy Parameters} \\
April Meeting of APS, Denver, Colorado, USA, 05/01/2004 -
05/04/2004.
\item {\em Open Charm Yield in 200 GeV d + Au Collisions at RHIC}
\\
2004 April Meeting of APS, Denver, Colorado, USA, 05/01/2004 -
05/04/2004.
\item {\em The Performance of a Prototype Multigap Resistive
Plate Chamber Time-Of-Flight Detector for the STAR Experiment}
(poster) \\
Quark Matter 2004, Oakland, California, USA, 01/11/2004 -
01/17/2004.
\item {\em Single Electron Spectra from d+Au and p+p collisions
at $\sqrt{s_{NN}}$ = 200 GeV} \\
2003 Fall Meeting of the Division of Nuclear Physics of APS,
Tucson, Arizona, USA, 10/28/2003 - 10/31/2003.
\end{itemize}

\vskip 0.1 in \Large{$\underline{\textbf{Publication List}}$}
\normalsize
\begin{itemize}
\item {\em Open charm yields in $d$ + Au collisions at
$\sqrt{s_{NN}}$ = 200 GeV} \\ J. Adams {\em et al.} (STAR
Collaboration), Phys. Rev. Lett. \textbf{94}, 062301(2005) \\
\emph{Principle authors:} X. Dong, L. Ruan, Z. Xu and H. Zhang.
\item {\em Resonance decay effects on anisotropy parameters} \\
X. Dong, S. Esumi, P. Sorensen, N. Xu and Z. Xu, Phys. Lett.
\textbf{B597}, 328(2004).
\item {\em Open charm production at RHIC - recent results from
STAR} \\
X. Dong {\em et al.} (for STAR Collaboration), Proceedings of 21st
Winter Workshop on Nuclear Dynamics.
\item {\em Improvement on the charge resolution with the average
of truncated dE/dx and optimized combination} \\
X. Dong, S.W. Ye, H.F. Chen, Z.P. Zhang and Z.Z. Xu, Journal of
University of Science and Technology of China Vol 8 (2002) (in
Chinese).
\item {\em Pion, kaon, proton and anti-proton transverse momentum
distributions from $p+p$ and $d$ + Au collisions at
$\sqrt{s_{NN}}$
= 200 GeV} \\
J. Adams {\em et al.} (STAR Collaboration), {\em arXiv:
nucl-ex/0309102} \\
\emph{Principle authors:} L. Ruan, X. Dong, F. Geurts, J. Wu and
Z. Xu.

\item {\em Azimuthal anisotropy and correlations at large
transverse momenta in $p+p$ and Au + Au collisions at
$\sqrt{s_{NN}}$ = 200 GeV} \\
J. Adams {\em et al.} (STAR Collaboration), Phys. Rev. Lett.
\textbf{93}, 252301(2004).
\item {\em Azimuthally sensitive HBT in Au + Au collisions at
$\sqrt{s_{NN}}$ = 200 GeV} \\
J. Adams {\em et al.} (STAR Collaboration), Phys. Rev. Lett.
\textbf{93}, 012301(2004).
\item {\em Multi-strange baryon production in Au-Au collisions
at $\sqrt{s_{NN}}$ = 130 GeV} \\
J. Adams {\em et al.} (STAR Collaboration), Phys. Rev. Lett.
\textbf{92}, 182301(2004).
\item {\em Cross Sections and Transverse Single-Spin Asymmetries
in Forward Neutral Pion Production from Proton Collisions at
$\sqrt{s}$ = 200 GeV} \\
J. Adams {\em et al.} (STAR Collaboration), Phys. Rev. Lett.
\textbf{92}, 171801(2004).
\item {\em Identified particle distributions in pp and Au + Au
collisions at $\sqrt{s_{NN}}$ = 200 GeV} \\
J. Adams {\em et al.} (STAR Collaboration), Phys. Rev. Lett.
\textbf{92}, 112301(2004).
\item {\em $\rho^0$ Production and Possible Modification in Au + Au
and $p+p$ Collisions at $\sqrt{s_{NN}}$ = 200 GeV} \\
J. Adams {\em et al.} (STAR Collaboration), Phys. Rev. Lett.
\textbf{92}, 092301(2004).
\item {\em Azimuthal anisotropy at the Relativistic Heavy Ion
Collider: the first and fourth harmonics} \\
J. Adams {\em et al.} (STAR Collaboration), Phys. Rev. Lett.
\textbf{92}, 062301(2004).
\item {\em Particle-type dependence of azimuthal anisotropy and
nuclear modification of particle production in Au+Au collisions at
$\sqrt{s_{NN}}$ = 200 GeV} \\
J. Adams {\em et al.} (STAR Collaboration), Phys. Rev. Lett.
\textbf{92}, 052302(2004).
\item {\em Pion-Kaon Correlations in Central Au + Au Collisions
at $\sqrt{s_{NN}}$ = 130 GeV} \\
J. Adams {\em et al.} (STAR Collaboration), Phys. Rev. Lett.
\textbf{91}, 262302(2003).
\item {\em Three-Pion Hanbury Brown-Twiss Correlations in
Relativistic Heavy-Ion Collisions from the STAR Experiment} \\
J. Adams {\em et al.} (STAR Collaboration), Phys. Rev. Lett.
\textbf{91}, 262301(2003).
\item {\em Transverse momentum and collision energy dependence of
high $p_T$ hadron suppression in Au+Au collisions at
ultrarelativistic energies} \\
J. Adams {\em et al.} (STAR Collaboration), Phys. Rev. Lett.
\textbf{91}, 172302(2003).
\item {\em Evidence from d+Au measurements for final-state
suppression of high $p_T$ hadrons in Au + Au collisions at RHIC}
\\
J. Adams {\em et al.} (STAR Collaboration), Phys. Rev. Lett.
\textbf{91}, 072304(2003).

\item {\em Pseudorapidity Asymmetry and Centrality Dependence
of Charged Hadron Spectra in d + Au Collisions at $\sqrt{s_{NN}}$)
= 200 GeV} \\
J. Adams {\em et al.} (STAR Collaboration), Phys. Rev. C
\textbf{70}, 064907(2004).
\item {\em Measurements of transverse energy distributions in
Au + Au collisions at $\sqrt{s_{NN}}$ = 200 GeV} \\
J. Adams {\em et al.} (STAR Collaboration), Phys. Rev. C
\textbf{70}, 054907(2004).
\item {\em Photon and neutral pion production in Au + Au
collisions at $\sqrt{s_{NN}}$ = 130 GeV} \\
J. Adams {\em et al.} (STAR Collaboration), Phys. Rev. C
\textbf{70}, 044902(2004).
\item {\em Centrality and pseudorapidity dependence of charged
hadron production at intermediate $p_T$ in Au + Au collisions at
$\sqrt{s_{NN}}$ = 130 GeV} \\
J. Adams {\em et al.} (STAR Collaboration), Phys. Rev. C
\textbf{70}, 044901(2004).
\item {\em Rapidity and Centrality Dependence of Proton and
Anti-proton Production from Au + Au Collisions at $\sqrt{s_{NN}}$
= 130 GeV} \\
J. Adams {\em et al.} (STAR Collaboration), Phys. Rev. C
\textbf{70}, 041901(2004).
\item {\em Production of $e^+e^-$ Pairs Accompanied by
Nuclear Dissociation in Ultra-Peripheral Heavy Ion Collision} \\
J. Adams {\em et al.} (STAR Collaboration), Phys. Rev. C
\textbf{70}, 031902(R)(2004).
\item {\em Transverse-momentum dependent modification of
dynamic texture in central Au + Au collisions at $\sqrt{s_{NN}}$
= 200 GeV} \\
J. Adams {\em et al.} (STAR Collaboration), Phys. Rev. C
\textbf{70}, 031901(R)(2004).
\item {\em Net charge fluctuations in Au + Au collisions at
$\sqrt{s_{NN}}$ = 130 GeV} \\
J. Adams {\em et al.} (STAR Collaboration), Phys. Rev. C
\textbf{68}, 044905(2003).

\end{itemize}

\bibliography{PhDthesis}
\bibliographystyle{uclathes}

\end{document}